%% file: main.tex
\documentclass[12pt]{report}
\usepackage{hyperref}

\usepackage[backend=biber,style=numeric,citestyle=phys]{biblatex}
\usepackage{titlesec}
\usepackage[a4paper,top=1.125in,bottom=1.125in,margin=1in]{geometry}
\usepackage{amsmath}
\usepackage[makeroom]{cancel}
\usepackage{graphicx}
\usepackage{algorithm}
\usepackage[noend]{algpseudocode}
\usepackage[labelfont=bf]{caption}
\usepackage{multirow}
\usepackage{booktabs}
\usepackage[separate-uncertainty=true]{siunitx}
\usepackage{placeins}
\usepackage[title, toc]{appendix}
\usepackage{tablefootnote}
\usepackage{longtable}
\usepackage{pifont}

\usepackage{listings}
\usepackage{color}
\usepackage{titlesec}
\usepackage[utf8]{inputenc}
\usepackage{fancyhdr}
\newcommand{\of}{OpenFOAM }
\newcommand{\IFM}{\emph{interFoam} }
\newcommand{\gIF}{\emph{gtawFoam} }
\newcommand{\MFIF}{\emph{multiphaseInterFoam} }
\newcommand{\CPIF}{\emph{compressibleInterFoam} }
\newcommand{\cMRF}{\emph{chtMultiRegionFoam} }
\newcommand{\IRMFIF}{\emph{icoReactingMultiphaseInterFoam} }
\newcommand{\ITPCF}{\emph{interThermalPhaseChangeFoam} }
\newcommand{\vsf}{\emph{volScalarField} }
\newcommand{\vvf}{\emph{volVectorField} } 
\newcommand{\ssf}{\emph{surfaceScalarField} }
\newcommand{\svf}{\emph{surfaceVectorField} }
\newcommand{\ds}{\emph{dimensionedScalar} }
\newcommand{\dv}{\emph{dimensionedVector} }
\newcommand{\fvOp}{\emph{fvOptions} }
\newcommand{\sFF}{\emph{swak4FOAM} }
\newcommand{\gBC}{\emph{groovyBC} }
\newcommand{\geoF}{\emph{geometricField} }
\newcommand{\numParagraph}[1]{\paragraph{#1}\mbox{}\\}

\begin{document}
\input{chapters/thesis-title}
% I think it is a good idea to footnote the case names of github or similar so people can identify what they are after.

\pagenumbering{roman}
\chapter*{Abstract}
\input{chapters/thesis-abstract}

\chapter*{Acknowledgements}
\input{chapters/thesis-acknowledgements}

\setcounter{secnumdepth}{4}
\setcounter{tocdepth}{3}
\tableofcontents
\listoffigures
\listoftables

\chapter{Introduction}
\pagenumbering{arabic}
\input{chapters/thesis-chapter-1}

\chapter{Previous work on GTAW}
\input{chapters/thesis-chapter-2}

\chapter{Numerical methods}
\input{chapters/thesis-chapter-3}

\chapter{Multiphysics solver}
\input{chapters/thesis-chapter-4}

\chapter{Benchmark cases}
\input{chapters/thesis-chapter-5}

\chapter{Ultra-thin-walled tube welding}
\input{chapters/thesis-chapter-6}

\chapter{Thin structure welding}
\input{chapters/thesis-chapter-7}

\chapter{Conclusion}
\input{chapters/thesis-chapter-8}

\printbibliography

\appendix
\chapter{Glossary}
\input{appendicies/thesis-appendix-glossary}
\chapter{Case details}
\input{appendicies/thesis-appendix-cases}
\chapter{Method for melt front extraction}
\input{appendicies/thesis-appendix-extraction}
\chapter{Incomplete fusion image}
\input{appendicies/thesis-appendix-incomplete}
\chapter{Code listings}
\label{ap-listings}
\input{appendicies/thesis-appendix-code}
\end{document}

%% file: chapters/thesis-title.tex
\begin{titlepage}
    \begin{center}
        \vspace*{1cm}
            
        \Huge
        \textbf{Multiphysics modelling of\\ Gas Tungsten Arc Welding on ultra-thin-walled titanium tubing}
            
        \vspace{1.5cm}
        
        \includegraphics[width=13cm]{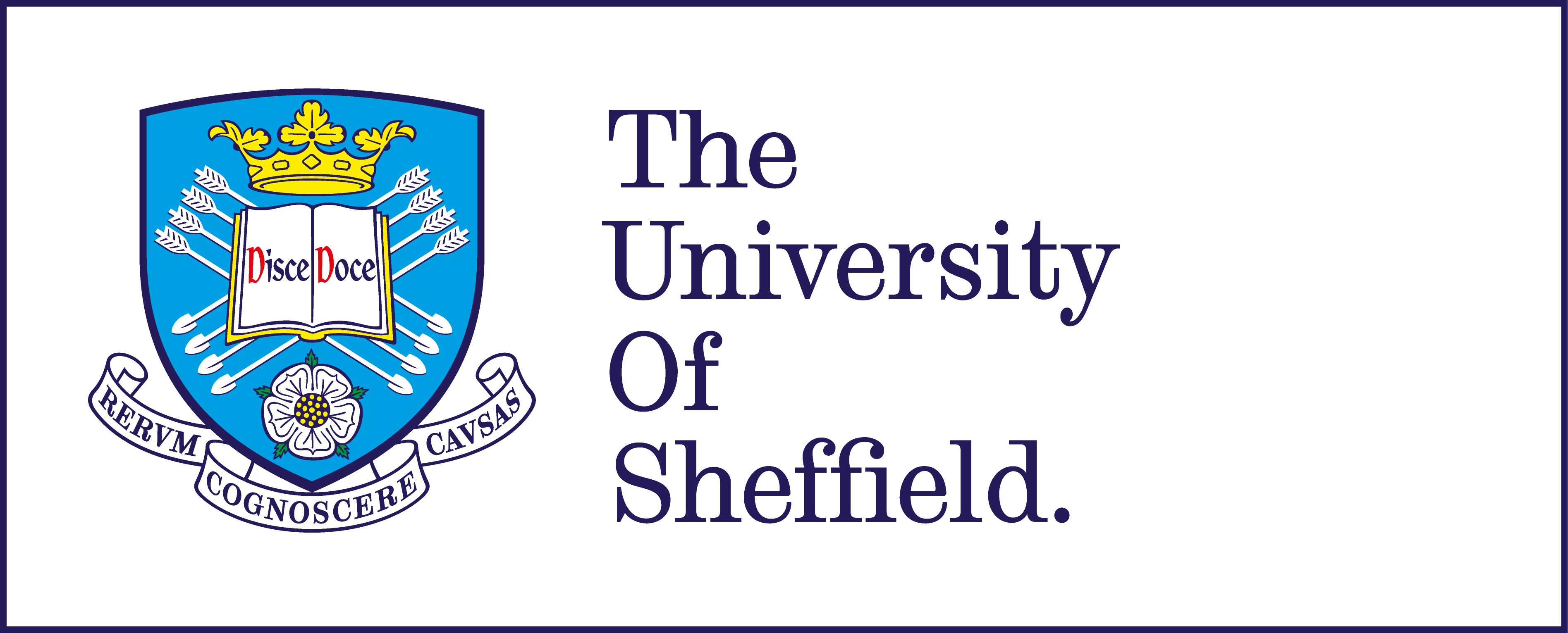}

        \vspace{1cm}
            
        \textbf{Will Yeadon}

        \vspace{1cm}
        \Large
        Department of Physics and Astronomy\\
        The University of Sheffield\\
        United Kingdom of Great Britain and Northern Ireland\\
            
       \vspace{2cm}
       A thesis presented for the degree of\\
       \emph{Doctor of Philosophy}

       \vspace{3cm}

       %\normalsize     
       September 2021
            
   \end{center}
\end{titlepage}

%% file: chapters/thesis-abstract.tex
\thispagestyle{plain}
This thesis presents a novel multiphysics solver, named \emph{gtawFoam}, for Gas Tungsten Arc Welding (GTAW) that is applied to simulate orbital GTAW on ultra-thin-walled titanium tubing. In this thesis, ultra-thin-walled tubing refers to tubing where the wall thicknesses are less than \SI{500}{\micro\meter}. Orbital welding of tubing with this wall thickness requires both a sufficient heat input to weld the tubing and an internal buttressing gas flow to ensure the tube retains its geometrical integrity. The specific use case is for the commercially pure grade 2 titanium tubing used in the ATLAS ITk cooling system which is \SI{2.275}{\milli\meter} outer diameter and \SI{300}{\micro\meter} wall thickness at the weld. 

The solver is created using the open source computational fluid dynamics library OpenFOAM and each component of the solver is benchmarked against an appropriate case. With the solver established, it is used to simulate a series of welding procedures that were performed experimentally on the aforementioned titanium tubing. Both the experimental and simulation results show a `goldilocks' region where the weld heat input and inner buttressing gas flow are moderated to a level where a fully penetrating weld is created but the geometric integrity of the tube is not compromised. 

\gIF is then used to simulate hypothetical tubing with larger and smaller wall thicknesses between \SI{250}{\micro\meter} and \SI{350}{\micro\meter}. The results suggest that the required buttressing gas pressure once achieved is relatively transferable between wall thickness changes but applying enough heat so as to achieve full penetration is critical. These results are then used to predict effective welding procedures for this hypothetical tubing. \gIF is subsequently applied to the welding of turbine blades. This includes the addition of multiple layers of filler metal to mimic additive manufacturing. The aerofoil shape of these blades include thin sections around their corners meaning this problem is analogous to ultra-thin-walled tube welding.  

\newpage
\thispagestyle{plain}
\vspace*{2.5cm}
{\Large \textbf{Publications}}
\vspace*{0.5cm}

Cooper, L., Crouvizier, M., Edwards, S., French, R., Gannaway, F., Kemp-Russell, P., Marin-Reyes, H., Mercer, I., Rendell-Read, A., Viehhauser, G. and \textbf{Yeadon, W}. `In situ micro gas tungsten constricted arc welding of ultra-thin walled 2.275 mm outer diameter grade 2 commercially pure titanium tubing'. Journal of Instrumentation, 15 (6). June 2020, DOI: 10.1088/1748-0221/15/06/P06022\\[12pt]
French, R., Marin-Reyes, H., \textbf{Yeadon, W}. `A Feasibility Study Comparing Two Commercial TIG Welding Machines for Deep Penetration' MATEC Web Conf. 269 01004, 2019, DOI: 10.1051/matecconf/201926901004\\[12pt]
French, R., \textbf{Yeadon, W}, Kapellmann, G. and Marin-Reyes, H., `Development of a Vision System for TIG Welding - A Work-in-Progress Study,' 2018 IEEE 23rd International Conference on Emerging Technologies and Factory Automation (ETFA), pp. 1193-1196, 2018, DOI: 10.1109/ETFA.2018.8502461.

\vspace*{0.5cm}
{\Large \textbf{Upcoming Publications (In preparation)}}
\vspace*{0.5cm}

\textbf{Yeadon, W.} `\emph{gtawFOAM}: A framework for Gas Tungsten Arc Welding additive manufacturing simulations'. SoftwareX
\newpage
\thispagestyle{plain}
\begin{center}
\vspace*{2.5cm}
\huge
\emph{The future's uncertain, \\and the end is always near\\}
\vspace{0.75cm}
The Doors
\end{center}

%% file: chapters/thesis-acknowledgements.tex
\thispagestyle{plain}
Firstly, I am indebted to the generosity of Prof. Iain Todd for taking me on as a student midway through my studies. His astute and creative comments and ideas helped me redefine seemingly intractable problems. Further, his strategising of a long term path ensured a cohesive thesis in spite of large unforeseen challenges; I am deeply thankful for this. Secondly, I thank Prof. Davide Costanzo for his role as my second supervisor, a role that perhaps proved larger than initially anticipated. His calm and steady hand ensured the successful completion of this thesis and I am thankful for this. Thirdly, I thank Dr. Al Buckley for his role as my personal tutor - checking up on me throughout the course of my studies.

This thesis would not have been possible without my initial supervisor Dr. Richard French. I am grateful for his conception and funding for this project without which I would not have had this opportunity. His deep practical intelligence and intuition solidified in me the importance of hands-on work. I also thank my old group, Enabling Sciences for Intelligent Manufacturing, and I wish Gabriel, Samantha, Bilal, Ben, Paul, Kieran, Jess and Sam all the best in the future and hope to see you again. I also thank the inhabitants of D44 for the memories - particularly the coffee and pizza related ones!

I thank my partner Lily for her continual love and support. I will always cherish the memories and laughter we shared during our studies and look forward to our future. Her outlook blessed me with grit and resilience that I will carry with me for the rest of my life. Finally, I thank my family for their inexhaustible kindness and support. Gran, I'll make sure I'll wear the `roundy hat'.

%% file: chapters/thesis-chapter-1.tex
\lhead{Chapter 1: Introduction}
\section{Background}
\subsection{Application}
\label{1-sec-ITk-intro}
This thesis addresses the challenge of modelling ultra-thin-walled tube welding for the ATLAS (\textbf{A} \textbf{T}oroidal \textbf{L}HC \textbf{A}pparatu\textbf{S}) ITk (\textbf{I}nner \textbf{T}rac\textbf{k}er) \cite{ITk-Strip-TDR}. Here, ultra-thin-walled tubing is a category of metal tubing where the wall thickness is less than $\SI{500}{\micro\meter}$. The ATLAS detector \cite{ATLAS-TDR} is a large ($46 \times 25 \times \SI{25}{\meter}$) general purpose particle physics detector at the Large Hadron Collider (LHC) at CERN in Geneva Switzerland. The ITk is a upgrade to the inner portion of the ATLAS detector which shall be the first ATLAS component to detect the products of particle collisions generated by the LHC. An embedded cooling system is required to cool the semiconductors that constitute some of the detecting mass of the ITk. However, all additional non-detecting mass within a particle physics detector decreases its efficacy (as non-detecting mass provides no information about particle tracks). Therefore, the minimum amount of non-detecting mass possible should be used. Not all non-detecting mass impacts a detector's efficacy equally thou with some materials having a more negative effect than others. This effect can be quantified in terms of radiation length. The radiation length is a measure of the mean length of a material in centimetres that will reduce the energy of an electron by a factor of $\frac{1}{e}$. Materials with a shorter radiation length impact the detector efficacy more than materials with longer radiation lengths. Thus the cooling system should be constructed with as little mass as possible and the material that is used should have a long radiation length. Additionally, the material used should be robust and have high reliability ensuring the cooling system lasts many years with limited maintenance access. 

To address these challenges, \SI{2.275}{\milli\meter} outer diameter and \SI{160}{\micro\meter} wall thickness commercially pure titanium tubing was chosen as the cooling tube material. With this geometry and material, the cooling system can effectively circulate the two phase liquid/vapour CO\textsubscript{2} coolant around the ATLAS ITk. However, with the complex curving geometry typical of a cooling system inevitably joints are required of which - to minimize leaks - welded joints are the optimum choice. Further, some of these welds must be able to be created in situ (inside of ATLAS) during detector construction as the separate parts of ATLAS are connected. This thesis presents simulation results for the computational modelling of these welded joints comparing the results to actual experimental results of the welding process used for the ITk cooling system. The results from the model are use to predict a welding procedure for hypothetical ultra-thin-walled tubing of different wall thicknesses to that used in the ITk.

The following sections in this chapter will introduce the general context of welding and the processes used in this thesis followed by a description of the specific challenges involved in ultra-thin-walled tube welding. After this, an overview of the rest of the thesis is given in Section \ref{1-sec-thesis-overview}. This chapter thus aims to elucidate the reader on the specific challenges involved in modelling ultra-thin-walled tube welding to emphasize the contributions of this thesis.

\subsection{Welding processes}
\subsubsection{Broad view}
\label{1-sec-weld-proc}
To weld anything, a user needs a source of heat to melt the metal and a method to prevent impurities. Welding is fundamentally different from brazing or bonding as it involves microstructural changes in the work piece through either completely melting it or through the coalescence of two work pieces at temperatures close to their melting points. Here, the metal that is being welded is referred to as the `work piece' although `work metal' or simply `work' are also commonly used. Creating a weld within the work piece of a sufficient size to ensure a strong joint can be challenging. Combining this with how the produced welds are often a point of mechanical failure leads to the common use of a reinforcing filler metal when welding. Here, a filler metal is combined with the work pieces during welding to add additional material to the weld improving its strength. When no filler metal is use, the process is categorized as autogenous welding. Further, a particular welding job can be categorized as homogenous if the work pieces and filler metal are of identical composition whereas it would be inhomogeneous if either the work pieces differ from each other or the filler metal differs from the work piece. 

Welding processes can be broadly split into those that form a melt pool in the work metal - fusion welding - and those which join two work pieces through coalescence - solid state welding. A common solid state welding technique is friction stir welding (FSW). Within fusion welding, popular techniques include: oxyacetylene welding, electron beam welding (EBW), laser beam welding (LBW) and Gas Tungsten Arc Welding (GTAW).

In FSW a non-consumable tool is rotated along an interface between two work pieces creating a weld. The rotating tool creates enough frictional heat to soften the work piece but not enough to form a melt pool. The rotating motion helps to mix this softened material to create a coalescence of the two materials resulting in a weld. The technical challenges involved in creating a custom friction stir welding setup capable of joining the ultra-thin-walled titanium tubing in the ITk are immense. Instead, welding the small work piece in the present application is much more suited to precision fusion welding. 

Oxyacetylene welding is a rudimentary fusion welding process involving the combustion of oxygen gas with a fuel source (traditionally acetylene). The resulting flame melts the work piece as well as providing limited removal of impurities. Whilst oxyacetylene welding is useful for many common welding jobs it is unsuitable for titanium welding leaving EBW, LBW and GTAW as candidates.

Electron beam welding involves the transfer of the kinetic energy of a beam of high energy electrons to the work to melt it. This bombardment of electrons can be focussed in a very small area to create a strong weld with a high depth to width ratio. However, a key disadvantage is that the EBW must be performed within a vacuum chamber. This physically limits the size and type of work pieces that can be welded.

Laser beam welding is a process whereby a high powered laser is used to melt the work piece. Nominally, the process works through electromagnetic radiant flux incident on the work piece forming a melt pool. However, the physics becomes quite complex when considering the reflectance and emittance of photons incident on a rough cathode \cite{laserPhysics}. Modern laser welding systems are capable of highly controlled heat input enabling the creation of intricate structures; this is particularly useful in additive manufacturing (AM). Even so, LBW has a similar issue to EBW in the present application in that it would be extremely challenging to create an LBW process capable of in situ production inside ATLAS.

The final candidate process is GTAW. Comparatively, EBW and LBW offer better weld properties in some circumstances than GTAW \cite{yunLianEBW}, however, they lack the advantage of in situ operation available to GTAW. Specifically, the equipment required for LBW and EBW could not physically fit into smaller areas for in situ production. In situ production is a critical requirement for the ATLAS ITk cooling system eliminating LBW and EBW as candidates. Further, GTAW based welding is considerably cheaper than EBW or LBW. GTAW produces precise, clean and strong joints on titanium and it is the industry standard for titanium joinery.    

\subsubsection{The GTAW process}
\label{1-sec-gtaw}
\begin{figure}[!htb]
\centering
\includegraphics[width=12.5cm]{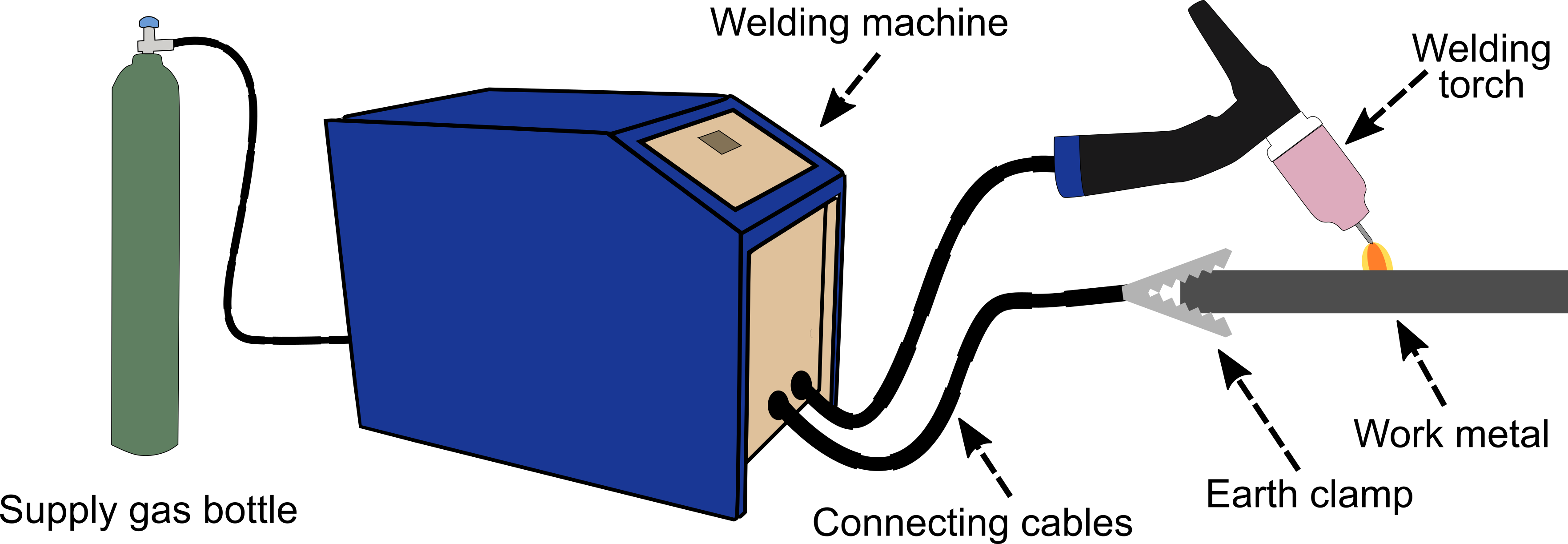}
\caption{Illustration of the GTAW process with the constituent equipment labelled.}
\label{1-fig-welding-process}
\end{figure}

Gas Tungsten Arc Welding (GTAW) is welding process involving the creation of an electric arc that conducts current into the work piece (providing the heat) enveloped by a supply of shielding gas (providing the method to prevent impurities). The process involves the use of a GTAW welding machine that draws power from the mains electricity supply to create a potential difference between the work piece connected to the welding machine by a metal clip (termed the 'earth clamp') and a welding torch connected to the welding machine via an electrical cable. In addition to supplying current to the welding torch, the welding machine also supplies shielding gas that is blown around the welding torch. The welding torch directs the delivery of the current to the work piece, it contains a non-consumable tungsten electrode as well as a shielding gas delivery manifold that pipes the shielding gas around the tungsten electrode guided by a surrounding a ceramic cup. Tungsten is used for the non-consumable electrode due to its high melting point. This tungsten electrode is typically doped with either lanthanum or thorium to improve conductivity. A diagram of the GTAW process is shown in Figure \ref{1-fig-welding-process}.

GTAW is a constant-current process meaning the welding machine provides a constant current irrespective of the voltage. In manual welding, this level of current is typically controlled by a foot pedal whereas in automated welding it would be controlled digitally. Often welding machines can provide both direct electrical current where charge always flows in the same direction and alternating electrical current where the direction of charge flow alternates. For direct current, the flow of charge can either be into the work piece or into the welding torch. When the flow of negative charge (free electrons) goes from the tungsten electrode to the work piece the process is termed `direct current electrode negative' (DCEN). All the welding procedures detailed in this thesis are DCEN. The voltage is determined by the gap between the electrode tip at one end of the arc and the work piece at the other; this interval is termed the `arc gap'. In order to initially generate the arc, a large initial arc strike voltage - typically \SI{1}{\kilo\volt} per \SI{1}{\milli\meter} - is required. With the GTAW arc created, a filler metal is often fed into the region where the arc meets the weld pool. Conventionally the addition of filler metal serves to enhance joint integrity. Although contemporaneously the filler metal is seeing increasing use as an additive material for GTAW based additive manufacturing. To create a strong joint, the choice of filler metal should match the base metal work piece (homogenous welding) but some niche applications allow for a filler metal different to the base metal to be used (hetrogenous welding).  

Noble gases are used for the shielding gas as they have very low chemical reactivity. Whilst Helium does see some use in GTAW, Argon is by far the most popular choice as it is comparatively a lot cheaper. Sufficient gas flow needs to be supplied so as to prevent the aforementioned impurities although too high of a gas flow will result in turbulent flow that can actually draw in impurities. When welding manually, the rate of gas flow is often tuned and judged by feel but it would more often be measured through flow meters in automated GTAW. For certain welding applications, a `purge gas' flow is used whereby the shielding gas is blown over the work piece for a few seconds before welding. This process helps to remove airborne impurities in the work piece's immediate atmosphere thus helping to limit impurities in the produced weld. Further, sometimes an additional `purge gas' flow is applied after welding to ensure an atmosphere free of impurities as the weld cools.

For a particular welding job, all the welding parameters mentioned thus far would be specified in a document termed a welding procedure specification (WPS). The WPS provides an outline for a welding practitioner to create a high quality joint for a particular work piece. Here, the welding parameters are the settings that a welding practitioner should use when welding. For instance, the WPS may specify a current of \SI{50}{\ampere} a voltage of \SI{5}{\volt}, a travel speed for the welding torch of \SI{5}{\milli\meter\per\second} and an arc gap of \SI{5}{\milli\meter} for a particular welding job. Mimicking these parameters should ensure a reasonable joint. However, due to the inevitable irregularity of the real work piece, following a WPS can't guarantee an excellent joint. For many manual welders the solution is to employ hard-won intuitions gained from experience to effectively adapt the welding procedure specification to changes in the work piece. The same process can not be applied to automated welding. Instead, an understanding of the role of each welding parameter and the physics of welding must be employed. 

A common intuition when considering the physics of welding is how the temperature of the GTAW arc affects the weld pool. However, given the main heat transfer mechanism in GTAW is joule heating \cite{gtawHeatTransfer}, the maximum temperature of a GTAW arc (\SI{>20000}{\kelvin}) is not particularly relevant. Despite this, the temperature gradient from the peak at the centre of the arc to the edge is important. This is because of the temperature dependence of surface tension of many metals. A gradient in temperature thus causes a gradient in surface tension that - due to the Marangoni effect - will cause the weld pool to flow a particular way. Although the Marangoni effect is typically the most impactful driving force on a weld pool \cite{weldDrive}, the arc pressure from the impinging plasma arc causes considerable warping of the weld pool surface. Further, the resulting shear stress from the impinging arc is sometimes categorized as a separate arc drag force. The remaining forces act solely within the weld pool: buoyancy and the Lorentz force. The latter of these is due to the current in the weld pool interacting with its self-induced magnetic field; this also applies for the current flow in the plasma arc. The buoyancy force occurs due to the archetypal reduction in density of liquid metals with increasing temperature causing a density gradient within the weld pool. 

\begin{figure}[!htb]
\centering
\includegraphics[width=12.5cm]{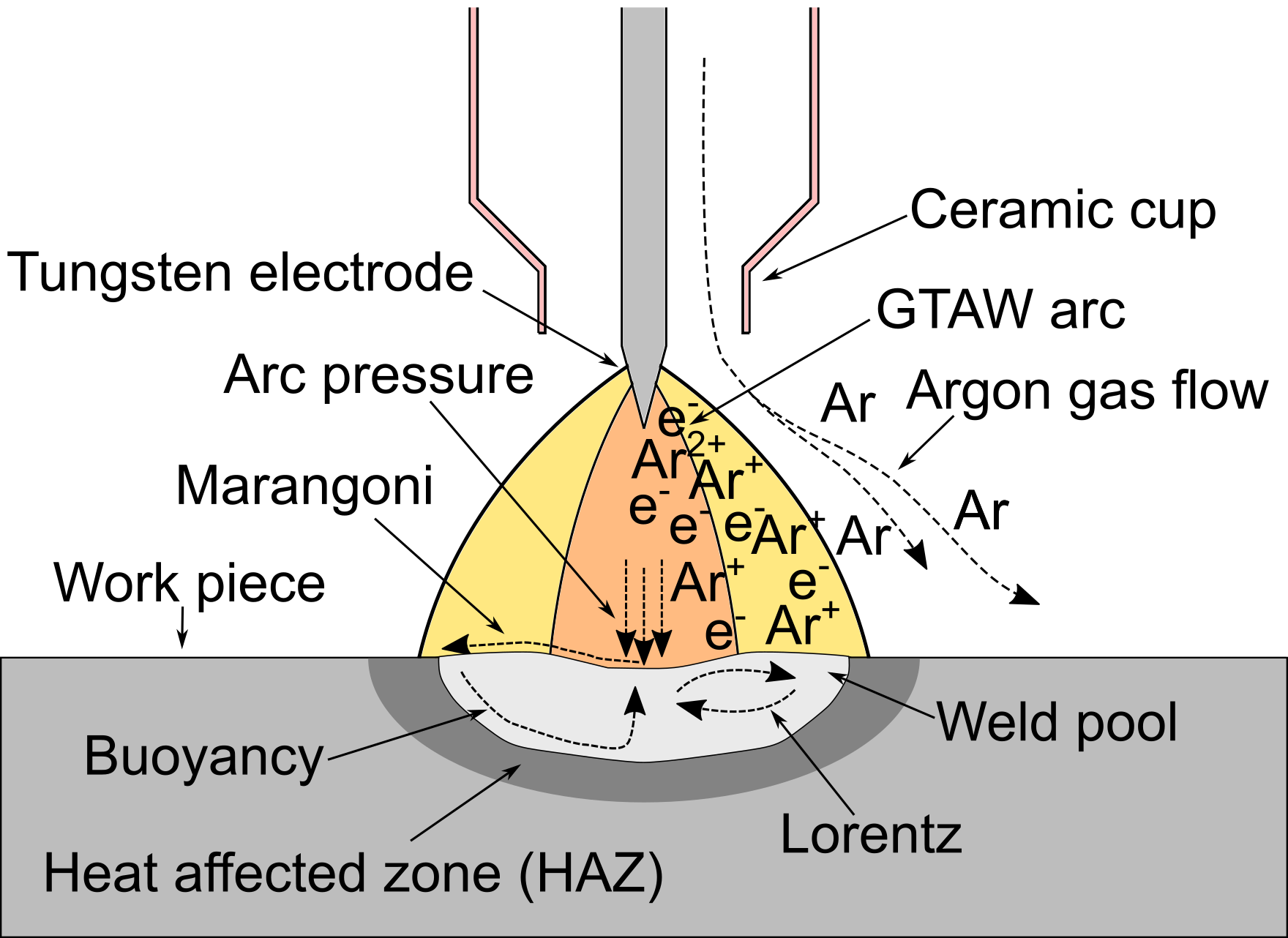}
\caption{Illustration of the autogenous GTAW process with the important elements highlighted. Note, all labels are symmetric but only one side of the features are illustrated/labelled to make the image more compact.}
\label{1-fig-gtaw}
\end{figure}

When it solidifies, the metal that previously constituted the weld pool will typically have considerably larger grains than the base metal. Although areas outside the weld pool (by definition) do not completely melt they are still affected by the inputted heat. The areas where the heat input is high enough for the microstructure of the material to be effected but not completely melted is termed the heat affected zone (HAZ). A labelled illustration of the GTAW process is shown in Figure \ref{1-fig-gtaw}.     

\subsection{Ultra-thin-walled tube welding}
\label{1-sec-ultra-thin-tube}
When welding tubes and pipes\footnote{NB the practical difference between tubes and pipes mainly comes down to ordering them from suppliers. Pipes will typically be ordered from their `Nominal Pipe Size' that specifies their internal diameter whereas tubes will be specified via an outer diameter and wall thickness. Although, this difference is not always clear cut. Also, some suppliers will classify hollow prisms with a polygon base as tubing (e.g. square tubing, rectangular tubing etc.) and hollow cylindrical structures as pipes.} a welding practitioner must ensure the welding heat input continually shifts angle and position to accommodate the geometry of the work piece. It is often impractical to rotate the work piece. Consequently the welding heat input must travel completely around the work piece instead; this process is termed `orbital welding'. As the heat input travels around from the top of the work piece to the underside there will be sections where the work piece is above the heat input. The practitioner must thus ensure that the heat input remains both continually sufficient to join the work piece and ideally consistent so that no potential failure points are created. This makes the situation more complex than the linear movement associated with joining sheets and plates.       

There is an added complication in orbital welding when the wall thickness of the work piece is particularly thin. In these situations full penetration is assured as the difference between the minimum heat input to initially form the weld pool and the heat input to fully penetrate the wall of the work piece is very small and instead, preventing burn through is the challenge. This would not be the case for a pipe with a wall thickness of say \SI{1}{\centi\meter} that could form a weld pool without burning through over a very broad range of heat inputs. This burn through prevention issue is exacerbated further in autogenous welding further reducing the volume of the liquid weld pool. In fact, a key quantity to consider in thin-walled tube welding is the surface to volume ratio and specifically the surface exposed to the shielding gas / atmosphere compared to the volume of the liquid metal. Here, from a phase geometry perspective, the surface is the boundary between the phase in question and everything else. However, the interface between the liquid metal and solid metal does not work in the same way as the interface between the liquid metal and atmosphere / shielding gas that constitutes the exposed surface. Thus, only the ratio of the liquid metal - gas interface against the liquid metal volume is of interest. As this ratio gradually increases with reductions in wall thickness, the challenges of thin-walled tube welding become even more acute. At very thin wall thicknesses it becomes vital to supply an internal gas flow to internally buttress the liquid weld pool and prevent it from collapsing inwards. The requirement of this internal buttressing force essentially forms a delimitation for a new category of orbital welding. To identify this new classification, the term ultra-thin-walled is introduced to categorize tubes with wall thicknesses less than $\SI{500}{\micro\meter}$. Below this approximate threshold which will change dependent on the specific material, an internal buttressing gas flow becomes a necessity. Ideally, a thicker tube wall would be employed to negate the challenges of ultra-thin-walled tube welding. However, there are uses of metal tubing that require both the thinnest possible wall thicknesses and welded joints.   

\begin{figure}[!htb]
\centering
\includegraphics[width=14.5cm]{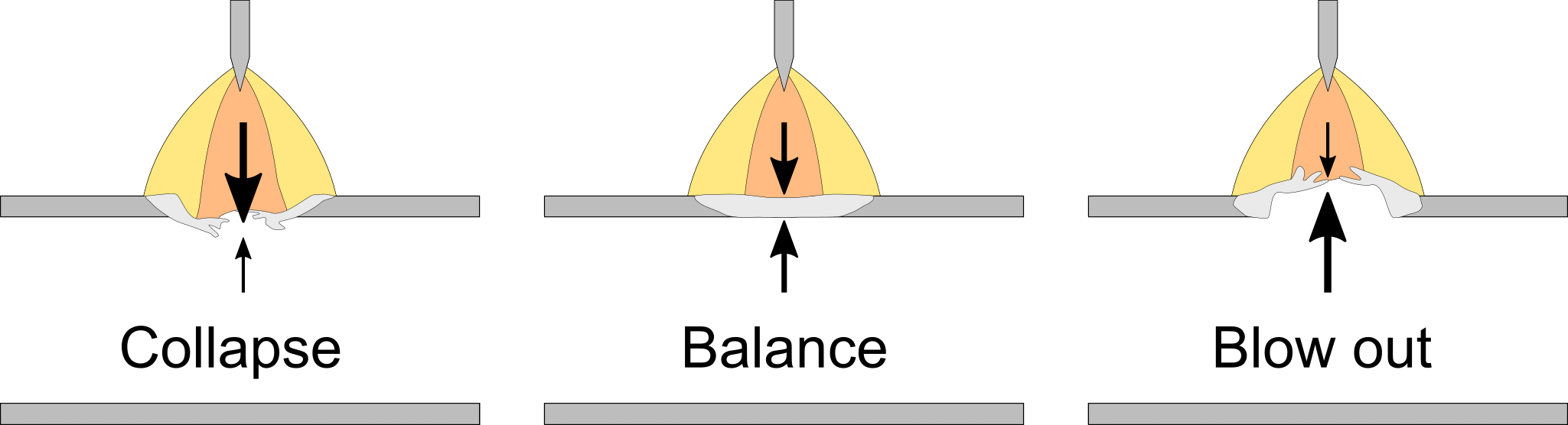}
\caption{Illustration of pressure balancing in ultra-thin-walled tube welding. The arc pressure pushes inwards whilst the buttressing internal gas pushes outwards; through balancing these two competing forces a consistent internal diameter can be achieved.}
\label{1-fig-tube-gtaw}
\end{figure}

To ensure consistent mass flow of coolant through a cooling circuit, tubing with a consistent internal diameter is required. Due to the aforementioned challenge of potential inwards collapse with ultra-thin-walled tube welding, tempering the internal gas flow so as to prevent collapse whilst also not causing blow out (where the internal pressure is too high) becomes challenging. These situations are illustrated in Figure \ref{1-fig-tube-gtaw}. Assessing the physics of this, it would appear that the liquid metal forms two competing Laplace pressures due to the inner pressurization and the outer arc pressure. However, taking an order of magnitude estimation of the arc pressure \cite{linEagarGtawPressure} with welding parameters that are known to successfully weld the tubes reveals that the arc pressures are considerably lower than the inner pressurizations. 

A possible solution to the pressure balance problem is perhaps the liquid weld solidifies before it collapses or blows out and the pressurization needs only maintain the weld pool for a few seconds. Understanding the surface tension of the liquid weld pool better may also provide a clue as to a solution. There could be some sort of measurement / setting error in the equipment used meaning the tubing is not actually under the pressurizations calculated. Also, the effects due to some forces such as gravity can be ruled out as the volume of the liquid weld pool is small (less than \SI{1}{\milli\meter\cubed}). This low effect of gravity is further proved by the lack of variation in the weld texture around the circumference of successfully welded tubes. These are just some potential solutions as to the author's knowledge, this problem has not been extensively addressed in the literature revealing a gap in the understanding of physics occurring during ultra-thin-walled tube welding.
 
A potential method to ensure a consistent internal diameter is to enhance the understanding of the process of ultra-thin-walled tube welding. This can be achieved through both simulation and experimental work. Simulation work comes with the standard advantages of software over hardware investigations: lower cost to run ``investigations'', no limit on the number of ``investigations'', trivial to change ``experimental'' parameters, and very high confidence in the values of the ``experimental'' parameters. Contrarily, modelling results are subject to modelling assumptions and choices and thus they may not accurately represent reality. By definition, hardware investigations do represent reality. Therefore if simulation work can help guide or compliment the hardware work then a full picture of ultra-thin-walled tube welding could be created. This would be the optimum method to pursue and was the initial plan for this thesis. However, whilst some hardware work was performed, the focus of this thesis turned\footnote{This was due to a combination of the COVID-19 pandemic, lack of access to equipment and an unavoidable change in supervision.} mainly to a software investigation. For this, the employment of multiphysics methods to simulate GTAW is required.

Finally, whilst ultra-thin-walled tube welding is somewhat of a niche procedure it is still potentially transferable to other welding challenges. One related challenge is in thin-structure welding. Here the term thin-structure refers to a metal structure with features comparable in size to the tube walls in ultra-thin-walled tube welding. The thin structures investigated in this thesis are aeroengine turbine blade tips. The aerofoil shape of turbine blades means that their structure becomes very thin around the corners and thus they are commensurate to ultra-thin-walled tubes. After these blades have been damaged during their service life layers of metal are welded onto them as part of a repair process. This welding repair process forms the transferable application of ultra-thin-walled tube welding.      

\section{Thesis Structure}
\label{1-sec-thesis-overview}
\subsection{Background}
Following this introductory chapter is a literature review that concludes with a list of objectives for this thesis. Then, Chapter 3 establishes the numerical methods used in this thesis enabling the reader to appreciate the contributions of this thesis present in the subsequent chapters. The chapter introduces computational fluid dynamics through showing how the Navier-Stokes equations are derived and solved in a domain. This is followed by a description of the program used for the simulations (OpenFOAM) as well as a description of the inbuilt solver that the simulations in this thesis are built upon (\emph{interFoam}). 

\subsection{The candidate's work}
With the background established, in the remainder of Chapter 3 a basic `toy' model is created as a first attempt to address the aforementioned objectives of this thesis. This transition into the candidate's work is emphasized. The limitations of this `toy' model are then outlined motivating the work in the remaining chapters. Chapter 4 is a detailed description of the computational solver developed to simulate ultra-thin-walled tube welding. The chapter details how the solver works as well as its implementation covering the overall algorithm as well as the individual features. Here, the implementation is detailed in a manner that an user familiar with \of could apply the model from the description. This is unless there is something unusual in which case it is highlighted specifically with code presented in the apt appendix. Chapter 5 is a series of systematic benchmarks that comprehensively gauges the efficacy of the computational solver. This section introduces a novel technique whereby the `stable' optimum values for one benchmark is found and subsequently applied to a series of other benchmarks that crucially test different aspects of the solver.

Chapter 6 answers the core question of this thesis through the presentation of a general case for ultra-thin-walled tube welding. This is undertaken through a combination of experimental and simulation work. The experimental work was undertaken as part of the development for the ATLAS ITK \cite{JINST} and was performed with the goal of rapidly achieving production capabilities rather than for the purpose of investigating the physics of ultra-thin-walled tube welding. The simulation work thus adapts to the available experimental data to benchmark against it before being extended to predict a general case. Chapter 7 demonstrates the simulation's flexibility as a general GTAW simulation tool. Finally, the contributions of this thesis are summarised in Chapter 8.    

All of the code used in this thesis can be found at \url{https://github.com/WillYeadon/thesis}. This is split into the multiphysics solver \emph{gtawFoam} in the folder `\emph{gtawFoam}', the individual \of cases in the folder `\emph{cases}', and the Python code for all of the plots in `\emph{thesis-python-scripts}'. In order to compile \emph{gtawFoam} a user needs a working version of \of 6. Further, to run some of the cases requires the add-on `\emph{groovyBC}' - these cases are indicated.   

%% file: chapters/thesis-chapter-2.tex
\lhead{Chapter 2: Previous work on GTAW}
\label{2-lit-review}
\section{Welding of titanium}
\label{2-sec-titanium-weld-review}
Due to the reactivity of titanium, a principal challenge when welding all forms of it is limiting embrittlement by hydrogen, nitrogen and oxygen absorption at the elevated temperatures of joinery \cite{wangWelsch}. Above \SI{500}{\celsius}, the oxidation resistance of titanium rapidly decreases allowing the aforementioned elements to dissolve interstitially creating microscopic hardening which increase the cracking susceptibility of the titanium. Successfully welded titanium ideally should have a bright silver colour on the weld surface. This lustre indicates a lack of contamination of the weld thus suggesting a good quality weld. When a titanium weld contains a little bit of contamination it has a characteristic light straw / gold colour. In general, this level of contamination is judged acceptable however in the present application the welds in ITk specifically require only the prior mentioned bright lustre to be judged acceptable. As the levels of contamination in a titanium weld increase, the light straw colour will turn to a dark straw colour and then a dark blue. When the colour of the weld changes from straw to blue the weld is rejected as a bad weld. Should the contamination levels continue to increase the colour will change to a lighter blue or purply colour, then a grey-blue, until turning powdery white. 

Depending on the application a range of alloying elements can be added to titanium to achieve the desired properties however in the present work only commercially pure titanium is of interest as it is the only type used for the ITk cooling system. Commercially pure titanium grades are $\alpha$ alloys of titanium with oxygen and iron as their primary alloying elements. Grade 1 commercially pure titanium has the lowest strength and highest formability of the commercially pure titanium grades whereas grade 4 has the lowest formability and highest strength. Commercially pure grade 2 (CP-2) titanium has properties between these two extremes making it moderately strong, corrosion resistant whilst having good cold forming properties and it is the grade used for the ITk cooling system. It is protected by a TiO$_2$ film, usually between \SIrange{50}{200}{\angstrom} thick, and is approximately 99.6\% pure titanium.    

To prevent impurities for GTAW welding on commercially pure titanium, previous reports have detailed the benefits of using three gas supply sources \cite{Anis2009, karpagaraj2019experimental} or using custom gas delivery tooling \cite{cpTiGTAW2} to ensure sufficient shielding gas coverage. As covered in Section \ref{1-fig-welding-process}, this coverage can be ensured through employing a pre-weld and post-weld purge gas flow. However, the cooling rate of a titanium weld effects its quality \cite{murthy1998phase} with lower cooling rates promoting granular-like grain morphologies that are conducive to good quality welds \cite{Lonardelli2007}. Orbital welding produces a three-dimensional heat flow into the work piece resulting in complex multidirectional grain morphologies. Limiting the excessive grain growth at higher heat inputs is key in producing an integral weld. The ultra-thin-walled tubes in the present work have limited neighbouring material thus the direction of the heat conduction is effectively limited to two dimensions. Thus, in the present work whilst substantial post-weld purge gas flow is applied, the flow rate is relatively low compared to typical GTAW flow rates.

% two to three for each sub section is fine.
\section{GTAW based thin structure titanium welding}
\label{2-sec-thin-struct-review}
\subsection{Broad view}
Lacking a widely accepted definition for thin-structures, this thesis uses the definition of 'thin' to be a structure with a minimum thickness of less than \SI{2}{\milli\meter}. This definition can then be extended to ultra-thin for structures with a minimum thickness of less than \SI{0.5}{\milli\meter}. `Structure' is specifically the work piece that is being welded. This definition is important as many modern additive manufacturing techniques can create very small parts where the dimension of importance is the thickness of each additive layer or `resolution' of the process. Yet irrespective of resolution the physics of welding on a small structure are the same whether that is a part of a single layer of welding or a multiple layer additive manufacture process. What matters is the thermal gradients and phase changes involved as a part of the entire process and the geometry of the structure has a large effect on this. Therefore, GTAW based thin structure titanium welding can be investigated in terms of thin sheet, thin tube, thin complex structures and finally ultra-thin structures.
 
%
% Very long already, just drop general thin structure and talk only of titanium
%
%\subsection{Thin structures}
%\subsubsection{Thin sheet}
%\subsubsection{Thin tube}
%\subsubsection{Thin complex}
%A popular research route for GTAW on thin complex structures is that of additive manufacturing (AM). In AM terms, GTAW is a direct energy deposition process where layers or solid filler metal are fed into the GTAW arc and welded on top of each other.  This is typically termed wire and arc additive manufacturing (WAAM).
\subsection{Thin titanium structures}
\subsubsection{Thin sheet}
\label{2-sec-ti-thin-sheet}
Using multiple shielding gas sources, Karpagaraj et al. investigated GTAW on \SI{2}{\milli\meter} commercially pure titanium sheet \cite{cpTiGTAW1}. Through analysing the experimental results, process parameters for \SI{1.6}{\milli\meter} commercially pure titanium sheet were predicted. Prediction of process parameters is exactly what is required for a general case of ultra-thin-walled tube welding and this study demonstrates the efficacy of an experimental only based approach. 

In addition to the pure experimental approach, Karpagaraj et al. investigated GTAW on \SI{1.6}{\milli\meter} Ti-6Al-4V sheet \cite{karpagaraj2020experimental} this time using the experimental work to validate a computational model. This methodology can be applied to the present work with experiment resulting used to calibrate a computational model. Furthermore, in a separate study Karpagaraj et al. \cite{karpagaraj2019experimental} sought to optimize both the area enveloped by the shielding gas and its post flow time for GTAW on \SI{1.6}{\milli\meter} and \SI{2}{\milli\meter} thick CP Ti sheet finding the control of shielding gas is a key parameter for titanium welding. 

Yunlian et al. \cite{yunLianEBW} compared the results of GTAW against LBW and EBW for \SI{500}{\micro\meter} commercially pure titanium sheets. The investigation found that overall EBW was the most suitable for welding the titanium sheets yet LBW achieve the smallest weld width and the least work piece deformation. GTAW performed relatively poorly compared to LBW and EBW indicating that potential they could be candidates for ultra-thin-walled tube welding but both EBW and LBW require bulky equipment which for the present application eliminates them.

\subsubsection{Thin tube}
\label{2-sec-ti-thin-tube}
The work of de Garcia et al. \cite{garcia2010advances} involved welding \SI{1}{\milli\meter} wall thickness, \SI{6}{\milli\meter} OD, titanium tubing for a satellite propulsion system using orbital GTAW with internal pressurization. Whilst this work does demonstrate the need for internal pressurization, a manual tacking process was used for tube alignment. This limits its applicability for the present work given this tacking process adds additional filler metal to the welds and possibly could affect the internal geometry of the tubing. 

\subsubsection{Thin complex}
\label{2-sec-ti-thin-complex}
Miyamoto et al. \cite{miyamoto1986} reported thin-walled CP-2 skelp production with wall thicknesses ranges from \SIrange{500}{700}{\micro\meter}. Here, multiple titanium sheets are continuously butt welded together using GTAW to create a skelp. However the forming of the skelp is performed through mechanical rollers so the contribution of GTAW in its creation remains as essentially typical butt welding thus this process is considerably simpler from a welding perspective. 

Work by Kumar et al. \cite{kumar2010} demonstrated tube-to-end-fitting welding of Ti-3Al-2.5V. The end fitting used has a shape reminiscent of the sleeves for the ATLAS ITk whereby the tubes are inserted into a socket. Yet the tubing had a wall thickness of \SI{1.7}{\milli\meter} thus is more robust against the pressure-balance concerns for ultra-thin-walled tube welding. Thus again there is limited applicability to ultra-thin-walled tube welding as the requisite internal pressurization means that it is a fundamentally different process.

\subsubsection{Ultra-thin titanium tube}
To the author's knowledge, reports on titanium ultra-thin-walled titanium tube welding are extremely limited. A plausibly close investigation was performed by Carvalho et al. \cite{carvalho2016} for laser welding of \SI{500}{\micro\meter}. Whilst not ultra-thin, it is still on the boundary. However, it was for longitudinal welding of CP-2 tubes. This process does not involve welding in a singular transverse plane thus the complex heat flow of orbital welding is not considered and this process is actually more similar to thin sheet welding. 

Whilst not titanium Liu et al. \cite{LiuThinWallTube} studied orbital GTAW on \SI{500}{\micro\meter} wall thickness 304 stainless steel tubing. Here, to eliminate deformations internal support structures were added inside of the tubing. The internal support structure proved successful illustrating the effectiveness of internal support from a structure or otherwise yet this was for tubing with \SI{500}{\micro\meter} wall thickness and it is reasonable to assume even more support would be required for thinner wall thicknesses. The introduction of support structures inside the stainless steel tubing in \cite{LiuThinWallTube} was possible as the tubing had an outer diameter of \SI{180}{\milli\meter}. In principle it may be possible to scale the support structures down to the \SI{2.275}{\milli\meter} outer diameter tubing used for the ITk yet the cooling system features a complex curving geometry making this impracticable to do. This therefore supports the method of using internal pressurization from gas flow to create internal support for orbital GTAW.

None of the research covered in this section is explicit about the requirement of the internal pressurization that is necessary for ultra-thin-walled tube welding. Therefore it has limited applicability beyond the standard ideas of the importance of shielding gas for welding titanium. Given this, there is a research gap for the present work.

\section{Multiphysics methods for GTAW}
% consider strucutre here with setting out the requirements to then compare against.
\label{2-sec-review}
\subsection{Broad view}
\label{2-sec-review-broad}
\begin{figure}[!htb]
\centering
\includegraphics[width=14.5cm]{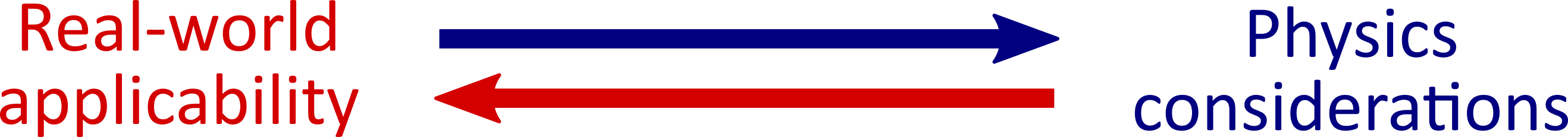}
\caption{Conceptual continuum of physics accuracy against real world applicability. Here, pure real world applicability works as an explicit welding procedure for a particular welding job to create a physical joint whereas a simulation prioritising physics accuracy would focus on simulating each physics process within GTAW.}
\label{2-fig-continuum}
\end{figure}

Simulating GTAW is a multiphysics problem. As detailed in Section \ref{1-sec-gtaw}, the GTAW process is a combination of magnetohydrodynamics, plasma physics and materials science. Due to this complexity, modelling efforts inevitably make trade-offs depending on the focus of the work. To illustrate this, it is helpful to consider a continuum between perfect physics accuracy where a complex multiphysics program simulates every physics process involved in GTAW and a real-world welding procedure specification that sets out exactly what welding parameters to use for a particular job. This is illustrated in Figure \ref{2-fig-continuum}. Simulating the flow of a finite number of electrons is fundamentally different from specifying the usage of a set amount of current for a user thus there is an inherent trade-off between physics accuracy and direct real world applicability. This difference manifests as a different focus of the work. For instance, work by Vel\'{a}zquez‑S\'{a}nchez et al. \cite{gtawHeatTransfer} presents a highly detailed representation of the GTAW process through simulating much of the physics involved with of goal of identifying the dominant heat transfer mechanisms in GTAW. Thus by definition it does not seek to provide a practitioner with a WPS for a particular welding job. Yet in the present work, the explicit goal is to simulate a practical WPS for welding hypothetical ultra-thin-walled tubing of different wall thicknesses to that used in the ITk. Therefore, on the continuum present in Figure \ref{2-fig-continuum} this work focuses more on real-work applicability over physics considerations.

% spacially $\mathcal{O}(\SI{\micro\meter})$ and temporally $\mathcal{O}(\SI{\milli\second})$
An ancillary to this continuum is the chosen scale of a simulation. A simulation focusing on the evolution of a liquid weld pool including its internal flow and the evolution of its surface requires both detailed information of the kinematic properties of the fluid and an apt temporal and spacial scale to capture these dynamics. For instance, the surface tension of the liquid weld pool with its surrounding atmosphere will cause the weld pool surface to curve. To model information about this curving requires a certain level of spatial and temporal granularity. Simulations that focus on capturing features of this size are termed mesoscale models. Compared to this, another simulation may look at the overall produced part including the microstructure, stresses and deformations. This is termed macroscale modelling \cite{lindgren2016am}.  Focusing on these elements does not require as small of a spatial and temporal granularity as mesoscale models thus freeing resources to better investigate microstructure, stresses and deformations. 

In a review article, Lindgren and Lundb\"{a}ck \cite{lindgrenReview} categorized this meso/macroscale modelling for particular welding based additive manufacture methods as either weld process modelling (WPM) for mesoscale and computational welding mechanics (CWM) for macroscale modelling. Given the scales involved are similar to ultra-thin-walled tube welding the CWM and WPM dichotomy can be used for the present work. Further, their review notes the common omission of the simulation of phase change in computational welding simulations of additive manufacturing where in lieu of simulating phase change various proxy methods can be used to assess microstructure, stresses and deformations. However in the present work to simulate the pressure balance of the liquid weld pool in ultra-thin-walled tube welding simulating phase change is a key requirement. Therefore, there is limited applicability of studies that use phase change proxies.

\subsection{Computational welding mechanics}
\label{2-sec-cwm}
As defined by Lindgren \cite{lindgrenNumericalModelling}, computational welding mechanics is a modelling process used to predict the thermal, material and mechanical effects of welding. CWM ignores fluid flow instead treating the liquid weld pool as a soft solid \cite{lindgren2016am}. This treatment is a key weakness for CWM from the perspective of predicting a general case for ultra-thin-walled tube welding as without modelling fluid flow the stability of the weld pool outlined in Section \ref{1-sec-ultra-thin-tube} cannot be predicted. This view is supported by Lindgren and Lundb\"{a}ck \cite{lindgrenReview} as they explicitly state how CWM cannot assist in selecting process parameters to obtain a stable process zone instead recommending using CWM for predicting deformations and microstructure. 

In a separate study, Lindgren and Lundb\"{a}ck \cite{lindgren2011metal} applied CWM to simulate ten layers of AM build up via GTAW on a Ti-6A-4V plate. Using the commercial software MSC Marc combined with a physical experiment using corresponding parameters, this work achieved an excellent match between the simulation and measurement for the temperature evolution across the work piece. The work also looked at distortions in the base Ti-6A-4V plate post welding finding overall reasonable agreement with simulation and measurement although the agreement deteriorated at later times. The close match of the temperature evolution - and to a certain extent the distortions - shows how CWM can work effectively. However, the process the work covers involves adding \SI{1.25}{\centi\meter} of \SI{1.14}{\milli\meter} wide metal filler wire into the weld pool every second. This is over an order of magnitude higher than the entire weld volume for the \SI{0.16}{\milli\meter} wall thickness CP-2 tubing for the ATLAS detector just in filler wire meaning the granularity requirements of \cite{lindgren2011metal} compared to the present work are not the same. CWM has also been used to investigate potential improvements to manufacturing processes. 

Fisk and Lundb\"{a}ck \cite{Lundback2012} used CWM to simulate GTAW on \SI{5}{\milli\meter} alloy 718 plate where filler metal was added into a pre-milled groove to mimic a repair process. Again using MSC Marc, this work investigated the possibility of replacing a global heat treatment process with an induction heating based local heat treatment. It found that there was no significant difference between the two processes. This demonstrates the use of CWM as an exploratory tool which is exactly the type of tool needed in the present application.

\begin{figure}[!htb]
\centering
\includegraphics[width=12cm]{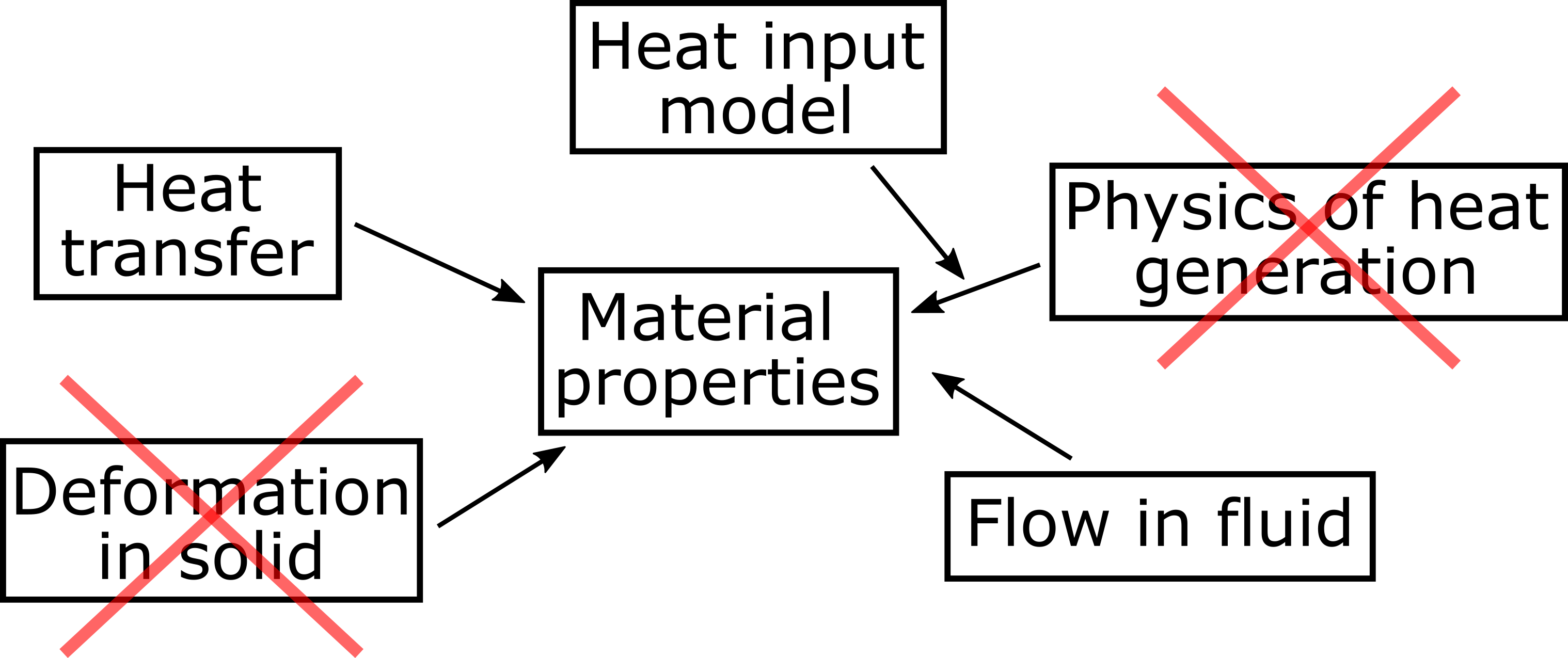}
\caption{Diagram showing the required modelling elements for the present application. The diagram has been adapted from \cite{lindgren2014computational} outlining the modelling elements in CWM. The key difference is that in \cite{lindgren2014computational} deformation in the solid is simulated and the flow in the fluid is not whereas in the present application the reverse is true.}
\label{2-fig-CWM}
\end{figure}

To evaluate the applicability of CWM to the present application, Lindgren's diagram in \cite{lindgren2014computational} can be adapted for a fluid-flow orientated model. In \cite{lindgren2014computational} for CWM to predict the material properties of a weld, heat input, heat transfer and deformations in the work piece must all be considered however the physics of heat generation and the flow in the liquid weld can be neglected. Yet in order to assess the challenges of ultra-thin-walled tube welding outlined in Section \ref{1-sec-ultra-thin-tube} it is vital that the fluid flow is simulated. Whereas simulating the deformations of the solid is not vital to understanding the pressure balance mechanism and nonetheless it would add considerable complexity to the model. The same is true for the physics of heat generation, ideally it would be simulated but it also adds additional complexity increasing the modelling challenge. In \cite{lindgren2014computational} the physics of heat generation is neglected in favour of the simpler weld heat input simulation, this same simplification can be applied to the present work. Therefore, through switching solid deformations and fluid flow in Lindgren's diagram the outline for the model in the present application can be established. This is shown in Figure \ref{2-fig-CWM}.

\subsection{Weld process modelling}
\label{2-sec-review-complex}
Historically, modelling the GTAW process has been split into modelling of the arc and modelling of the weld pool. Modelling the GTAW arc accurately requires a detailed representation of the cathode (the electrode in DCEN), the plasma column (the highly visible portion of the electric arc) and the cathode layer (the sources of the various electron emissions) \cite{choquetCathodeReview}. Whilst some issues still remain \cite{murphy2015perspective, choquetPhysicsReview}, in recent years there have been multiple sophisticated models of the GTAW arc that elucidate the plasma physics involved \cite{gtawArcSim1, gtawArcSim2, gtawArcSim3}. These accurate GTAW arc models enable a understanding of the heat input into the work piece. However, even with this understanding an accurate model of the evolution of the weld pool is required. 

Weld pool modelling or thermal fluid dynamics focuses on the formation and evolution of the liquid weld pool. This involves simulating not only the thermal fields and the transition from solid to liquid in the work piece but also the many oft competing forces acting on the liquid weld pool highlighted in Figure \ref{1-fig-gtaw}. The relative strength of these competing forces was categorized by Han et al. \cite{weldDrive} finding that the Marangoni force is dominant followed by the arc pressure, arc drag force, electromagnetic force and finally buoyancy. However the dominance of the Marangoni effect is dependent on the surface tension to temperature coefficient of the material. Steel, which is used in \cite{weldDrive}, typically not only has high surface tension to temperature coefficient but the sign of this coefficient has a strong effect on its weldability. At a critical sulfur concentration - $\approx \SI{40}{ppm}$ - the previously negative surface tension to temperature coefficient will become positive. A positive coefficent will help create a narrower and deeper and hence better weld thus improving weldability. This topic is explored in detail in \cite{millsMarangoni} and \cite{aidunMarangoni}. Further, weld pool simulation studies investigating the Marangoni effect specifically have been performed. Mishra et al. \cite{mishra2008} showed the well known change in weld pool shape with sulfur concentration in steel due to the Marangoni effect and Saldi \cite{saldi2012marangoni} investigated the Marangoni effect during laser welding of various steel grades in a thesis on the topic.

Fully coupled models involve both the arc and the weld pool being simulated concurrently. In an influential review article on the subject, Tanaka and Lowke \cite{tanakaPlasmaPhysics} showed how weld profiles can be predicted from details about the arc. The work showed that due to the aforementioned relative strength of the Marangoni effect an accurate description of the arc can accurately predict weld profiles. Although, the work lacked a free surface with the weld pool remaining flat. However, fully coupled GTAW models have been used to probe surface deformations dependent on the welding procedure. In \cite{WangCoupled}, Wang and Lu show how the effects of the Lorentz force, the Marangoni effect, and the arc pressure change as the current - both peak value and pulsing rate - is varied. In a thesis on the subject, Traidia \cite{traidiaThesis} took a different `hybrid' approach whereby a 2D model was created for stationary simulation of a GTAW arc to calculate a top surface boundary condition. This boundary condition was then applied to a full 3D model that calculated the surface deformation. 

\subsection{Finite element and finite volume methods}
\label{2-sec-femfvm}
\subsubsection{Broad view}
The finite element method (FEM) is a class of numerical techniques whereby a partial differential equation is discretized into a series of elements. These elements are then used to form a large system of linear equations that can be solved using various mathematical techniques. The benefit of FEM is that this discretization into finite elements can be done on physical geometries and thus can solve many real-world engineering problems. For instance, a flat beam fixed at one end with a downwards force applied on the opposite end could be represented as a series of two dimensional line elements. Using appropriate physics equations FEM could then be used to calculate the bending of this beam. It is a popular method for solid mechanics problems although in principle can be used for fluid mechanics.

Conversely, the finite volume method (FVM) relies on discretizing the domain into a series of control volumes. Compared to FEM elements, these control volumes have finite surface areas and volumes and the flux through these cell faces can be combined with physical conservation laws which enable complex fluid flow to be simulated. Due to this, FVM is typically used over FEM for the simulation of fluids as it can capture free surface flow more effectively. The aforementioned studies of Lindgren \cite{lindgrenNumericalModelling, lindgren2016am, lindgren2011metal} are all finite element modelling based. Whilst this may prove apt for a CWM focus, there is limited applicability for the WPM required in the present work.

\subsubsection{Ultra-thin-walled tube welding simulation research}
\label{2-sec-review-tube}
The simulation of ultra-thin-walled tube welding is not the most popular research route yet there are still some related FEM investigations. For instance FEM is commonly used to investigate the residual stresses created during GTAW pipe welding. Obeid et al. \cite{obeidOrbital} used ABAQUS to simulate orbital GTAW on \SI{6}{\milli\meter} AISI 304 pipe. Further Ravisankar et al. used another program, SYSWELD, to simulate orbital pipe welding on \SI{2.5}{\milli\meter} thick AISI 304 stainless steel \cite{sysweld}. Whilst these results do feature details on mechanical deformation, only the heat transfer equation is solved for thus this does not address the liquid metal flow critical for ultra-thin-walled tube welding which a FVM method would be able to achieve. Despite this, ABAQUS remains popular for this vein of research shown by recent results from Asadi et al. \cite{asaidOrbital} investigating the effects of welding parameters on the residual stresses created during GTAW pipe welding also on \SI{6}{\milli\meter} AISI 304 pipe. The efficacy of ABAQUS was demonstrated in \cite{asaidOrbital} by the reasonable match to the complementary experiment performed in the same study.

Other commercial software such as ANSYS have also been used successfully for FEM orbital welding simulation shown by Singh and Pradhan \cite{singhPradhanOrbital} where \SI{1.5}{\milli\meter} thick AISI 316 L pipe was used. The ANSYS package features ANSYS Fluent which is a FVM program yet this is not commonly used for ultra-thin-walled tube welding simulation. ANSYS was also used by Purmohamad et al. \cite{purmohamadOrbital} for GTAW on \SI{5.54}{\milli\meter} thick Incoloy 800 pipe. Here, `element birth and death' was employed to simulate the addition of filler metal but again only the temperature equation was solved for limited the efficacy in an ultra-thin-walled tube welding context. 

Without clear phase change and a free surface, it is difficult to apply this FEM research to ultra-thin-walled tube welding where identifying collapsing or bursting conditions is critical. Yet the FVM method is adept at simulating free surface flow and phase change thus would be able to achieve this. Therefore from combining the present section with Section \ref{2-sec-review-complex} it is concluded that a FVM based simulation is the most apt for ultra-thin-walled tube welding simulation.

\subsection{Heat source modelling}
\label{2-sec-hs}
\subsubsection{Broad view}
\label{2-sec-hs-bv}
Simulating the heat transfer in GTAW is a complex multiphysics problem. The important heat transfer processes associated with GTAW are: joule heating from the electric current flow and associated losses, thermal conduction, convective flow within the liquid weld pool and phase change from liquid to solid (and vice versa) \cite{murphy2018heatTransBook}. The most impactful of these processes is joule heating from the electric current flow \cite{murphy2015perspective}. Therefore, simulating the heat transfer from a GTAW arc requires a strong focus on plasma physics to capture the electromagnetic field properties of the arc during welding. For instance, understanding the electron emission mechanisms is key in determining the energy distribution of ions and electrons in the plasma column and hence its conductivity. This is before any interaction between the arc plasma and work piece is considered. This modelling approach is achievable with appropriate key assumptions \cite{choquetPhysicsReview}, yet adds considerable complexity beyond CWM and WPM. Instead, as shown previously in Figure \ref{2-fig-CWM}, heat input can be simulated in lieu of the physics of heat generation.

\begin{figure}[!htb]
\centering
\includegraphics[width=4.5cm]{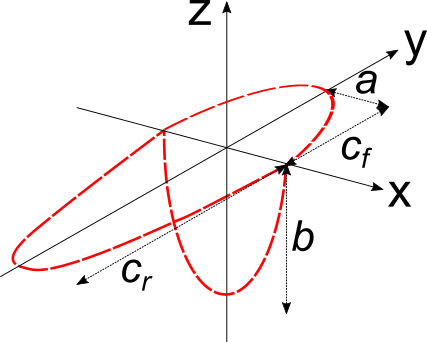}
\caption{Diagram showing the parameters for Goldak-type heat source models.}
\label{2-fig-goldak}
\end{figure}

One of the most widely used volumetric heat source models is the double-ellipsoid proposed by Goldak et al. \cite{goldak1984}. This heat source model involves two volumetric ellipsoid regions typically a front ellipsoid and a rear ellipsoid where the fraction of heat input is split as $f_f$ in the front ellipsoid and $f_r$ in the rear ellipsoid. Using the convention $f_r = 1 - f_f$, the heat input $q_i$ from each ellipsoids can be defined as 

\begin{equation}
\label{2-eq-goldak}
q_i = f_i \frac{q_{src} 12 \sqrt(3)}{abc_i\pi^\frac{3}{2}} \cdot exp\left( 3\left(\frac{x}{a}\right)^2 - 3\left(\frac{y}{c_i}\right)^2 - 3\left(\frac{z}{b}\right)^2\right)
\end{equation}

for heat inputs $q_i, f_i, c_i$ where $i$ is either the front ellipsoid $f$ or the rear ellipsoid $r$ and $q_{src}$ is the heat input from the source. $a, b$ and $c_i$ define the depth, width and length of the ellipsoid and $x, y, z$ the coordinate system it is defined upon. These values are illustrated in Figure \ref{2-fig-goldak}. Note $q_{src}$ is equal to the power input (which for GTAW is voltage multiplied by the welding current) scaled by an efficiency value $\eta$. The apt value for $\eta$ is discussed in the following section. 

A key advantage of Goldak-type heat sources is their adaptability; through altering the shape and number of ellipsoids used LBW \cite{goldakLBW}, EBW \cite{goldakEBW} and K-TIG \cite{goldakKTIG2017} have all been modelled with a Goldak-type source. For GTAW, the key assumption with Goldak-type heat sources is the neglecting of physics based mechanisms for heat input - such as the aforementioned joule heating - in favour of a generic heat added term. With this assumption Goldak-type heat sources serve as literal source terms in the energy transport equation used in the CWM/WPM model. This is why \cite{goldakLBW, goldakEBW, goldakKTIG2017} can all use Goldak-type heat sources despite the dramatic differences in the process as the heat sources are just source terms. Therefore, whilst it may be apt in some circumstance to characterize slight modifications to equation \ref{2-eq-goldak} as new or distinct heat source models \cite{amHeatSourceReview}, in the case of GTAW they are effectively all Goldak-type. 

In addition to Goldak, fundamentally different assumptions about the GTAW arc can be made such as modelling it as a point source of inputted heat as is the case in the historically popular Rosenthal equation. Similarly, some researchers opt to use a boundary condition based GTAW heat source \cite{boundaryHS, boundaryHS2} whereby an Gaussian heat distribution is used as either a fixed value or fixed gradient on a simulation domain boundary. However, these assumptions are less representative of how GTAW melts a work piece thus often fail to achieve as good agreement with experiment as Goldak-type heat sources do. Also, heat source modelling as a concept only makes sense when the physics based modelling detailed in Section \ref{2-sec-review-complex} is neglected and a heat source term is introduced. Combining this with how Goldak-type heat sources perform better than those with different assumptions means for this thesis only Goldak-type models need be discussed. 

Finally, note the second ellipsoid can be dropped from equation \ref{2-eq-goldak} and the heat input treated as Gaussian in the $xy$ plane and linear in the $z$ plane. This is the approach taken in \cite{LundbackXYLinZ} for EBW and \cite{paraboloid-paper} for GTAW. This modification reduces the geometric parameters ($c_f$ and $c_r$ to just $c$) which is a key advantage when working in a scenario where the exact weld dimensions cannot be calculated easily as is the case in the present work.

\subsubsection{Efficiency value}
The electrical power produced by GTAW is simply the welding current multiplied by the voltage. However, not all of this power will go directly into the work piece as heat as much of it is lost in the arc plasma by lateral radiation or convection. Therefore, the heat actually inputted into the work piece $Q_{in}$ is modelled as the electrical power $P = IV$ scaled by an efficiency $\eta$. Strictly speaking, given the non-linearities of the various electron emission mechanisms in GTAW this efficiency value will depend on the current magnitude. However as research has found this effect to be small \cite{dupont1995thermal} when modelling it is reasonable to just use one efficiency value. 

In principle, the choice for this efficiency value would be found from experiment. Yet the combination of variety in the work piece composition, the work piece geometry and the shielding gas used amongst other factors means efficiency values found through experiment vary considerably. To illustrate this compare how the efficiency of DCEN GTAW with argon gas and a flat work piece is measured as 0.70 by Gonzalez et al \cite{effGonzalez2007} with a water cooled copper work piece comped to Cantin and Francis who measured 0.78 with an aluminium work piece \cite{effCantin2005}. This is a difference of over 10\% just with a work piece composition change. However, from analysing multiple studies Nils Stenbacka found an average value of the measured efficiency for DCEN GTAW to be $0.77 \pm 0.07$ and that this value is valid for a wide variety of work pieces \cite{stenbackaEff}.

An alternate view is to fit the efficiency value via an experimental investigation. Farias et al. \cite{fariasEff} conducted a study where both butt joint and lap joint welds for AISI 1020 carbon steel and AISI 304 stainless steel were performed experimentally with thermocouples embedded into the work pieces. The temperature evolution and weld cross section were then used to find a set of parameters for both the efficiency and the other Goldak heat source parameters in equation \ref{2-eq-goldak} that resulted in the closest match between simulation and experiment for a FEM simulation using a Goldak-type heat source. For butt joints on the AISI 1020 carbon steel an efficiency of merely 59\% was best whereas for AISI 304 69\% proved the optimum. Additionally, for the lap joints the efficiencies were 79\% and 77.5\% for the AISI 1020 and the AISI 304 respectively.

The values for efficiency used for Goldak-type heat sources in a selection of investigations are shown in Table \ref{2-tab-eff-sim}. Here the average value is 70\% with a standard deviation of 10\%. Note in the origional study Goldak et al. \cite{goldak1984} used an efficiency of 95\% but this is omitted as it is quite out of step with modern experimental measurements and simulations.

\begin{table}[h]
\caption{\label{2-tab-eff-sim}Summary of the value used for the efficiency of the autogenous GTAW process for models using Goldak-type heat sources.} 
\centering
\begin{tabular}{c c c}
\toprule
Reference & Efficiency value & Work piece \\
\hline
%Lindgren and Lundb\"{a}ck \cite{lindgren2011metal} & 0.58 & Titanium alloy \\ This one has filler metal
Asadi et al. \cite{asaidOrbital}\rule{0pt}{2.6ex} & 0.80 & AISI 304 \\
Farias et al. $a)$ \cite{fariasEff} & 0.59 & AISI 1020 \\
Farias et al. $b)$ \cite{fariasEff} & 0.69 & AISI 304 \\
Fisk and Lundb\"{a}ck \cite{Lundback2012} & 0.75 & Alloy 718 \\
Karpagaraj et al. \cite{karpagaraj2019experimental} & 0.50 & CP Titanium \\
Malik et al. \cite{malik2008analysis} & 0.80 & AH36 Steel \\
Obeid et al. \cite{obeidOrbital} & 0.70 & AISI 304 \\
Ravisankar et al. \cite{sysweld} & 0.80 & AISI 304 \\
\hline
Average\rule{0pt}{2.6ex} & $0.70 \pm 0.10$ & - \\
\bottomrule
\end{tabular}
\end{table}

Given this overview it is the view of the author that the choice of what value to use for efficiency is - within reason - up to the practitioner. The broad variety of efficiencies commonly used in simulation research combined with the variety of efficiencies measured experimentally means that any value for efficiency from 60\% to 80\% could in principle be supported with previous research. Therefore it is up to the practitioner to pick an apt value for their application.  

\subsubsection{Additive manufacturing}
When GTAW includes the use of a filler metal, some of the heat input from the GTAW arc is used to melt it. A simple modification to Goldak-type heat sources to account for this is to reduce the efficiency value. For instance, Lindgren and Lundb\"{a}ck \cite{lindgren2011metal} used an $\eta$ value of 0.58 for GTAW based AM build up on a titanium alloy. Here multiple layers of filler metal were welded on top of each other creating the AM part. However, geometrically as layers are welded on top of each other and the base metal work piece moves physically further away from the GTAW heat input a different amount of cooling takes place. Given how Goldak-type heat sources are literately just source terms in energy equations simply reducing the $\eta$ value is functionally equivalent to increasing the cooling.

However, simply just changing the $\eta$ value does not simulate the melting of the filler metal. Looking back to Figure \ref{2-fig-goldak} shows the Goldak-type heat inputted into a flat work piece. Mathematically equation \ref{2-eq-goldak} is reflected in the $xy$ plane thus it defines a full ellipsoid. An idea to account for the filler metal is to model the full ellipsoid and have the added filler metal melt within it as was performed by Montevecchi et al. \cite{heatSource50WAAM}. Here, the top and bottom halves of the ellipsoid were modelled to take 50\% of the total inputted heat e.g. $Q_{in, work} = \frac{Q_{in, total}}{2} = \frac{\eta I V}{2}$ with the 50\% value coming from prior research \cite{dupont1995thermal}. This is a different situation to \cite{lindgren2011metal} were the $\eta$ value was simply reduced. 

When considering implementation of \cite{heatSource50WAAM} it requires continuous definitions of what is filler metal and what is base metal. As an example, when the cold filler metal is melted by the arc, at what point does is change into base metal? The finite element modelling in \cite{heatSource50WAAM} allows for dedicated filler metal and base metal elements negating this concern yet with the finite volume method the cells cannot be tagged in the same manner. As discussed in Section \ref{2-sec-femfvm} the free surface modelling capabilities FVM provides are key for ultra-thin-walled tube modelling hence to implement this method in FVM a feature that identifies the volumes with filler metal at each time step would be required. The Lindgren and Lundb\"{a}ck investigation \cite{lindgren2011metal} would require no such feature and thus would be considerably easier to implement than the Montevecchi et al. investigation \cite{heatSource50WAAM} as only changing the $\eta$ value would be required.

Finally once an apt efficiency value is established, in principle it can be used to predict the heat input during AM and thus the weld pool size. By analysing the heat input into both H13 and 316 L steel additive substrate, Pinkerton and Li \cite{Pinkerton_2004} were able to successfully achieve this and predict weld pool geometry. However, this study was for laser welding which has a different heat input mechanism than GTAW.

\subsection{Computational software}
\label{2-sec-review-cs}
\subsubsection{Broad view}
There are many different computational software packages available to simulate GTAW. Broadly, these can be split into commercial software, open-source non-profit software and complete custom software created from scratch. Much of the commercial software mentioned thus far is FEM based such as ABAQUS \cite{obeidOrbital, asaidOrbital}, SYSWELD \cite{sysweld} and MSC Marc \cite{lindgren2011metal, Lundback2012}. However there are two prominent FVM software packages ANSYS Fluent and COMSOL. At the time of evaluation (2019), there was not an obvious candidate for which strong in-house knowledge was present for thus the programs were evaluated on their perceived merit and accessibility. Given the author was familiar with C++, \of emerged as a potential good candidate to investigate. As it is open source, \of came with the advantage of letting the user customize the program to suit their needs - something (closed source) commercial software limits. When compared with ANSYS FLUENT over 6 weeks considerably more progress\footnote{NB in no small part due to the fantastic free \of course and resources maintained by H\r{a}kan Nilsson \cite{hakanOfCourse}.} was made with \of thus it was chosen as the sole simulation program.    

\subsubsection{OpenFOAM}
\label{2-sec-review-of}
OpenFOAM, the program used in this thesis, is an open source computational fluid dynamics program that has been used for GTAW simulation in multiple capacities. Compared to commercial welding simulation software, the open source nature of \of means that the full details of exactly how the simulation works are available for anyone with the source code to inspect. Researchers from Chalmers University have published many GTAW simualtions with \of investigating a variety of aspects of the process but principally focusing on the arc. For example, Choquet has been involved in a large body of research related to GTAW arc simulation in \of \cite{choquet2011electric, choquet2016effect, choquet2018coupling}. Further, in a thesis on the topic, Sass-tisovskaya \cite{tisovskayaThesis} used \of to simulate a GTAW arc for an investigation into tandem GTAW. Tandem GTAW is a GTAW variation where two electrodes are used.

A collaborator of Choquet, Shirvan produced a thesis \cite{shirvanThesis} demonstrating a sophisticated treatment of the GTAW arc in \of coupled to a solid region. However, the coupling involved splitting the domain between a fluid section (gas) and solid section (work piece) with different equations in each section. Shirvan's work is an example of a solver based off of the inbuilt Conjugate Heat Transfer solver group in OpenFOAM. In this thesis, the multiphase group of solvers are used instead. Given these regions are set at the start of the simulation, it is difficult to capture the liquid weld pool surface evolution in a situation like ultra-thin-walled tube welding.  

In addition to arc physics, \of has been used to investigate weld pool evolution. The prominent aforementioned thesis by Saldi \cite{saldi2012marangoni} used \of to investigate the Marangoni effect in laser welding. The thesis demonstrated a variety of weld pool shapes created using different parameters illustrating the potential efficacy of \of for welding simulations. A strong aspect of the thesis is how the solver is benchmarked against various test cases to validate it. In fact, the present work benchmarks against the Marangoni test cases used by Saldi \cite{saldi2012marangoni}. 

\section{Research gaps for the GTAW simulation}
\label{2-sec-novelty}
\subsection{Experimental work}
\label{2-sec-gap-exp}
As covered in Section \ref{2-sec-thin-struct-review}, there is a moderate amount of research focusing on thin-structure titanium GTAW. However for ultra-thin structures there is a clear paucity of research focussing on preserving a specific geometry in the work pieces. The closest piece of work is Liu et al. \cite{LiuThinWallTube} yet as previously covered a physical internal support structure could not be created for the ITk cooling system. Further, the tubing in \cite{LiuThinWallTube} is 304 stainless steel which does not have the same shielding gas requirements as titanium \cite{karpagaraj2019experimental}.

As outlined in Chapter 1, GTAW is the clear optimum method for welding of the ultra-thin-walled CP-2 tubing in the ITk given the quality of the welds it can produce combined with its small portable equipment size. Yet to the author's knowledge, there is no previous research that addressed the combined requirements of a tube geometry, a CP-2 titanium material, an orbital GTAW process and an ultra-thin $\leq$ \SI{500}{\micro\meter} wall thickness. Therefore, purely from an experimental perspective, successfully welding the ATLAS ITk cooling system tubes would demonstrate a novel advance. The requirements for this advance against previous research is compared in Table \ref{2-tab-exp-summary}.

\begin{table}[h]
\caption{\label{2-tab-exp-summary}Summary of the experimental work covered compared to the target experimental work to create a WPS for the ITk cooling system.} 
\centering
\begin{tabular}{l c c c c}
\toprule
Investigation & Geometry & Material & Process & Wall thickness \\
\hline
Target\rule{0pt}{2.6ex} & Tube & Titanium & Orbital GTAW & $\leq$ \SI{500}{\micro\meter} \\
\hline
Asaid\rule{0pt}{2.6ex} \cite{asaidOrbital} & \ding{51} & \ding{55} & \ding{51} & \ding{55} \\
Carvalho et al. \cite{carvalho2016} & \ding{55} & \ding{51} & \ding{55} & \ding{51} \\
de Garcia et al. \cite{garcia2010advances} & \ding{51} & \ding{51} & \ding{51} & \ding{55} \\
Karpagaraj et al. $a)$ \cite{cpTiGTAW1} & \ding{55} & \ding{51} & \ding{55} & \ding{55} \\
Karpagaraj et al. $b)$ \cite{karpagaraj2019experimental} & \ding{55} & \ding{51} & \ding{55} & \ding{55} \\
Karpagaraj et al. $c)$ \cite{karpagaraj2020experimental} & \ding{55} & \ding{51} & \ding{55} & \ding{55} \\
Kumar et al. \cite{kumar2010} & \ding{51} & \ding{51} & \ding{51} & \ding{55} \\
Liu et al. \cite{LiuThinWallTube} & \ding{51} & \ding{55} & \ding{51} & \ding{51} \\
Miyamoto et al. \cite{miyamoto1986} & \ding{55} & \ding{51} & \ding{55} & \ding{51} \\
Yunlian et al. \cite{yunLianEBW} & \ding{55} & \ding{51} & \ding{55} & \ding{51} \\
\bottomrule
\end{tabular}
\end{table}

\FloatBarrier
\subsection{Simulation work}
\label{2-sec-gap-sim}
Broadly, the solvers described in Sections \ref{2-sec-cwm}, \ref{2-sec-femfvm}, \ref{2-sec-hs} and \ref{2-sec-review-cs} seek to elucidate the GTAW process. Whether by arc modelling, weld pool modelling, or coupled modelling solvers are typically used to investigate weld pool evolution in controlled situations. This means situations where the results are either known or the domain / case is set up in such a way as to facilitate investigations. For instance, the work by Vel{\'a}zquez-S{\'a}nchez et al. \cite{gtawArcSim1} presents a richly detailed and sophisticated understanding of the temperature profile of a welding arc. But, this is for a stationary arc in a steady state simulated in two dimensions. This idealized situation facilitates a meticulous investigation of the GTAW process but by definition its focus is away from a real world scenario. This is a fundamentally different requirement than simulating an unseen and relatively unusual welding process. Simulating welding on ultra-thin-walled tubing with the express intention of predicting a general case will mean less of a focus on the welding physics. This is shown by the solvers featured in Section \ref{2-sec-review-tube}. Here the focus on orbital pipe welding necessitates a pivot away from a welding physics only perspective. However, there are still core welding physics requirements that need to be incorporated in any simulation of ultra-thin-walled tube welding. 

There is a omitted issue with the verification of many models. This is that in many welding simulations either no benchmarking is performed or only one benchmark is used. Here benchmarking refers to assessing the performance of a simulation against known (e.g. experimental) results. This is an issue as it is possible for a simulation to be designed to perform very well on a particular task but this same simulation performs poorly on others. As there is somewhat limited space in a research article it is understandable that often only one or even no benchmarks are used. However, there is a research gap for a solver to be benchmarked against a plethora of investigations. 

Thus in order to simulate ultra-thin-walled tube welding, a simulation must conform to various requirements. It must be well benchmarked for credible results, it must be three phase to capture the weld pool breaking. Further, it must have a moving GTAW source to enable the full weld to be simulated and check whether the weld will collapse at any point. Finally, it should be capable of simulating the complex geometry of a ultra-thin-walled tubed. A summary of the models mentioned thus far and whether they conform to these requirements is shown in Table \ref{2-tab-model-summary}. This table shows that there is a clear research gap for ultra-thin-walled tube welding research.

% tick   \ding{51}
% cross \ding{55} 
\begin{table}[h]
\caption{\label{2-tab-model-summary}Summary of the multiphysics models covered in Section \ref{2-sec-review} compared to a target model for ultra-thin-walled tube welding simulation.} 
\centering
\begin{tabular}{l c c c c c}
\toprule
Model & \begin{tabular}{@{}c@{}}Bench-\\marks\\\end{tabular} & \begin{tabular}{@{}c@{}}Three\\phase\\\end{tabular} & \begin{tabular}{@{}c@{}}Moving\\GTAW source\\\end{tabular} & \begin{tabular}{@{}c@{}}Melting \&\\ Solidification\\\end{tabular} &\begin{tabular}{@{}c@{}}Geometry\\\end{tabular} \\
\hline
Target\rule{0pt}{2.6ex} & $\geq 5$ & \ding{51} & \ding{51} & \ding{51} & Tube \\
\hline
Asaid\rule{0pt}{2.6ex} \cite{asaidOrbital} & 0 & \ding{55} & \ding{51} & \ding{55} & Tube \\
Choquet \cite{choquet2011electric} & 0 & \ding{55} & \ding{55} & \ding{55} & Plane \\
Han \cite{weldDrive} & 0 & \ding{51} & \ding{55} & \ding{51} & Plane \\
Liang \cite{gtawArcSim2} & 0 & \ding{55} & \ding{55} & \ding{55} & Plane \\
Mishra \cite{mishra2008} & 0 & \ding{55} & \ding{55} & \ding{51} & Plane \\
Obeid \cite{obeidOrbital} & 0 & \ding{55} & \ding{51} & \ding{55} & Tube \\
Purmohamad \cite{purmohamadOrbital} & 0 & \ding{55} & \ding{51} & \ding{51} & Tube \\
Singh \cite{singhPradhanOrbital} & 0 & \ding{55} & \ding{51} & \ding{55} & Tube \\
Tanaka \cite{tanakaPlasmaPhysics} & 0 & \ding{51} & \ding{55} & \ding{51} & Plane \\
Traidia \cite{traidiaThesis} & 0 & \ding{51} & \ding{51} & \ding{51} & Plane \\
Uhrlandt \cite{gtawArcSim3} & 0 & \ding{55} & \ding{55} & \ding{55} & Plane \\
Vel{\'a}zquez-S{\'a}nchez \cite{gtawArcSim1} & 0 & \ding{55} & \ding{55} & \ding{55} & Plane \\
Wang \cite{WangCoupled} & 0 & \ding{51} & \ding{55} & \ding{51} & Plane \\
Saldi \cite{saldi2012marangoni} & $\geq 5$ & \ding{51} & \ding{55} & \ding{51} & Plane \\
Sass-tisovskaya \cite{tisovskayaThesis} & 2 & \ding{55} & \ding{55} & \ding{55} & Plane \\
Shirvan \cite{shirvanThesis} & 1 & \ding{55} & \ding{55} & \ding{55} & Plane \\
\bottomrule
\end{tabular}
\end{table}

\FloatBarrier
\subsection{Novelty of thesis}
% This bit is the methodology
Given the research gaps highlighted in Sections \ref{2-sec-gap-exp} and \ref{2-sec-gap-sim} it is clear that there is a research gap for both an experimental based investigation into ultra-thin-walled tube welding and, separately, a simulation based investigation. Yet fulfilling the targets in both Table \ref{2-tab-exp-summary} and Table \ref{2-tab-model-summary} to create an experimental and a simulation study together would demonstrate a strong research novelty and a valuable addition to the field. The efficacy of a joint modelling and experimental approach to predict process parameters has been demonstrated in prior mentioned studies \cite{cpTiGTAW1, asaidOrbital} which further underlines the viability of this research novelty. 

The present work can also be compared to two fully coupled WPM studies, Wang and Lu \cite{WangCoupled} and Traidia \cite{traidiaThesis}. In the thesis by Traidia an idealized situation is used to simulate a `full picture' of the arc and this information is then used for a simulation of a specific joint. In \cite{traidiaThesis}, this was for narrow gap welding and the approach proved successful. However, the complex weld pool dynamics in ultra-thin-walled tube welding shown in Figure \ref{1-fig-tube-gtaw} is resistive to this methodology as the arc and weld pool dynamics at the low amperages and tube wall thicknesses are unknown impeding assessment of the simulation. This is further exacerbated by the fact the orbital weld head completely encloses the tubing preventing any imaging of the process. In fact, a challenge with predicting a process that has limited experimental data is that the critical elements are unknown in advance. For instance, for thermophysical properties, a simulation can use either a single value for a phase, a series of `step' values, or a function. Some processes may be relatively robust against the choice of implementation for a particular thermophysical property but for others the implementation of a particular property may be critical. For these reasons the technique used by Traidia cannot be directly applied to ultra-thin-walled tube simulation and in fact the introduction of the required predictive methodology will be a novelty over Traidia.    

Wang and Lu \cite{WangCoupled} present an interesting simulation study for pulsed current GTAW on AISI 304 stainless steel. Considering a current pulsing between \SI{80}{\ampere} and \SI{400}{\ampere} with a \SI{10}{\hertz} pulsing frequency, their results show how simulations can be used to isolated the separate mechanisms with GTAW. However, this work sits in the same category as the aforementioned work by Vel\'{a}zquez‑S\'{a}nchez et al. \cite{gtawHeatTransfer} where a simpler spot weld situation is used to allow for sophisticated mathematical investigation. The present work has a moving GTAW arc in a complex geometry meaning that - as shown in Figure \ref{2-fig-continuum} - Physics accuracy is sacrificed for real-world applicability.

Additionally, both of these works use flat plane geometries; the present work presents a novel solution to simulating tube welding through a `flattening' process. Further, as neither work benchmarks their simulations to improve upon this the present work uses a more robust approach whereby many benchmarks are incorporated. Finally, even with the simulation novelties compared to the work in \cite{WangCoupled} and \cite{traidiaThesis} neither presents experimental analysis. The present work includes both simulation and experimental results.
%An anology here is the aforementioned comparison for AM heat source efficiency between Lindgren and Lundb\"{a}ck \cite{lindgren2011metal} and Montevecchi et al. \cite{heatSource50WAAM} where 
% Mention FEA / Lindgren’s work again
% Detail how approach is robust
% This bit is the application

Overall, to best investigate the ultra-thin-walled tubing for the ATLAS ITk both a simulation and an experimental approach should be used. The simulation must conform to various requirements. It must be well benchmarked for credible results, it must be three phase to capture the weld pool breaking. Further, it must have a moving GTAW source to enable the full weld to be simulated and check whether the weld will collapse at any point. Finally, it should successfully predict the required welds for the novel application of the ITk cooling system. Even in the hypothetical absence of novel simulation and experimental methodologies, investigating a procedure for this unique ultra-thin-walled tube welding challenge is a novel contribution.
%Revise with exp details
%Methodology, robustness of approach, applications ..
\FloatBarrier
\section{Objectives}
\label{2-sec-objectives}
\begin{enumerate}
    \item The main aim of this thesis is to create a general case for ultra-thin-walled tube welding. Specifically, this means some sort of model that is rules based, formulaic, or otherwise that can predict whether a particular welding procedure will be successful in welding ultra-thin-walled tubing. This will allow a user to identify a required procedure for welding \emph{any} ultra-thin-walled metal tubing in advance saving them time and money.
    \item An ancillary objective is to create a well documented open source GTAW simulation tool. Many multiphysics simulation programs work as effective black boxes where the simulation tools simply state `solves the Navier-Stokes equations' without any details on the implementation mathematics potentially obscuring key design choices. Further, many simulations rely on being tuned to a particular case. Whilst this may work for simpler geometries, this lack of robustness limits confidence in the novel ultra-thin-walled tube welding simulation described in the present work. In both of these situations a user is left with limited predictability. This narrow scope curtails them to post-hoc tools that show how an experiment worked rather than predicting whether it will work. A well documented tool allows the user to understand exactly how it works and an open source tool allows the user to change it to fit their needs. Thus this combination will allow a user a basis to both understand their experiment and crucially predict what will happen in an experiment.
    \item The final objective is to better elucidate and optimize the enthalpy-porosity method for GTAW simulation. Typical enthalpy-porosity models include some arbitrary computational constants such as the Darcy constant. As mentioned in the previous objective these constants are often chosen purely to fit the case at hand. Therefore systemically addressing standard questions such as `what is the best value for the Darcy constant' or `what is the best formulation of enthalpy to use' allows a user to better implemented the aforementioned GTAW simulation tool.  
\end{enumerate}

\section{Scope}
\label{2-sec-scope}
\subsection{Included scope}
The research covered in this thesis focuses on understanding the evolution of the weld pool during the GTAW process particularly on ultra-thin-walled tubing. The core undertaking to achieve this is through a custom built multiphysics simulation program. The program is created through extending an open source C++ toolbox: OpenFOAM. \of is an extensive program so only the parts relevant to the custom solver are considered in-scope. There are simulations covering melting, solidification, buoyancy driven flow, Marangoni driven flow as well as multiple GTAW simulations. The materials simulated included gallium, tin, aluminium, water, bismuth, 304 stainless steel, 316 stainless steel and commercially pure grade 2 titanium. Additionally, some limited experimental work is present although this was impacted due to unforeseen circumstances. The author was involved in the experimental work in Chapter 5 but the experimental results from Chapter 6 were created by other researchers.

Throughout the development process a variety of features were added to achieve the main objective of finding a general case for ultra-thin-walled tube welding. Therefore, the in-scope boundary was established through which tools enabled the achievement of this goal. Given this was unknown in advance, this resulted in an ad hoc addition of features some of which do not have a large effect on the final results for ultra-thin-walled tube welding. However, these features are largely still included as - with the second objective in mind - they provide functionality which may be vital for an unexpected application.   

\subsection{Excluded scope}
Due to the complexity of simulating the GTAW process certain aspects aspects necessarily are neglected to allow a stronger focus on others. For example, the arc physics of the GTAW arc is not covered at all in this thesis as it is a discipline within itself - these arc physics modelling efforts are described extensively by Murphy et al. in \cite{murphyArcModellingReview}. Although, as a proxy for this, the resultant heat from the GTAW arc is modelled. Similarly, arc pressure effects have to be separately mimicked. Whilst it is implemented in \emph{gtawFoam}, the Lorentz force is not modelled due to a combination of it having a small effect when tested and its incompatibility with the volumetric heat source in some situations. Also, materials are modelled as phases with no further granularity. This means the amount of an alloy material such as 316 stainless steel would be stored effectively as a single scalar value at the centre of each mesh cell devoid of any details about the grain structure. This is relevant in GTAW as it means the HAZ will be uncharacterised and can only be estimated through looking at the maximum temperature that the non-melted region was subject to. Finally, given the extensive mathematics behind computational fluid dynamics and its many applications, only the ideas and derivations deemed relevant for the reader to understand the contributions of this thesis are included. Some theses opt to feature extensive derivations of general forms for all fluids however as only incompressible Newtonian fluids are used in this thesis they are judged the only kind relevant to be detailed.

%% file: chapters/thesis-chapter-3.tex
\lhead{Chapter 3: Numerical methods}
\section{Introduction}
This chapter details numerical methods used to model welding within this thesis. Its purpose is to cover the necessary background information to contextualize the following chapters. Initially, computational fluid dynamics, CFD, is introduced through a basic case of a single phase incompressible Newtonian fluid along with a description of its implementation in the program used for the simulations in this thesis - OpenFOAM. Then, to show how \of works in practice, a popular OpenFOAM solver is described - \emph{interFoam}. With this established, a basic `toy' model is introduced as a rudimentary attempt to simulate GTAW using CFD. Finally, this 'toy' model is assessed to investigate what is required for a bespoke model.   

\section{Computational Fluid Dynamics}
\subsection{Overview}
Sitting at the confluence between maths, Physics, and computer science, Computational Fluid Dynamics (CFD) is - as the name suggests - a broad class of computational methods to solve fluid dynamic problems. It can be broken down into problem formulation from a Physics perspective, solving the problem with numerical methods, and analysing the results. Taking a continuum mechanics example, the Physics of the conduction of heat through a three dimensional solid can be modelled with the heat equation $\partial_t T = \alpha_D \nabla^2 T$. This heat equation can then be imposed on a domain with specific boundary conditions and could be solved for using the conjugate gradient method to find solutions at various times. The temperature at a particularly point could then be plotted using a program of the users choice. This is essentially the process for most CFD problems. However, when dealing with fluids the equations are more complex. Therefore the description of CFD will first look at a simple case of a single phase incompressible Newtonian fluid. Starting from the laws of conservation of mass and momentum the governing equations of which - the Navier-Stokes - can be derived.

\subsection{Navier-Stokes}
\label{2-sec-NS}
The Navier-Stokes equations are the governing equations of fluid dynamics. For an incompressible Newtonian fluid they are written as two equations: firstly the conservation of mass (equation \ref{2-eq-NS-com}) and secondly the conservation of momentum (equation \ref{2-eq-NS}).

\begin{equation}
\label{2-eq-NS-com}
\nabla \cdot \hat{U} = 0
\end{equation}

\begin{equation}
\label{2-eq-NS}
\rho \cdot \left( \partial_t \hat{U} + \hat{U} \cdot \nabla \hat{U} \right) = -\nabla p + \rho \hat{F} + \mu \nabla^2 \hat{U}   
\end{equation}

\begin{figure}[!htb]
\centering
\includegraphics[width=14cm]{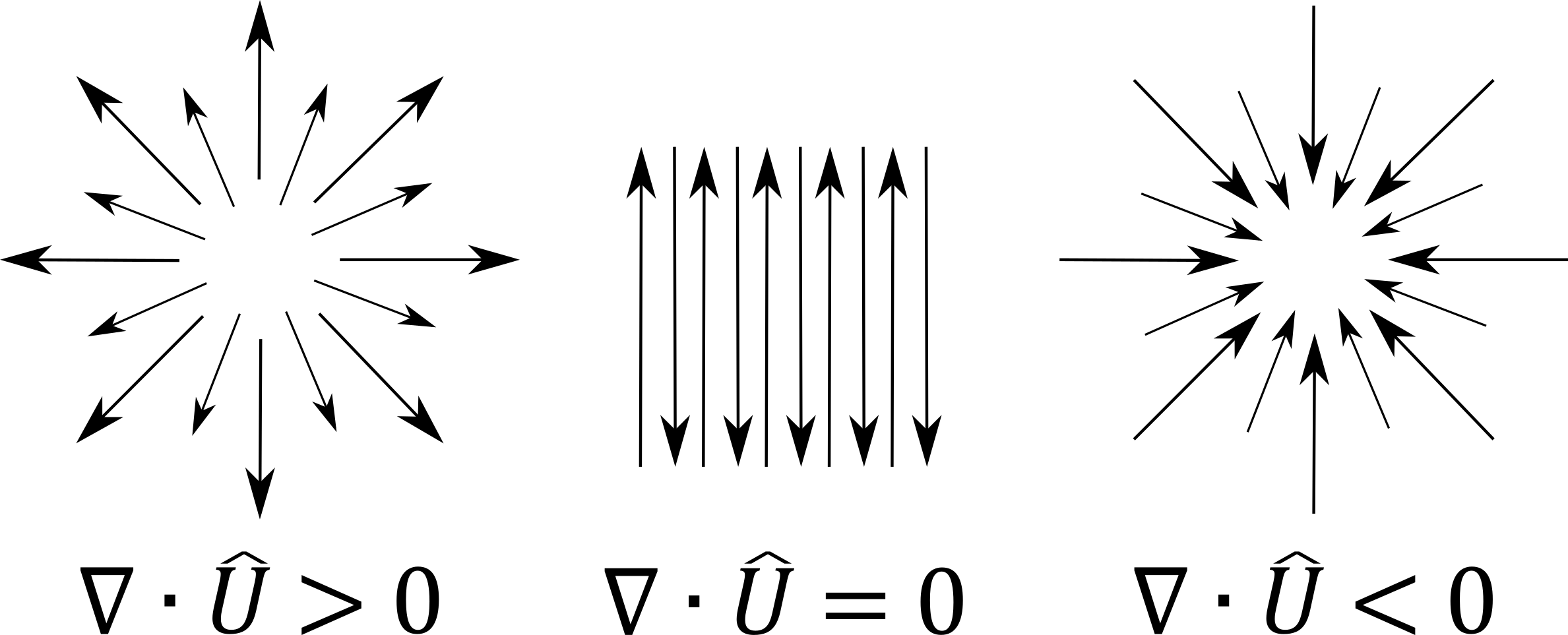}
\caption{Visualization of the divergence of various vector fields. Note only $\nabla \cdot \hat{U} = 0$ results in no `creation' or `destruction'.}
\label{2-fig-div}
\end{figure}

Here: $\rho$ is the density, $\hat{U}$ is the flow velocity, $t$ is time, $\hat{F}$ is a general external force term and $\mu$ is dynamic viscosity. To illustrate how equation \ref{2-eq-NS-com} ensures the conservation of mass the divergence of various vector fields is shown in Figure \ref{2-fig-div}. Here, for a physical substance when $\nabla \cdot \hat{U} \neq 0$ there are areas in which fluid will either need to be destroyed $\nabla \cdot \hat{U} < 0$ or created $\nabla \cdot \hat{U} > 0$. This is because in the centre of a inward flow ($\nabla \cdot \hat{U} < 0$) the amount fluid would have to decrease (else wise it would tend to infinity at the centre). Similarly in middle of an outflow situation ($\nabla \cdot \hat{U} > 0$) fluid would have to be created. Thus when $\nabla \cdot \hat{U} = 0$ conservation of mass is ensured as there is no increase or decrease. Also, note this is for an incompressible fluid. More rigorously the mass continuity equation $\partial_t \rho + \nabla \cdot (\rho \hat{U}) = 0$ is used. Expanding this out gives $\partial_t \rho + \nabla \rho \cdot \hat{U} + \rho (\nabla \cdot \hat{U}) = 0$. With incompressible fluids by definition $\partial_t \rho = 0$ and $\nabla \rho = 0$ thus $\nabla \cdot \hat{U} = 0$ is recovered.

Equation \ref{2-eq-NS} is essentially Newton's second law, $\hat{F} = m\hat{a}$ for $\hat{F}$ on mass $m$ with acceleration $\hat{a}$. However, it is written as $m\hat{a} = \hat{F}$ with the left hand side (LHS) equalling $\sum \text{mass} \times \text{acceleration}$ and right hand side (RHS) equalling $\sum \text{Force}$. To get from Newton's second law to equation \ref{2-eq-NS}, on the LHS the substitution $m = \rho \delta V$ is made where $\delta V$ is an infinitesimal cubic volume element. In three dimensions, the velocity of this fluid volume element $\delta V$ is a function of time and its position in time $\hat{U}(x(t), y(t), z(t), t)$ e.g. the motion of $\delta V$ and the bulk motion of the fluid. The intuition here is measuring the concentration of sediment, $S$, in a flowing river. An experimenter can stand at a fixed point in the river and measure the change in sentiment concentration with time, $S(t)$, or they could get in a raft and float down the river and measure $S(t)$ as they go. An experimenter on a raft has to agree about $S(t)$ with an experimenter at a fixed point when they intercept (in space and time) thus $S(t)$ is a function of $t$ and $\hat{U}(x(t), y(t), z(t))$. Instead of measuring the change in sediment concentration the experimenter could be measuring any property of a river such as its velocity. Therefore the velocity is also a function of time and position (which is itself a function of time!). Acceleration is the change in velocity over time - $\frac{d \hat{U}}{d t}$ - therefore acceleration can be written as

\begin{equation}
\label{2-eq-acceleration}
\begin{split}
a &= \frac{d}{d t} \left( \hat{U}(x(t), y(t), z(t), t) \right) \\
  &= \frac{\partial \hat{U}}{\partial x}\frac{\partial x}{\partial t} + \frac{\partial \hat{U}}{\partial y}\frac{\partial y}{\partial t} + \frac{\partial \hat{U}}{\partial z}\frac{\partial z}{\partial t} + \frac{\partial \hat{U}}{\partial t} \\
  &= \hat{U} \cdot \left(\frac{\partial \hat{U}}{\partial x} + \frac{\partial \hat{U}}{\partial y} + \frac{\partial \hat{U}}{\partial z}\right) + \frac{\partial \hat{U}}{\partial t} \\
  &= \hat{U} \cdot (\nabla \hat{U}) + \frac{\partial \hat{U}}{\partial t} \\
  &= \frac{\partial \hat{U}}{\partial t} + \hat{U} \cdot \nabla \hat{U}.
\end{split}
\end{equation}

Combining acceleration with the aforementioned density $\rho$ creates the LHS of equation \ref{2-eq-NS}. Interesting equation \ref{2-eq-acceleration} is actually the specific case of the velocity of the infinitesimal element $\delta V$ within the bulk motion of \textbf{$\hat{U}$} . It can be generalized to a fluid property $\phi$ in a bulk velocity $\hat{U}$ as the material derivative written as

\begin{equation}
\label{2-eq-material}
\frac{D \phi}{D t} =  \frac{\partial \phi}{\partial t} + \hat{U} \cdot \nabla \phi. 
\end{equation}

\begin{figure}[!htb]
\centering
\includegraphics[width=10cm]{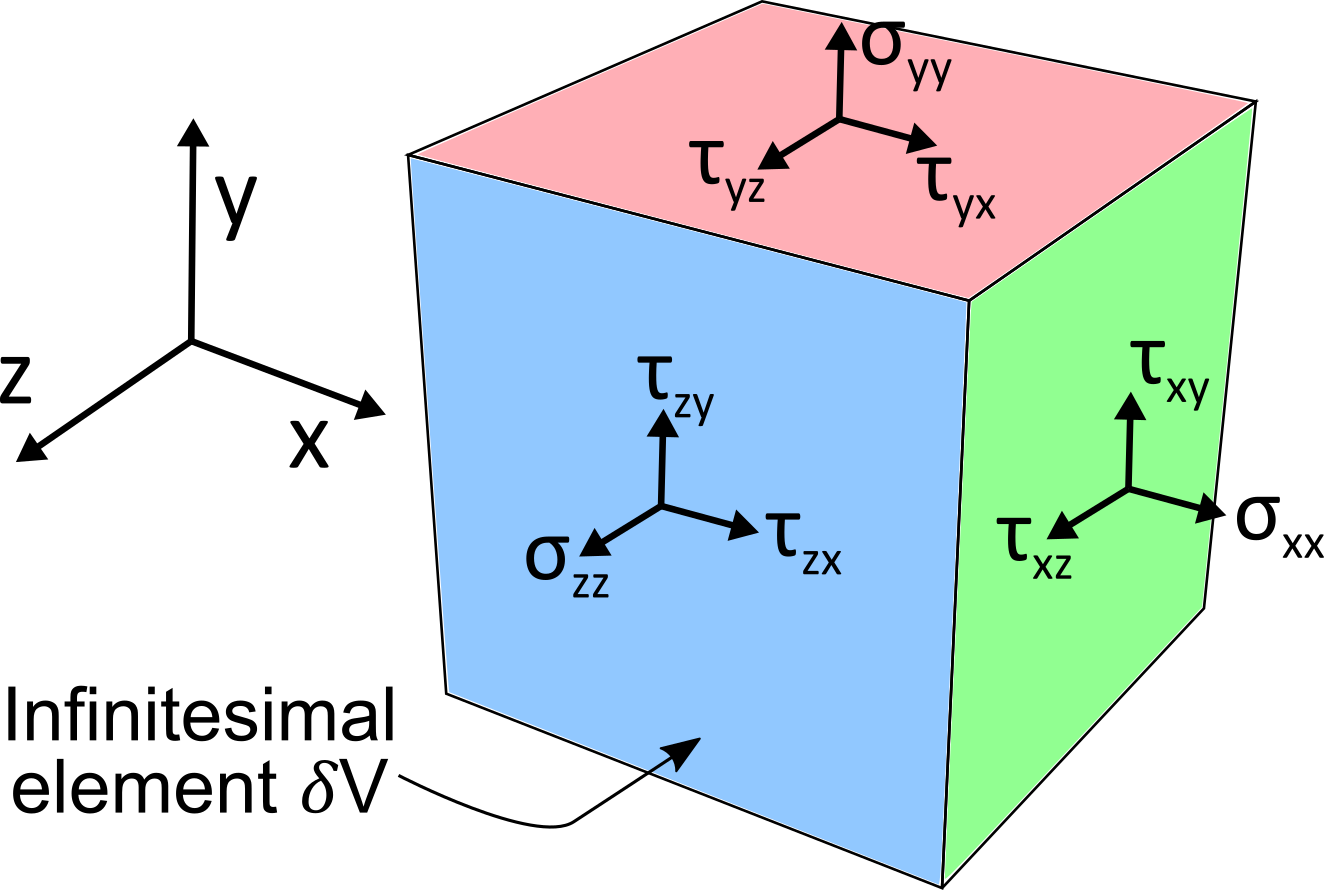}
\caption{Stress tensor elements on an infinitesimal element $\delta V$.}
\label{2-fig-stress}
\end{figure}

The RHS of equation \ref{2-eq-NS} is the sum of each contributing force acting on the element $\delta V$. This can be broken down into internal forces denoted as $\nabla \cdot \sigma$ where $\sigma$ is the stress tensor and external forces denoted as $\rho \hat{F}$ where $\hat{F}$ is a general external force. Within this chapter the only external force of interest is gravity thus $\rho \hat{F} = \rho \hat{g}$ where $\hat{g}$ is acceleration due to gravity. However for flexibility in equation \ref{2-eq-NS} the external force term is left as $\rho \hat{F}$. For the internal forces acting on the cubic element $\delta V$, as shown in Figure \ref{2-fig-stress}, it can be mechanically stressed in nine different directions. This can be used to form a tensor as

\begin{equation}
\label{2-eq-stress-tensor}
\sigma_{ij} = 
\begin{bmatrix}
\sigma_{xx} & \tau_{xy} & \tau_{xz}\\
\tau_{yx} & \sigma_{yy} & \tau_{yz}\\
\tau_{zx} & \tau_{zy} & \sigma_{zz}
\end{bmatrix}
\end{equation}

with normal stresses $\sigma$ when $i = j$ and shear stress $\tau$ when $i \neq j$. The type of flow will determine the values for $\sigma_{ij}$. For the simple case of inviscid flow (where viscosity $\mu = 0$) all the viscosity contributions to $\tau_{ij} = 0$. Therefore, the remaining elements with normal stresses in $\sigma_{ij}$ are equal to the force $\hat{F}$ on each face (of $\delta V$) over the area $\delta \hat{A}$ of each face - by definition, this is the pressure $p$: 

\begin{equation}
\sigma = \frac{\hat{F}}{\delta \hat{A}} = p.    
\end{equation}

\begin{figure}[!htb]
\centering
\includegraphics[width=10cm]{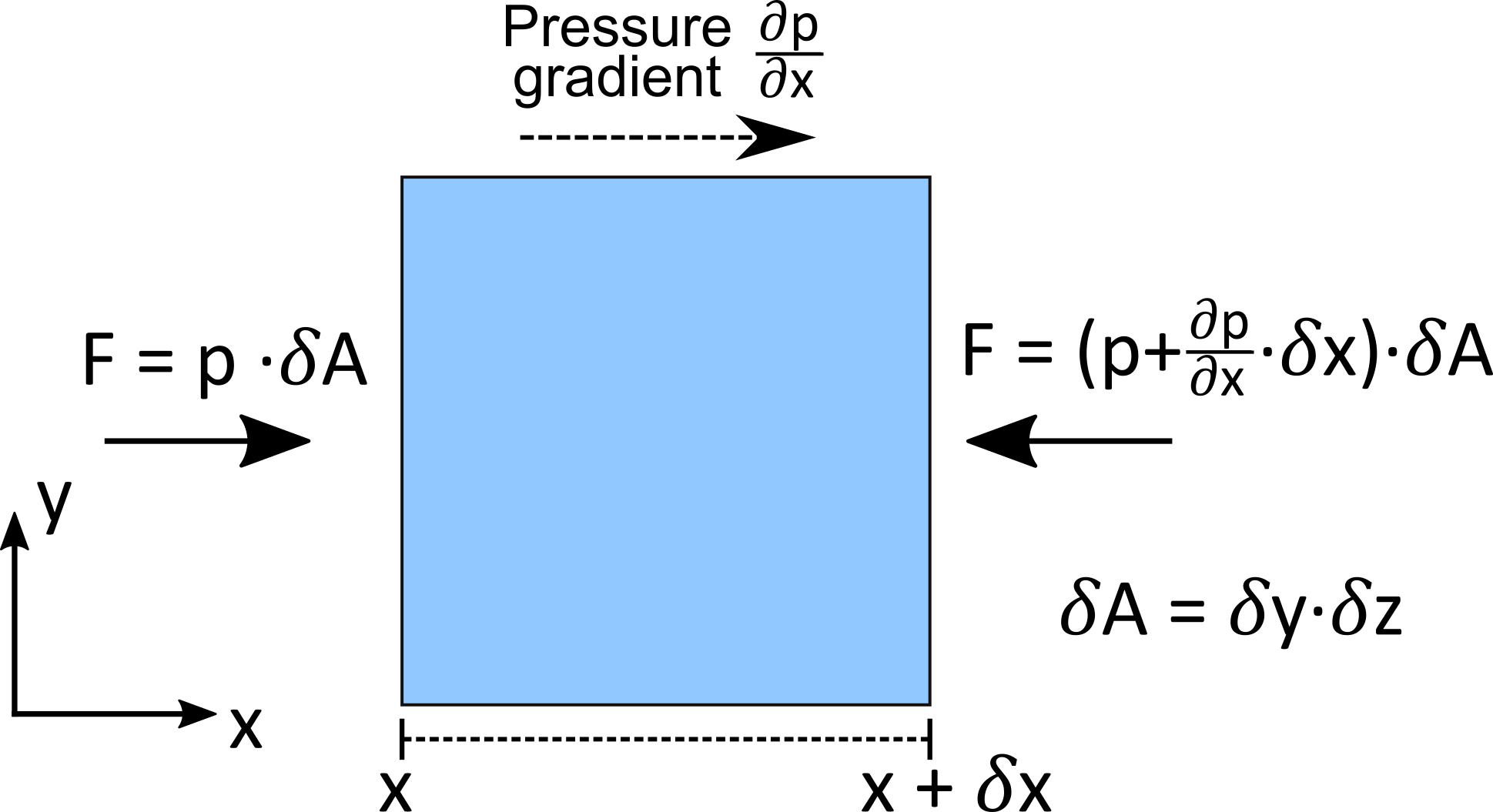}
\caption{Pressures on opposite faces of the infinitesimal element $\delta V$ (where each face has an area $\delta A$) shown in Figure \ref{2-fig-stress} for a pressure gradient increasing with the positive x direction.}
\label{2-fig-pressure}
\end{figure}

As shown in Figure \ref{2-fig-pressure}, for a pressure gradient $\frac{\partial p}{\partial x}$ in the $x$ direction ($\hat{i}$) the faces of the infinitesimal element $\delta V$ will have different pressures. The pressure over $\delta V$ in Figure \ref{2-fig-pressure} can thus be expressed as $(p|_{x} - p|_{x + \delta x}) \cdot \delta y \cdot \delta z = \frac{\partial p}{\partial x} \cdot \delta V$. Through extending this for pressure gradients in the $y$ ($\hat{j}$) and $z$ ($\hat{k}$) directions the pressure over $\delta V$ is expressed as

\begin{equation}
\left(\frac{\partial p}{\partial x}\hat{i} + \frac{\partial p}{\partial y}\hat{j} + \frac{\partial p}{\partial z}\hat{k}\right) \cdot \delta V = -\nabla p \cdot \delta V
\end{equation}

which gives the $-\nabla p$ term Navier-Stokes in equation \ref{2-eq-NS}. This can be written as $\nabla \cdot \sigma = \nabla \cdot (-p)$ however as the off diagonal terms are all equal to zero to avoid ambiguity it is written as $-\nabla p$. Given the pressure term just derived is useful for inviscid flow (and fluid at rest) it can be separated out from equation \ref{2-eq-stress-tensor} creating the deviatoric stress tensor shown in equation \ref{2-eq-deviatoric-tensor}. 

\begin{equation}
\label{2-eq-deviatoric-tensor}
\tau_{ij} = 
\begin{bmatrix}
\sigma_{xx} + p & \tau_{xy} & \tau_{xz}\\
\tau_{yx} & \sigma_{yy} + p & \tau_{yz}\\
\tau_{zx} & \tau_{zy} & \sigma_{zz} + p
\end{bmatrix} =
\begin{bmatrix}
\tau_{xx} & \tau_{xy} & \tau_{xz}\\
\tau_{yx} & \tau_{yy} & \tau_{yz}\\
\tau_{zx} & \tau_{zy} & \tau_{zz}
\end{bmatrix}
\end{equation}

\begin{figure}[!htb]
\centering
\includegraphics[width=10cm]{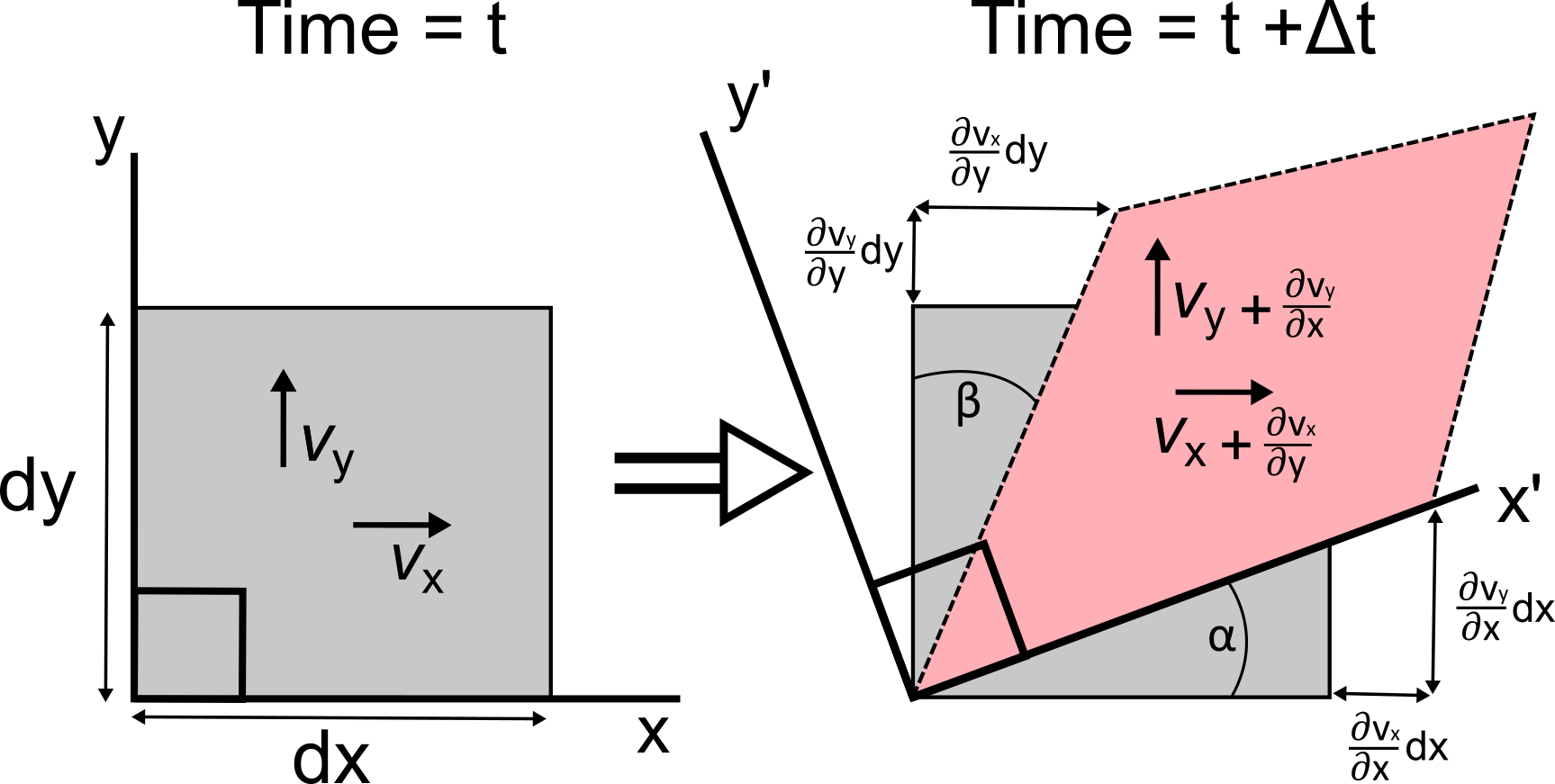}
\caption{Induced (\emph{engineering}) strain $\gamma_{xy}$ characterized by the sum of the angles $\alpha$ and $\beta$ on the surface of a fluid element $\delta V$ due to shear stresses. Note the primed axis are included to represent a situation of $\alpha - \beta$ e.g. rotation.}
\label{2-fig-strain}
\end{figure}

The deviatoric stress tensor can be combined with Stokes' stress constitutive equation $\tau = 2 \mu \epsilon_{ij}$ where $\mu$ is the dynamic viscosity and $\epsilon_{ij}$ the rate of strain tensor. For the viscosity in this thesis, only incompressible Newtonian fluids are considered. These are fluids that obey Newton's law of viscosity given by

\begin{equation}
\label{2-eq-newton-visc}
\tau = \mu \frac{\partial \hat{v}}{\partial \hat{y}} 
\end{equation}

for a fluid with dynamic viscosity $\mu$ and velocity gradient $\frac{\partial \hat{v}}{\partial \hat{y}}$ in direction parallel to the shear and displacement perpendicular to it Navier-Stokes. This is sometimes written with a negative sign which comes from the definition of viscosity whereby it opposes an force generated by the velocity gradient. The rate of strain tensor can be found through inspecting an infinitesimal strain element undergoing an applied shear stress from time $t$ to $t + \Delta t$. Here, as shown in Figure \ref{2-fig-strain}, the stain induced from this stress can be characterized through angles $\alpha$ and $\beta$ which when small  
\begin{align}
tan(\alpha) \approx \alpha = \frac{\partial \hat{v_y}}{\partial \hat{x}} && tan(\beta) \approx \beta = \frac{\partial \hat{v_x}}{\partial \hat{y}}.
\end{align}

The induced shear strain for Figure \ref{2-fig-strain} is $\gamma_{xy} = \alpha + \beta$. Here $\gamma_{xy}$ is technically the \emph{engineering} shear strain which for historical reasons is define as double the \emph{tensorial} shear strain $\epsilon$ that features in  Stokes' stress constitutive equation. For brevity the \emph{tensorial} shear strain will be refereed to simply as the shear strain or just the strain. The shear strain is thus $\epsilon_{yx} = \frac{1}{2} \left(\frac{\partial \hat{v_y}}{\partial \hat{x}} + \frac{\partial \hat{v_x}}{\partial \hat{y}}\right)$. Through characterizing the change from time $t$ to $t + \Delta t$ for all combinations of $x$, $y$ and $z$ in Figure \ref{2-fig-stress} the strain rate tensor $\epsilon_{ij}$ in equation \ref{2-eq-strain-rate-tensor} is created. Here, for matching indices the aforementioned $\frac{1}{2}$ factor cancels with the $2$ generated from identical differentials. The deviatoric stress tensor can thus be recovered using the above-mentioned Stokes' stress constitutive equation $\tau = 2 \mu \epsilon$. 

\begin{equation}
\label{2-eq-strain-rate-tensor}
\begin{split}
\epsilon_{ij} &=  
\begin{bmatrix}
\frac{\partial \hat{v_x}}{\partial \hat{x}} & \frac{1}{2} \left(\frac{\partial \hat{v_y}}{\partial \hat{x}} + \frac{\partial \hat{v_x}}{\partial \hat{y}} \right) & \frac{1}{2} \left(\frac{\partial \hat{v_z}}{\partial \hat{x}} + \frac{\partial \hat{v_x}}{\partial \hat{z}} \right)\\
\frac{1}{2} \left(\frac{\partial \hat{v_x}}{\partial \hat{y}} + \frac{\partial \hat{v_y}}{\partial \hat{x}} \right) & \frac{\partial \hat{v_y}}{\partial \hat{y}} & \frac{1}{2} \left(\frac{\partial \hat{v_z}}{\partial \hat{y}} + \frac{\partial \hat{v_y}}{\partial \hat{z}} \right)\\
\frac{1}{2} \left(\frac{\partial \hat{v_x}}{\partial \hat{z}} + \frac{\partial \hat{v_z}}{\partial \hat{x}} \right) & \frac{1}{2} \left(\frac{\partial \hat{v_y}}{\partial \hat{z}} + \frac{\partial \hat{v_z}}{\partial \hat{y}} \right) & \frac{\partial \hat{v_z}}{\partial \hat{z}}
\end{bmatrix} \\
&= \frac{1}{2}\left(\nabla \hat{U} + (\nabla \hat{U})^T\right) \\
\tau_{ij} &= 2 \mu \epsilon_{ij} \\
 &= \mu \left(\nabla \hat{U} + (\nabla \hat{U})^T\right)
\end{split}
\end{equation}

% Maybe use diffeent \hat{U} for bulk and dV?
Finally, using the same argument as for pressure (shown in Figure \ref{2-fig-pressure}), due to the incompressible flow assumption the divergence of the final result from equation \ref{2-eq-strain-rate-tensor} can be written as $\nabla \cdot \tau_{ij} = \mu \nabla \cdot \left(\nabla \hat{U} + (\nabla \hat{U})^T\right) = \mu \nabla^2 \hat{U}$. The final term of which is combined with $- \nabla p$ to create $\nabla \cdot \sigma = - \nabla p + \mu \nabla^2 \hat{U}$ which finishes the RHS of equation \ref{2-eq-NS}. This completes the Navier-Stokes equations. So far an infinitesimal element $\delta V$ has been used. Yet to solve the Navier-Stokes equations computationally requires a finite number of operations. Therefore the infinitesimal element $\delta V$ must be converted into a finite element with an actual volume.  
\subsection{Mesh}
\label{2-sec-mesh}
\subsubsection{Overview}
Meshing is the process whereby a domain is split into small cells. The principal purpose of meshing in CFD is to enable the discretization of the relevant equations (e.g. the Navier-Stokes) for the CFD case. Through discretization, the equations in the case can be represented as a system of linear equations that can be solved using numerical methods such as the ones covered in Section \ref{2-sec-numerical-methods}. The cells created through meshing feature a cell centre and cell faces. The cell centre defines a discrete position for an average value of a field whilst the cell faces enable identification of neighbours. Neighbours can be either boundaries that specify a discrete condition for the discretization or other cells that contain their own values for the aforementioned system of linear equations.     

\begin{figure}[!htb]
\centering
\includegraphics[width=14cm]{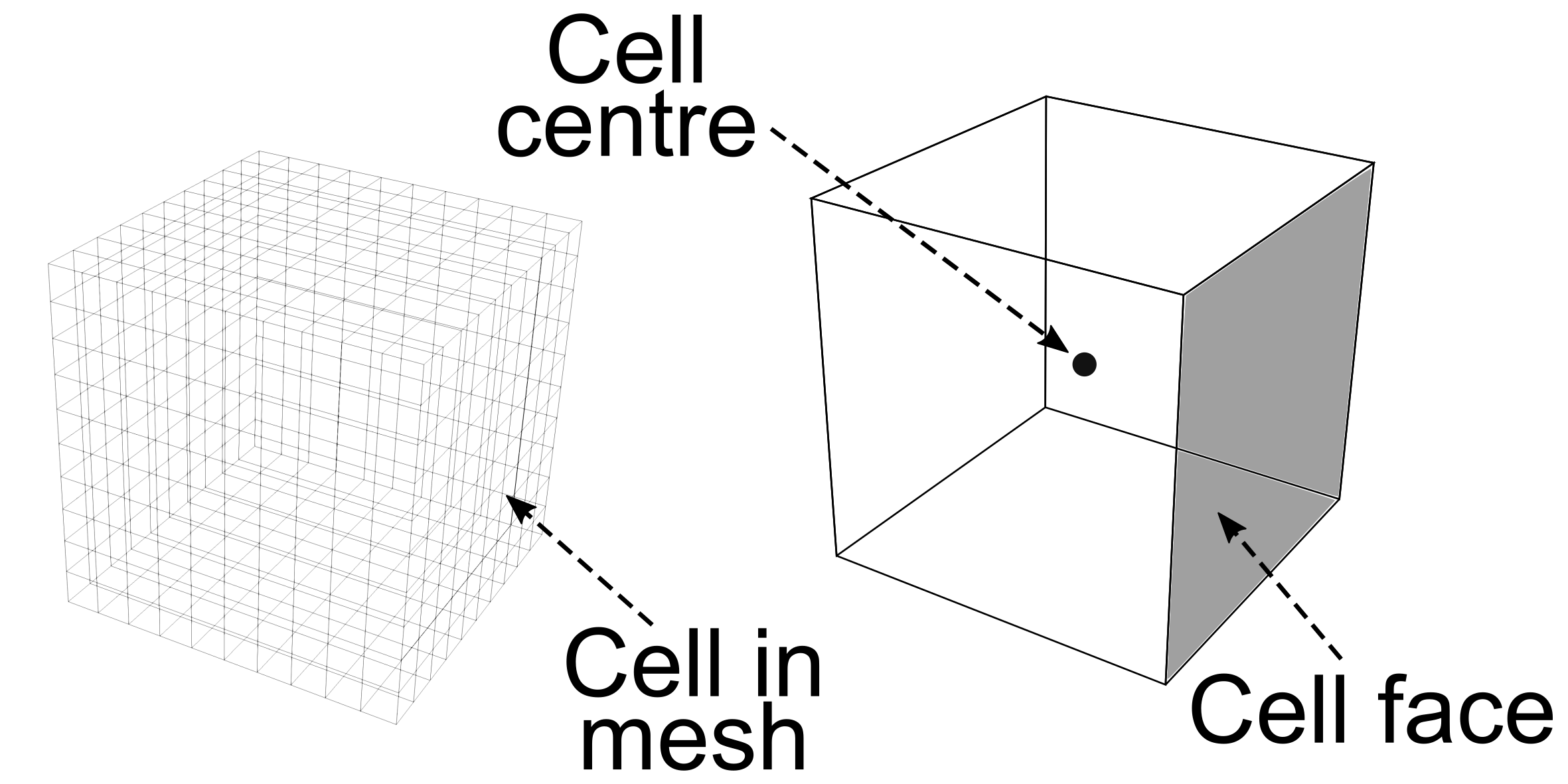}
\caption{Visualization of a $10 \times 10$ cell cubic mesh (left) next to an individual cubic mesh cell with the cell centre and cell face highlighted (right). Note cubic cells are used purely for illustration purposes mesh cells can - and typically are - complex polyhedrons. Further, meshes in \of can have curved edges between the points providing that each cell face only matches face to face with one and only one other cell face.}
\label{2-fig-mesh}
\end{figure}

The actual mathematical process of meshing is out of scope for this thesis however an example of meshing is shown in Figure \ref{2-fig-mesh}. Here are cubic domain is split into many smaller cubic cells each featuring a cell centre and six cell faces. Cubic cells are shown for simplicity but many polyhedrons can be used with these same general principals. Although equilateral cells are preferred, in principal highly complex shapes could be used for mesh cells.   

%\subsubsection{Finite Volume Method}
%\label{2-sec-fvm}

\subsubsection{Mesh sensitivity}
In order to capture the Physics of a CFD case accurately, the mesh cells (and thus control volumes) that it is calculated upon need to be sufficiently small so as to avoid coarsely averaging away details. In general, the smaller the average mesh cell (e.g. the finer the mesh) used in a CFD case the more computationally intensive it will be to calculate fields on it. This is due simply to there being more mesh cells in a finer mesh. There is thus a trade off between accuracy of the solution and time taken to achieve it. Also, when taken to an extreme an excessively fine mesh can cause some issues with convergence but in this thesis this is never a problem.

In general the results of most CFD cases will improve with a finer mesh up to a point whereby there is diminishing additional gain in accuracy. The user must make a judgement what mesh resolution to use - this assessment is called a mesh sensitivity analysis. For this thesis, it was found that the required `fine enough' mesh resolution was a few thousands of cells for 2D cases and a few tens of thousands for 3D cases with the main issue being increasing computation time. Fine meshes were used especially in 2D cases where the total number of cells is low but the mesh resolutions were not artificially inflated for no reason. Secondly, in 3D for cases where there are $>\num{5e5}$ cells, computation times can get quite ($\approx$ hours) long for long simulated times and the field objects the program writes can get quite ($\approx$ GB) big. There needs to be a good reason to use a fine mesh of which there rarely was in this thesis as the cases could often be reformulated. Given this, all the meshes used in this thesis are set to be comfortably finer than the `fine enough' floor and formal mesh sensitivity analyses are omitted.     

%\subsubsection{Dynamic Meshing}
%The CFD program used in this thesis - \of - includes a dynamic meshing feature. Here, the mesh can be updated during the simulation. Whilst dynamic meshing can be used to ... 

\subsection{Finite Volume Method}
\subsubsection{Overview}
\label{2-sec-numerical-methods}
With meshing and the Navier-Stokes equations established the next step is to solve them on a mesh. In this thesis, the technique used to do this is the finite volume method (FVM) \cite{jasakThesis}. Here, the domain is discretized into a series of finite control volumes which for this thesis\footnote{Control volumes are technically a mathematical abstraction but their implementation in this thesis is done via meshing.} are equivalent to the mesh cells covered in Section \ref{2-sec-mesh}. In the case of Navier-Stokes, the evolution of the values for momentum of a fluid in these control volumes would then be solved for using a series of mathematical techniques. However, the description of the FVM is typically abstracted in terms of a general fluid property $\phi$ per unit mass. Here, within the control volume, $\phi$ will evolve as a combination of time (temporal change), transport of $\phi$ through the control volume faces (via convection or diffusion), and the creation / destruction of $\phi$ inside the control volume (source / sink). Mathematically, this is expressed as: 

\begin{equation}
\label{2-eq-CV-words}
\parbox{2cm}{\centering temporal change} \ + \ \sum_{\text{Through faces}} \left(\parbox{2.5cm}{\centering transport by convection} \ + \ \parbox{2.5cm}{\centering transport by diffusion}\right) \ = \ \parbox{2cm}{\centering Source / Sink}.
\end{equation}

Given $\phi$ is per unit mass, its temporal evolution will be the rate of change of $\phi \times \text{mass} \equiv \partial_t (\rho \phi)$. Further, from the aforementioned material derivative, the mass flux of $\phi$ due to transport by convection with velocity $\hat{U}$ can be introduced as $\nabla \cdot (\rho  \textbf{U} \phi)$. Here, \textbf{U} is the cell face flux - accounting for the summation term in equation \ref{2-eq-CV-words}. From Fick's second law of diffusion the term $\nabla \cdot \Gamma_{\phi} \nabla \phi$ is (where $\Gamma_{\phi}$ is the diffusion constant $\equiv D$) introduced which as $\phi$ is per unit mass is also multiplied by $\rho$. Finally terms that cannot be written as temporal, convection or diffusion terms are implemented as sources or sinks which are combined into a general term $S_{\phi} (\phi)$. Combining all these terms yields the scalar transport equation (sometimes also called the convection-diffusion or advection-diffusion equation) for a general fluid property of $\phi$ per unit volume as

\begin{equation}
\label{2-eq-scalar-transport}
\partial_t (\rho \phi) + \nabla \cdot (\rho  \textbf{U}   \phi) + \nabla \cdot (\rho \Gamma_{\phi} \nabla \phi) = S_{\phi} (\phi).
\end{equation}

The scalar transport equation can then be applied to specific properties. For instance, the evolution of enthalpy $H$ could be written using the substitutions $\Gamma = \alpha_{D}$ creating the equation. Here, thermal diffusivity $\alpha_{D} = \frac{k}{\rho c_p}$ where $k$ is the thermal conductivity and $c_p$ the specific heat capacity and enthalpy $H = c_p T$ where $T$ is temperature creating the diffusion term in equation \ref{2-eq-enthalpy-transport}. 

\begin{equation}
\label{2-eq-enthalpy-transport}
\rho \frac{\partial H}{\partial t} + \nabla \cdot (\boldsymbol{\rho U} H) = \nabla \cdot (k\nabla T).
\end{equation}

Here, the \of convention of incorporating $\rho$ into the cell face flux is used to create $\boldsymbol{\rho U}$, the cell face mass flux. To recover equation \ref{2-eq-NS}, the substitutions $\phi = \hat{U}$, $\Gamma_{\phi} = \frac{\mu}{\rho}$ and $S_{\phi}(\phi) = (\rho \hat{F} -\nabla p)$ are made. This shows the flexibility of the scalar transport equation. The next step in the FVM is to systematically solve it through numerical schemes and algorithms.
% \footnote{This simple recovery is the case for the incompressible Newtonian fluid form of the Navier-Stokes equations - in other forms this it more complex.}

\subsubsection{Numerical schemes}
\label{2-sec-numerical-schemes}
In this thesis, numerical schemes are the methods employed by \of to solve equation \ref{2-eq-CV-words}. In equation \ref{2-eq-CV-words}, the transport by convection and diffusion is evaluated at cell faces rather than cell centres. However, \of stores all values at cell centres rather than cell faces. Thus in order to solve forms of equation \ref{2-eq-CV-words} the cell centre values need to be interpolated onto the cell faces; these methods are termed interpolation schemes. The clear starting point is to simply linearly interpolate between adjacent cell faces - termed the central differencing scheme. This scheme can potentially lead to unbounded solutions where the calculated values oscillate however in \of this linear interpolation is used in almost every case. If required, an alternative is to simply take the cell centre value of the current cell in the positive axial direction. This means that the right face of the cell will have the cell centre value of the cell and the left face will have the cell centre value of the cell to the left of the current cell. This method of interpolation is termed `upwind' and thus this is the upwind differencing scheme. The advantage of this scheme is that unlike the central differencing scheme the calculated solution will definitely be bounded. There a even more sophisticated schemes such as the van Leer scheme \cite{vanLeerScheme}. In this complex scheme, a series of values are created to step between adjacent cell centres that limit the slope between them and ensure boundedness. This is useful for a scalar field that should always be bounded such as a volume of fluid phase fraction field that needs to remain between 0 and 1. 

With the cell centre and cell face values established, additional numerical schemes are required for integrating over the time steps, finding the gradients, calculating the divergences and calculating the laplacians if required. The specific schemes occasionally vary dependent on the case but in this thesis the Euler implicit time scheme\footnote{For clarity this is simply $\frac{\partial \phi}{\partial t} = \frac{\phi - \phi_{0}}{\Delta t}$.} is used for time, and `Gauss linear' method is used for the rest. Here `Gauss' refers to the standard Gaussian integration over the cell faces \cite{jasakThesis} and `linear' the aforementioned linear interpolation method. However, for divergence `Gauss linearUpwind' is used  for $\nabla \cdot \rho  \textbf{U}   \phi$ and `Gauss vanleer' for the divergence terms in the $\alpha$ equation. These schemes use the above mentioned upwind and van Leer interpolations. A full list and description of the numerical schemes available in \of can be found in `Numerical Schemes' portion of the \of 6 user guide \cite{of6UG}.
 
\subsubsection{Solution and algorithms}
\numParagraph{Overview}
Once the equations have been discretized, a series a algorithms are applied to iteratively solve them. The discretized equations are essentially a system of algebraic linear equations of form $\textbf{A} \phi = \textbf{b}$ thus they can be solved for using guess and correct methods. As the word `solver' is already used in the \of context to refer to the entire executable programs that solve all the equations on a domain the specific steps that solve these individual systems of linear equations are commonly referred to as `linear solvers'.  

\of features numerous inbuilt linear solvers which like all other elements of \of can be modified by the user to suit their needs. In this thesis, the inbuilt linear solvers used are: `Geometric Agglomerated Algebraic multiGrid (GAMG)' solver, the `Preconditioned Conjugate Gradient (PCG)' solver, the `Preconditioned Bi-conjugate Gradient (PBiCG)' solver and the `Smooth Solver'. Additionally, to improve convergence, many solvers include preconditioning. Here, preconditioning refers to applying a preconditioner $P$ to a linear equation such that $P^{-1}\textbf{A} \phi = P^{-1}\textbf{b}$ before applying the main linear solver. The purpose of this step is to improve convergence. In this thesis the preconditioners used are: the GAMG preconditioner, the Diagonal-based Incomplete Cholesky (DIC) preconditioner, and the Diagonal-based Incomplete LU (DILU) preconditioner. Further to this, the solvers can include `smoothing' a process where sparse linear solution methods such as Gauss Sidel are used to `smooth out' the error in some steps. It is variations of the Gauss Sidel method that are used as smoothers in this thesis. 

The individual steps used in the above mentioned linear solvers can be found through either inspecting the source code or from \cite{saad2003iterative} and \cite{of6UG}. There is no optimum choice out of these solvers and the choices used in this thesis vary depending on the case. For this thesis the exact choices per case are in the `fvSchemes.H' files in the case directory on GitHub. This is due to the specificness to \of and the length of the files. Finally, in addition to these linear solvers there are two algorithms of note employed: PIMPLE for the pressure-velocity coupling and, in the case of volume of fluid, MULES which shall be described in the relevant sections.

\numParagraph{PIMPLE}
PIMPLE is a very common CFD algorithm that employs both the Semi Implicit Method for Pressure-Linked Equations (SIMPLE) algorithm and the Pressure Implicit with Splitting Operators (PISO) algorithm. Inspecting equation \ref{2-eq-NS} reveals that there are four unknowns: the three velocity components $\hat{U_x}$, $\hat{U_y}$, and $\hat{U_z}$, and the pressure. To solve this, the conservation of mass equation (equation \ref{2-eq-NS-com}) can be applied to both sides of equation \ref{2-eq-NS} leaving a Poisson equation for pressure. These two algorithms work through solving for $\hat{U_x}$, $\hat{U_y}$, and $\hat{U_z}$ sequentially to create a `guess' for the velocity field. This guess can then be used to calculate the mass flux at the cell faces. Next the pressure equation is solved with this guess and the mass fluxes and velocities are then corrected with the new pressure field followed by an boundary field update. An extensive description of the PIMPLE, PISO and SIMPLE algorithms are detailed in \cite{holzmann2016mathematics}.

\subsection{OpenFOAM}
\subsubsection{Overview}
%\subsubsection{Why OpenFOAM?} - Put in Introduction
\of - \underline{OP}en source \underline{F}ield \underline{O}peration and \underline{M}anipulation - is an open source CFD program written in C++. Due to a development fork, there are two main versions of \of available: \of ESI and \of Foundation. The ESI versions of \of are released biannually whereby the December 2020 version is v2012 and the June 2021 version is v2106. Whereas, the \of Foundation versions are released every year with a simple number so the 2021 version is \of 9. This is addition to other project forks such as foam-extend. Therefore, there are many programs that are plausibly referred to as \of that have different features. In this thesis, the \of version used is \of 6 released in July 2018\footnote{For the reader interested in writing solvers for a project in \of it is the recommendation of the author to pick the latest full \of release and stick with it.}. Given most of the dependencies in the full model presented in this thesis have been re-written with new names from the inbuilt versions it should work with most other \of versions. However, this has not been extensively tested. \of is a very powerful program with lots of applications. To fully understand the capabilities of \of 6 see the user guide \cite{of6UG} and to see the up to date features in \of see both https://www.openfoam.com/ for the ESI version and https://www.openfoam.org/ for the foundation version.       

The following sections detail specific information about \of relevant to its use in this thesis. There are a large amount of classes and modules available in \of with a huge amount of functionality. Given that detailed documentation and source code repositories are available online, within this thesis only a small subset of classes relevant to understanding the Physics are included. Further, modules are described on an ad hoc basis e.g. the boundary conditions module isn't detailed but a description of a specific boundary condition would be. Finally, descriptions of computational details such as the object registry are largely neglected. 

\subsubsection{Fields and properties in OpenFOAM}
The main unit of field description in \of are instances of the \geoF class. Amongst other members, the \geoF class contains information about the internal field, the boundary field, and the dimensions of an \of field. The internal field essentially contains a list of values with each value corresponding to a mesh element such as the cell centre of a mesh cell. The boundary field contains information about the cell faces on the boundary of the domain; for instance cells with a fixed value boundary face will have values specified for those boundaries. Dimensions in OpenFOAM serve as a convenient a method to check the physical validity of calculations preventing operations such as $\SI{1}{\kilogram} + \SI{1}{\meter}$. The four main \emph{geometricFields} used in this thesis are: 

\begin{itemize}
    \item \vsf which are the simplest \geoF to understand. Their internal field essentially contains a list of scalar values with each value corresponding to the value with a mesh cell centre. For example, a temperature field may have a value of \SI{300}{\kelvin} in the centre of cell number 10, a value of \SI{300.5}{\kelvin} in the centre of cell number 11 etc. In general, their boundary fields are specified at run time - although in some cases they are specified in the class constructor. 

    \item \vvf work in a similar way to \vsf except their internal field contains vectors for the centre of cells. For example a velocity field may have a value of (0 0 1) \si{\meter\per\second} in cell centre number 5, a value of (0 0 2) \si{\meter\per\second} in cell centre number 6 etc.  

    \item \ssf is a cell face value for a scalar field. Surface fields are all defined on cell faces. The most common use of them in this thesis is the (mass) flow through the cell faces (rho) phi with symbol(s) ($\rho$)$\phi$. 

    \item \svf are another type of surface field whose vectors use cell face vectors. A typical examples is the face area vector is $S_f$ which is the area of the face combined with the face normal vector.  

\end{itemize}

Dimensions in OpenFOAM are expressed as a \emph{dimensionSet} as [\# \# \# \# \# \# \#] which is a list of scalars where a scalar's position correspond to the properties in Table \ref{2-tab-dimensions}. The first scalar corresponds to property number 1 - mass, the second property 2 - length etc. The value of the scalar corresponds to its power in the units so \SI{1}{\meter} would be 1 whereas \SI{1}{\meter\cubed} would be 3. The same holds for negative powers so \SI{1}{\per\kilogram} would be -1. As an example $\SI{1}{\newton} = \SI{1}{\kilogram\meter\per\second\squared}$ would be written as [1 1 -2 0 0 0 0]. By default operations in \of undergo dimension checking whereby the dimensions of an operation are checked to be equal before the operation is performed.   

\begin{table}[ht]
\setlength{\tabcolsep}{10pt}
    \centering
    \caption{\label{2-tab-dimensions} Dimensions in OpenFOAM ordered by their position in the [\# \# \# \# \# \# \#] \emph{dimensionSet}.}
    \begin{tabular}{l l l}
    \toprule
    Position \# & Property & SI Unit \\
    \hline
    1\rule{0pt}{2.6ex} & Mass & kilogram \si{\kilogram} \\
    2 & Length & metre \si{\meter} \\
    3 & Time & second \si{\second} \\
    4 & Temperature & Kelvin \si{\kelvin} \\
    5 & Quantity & mole \si{\mole} \\
    6 & Current & ampere \si{\ampere} \\
    7 & Luminosity & candela \si{\candela} \\    
    \bottomrule
    \end{tabular}
\end{table}

In addition to \geoF there are also \emph{dimensionedTypes}. These are typically used to define or modify a \emph{geometricField}. For example, in the aforementioned heat equation $\partial_t T = \alpha_D \nabla^2 T$ whilst the temperature $T$ is a \geoF the thermal diffusivity $\alpha_D$ is a \emph{dimensionedType}. The two relevant \emph{dimensionedTypes} in this thesis are:

\begin{itemize}
    \item \ds contain simply a scalar value and an associated dimensions. For instance \SI{1}{\meter} can be expressed as a \ds of value 1 and dimensions [1 0 0 0 0 0 0].
    \item \dv contain a vector and associated dimensions. A typical usage is defining acceleration due to gravity where -9.81 \si{\meter\per\second\squared} can be expressed as vector (0 0 -9.81) and dimensions [0 1 -2 0 0 0 0].
\end{itemize}

In this thesis, \emph{geometricFields} are referred to as fields and are what are numerically solved for. Whereas \emph{dimensionedTypes} are termed properties which in general modify or define fields. For instance a fluid may have a velocity and a pressure field of which the numerical solutions for depend on its viscosity and density properties. There is some flexibility with the usage of these terms whereby some properties are turned into fields such as when the density of a phase is made temperature dependent. However, the distinction is still drawn to aid in the descriptions of the solver calculations. 

\subsubsection{Equations in OpenFOAM}
One of the key features of \of is its ability to represent the equations within a solver's code in a human readable fashion. For instance, the solver \emph{laplacianFoam} that solves the diffusion equation for temperature (e.g. thermal diffusion in a solid)

\begin{equation}
\label{2-eq-laplacianFoam}    
\partial_t T = \alpha_{D} \nabla^2 T
\end{equation}

is written in \of as:

%[caption = {Implmentation of equation}, label = {3-code-T-eqn}]
\lstset{frame = none}
\begin{lstlisting}  
    fvScalarMatrix TEqn
    (
        fvm::ddt(T)
        ==
        fvm::laplacian(alpha_diffusion, T)
    );
\end{lstlisting}

where thermal diffusivity $\alpha_{D}$ is `alpha\_diffusion'. In this situation T is a \vsf which the \emph{fvScalarMatrix} solves for using the specified numerical method. Conversely, alpha\_diffusion is a \ds that is a constant in the equation. Adding a source term to this equation is as simple as inserting a `+ Source' on the line after the laplacian (assuming `Source' is already defined).    
The `laplacian' and `ddt' terms in \emph{laplacianFoam} are preceded by `fvm' - followed by the C++ scope resolution operator `::' - which stands for `finite volume method'. This is the \of implementation of the finite volume method covered in Section \ref{2-sec-numerical-methods}, for a more in depth description see one of the original \of authors thesis on the topic \cite{jasakThesis}. 

\subsubsection{Solvers in OpenFOAM}
Solvers in \of are executable files that solve one or more equations relevant to the specified case. The solver source directories can typically be split into a \emph{solverName.C} file containing (amongst other dependencies) a \emph{createFields.H} header file that defines all of the fields and properties, and one or more \emph{fieldEqn.H} header files that handle the equation for a specific field. For instance, the source directory for \emph{simpleFoam} contains the main \emph{simpleFoam.C} file, a \emph{createFields.H} file that defines various fields, and \emph{pEqn.H} and \emph{UEqn.H} files that define the pressure and velocity equations that \emph{simpleFoam} solves. It is important to note that the role of these three files types is often conflated; some \emph{solverName.C} contain equations and \emph{fieldEqn.H} header files often contain definitions of fields and properties. Also, it is frequently convenient to slice off various field and property definitions from \emph{createFields.H} into other header files. A common case of this is when adding a feature a user may elect to create a \emph{featureFields.H} file such as \emph{temperatureFields.H} or \emph{electromagFields.H}.   

Beyond these three core elements solvers contain calls to versions of specific \of header files required to create the solver. For example the \emph{fvCFD.H} header file is typically included in the first few lines as it features calls to a basket of header files such as \emph{fvMesh.H} for the mesh and \emph{Time.H} for the time. Importantly, some of these header files include constructors for classes required for the solver. A case in point is the \emph{twoPhaseMixture} transport model that is necessary for the \IFM solver. One of the unique elements of the novel solver presented in Chapter 3 is that as many of the inbuilt classes in \of are insufficient to simulate ultra-thin-walled tube welding custom ones have been created.  

Finally it should be noted that the solver directories all contain instructional files for the \of \emph{wmake} utility that compiles the solver into an executable. Additionally many inbuilt \of solvers include enhanced versions of themselves within their solver directory. As an example a version of \emph{simpleFoam} with added porosity treatment - \emph{porousSimpleFoam} - is included in the above mentioned \emph{simpleFoam} directory.

\subsubsection{Cases in OpenFOAM}
In \of simulations are run via a case. In essence, this is a folder that includes all the necessary folders and files for a solver to run a simulation. Principally, this includes a mesh, initial conditions, and numerical instructions for the solver. With these requirements fulfilled an \of solver can then run the case directory and write the calculated values at specified times within it. By default the start time will be \SI{0}{\second} so for a case specified to write every \SI{0.1}{\second} for \SI{0.5}{\second} the written time folders will be 0.1, 0.2, 0.3, 0.4, and 0.5. Combined with the initial directory 0 the contents of time folders can be analysed using a program such as ParaView \cite{paraView} which is the program used for the case visualizations in this thesis.

The mesh is defined by a dictionary file in a \emph{system} subfolder within the case named \emph{blockMeshDict}. This file provides instruction for the \emph{blockMesh} utility to create the mesh. Within this thesis, only the simplest type of mesh in \of is used - a \emph{polymesh} - which is stored in the \emph{constant/polymesh} subfolder. The syntax details of \emph{blockMeshDict} are omitted as they are covered extensively in the user guide \cite{of6UG}. For the purpose of this thesis, the only requirement is that the reader be aware \emph{blockMeshDict} defines the mesh and \emph{blockMesh} creates it.  

\begin{figure}[!htb]
\centering
\includegraphics[width=8cm]{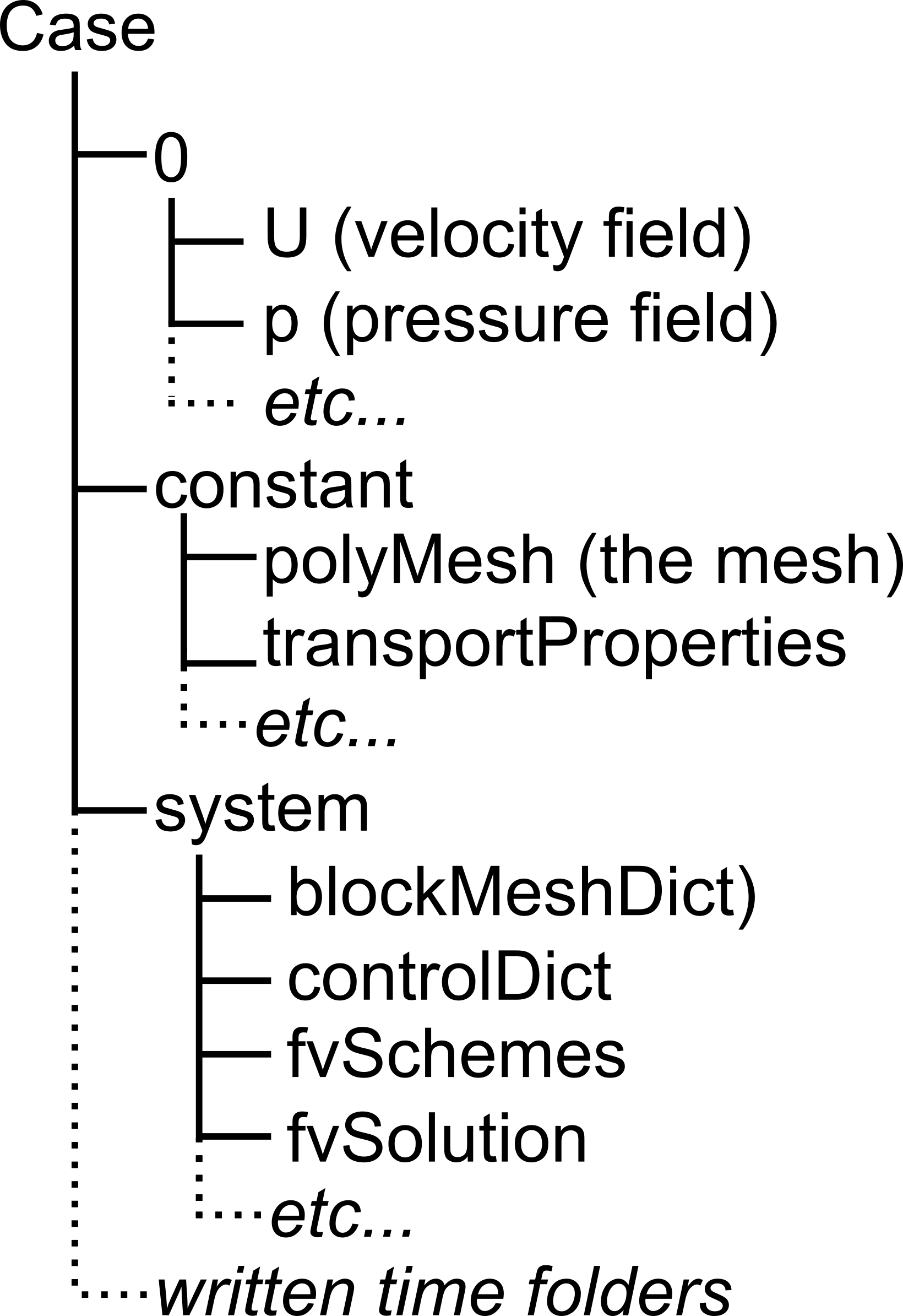}
\caption{Illustration of the structure of a case folder in OpenFOAM. Bracketed text are descriptions of the file added to improve the illustration. Dotted lines and `... \emph{etc}' indicate various other files dependent on the specific case.}
\label{2-fig-of-case}
\end{figure}

The initial conditions are split into fields\footnote{Technically it is input/output types with specific read options that are stored in the \emph{0}/ subfolder.} that are stored in a \emph{0}/ subfolder and properties that are store in a \emph{constant} subfolder. Examples of fields include pressure, temperature, and velocity whereas examples of properties include density, kinematic viscosity, and properties such as the gravitational field strength. For clarity, the properties of a case for a particular solver are usually organised in seprate \emph{xProperties} files. For instance, transport properties are stored in a \emph{transportProperties} file in the \emph{constant} subfolder. In addition to the \emph{blockMeshDict} file, the \emph{system} folder also contains computational instructions for the solver. For instance, the numerical methods used in the case are defined in two files: \emph{fvSchemes} for the numerical schemes and \emph{fvSolution} for the algorithms. Further a \emph{controlDict} file is also included that details information such as the start and end times as well as the time step for the solution. Figure \ref{2-fig-of-case} shows the outline of a case structure.

\section{Interfoam solver}
\subsection{Overview}
\IFM is an inbuilt two phase volume of fluid solver for two immiscible, isothermal and incompressible fluids. This mean the fluids will always stay separate (immiscible), their temperature will not change (isothermal) and their density remains constant (incompressible). Described in detail in \cite{interFoamDscription, interFoamEvaluation}, \IFM is mostly used for wave simulations in marine-related applications \cite{interfoamApp1, interfoamApp2, interfoamApp3}. However, for the present work the key attribute of \IFM is that in addition to the aforementioned Navier-Stokes equations \IFM uses the volume of fluid method.   

\subsection{Volume Of Fluid}
\label{2-sec-vof}
\begin{figure}[!htb]
\centering
\includegraphics[width=7.5cm]{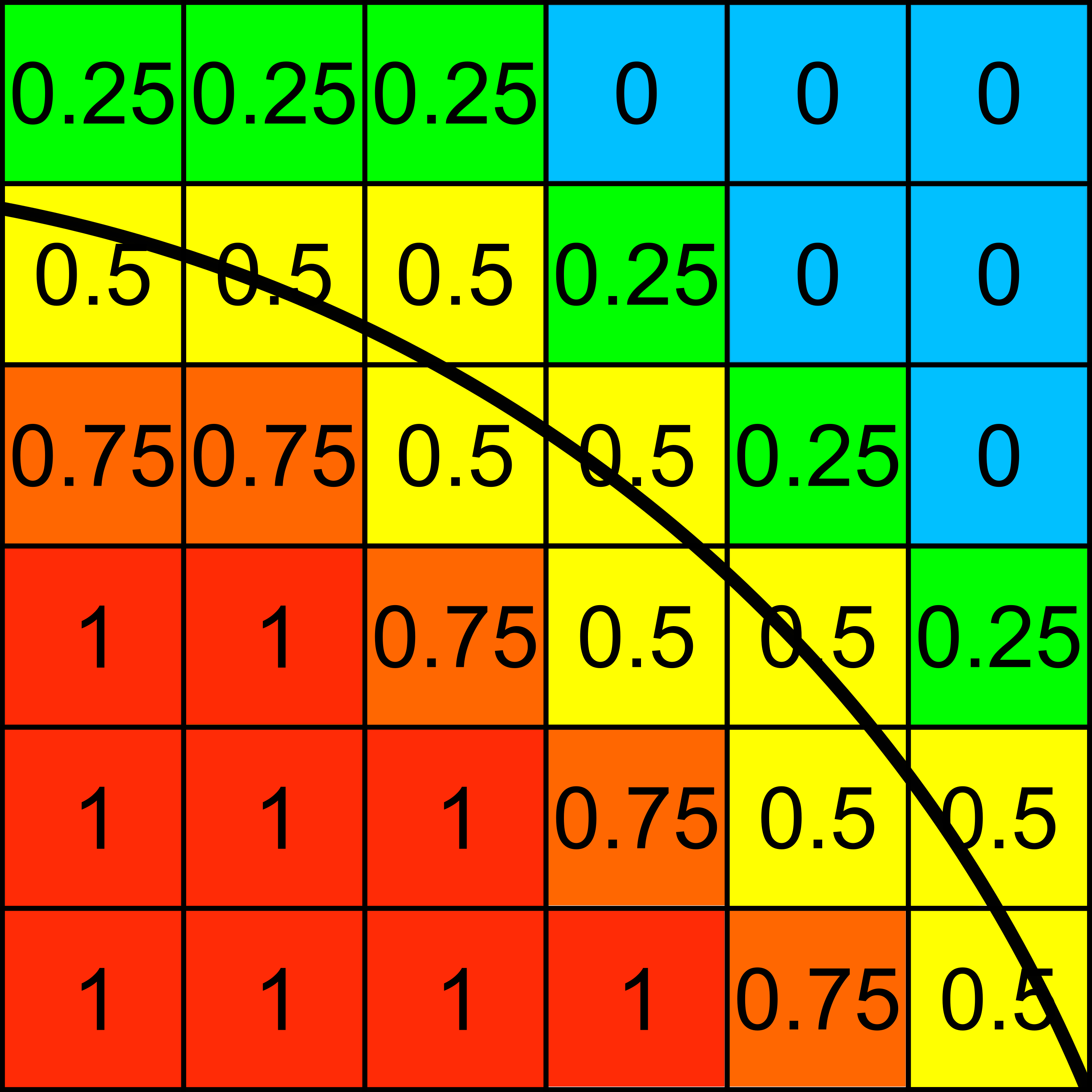}
\caption{Representation of the interface capturing method in \IFM with arbitrary phase fractions between 0 and 1. The cells defined as the interface are shown in yellow.}
\label{2-fig-IF-interface}
\end{figure}

The volume of fluid (VoF) method \cite{vofOrig} involves the introduction of a scalar field $\alpha_i$ that represents the volume of a mesh cell occupied by a phase `i'. A mesh cell containing 100\% phase `i' will have $\alpha_i = 1$ whereas a mesh cell with 0\% phase `i' will have $\alpha_i = 0$. Cells containing somewhere between 0\% and 100\% of phase `i' will have $0 < \alpha_i < 1$. With two phases `i' and `j', the VoF method results in $\alpha_j = 1.0 - \alpha_i$ where cells with $0 < \alpha_i < 1$ form the interface between the phases. To create a better visualizations, in this thesis the interface is defined as the cells with $\alpha_i = 0.5$; references to `melt front' are referring to these cells. To avoid excessive gradient between cells, \of automatically blends the values at the interface. An illustration of this blending in \IFM is shown in Figure \ref{2-fig-IF-interface}. 

Once these cells are defined, their evolution is governed by the standard material derivative: 

\begin{equation}
\label{2-eq-alpha-eq}
\partial_t \alpha_i + \nabla \cdot  \textbf{U}   \alpha_i = 0.    
\end{equation}

\noindent To enhance the results an artificial compression term can be added to equation \ref{2-eq-alpha-eq}. This term is introduced to create interface compression and features a relative velocity $\textbf{U\textsubscript{r}} = \textbf{U\textsubscript{j}} - \textbf{U\textsubscript{i}}$. Adding this term to equation \ref{2-eq-alpha-eq} gives:

\begin{equation}
\label{2-eq-full-alpha-eq}
\partial_t \alpha_i + \nabla \cdot  \textbf{U}   \alpha_i + \nabla \cdot (\alpha_j \alpha_i \textbf{U\textsubscript{r}}) = 0.    
\end{equation}

As \IFM is a two phase solver the formula $\alpha_i = 1.0 - \alpha_j$ can be used for the other phase. Therefore, only one phase needs to be solved for. The relevant transport properties - viscosity and density - can then be distributed using the phase fraction so for density $\rho = \alpha_i \rho_i + \alpha_j \rho_j$. This distribution means that the aforementioned Navier-Stokes equation automatically uses the apt transport properties depending on the phase fraction for a particular cell. At run time, the phase fraction equation is solved first with the MULES algorithm followed by equation \ref{2-eq-NS} using the PIMPLE algorithm. 

\subsection{MULES}
Multidimensional Universal Limit for Explicit Solution (MULES) is an algorithm inbuilt in \of employed specifically to ensure the phase fraction in VoF remains bounded. At the phase boundary, there is a sharp discontinuity between where the phase fraction $\alpha_i = 1$ and where $\alpha_i = 0$. The van Leer scheme and similar detailed in Section \ref{2-sec-numerical-schemes} help to deal with this sharp discontinuity on the faces and MULES serves to ensure its boundedness throughout the time steps. An detailed description of this algorithm is given in a thesis by M{\'a}rquez Dami{\'a}n \cite{marquezThesis}.

\subsection{Dam break}
\begin{figure}[!htb]
\centering
\includegraphics[width=10.5cm]{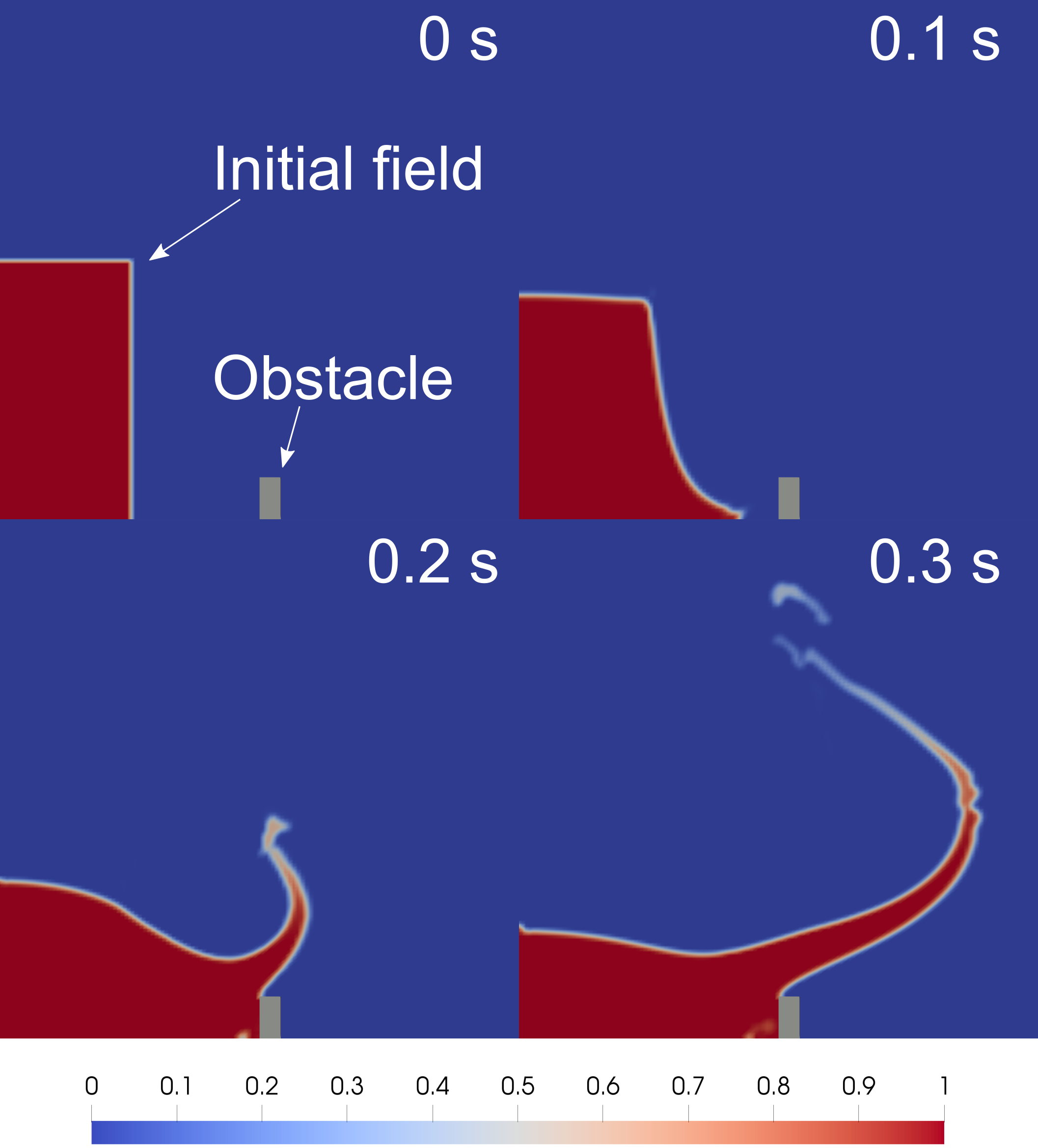}
\caption{Time evolution of the \IFM tutorial case damBreak. The upper left quadrant shows the initial condition of the water phase \vsf for $t = \SI{0}{\second}$. At $t = t + \Delta t$ the water phase fraction of the mesh cells begins to evolve in accordance with the model detailed in Section \ref{2-sec-vof}. The color bar at the bottom of the image shows the colour scheme for the water phase fraction.}
\label{2-fig-damBreak}
\end{figure}

A built in tutorial case for the \IFM involves an instantaneous breaking of a dam causing water to slosh around in a 2D domain. The two phases involved are water and air. Relevant transport properties need to be specified: the kinematic viscosity, $\nu$, and the density, $\rho$ of the water and of the air and, a single surface tension coefficient, $\sigma$, between the water and the air. The results from the tutorial are shown in Figure \ref{2-fig-damBreak}. Cases like the tutorial case where a liquid phase flows around objects are the types of cases that \IFM was built to solve.  

\FloatBarrier
\newpage
\begin{center}
\hrulefill 
{\\ \Large \emph{The candidate's work} \\ }
\hrulefill
\end{center}

\section{Toy model}
\subsection{Overview}
\label{2-sec-toy}
With the `tools' for welding simulation established, a rudimentary `toy' model can be created. The aim of this 'toy' model is to create a minimum viable model for GTAW simulation. To achieve this, starting with \IFM as a base model, the following four features need to be added:  

\begin{itemize}
    \item Make one of the phases act like a solid through preventing it from moving around like the phases in Figure \ref{2-fig-damBreak}.
    \item Melting and solidification; this takes the form of phase change from the solid phase to the liquid phase and vice versa.
    \item Introduce a temperature field to identify the regions requiring phase change.
    \item Add a GTAW heat source term to create changes in the temperature field. Combined with the other features this enables modelling of the GTAW process.
\end{itemize}

With these features in place an initial GTAW simulation model will be created. The following subsections briefly detail the implementations of these features. To avoid redundant material and to keep the scope of this chapter focused only broad outlines of the features are provided with the full details of the implementations in the full model covered in Chapter 3. 

% either class or createFields

\subsection{Phase modelling}
\label{2-sec-toy phase}
The VoF method native to \IFM allows the two phases to have different properties. Here, each phase can be assigned a general phase property $\psi$ as $\psi = \psi_1\alpha_1 + \psi_2\alpha_2$. With only two phases, the presence of one phase necessitates the absence of the other e.g. $\alpha_2 = 1.0 - \alpha_1$. This property can be used to model a solid phase. An artificial momentum sink term such as $C_{sink}$ can be added to mesh cells containing the solid fraction preventing them from moving and thus mimicking a solid. This is a Darcy source term of the form suggested in \cite{voller1987darcy}. Chapter 3 covers this in much greater detail so it is covered only briefly here. Given the cell gradients of the phase fractions shown in Figure \ref{2-fig-IF-interface} the cells effected by the solid momentum sink term need to be identified more distinctly. Hence a solid momentum sink term, $S_{sink}$, can be written as 

\begin{equation}
\label{2-eq-darcy}
S_{sink} = C_{sink} (1.0 - \alpha_1) \xrightarrow{}  C_{sink} \frac{(1.0 - \alpha_1)^2}{\alpha_1 + \epsilon}
\end{equation}

where $\epsilon = 10^{-3}$ to prevent division by zero. Here phase one $\alpha_1$ is the liquid phase and $\alpha_2$ is the solid phase. Given this, $S_{sink}$ will therefore $= C_{sink}$ in solid regions ($\alpha_1 = 0$) and will have a value of $0$ in liquid regions ($\alpha_1 = 1$). For this toy model $C_{sink} = \num{5e9}$; Chapter 4 features an in depth investigation of the optimum value for this term. The $S_{sink}$ term can be added to the inbuilt momentum equation to simulate a solid phase addressing the first `toy' model requirement.

In addition to simulating a solid phase, the `toy' model needs to be able to simulate the change of one phase into another. This takes the form of an additional source term $S_{pc}$ to a phase fraction transport equation (that `creates' more of a phase through acting as a source). Again, the property of one phase necessitating the absence of another can be used as an increase is one phase necessarily accounts for a decrease in the other when it is updated as $\alpha_2 = 1.0 - \alpha_1$. With this, the $S_{pc}$ term needs to only be present in regions where melting should occur. To focus the scope of this chapter, the equation for $S_{pc}$ is simply presented as 

\begin{equation}
\label{2-eq-pc}
    S_{pc} = C_{pc}\left(\frac{(T - T_{melt})c_p}{L}\right)
\end{equation}

for a material with melting temperature $T_{melt}$, heat capacity $c_p$ and latent heat $L$. The full details of this term are presented in Chapter 3 with the choice for a value of the computational term $C_{pc}$ presented in Chapter 4. This $S_{pc}$ term is added to the liquid phase transport equation fulfilling the second `toy' model requirement.

\subsection{Energy equation addition}
With the phases established, the next step is to introduce a method to simulate the temperature of the phases; this is achieve through simulating heat transfer. Heat transfer within a material is heavily dependent on its thermophysical properties. By default \IFM only includes density and kinematic viscosity properties for the phases. Thus to simulate heat transfer additional properties need to be added to each phase namely heat capacity $c_p$ and thermal conductivity $k$. Additionally, values for the melting temperature and latent heat need to be available within the solver. For the `toy' model, these can simply be added as \ds to the \emph{createFields.H} file present in \IFM by default. Then, they can be initialised at run time through entries in the \emph{transportProperties} dictionary in the \emph{constant} subfolder. Note, it may be more appropriate to put these values in a separate `\emph{thermalProperties}' dictionary. However, it is simpler to just add these values as additional properties in the already present \emph{transportProperties} dictionary.

To use these thermophysical properties an appropriate heat transfer equation from which the temperature field is calculated needs to be added to the solver. Here, the transport equation for enthalpy for a fluid with zero viscous dissipation can be introduced as 

\begin{equation}
\label{2-eq-enthalpy-transport-2}
\rho \frac{\partial H}{\partial t} + \nabla \cdot (\boldsymbol{\rho U} H) = \nabla \cdot (k\nabla T).
\end{equation}

Due to latent heat, the enthalpy will change between when there is phase change between the solid and liquid phases. The basis of enthalpy-porosity type models is to split the enthalpy of the liquid phase between a `sensible enthalpy' term and a latent heat term. This allows the enthalpy within the two phases to be described as  

\begin{subequations}
\label{2-eq-enthalpy-int}
\begin{eqnarray}
&& H = \int c_p dT \\
&& H = \begin{cases}
      c_{p1} \cdot T + L & \text{liquid phase} \\
      c_{p2} \cdot T  & \text{solid phase}
    \end{cases}
\end{eqnarray}
\end{subequations}

Shown in equation \ref{2-eq-energy}, equations \ref{2-eq-enthalpy-transport-2} and \ref{2-eq-enthalpy-int} can be combined to create a heat transfer equation for the `toy' model. Here, for brevity, given each phase has its own thermophysical properties the substitution $(\alpha_1 \rho_1 c_{p1} + \alpha_2 \rho_2 c_{p2}) = \rho c_{p}$ is used. Note the $\boldsymbol{\rho U c_p}$ term is found through the use of the \emph{fvc::interpolate} utility as $\boldsymbol{\rho U c_p} = \text{fvc::interpolate}(\rho c_p)$; this is discussed in more detail in Chapter 3.  

\begin{equation}
\label{2-eq-energy}
\rho c_p \frac{\partial T}{\partial t} + \nabla \cdot (\boldsymbol{\rho U c_p} T) - \nabla \cdot (k\nabla T) = - S_{Latent}.
\end{equation}

To model temperature accurately, it is necessary to take into account the effect temperature has on other properties. For instance, density can vary considerably with temperature which can create density gradients within a phase and cause buoyancy-driven flow. To implement this in the `toy' model, the Boussinesq approximation is used whereby 

\begin{equation}
\label{2-eq-bouss}
\rho_{liquid}(T) = \rho_{ref}(1.0 - \alpha_{liquid}\beta(T - T_{melt}))
\end{equation}

for a liquid with reference density $\rho_{ref}$, melting point $T_{melt}$ and coefficient of volumetric expansion $\beta$. With the additional temperature equation, the third required feature for the `toy' model has been added.   

\subsection{GTAW term}
To model the GTAW process, a heat source term to mimic the heat input in GTAW needs to be added to equation \ref{2-eq-energy}. The implementation on this term is a little convoluted so to focus the scope of this chapter it is termed simply as a black box `$Q_{GTAW}$' with the details covered in Chapter 3. Essentially, this term applies additional heat to mesh cells that fall within a defined heated / welded region. Recall how the latent heat term is negative to capture how additional heat needs to be added for the solved variable ($T$) to increase when phase change occurs. The $Q_{GTAW}$ is positive as it adds additional heat to the system. With this addition term, equation \ref{2-eq-energy} is transformed into equation \ref{2-eq-energy-GTAW} as

\begin{equation}
\label{2-eq-energy-GTAW}
\rho c_p \frac{\partial T}{\partial t} + \nabla \cdot (\boldsymbol{\rho U c_p} T) - \nabla \cdot (k\nabla T) = - S_{Latent} + Q_{GTAW}.
\end{equation} 

This can then be combined with a equation \ref{2-eq-NS} that employs equation \ref{2-eq-darcy} and equation \ref{2-eq-bouss} and a modified equation \ref{2-eq-full-alpha-eq} with \ref{2-eq-pc} as a source term to create the full `toy' model.

\section{Benchmark cases for toy model}
\label{2-sec-benchmark-cases}
\subsection{Melting benchmark}
\begin{figure}[!h]
\centering
\includegraphics[width=8cm]{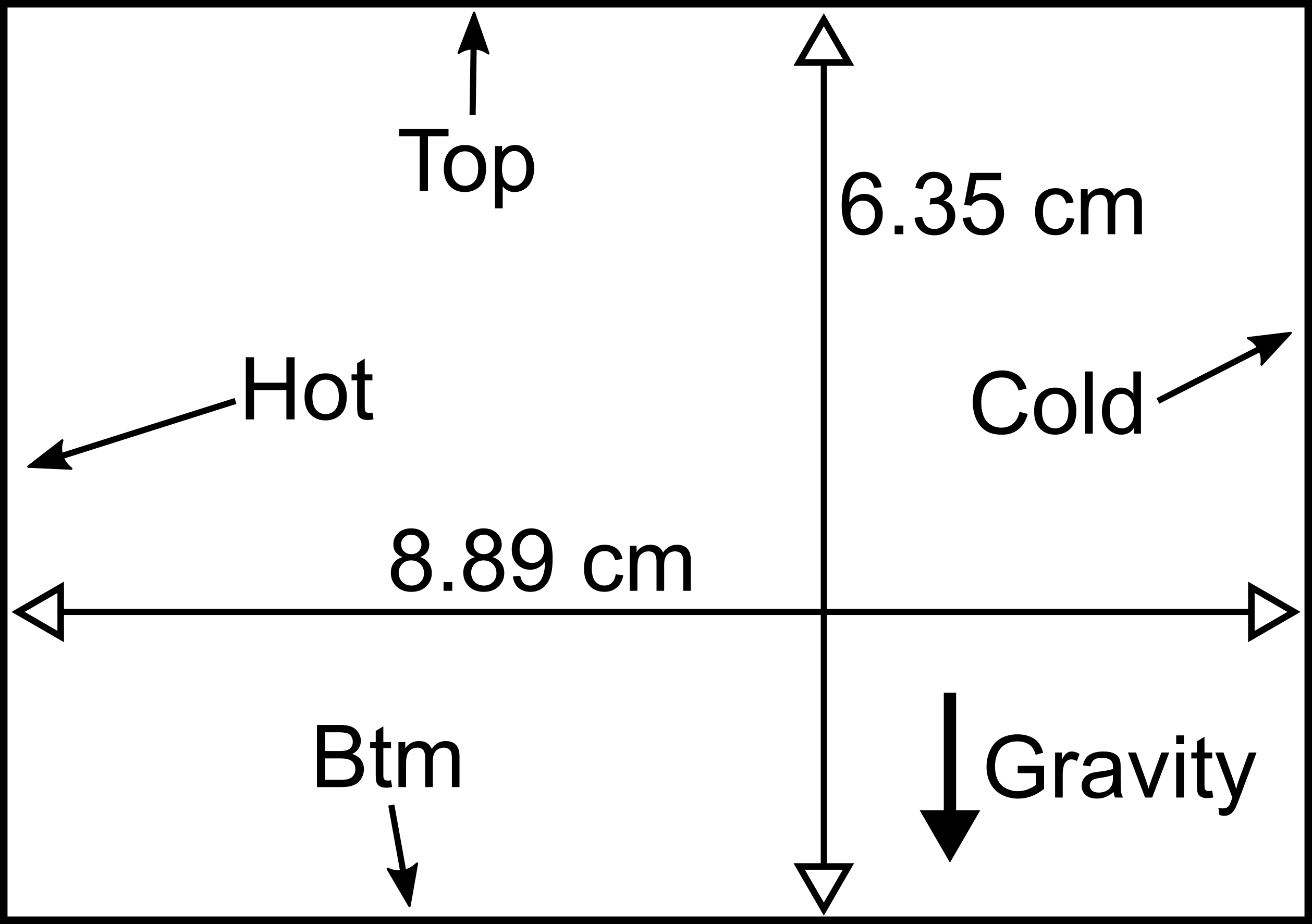}
\caption{2D computational domain used for gallium melting simulations.}
\label{4-fig-ga-melt-domain}
\end{figure}

In order to assess the quality of the 'toy' model, a benchmark case of the melting of gallium in a rectangular cavity by Gau and Viskanta \cite{GauGaMeltExp} is used. Particularly with this enthalpy-porosity type models, benchmarking with the melting of gallium is very popular \cite{brentExp1, brentExp2, brentExp3, brentExp4, rosler2011brentStep, brentUpdateCFD1, brentUpdateCFD2} so it is an apt benchmark to contextualize the 'toy' model. In Gau and Viskanta's experiment, a $8.89 \times 6.35 \times 3.81$ \si{\cm} block of gallium was differentially heated with a hot side, $T_{hot}$, at \SI{311}{\kelvin} and a cold side, $T_{cold}$, at \SI{301.3}{\kelvin}. As the melting point of gallium is \SI{302.93}{\kelvin}, the gallium changes phase as the temperature changes across the domain in accordance with equation \ref{2-eq-energy}. All other sides were insulated so as to be adiabatic. To run a simulation of this experiment in \of the domain outlined in Figure \ref{4-fig-ga-melt-domain} is used. The case is two dimensional but as \of requires three dimensions the domain is one cell thick with \emph{empty} boundaries in the third dimension. An $84 \times 64$ cell mesh was used. The boundary conditions used are shown in Table \ref{4-tab-bc} and the thermophysical properties used are shown in Table \ref{ap-tab-toy-ga-TPP}.   
% comment on the fact sim is 2d exp is 3d

\begin{table}[!h]
\caption{\label{4-tab-Ga-Case-Prop}Boundary conditions for gallium melting benchmark case.} 
\centering
\begin{tabular}{l l l l l l l}
\toprule
Boundary & Velocity & Pressure & Temperature & $\alpha_1$ & $\alpha_2$ & $\alpha_3$ \\
\hline
Top & \emph{PIOV} & \emph{TP} & $\partial_n = 0$ & $\partial_n = 0$ & $\partial_n = 0$ & $\partial_n = 0$ \\
Hot & \emph{NS} & \emph{FFP} & $T_{hot} = \SI{311}{\kelvin}$ & $\partial_n = 0$ & $\partial_n = 0$ & $\partial_n = 0$ \\
Cold & \emph{NS} & \emph{FFP} & $T_{cold} = \SI{301.3}{\kelvin}$ & $\partial_n = 0$ & $\partial_n = 0$ & \emph{IO} \\
Btm & \emph{NS} & \emph{FFP} & $\partial_n = 0$ & $\partial_n = 0$ & $\partial_n = 0$ & $\partial_n = 0$ \\
\bottomrule
\end{tabular}
%\vspace{-1cm}
\end{table}

\begin{table}[!h]
\caption{\label{ap-tab-toy-ga-TPP}Thermophysical properties used for the toy gallium melting case.}
\centering
\begin{tabular}{l l l l}
%\begin{longtable}{l l l l}
%\caption{Thermophysical properties used for the toy gallium melting case.}
\label{ap-tab-toy-ga-TPP} \\
\toprule
Phase & Property & Value & Units\\
\hline
$\alpha_1$\rule{0pt}{2.6ex} & Density, $\rho_1$ & 6092 & \si{\kilogram\per\metre\cubed} \\
 & Specific Heat Capacity, $c_{p,1}$ & 381.5 & \si{\metre\squared\per\second\squared\per\kelvin} \\
 & Thermal Conductivity, $k_1$ & 32 & \si{\kilogram\metre\per\second\cubed\per\kelvin} \\
 & Kinematic Viscosity, $\nu_1$ & \num{2.97e-07} & \si{\metre\squared\per\second} \\
\hline
$\alpha_2$\rule{0pt}{2.6ex} & Density, $\rho_2$ & 6092 & \si{\kilogram\per\metre\cubed} \\
 & Specific Heat Capacity, $c_{p,2}$ & 381.5 & \si{\metre\squared\per\second\squared\per\kelvin} \\
 & Volumetric Thermal Expansion Coefficient, $\beta$ & \num{1.2e-4} & \si{\per\kelvin} \\
 & Thermal Conductivity, $k_2$ & 32 & \si{\kilogram\metre\per\second\cubed\per\kelvin} \\ 
 & Melting Point, $T_{melt}$ & 302.93 & \si{\kelvin} \\
 & Reference Temperature, $T_{ref}$ & 302.78 & \si{\kelvin} \\
 & Latent Heat of Fusion, $L_f$ & \num{8.016e4} & \si{\metre\squared\per\second\squared} \\
 & Kinematic Viscosity, $\nu_2$ & \num{2.97e-07} & \si{\metre\squared\per\second} \\
%\hline
%All & Specific Heat Capacity & 381.5 & \si{\kilogram\per\metre\cubed} \\
\bottomrule
%\end{longtable}
\end{tabular}
\end{table}

\begin{figure}[!h]
\centering
\includegraphics[width=10.5cm]{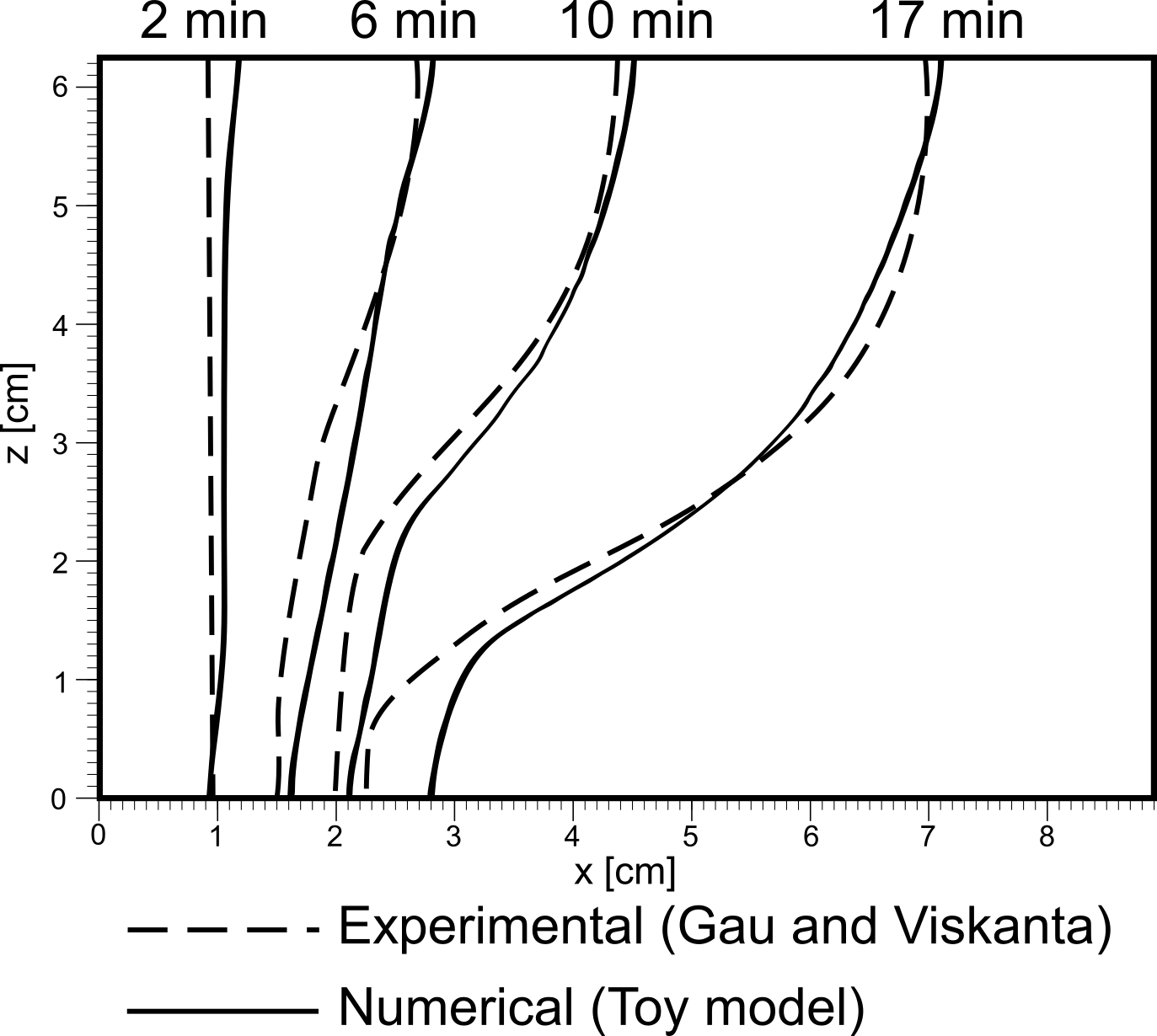}
\caption{Experimental results from Gau and Viskanta \cite{GauGaMeltExp} (dashed line) against 'toy' model results (solid line) showing the melt front at a selection of times for the melting of gallium. $x$ and $z$ axis show the distance from the further bottom left in Figure \ref{4-fig-ga-melt-domain}}
\label{2-fig-toy-ga-bench}
\end{figure}

Figure \ref{2-fig-toy-ga-bench} shows the `toy' model matches the experimental results well. Here, as the temperature increases from left to right in accordance with equation \ref{2-eq-energy} the phase change term described in equation \ref{2-eq-pc} becomes positive causing the phase fraction composition to change. Note how the melting is uneven between the top and bottom of the domain. This is due to the density variations governed by equation \ref{2-eq-bouss}.

\subsection{GTAW benchmark}
Given the `toy' model successfully simulates the melting of gallium, the next step is to investigate its efficacy at simulating welding. Here the heat source is the $Q_{GTAW}$ term rather than the boundary as is the case with the melting of gallium. Given that the toy model includes just two phases, only autogenous welding can be simulated. The chosen benchmark involves autogenous welding on \SI{5}{\milli\meter} thick 304 stainless steel plate. The benchmark uses a current of \SI{150}{\ampere} and a voltage of \SI{12.6}{\volt} to weld at \SI{1}{\meter\per\minute}. The full case details are presented in Appendix B.2. Figure \ref{2-fig-toy-weld} shows a comparison between the weld predicted from simulation compared to the experimental results. Here, the match between the simulation and experimental results is close indicating the `toy' model is successful in simulating this welding case.   

\begin{figure}[!htbp]
\centering
\includegraphics[width=12.5cm]{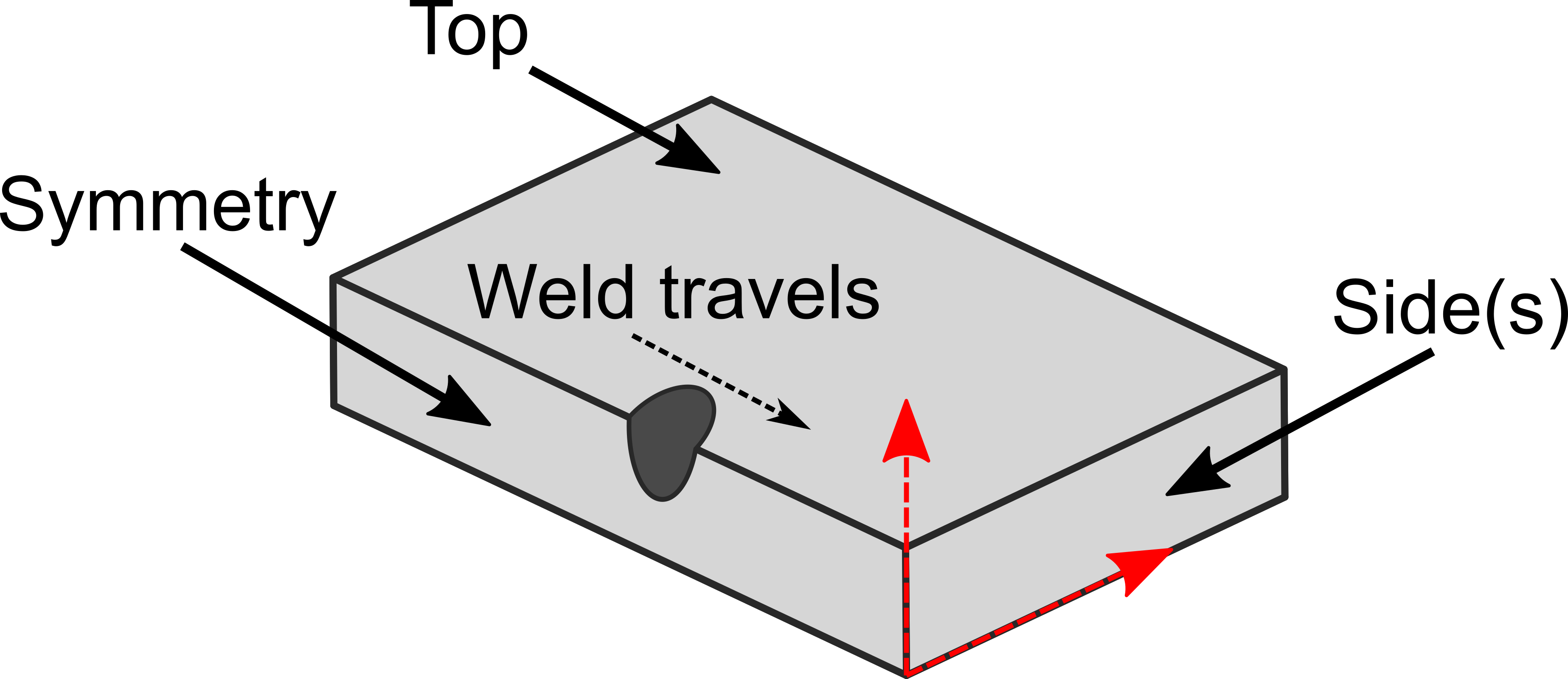}
\caption{3D domain used for the `toy' model GTAW benchmark. The boundary marked symmetry uses the symmetry plane condition which allows the simulation to only have to run half of a full weld. All boundaries not marked are grouped into the `side' boundary. The two red arrows show the perspective of Figure \ref{2-fig-toy-weld}}
\label{2-fig-toy-domain}
\end{figure}

\begin{figure}[!htb]
\centering
\includegraphics[width=10.5cm]{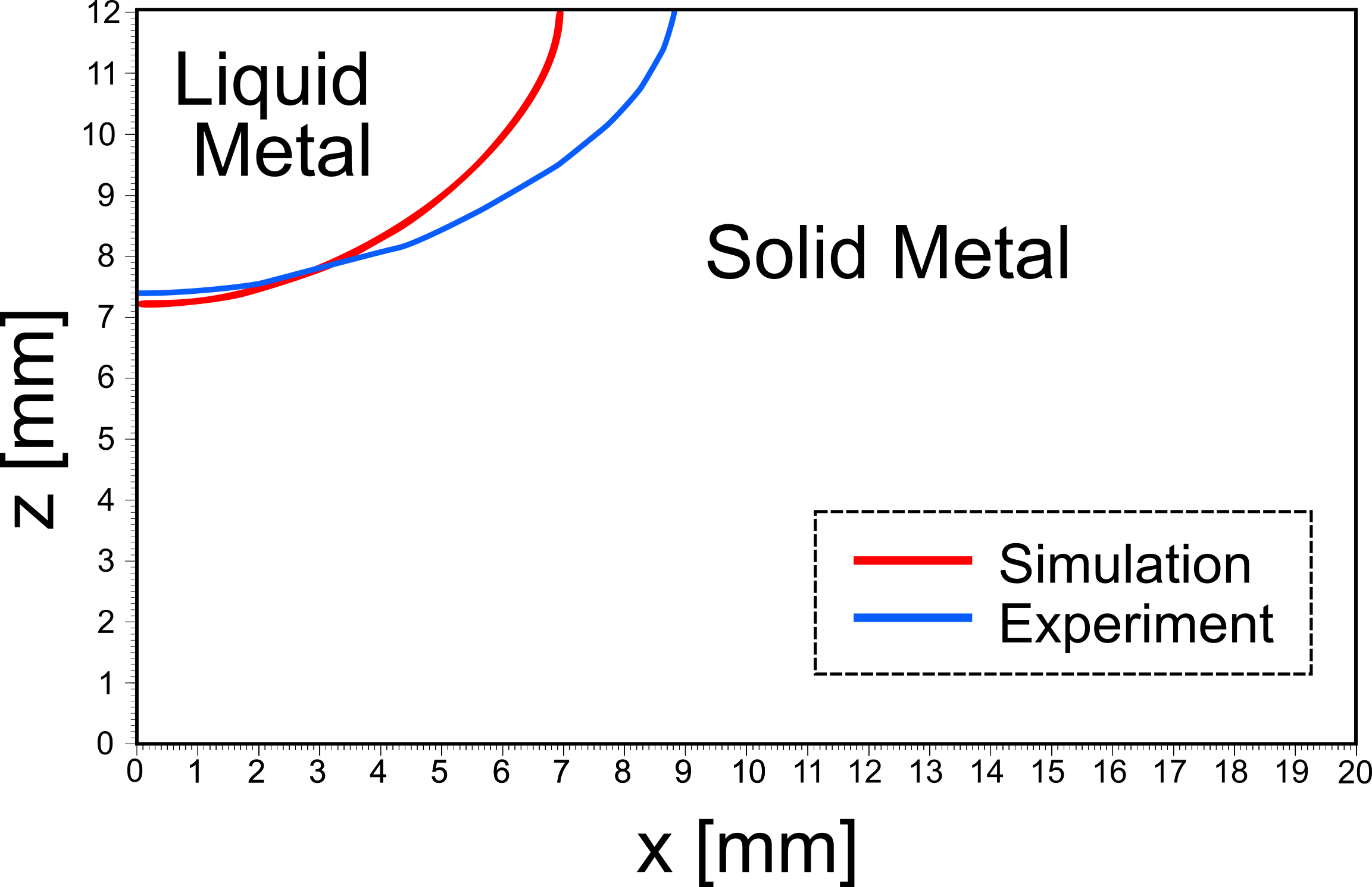}
\caption{Experimental (blue) vs simulation results (red) for the toy model welding. Here, the $x$ and $z$ axis show the distance from the furtherest bottom left of Figure \ref{2-fig-toy-domain} indicated by the two red arrows.} 
\label{2-fig-toy-weld}
\end{figure}

\section{Model limitations}
The `toy' model only has two phases and therefore it is unable to simulate both melting (solid and liquid) and free surface deformation (liquid and gas) simultaneously. Given accurately simulating both of these elements is required to capture the time evolution of a weld pool an additional (third) phase is required. Further, whilst the results presented in Section \ref{2-sec-benchmark-cases} may look good, they are a little bit deceiving. This is because the `toy' model has been designed to get the good results. As detailed throughout this chapter, there are many specifications within an \of case  - notwithstanding the thermophysical properties of the materials used - that affect the solution. By choosing the `correct' specifications and an apt case good results can be achieved. However, it does not then follow that the solver is effective\footnote{As an interesting side note, this is essentially almost a ludic fallacy \cite{taleb2007black} in a CFD context. The entire solver and case can be designed to perform a melting benchmark exceptionally well but this does not ensure the solver can predict the melting of anything else other than the benchmark case. All it actually does is show that the `toy' model performs well within the `game' case and not that is predicts real-world GTAW e.g. the ludic fallacy! As will be explored in detail in Chapter 4, it is the view of the author that the prediction abilities of a solver is synonymous with its efficacy.}. A good solver should perform well in a variety of situations. Thus to assess this it needs to be tested in a variety of situations. This becomes an issue whenever any additional solver elements are added to allow the solver to work for particular situations that then invalidate previous tests. Chief among these additional elements is the introduction of a third phase. Thus, there are two issues that need to be resolved.

\begin{enumerate}
    \item With only two phases (solid and liquid) there is no free surface capable of deformation. Surface deformation is a vital simulation element for simulating ultra-thin-walled tube welding. This is because unless specific parameters are used a GTAW arc can easily melt through a thin wall. Assessing whether the melt pool will burst / melt through requires assessing surface deformation. Therefore, a third (gas) phase is needed to fully simulate ultra-thin-walled tube welding.
    \item The `toy' model has been optimized for a specific benchmark whereby certain decisions on how and what to model were made. The results presented in \cite{GauGaMeltExp} are from one experiment with one set of equipment from 1986. Matching these experimental results \emph{too} closely actually indicates possible over-fitting / forcing of the results. Therefore, claims about the efficacy of a solver are unsupported without a far more robust set of benchmarks.
\end{enumerate}

\section{Chapter summary}
This chapter introduces the concepts and ideas required to understand the contributions of this thesis. The program used for the simulations - \of - features an inbuilt solver - \IFM - that provides a foundation on which to build a GTAW simulation. Limited modifications to \IFM enables the creation of a `toy' model that allows for reasonable results to be achieved. However, there are two outstanding issues with the `toy' model: the lack of a third phase and the lack of a robust set of benchmarks. The former will be addressed in Chapter 3 and the latter in Chapter 4.

%% file: chapters/thesis-chapter-4.tex
\lhead{Chapter 5: Multiphysics solver}
\section{Introduction}
This chapter details the full multiphysics solver used in this thesis. `Solver' is used to refer to the entire program including all of the custom classes and the separate additional modules. Whereas, `algorithm' is used to refer to the computational process that returns data structures for specified times given a mesh and initial boundary conditions. `Equation' is used interchangeably between the mathematical statements and the implementation in both singular computational statements and their \emph{fvMatrix} class. Initially the solver structure and algorithm is detailed followed by the OpenFOAM C++ class structure used to create the solver. Then, the numerical structure of the requisite equations used in the algorithm is covered in Section \ref{3-num-struct}. This arrangement is used to enable the abstraction of the problem to be detailed first so the motive for the specific choices made is clearer. Finally, small factors $\mathcal{O}(10^{-8})$ are often included in the denominators of various equations to prevent floating point exceptions. These are omitted for brevity unless particularly relevant. The full code for the solver is available on GitHub at \url{https://github.com/WillYeadon/thesis} in the \gIF directory. 

\section{Solver overview}
\subsection{Background}
The solver was designed with the intention of simulating ultra-thin-walled tube welding. In order to achieve this, the solver needs to be able to handle:

\begin{itemize}
    \item Three phases (solid, liquid and gas)  
    \item Motion of the phases (Navier-Stokes)
    \item Phase change between solid and liquid phase (melting and solidification)
    \item Moving GTAW heat source
    \item Accurate prediction (measured through benchmarks)
\end{itemize}

This blends two kinds of CFD problems: free surface flow and heat transfer. Calculating the free surface flow is key to the understanding of whether the weld pool will be stable as it rotates around the ultra-thin-walled tube. Whereas, the calculation of the heat transfer is needed to understand the size and shape of the weld pool. This combination makes popular inbuilt solvers for multiphase flow such as \IFM and \MFIF or for heat transfer such as \cMRF insufficient. Further, as discussed in Chapter 2, limited modifications to the inbuilt \IFM solver are also inadequate. Additionally, to explore a general case for ultra-thin-walled tube welding it is imperative that the solver matches the experimental results detailed in Chapter 6 so that there can be confidence in the solver's predictions. To the authors knowledge, the best publicly available candidate \of solvers are \emph{compressibleInterFoam}, \ITPCF \cite{interThermalPhaseChangeFoam} and \IRMFIF\cite{icoReactingMultiphaseInterFoam} all of which are volume of fluid (VoF) based solvers. 

\CPIF is a two phase inbuilt solver for two compressible non-isothermal fluids. Given the fluids are non-isothermal, the solver includes a temperature equation. However, as covered in Chapter 2, adding a basic temperature equation to a solver is relatively straightforward so this is not much of an advantage. Secondly the solver is built around gas-liquid cases where there are large temperature dependent changes in the phases e.g. an underwater explosion. As such, the inbuilt equations include many extra terms to fulfil this scope. Thus to extend it to a three phase solid-liquid-gas welding simulation would require considerable reworking to the point whereby it is simpler to build up a new solver from \emph{interFoam}.

\ITPCF is a user created two phase solver which also requires a third phase extension to simulate solid-liquid-gas interaction. Further, the solver is focused on liquid-gas phase change and so would also require a new phase change model for solid to liquid. With these two points considered, modifying the solver presents largely the same challenge as starting from \IFM without the same extensive body of knowledge available online. Therefore it was not chosen. 

\IRMFIF is an inbuilt solver available in the \emph{ESI} version of \of and is the best candidate as it can handle multiple phases with phase change and features a separate inbuilt heat source model. This solver was initially selected however it did not perform as expected for various test cases. This issue was exacerbated by the lack of documentation available. With a hindsight assessment there is definite scope for using this solver for welding simulation yet at the time of evaluation it was judged to be easier to instead use it as an influence for the development of a separate solver. 

Given the problems outlined with these solvers, the development of the custom solver covered in this chapter was deemed necessary. Using the skeleton of the \IFM solver, the constituent classes were replaced with custom `\emph{gtaw}' versions to create a multiphysics solver capable of simulating ultra-thin-walled tube welding: \emph{gtawFoam}. This solver was written by the author to simulate the welding of ultra-thin-walled tubing.  

\subsection{Solver algorithm}
\label{3-solver-algorithm}
% Explain the overall simplified version
The core solver runs a series of files that solve specific field objects (e.g. the alpha equation solves the volume fraction) and update the field objects each time step with the results from those calculations. The structure is similar to the inbuilt \IFM solver but the constituent parts are all replaced. As such, the solver uses the PIMPLE algorithm discussed in Chapter 2 to solve the pressure-velocity coupling. However, the temperature equation is calculated outside of the main PIMPLE loop; it is one-way coupled to the other equations. Some research has advocated the temperature equation should be included inside the main PIMPLE loop \cite{faden2019optimum} yet during development of the solver, one-way coupling was found to be superior. This mismatch is probably due to the handling of the liquid fraction - this is detailed further in Section \ref{3-alpha phase-change}. The full solver algorithm is shown in Algorithm \ref{3-alg-solver}.

\begin{algorithm}
\caption{Solver Algorithm}
\begin{algorithmic}[1]
    \State Create field objects 
    \While{$\mbox{t}<\mbox{t}_{end}$}
        \While{PIMPLE}
            \State Do alpha equation
            \State Do mixture update
            \State Do velocity equation
            \While{Pressure Correction}
                \State Do pressure correction
            \EndWhile
        \EndWhile
    \State Do temperature equation
    \If{t == write control}{
        write field objects
        } \EndIf
    \State t += $\Delta$t
    \EndWhile
\end{algorithmic}
\label{3-alg-solver}
\end{algorithm}

\section{Class structure}
\label{3-class-structure}
\begin{figure}[p!]
\includegraphics[width=15cm]{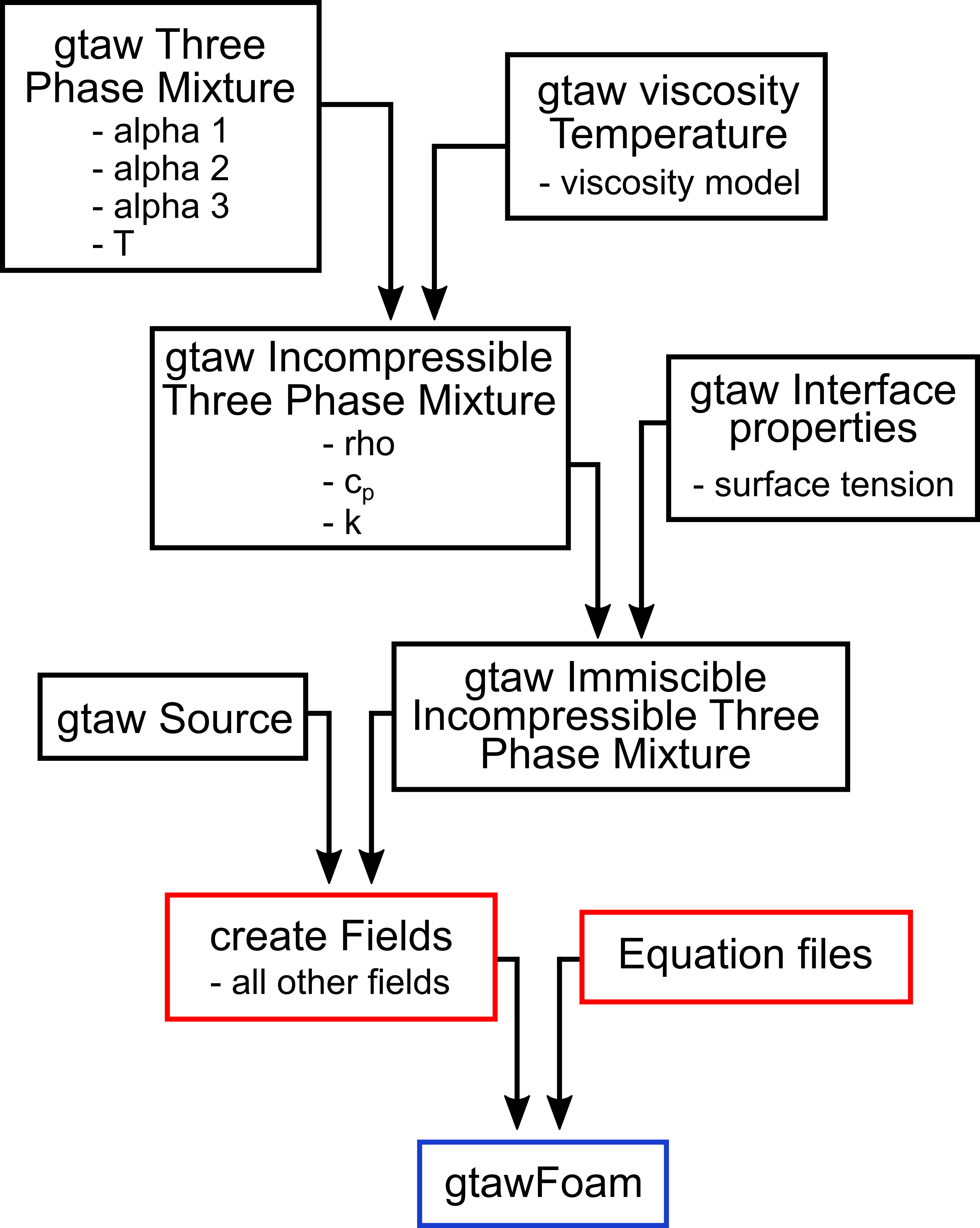}
\caption{Full structure of the solver elements for \emph{gtawFoam}. Classes are surrounded by a black box, header files are in a red box and the final executable is in a blue box. The black lines indicate a has-a relationship e.g. \emph{gtawIncompressibleThreePhaseMixture} has-a \emph{gtawThreePhaseMixture}. The dash points are the important fields within the solver elements. This structure enables reusability for instance, \emph{gtawSource} could be replaced by an apt `\emph{laserSource}' or similar and the solver would continue to work.}
\label{3-fig-class-structure}
\end{figure}

The solver uses the same class structure as \IFM and as such has strong modularity. As shown in Figure \ref{3-fig-class-structure}, classes such as \emph{gtawSource} could be replaced with `\emph{laserSource}' to adapt the solver to a specific requirement (as the present solver has been adapted from \emph{interFoam}). Inbuilt classes in \of are well tested and should preferably be used. However, due to the requirements of \gIF eventually all constituent classes in \IFM were replaced. 

Initially, \emph{twoPhaseMixture} is replaced with \emph{gtawThreePhaseMixture} which includes a temperature field and three phases. As subsequent classes `have' the \emph{gtawThreePhaseMixture} class a singular temperature object is always accessed and temperature dependent thermophysical properties can therefore be implmented. This class, combined with a custom temperature dependent viscosity model detailed in Section \ref{3-temp-dep}, is used to construct  \emph{gtawIncompressibleThreePhaseMixture}. Compared to the inbuilt \emph{incompressibleTwoPhaseMixture} which simply contains two \emph{dimensionedScalars} - $\rho_1$ and $\rho_2$ - \emph{gtawIncompressibleThreePhaseMixture} contains a suite of thermophysical properties (density, specific heat capacity, thermal conductivity etc...). \of lacks a database of common materials and therefore a user often has to look up densities and viscosities for specific materials. \gIF solves this by including a header file with the thermophysical properties of commonly welded materials thus users simply need to specify a string argument such as `titanium' to retrieve the required properties. 

Detailed in Section \ref{3-momentum-section}, the interface treatment is modified from \IFM to better suit welding simulation and so the interface properties class is modified to create \emph{gtawInterfaceProperties}. This is combined with \emph{gtawIncompressibleThreePhaseMixture} to make the `mixture' object which is a instance of the class \emph{gtawImmiscibleIncompressibleThreePhaseMixture}. This `mixture' object has all the member variables required to calculate solid-liquid-gas flow. This is combined with a heat source class, \emph{gtawSource}, which is detailed in Section \ref{3-GTAW-source-term} and initialized in a header field `\emph{createFields.H}'. With the field objects created, individual header files are used for the equations shown in Algorithm \ref{3-alg-solver} and are compiled (with other dependencies) into an executable that solves the equations for a particular case. 

\section{Numerical structure}
\subsection{Overview}
\label{3-num-struct}
This section comprehensively covers the features of the multiphysics solver used in the rest of this thesis. As the full code is available on GitHub, code snippets are only included when particularly relevant or non-trivial. For brevity, the separate fvOptions module and custom boundary conditions for the energy equation are treated as part of the solver despite technically being separate programs. The algorithm is covered in its run order as detailed in Section \ref{3-solver-algorithm}.

Throughout the solver, various computational constants (written $C_x$) are introduced to improve solver accuracy. In many CFD simulations it is possible to `force' the simulation to produce the desired results through appropriate tuning of these computational constants. However, aggressively tuning these parameters to specific cases - locally optimizing them - can lead to poor performance for other (global) cases. Therefore computational constants are only introduced when absolutely necessary and where relevant to a broad class of problems. The optimization of the values for these parameters are covered in Chapter 5.
\subsection{Alpha equation}
\label{3-alpha-section}
Similarly to the solver detailed in Chapter 2, the multiphysics solver used a volume of fluid method to calculated each phase fraction. However, with three phases the method of calculating one phase, $\alpha_{i}$, and using it to define the other as $\alpha_{j} = 1 - \alpha_{i}$ no longer works and each phase must be calculated. Starting the with standard scalar transport equation the three phases are defined as     

\begin{subequations}
\label{3-alpha-i-transport}
\begin{gather}
\sum_{i=1}^3 \alpha_{i} = 1 \\
\label{3-alpha-i-transport-B}
\partial_{t}\alpha_{i} + \nabla \cdot \boldsymbol{\alpha_{i} U_{\alpha}}  = S_{\alpha_{i}}.
\end{gather}
\end{subequations}

As with the base \IFM solver, artificial interface compression is required for the solver to work effectively. As detailed in Chapter 2, the interface compression in \IFM is found through introducing a relative flux $U_{c_{ij}} = U_{i} - U_{j}$ and rewriting equation \ref{3-alpha-i-transport-B} as  

\begin{equation}
\partial_{t}\alpha_{i} + \nabla \cdot \boldsymbol{\alpha_{i} U_{\alpha}} + \nabla \cdot (\alpha_i \alpha_j \boldsymbol{U_{c_{ij}}}) = S_{\alpha_{i}}.
\end{equation}

However, for three phases the compression fluxes for $\alpha$ pairs $\alpha_{12}$, $\alpha_{13}$ and $\alpha_{23}$ (where $\alpha_{ij} \equiv \alpha_{ji}$) are required. Importantly, the pairs are redundant and thus must only be applied once each in the $\alpha$ equations for all phases. Applying the compression flux to $\alpha$ pair $\alpha_{12}$ gives  

\begin{equation}
\partial_{t}\alpha_{1} + \nabla \cdot \boldsymbol{\alpha_{1} U} + \nabla \cdot (\alpha_{1} \alpha_{2} \boldsymbol{U_{c_{12}}}) + \nabla \cdot (\alpha_{1} \alpha_{3} \boldsymbol{U_{c_{13}}})  = 0.    
\end{equation}

This compression flux is discretized in the same way as \IFM \cite{interFoamDscription} except is of a general form 

\begin{equation}
\boldsymbol{U_{c_{ij}}} = \hat{n}_{ij} \cdot min[C_{\alpha} \cdot \frac{|\boldsymbol{U}|}{|\boldsymbol{S_f}|},  max\left(\frac{|\boldsymbol{U}|}{|\boldsymbol{S_f}|}\right)]     
\end{equation}

for total flux $\boldsymbol{U}$, surface vector $\boldsymbol{S_f}$ and interface compression factor $C_{\alpha}$. The surface normal $\hat{n}_{ij}$ for a phase pair $\alpha_{ij}$ is given by

\begin{subequations}
\begin{gather}
\hat{n}_{ij} = \frac{\alpha_j \nabla \alpha_{i} - \alpha_i \nabla \alpha_{j}}{|\alpha_j \nabla \alpha_{i} - \alpha_i \nabla \alpha_{j}|}.    
\end{gather}
\end{subequations}

For implementation, all the $\boldsymbol{\alpha U}$ terms need to be combined into one variable for each phase. It was found that iterating over an array of \emph{surfaceScalarFields} was the best solution for this.

Phase change is handled through incorporating an explicit source term $S_{\alpha_{i}}$. Detailed in Section \ref{3-energy-section} the values for the melting, $S_{\alpha_{3} \rightarrow{} \alpha_{2}}$, term is calculated from the energy equation. With a zero sum material change in melting and solidification by definition $S_{\alpha_{3} \rightarrow{} \alpha_{2}} = - S_{\alpha_{2} \rightarrow{} \alpha_{3}}$. This generates the phase change source terms for the solid and liquid phase. For brevity the notation $S_{\alpha_{3} \rightarrow{} \alpha_{2}} \equiv S_{\alpha_{2}}$ and $S_{\alpha_{2} \rightarrow{} \alpha_{3}} \equiv S_{\alpha_{3}}$ is used. Further, in this thesis, the terms $\alpha_2 \equiv \text{liquid}$ or liquid metal and $\alpha_3 \equiv \text{metal}$ or solid metal. Combining equation \ref{3-alpha-i-transport}, interface compression and introducing source terms for phase changes gives a final alpha equation of 
% This needs to match whatever term you used for alphaPhi
\begin{equation}
\label{3-full-alphas-equation}
\begin{gathered}
    \partial_{t}\alpha_{1} + \nabla \cdot \boldsymbol{\alpha_{1} U} + \nabla \cdot (\alpha_{1} \alpha_{2} \boldsymbol{U_{c_{12}}}) + \nabla \cdot (\alpha_{1} \alpha_{3} \boldsymbol{U_{c_{13}}})  = 0 \\ 
    \partial_{t}\alpha_{2} + \nabla \cdot \boldsymbol{\alpha_{2} U}  + \nabla \cdot (\alpha_{2} \alpha_{3} \boldsymbol{U_{c_{23}}}) = S_{\alpha_{2}} \\
    \partial_{t}\alpha_{3} + \nabla \cdot \boldsymbol{\alpha_{3} U}  = S_{\alpha_{3}}. \\
\end{gathered}
\end{equation}

%Equation \ref{3-full-alphas-equation} is solved with the MULES algorithm as detailed in Chapter 2. 

    \subsection{Dynamic meshing on alpha fraction}
    To better capture the interface between the phases, dynamic mesh refinement can be applied. However, with phase changes where for some time steps the total phase fraction in the entire domain may be $\mathcal{O}(10^{-6})$ (e.g. when melting/solidification begins) mesh refinement can crash the solver. To resolve this, refinement is performed on a `shadow' phase fraction $\alpha_{dyn}$ that only $\neq 0$ when the main phase fraction occupies enough cells to avoid crashing. This is implemented as       

    \begin{equation}
    \label{3-dyn-mesh}
    \alpha_{dyn} = \begin{cases}
        \alpha_i, & \text{if}\ \frac{\sum \alpha_i}{n_{cells}} > C_{dyn}\\
        0, & \text{otherwise}        
        \end{cases}
    \end{equation}    
    % mention that it is a trigger e.g. when solidifying is okay
    where dynamic meshing is applied to phase $\alpha_i$ through refining over $\alpha_{dyn}$ in a mesh with $n_{cells}$ cells. The value of $C_{dyn}$ is highly variable and depends on the specific case but in general it needs to be high enough that $\alpha_{dyn} = 1$ in at least 10 cells. Thus with equation \ref{3-dyn-mesh}, dynamic meshing can be used for mesh refinement at phase interfaces. The \of version used in this thesis (version 6) by default only features dynamic mesh refinement in three dimensions so a user contributed 2D dynamic mesh refinement for 2D cases is used\footnote{I use Luca Cornolti's 2D dynamic mesh refinement code for \of v6 from 2018. However in the years proceeding its release there has been a strong effort from the \of community in developing this feature further (see \cite{2D-mesh-rettenmair}). Thus, it is recommended to use a more updated version.}. 
\subsection{Momentum equation}
\subsubsection{Overview}
\label{3-momentum-section}
The momentum equation used in the solver is essentially the incompressible Navier-Stokes equation with a few additional source terms. As covered in detail in Chapter 2, the momentum equation in the inbuilt \IFM solver reads 

\begin{subequations}
\label{3-IF-momentum-eq}
\begin{gather}
\partial_t \rho \hat{U} + \nabla \cdot \boldsymbol{\rho U} \hat{U} - \nabla \cdot \hat{T}  = -\nabla p_{rgh} + (g \cdot \hat{x}) \nabla \rho  + F_{ST}  + \mu \nabla^2 \hat{U}  \\
\label{3-IF-momentum-eq-B}
\text{where} \ p_{rgh} = p - \rho(g \cdot \hat{x}) \\
\label{3-IF-momentum-eq-C}
\text{and} \ F_{ST} = \sigma \kappa \nabla \alpha_{i} \ \text{where} \ \kappa = - \nabla \cdot \hat{n} = - \nabla \cdot \left( \frac{\alpha_{i}}{|\alpha_{i}|} \right).
\end{gather}
\end{subequations}

Here, the pressure is transformed into $p_{rgh}$ which removes the hydrostatic element of the pressure as shown in equation \ref{3-IF-momentum-eq-B}. This is to ease boundary specifications. The turbulent stress tensor, $\hat{T}$ is never used for the present application as all simulations are set to smooth laminar flow so it is dropped from the final momentum equation.

$F_{ST}$ is a source term to represent the effects of surface tension. However, the interface between the phases is not explicitly tracked and thus its position is unknown. This well known issue is overcome through the Continuum Surface Force (CSF) model proposed in \cite{brackbillCSF} shown in equation \ref{3-IF-momentum-eq-C} where $\kappa$ is the curvature of the interface and $\hat{n}$ the unit normal. As there are only two phases in the inbuilt \IFM solver, the surface tension term only considers one alpha phase - written $\alpha_{i}$. With three phases this needs to be extended for alpha pairs $\alpha_1-\alpha_2$, $\alpha_1-\alpha_3$ and $\alpha_2-\alpha_3$. However, only the gas-liquid ($\alpha_1-\alpha_2$) surface tension is of interested. The solution for this in the inbuilt \MFIF solver for CSF is to transform equation  \ref{3-IF-momentum-eq-C} into   

\begin{subequations}
\begin{gather}
F_{ST} = \sigma_{ij} \kappa_{ij} \cdot \left(\alpha_j \nabla \alpha_{i} - \alpha_i \nabla \alpha_{j}\right)  \\ \text{where} \ \kappa_{ij} = - \nabla \cdot \hat{n}_{ij} = - \nabla \cdot \left( \frac{\alpha_j \nabla \alpha_{i} - \alpha_i \nabla \alpha_{j}}{|\alpha_j \nabla \alpha_{i} - \alpha_i \nabla \alpha_{j}|} \right).    
\end{gather}
\end{subequations}

for a phase pair $\alpha_i-\alpha_j$. However, this was found to have considerable negative impact on solver performance. The solution here is to treat $\sigma \kappa$ in equation \ref{3-IF-momentum-eq-C} as a \vsf and multiply it by $1 - \alpha_3$ so that the surface tension for will be equal to zero in the solid region and the fast two phase treatment can still be used. The final equation for surface tension is therefore 
\begin{equation}
    \ F_{ST} = \sigma_{12} (1 - \alpha_3) \kappa \nabla \alpha_{1} \ \text{where} \ \kappa = - \nabla \cdot \hat{n}_{12} = - \nabla \cdot \left( \frac{\alpha_{1}}{|\alpha_{1}|} \right).
\end{equation}

Given $\hat{n}_{12}$ is computed for the surface tension in the \emph{gtawInterfaceProperties} class, the $\hat{n}_{13}$ and $\hat{n}_{23}$ used for interface compression are also included as member functions of the class. With the three phase surface tension modification implemented, to transform equation \ref{3-IF-momentum-eq} into one useful for GTAW simulation three additional terms are added: Buoyancy, Darcy and Marangoni. The momentum equation is thus rewritten as

\begin{equation}
\label{3-momentum-eq}
\partial_t \rho \hat{U} + \nabla \cdot \boldsymbol{\rho U} \hat{U} + S_{m}  = -\nabla p_{rgh} + (g \cdot \hat{x}) \nabla \rho  + F_{ST} + S_{d} + S_{buoyancy}.
\end{equation}

\subsubsection{Lorentz Force}
As mentioned in the excluded scope section in the introduction, the Lorentz force is omitted in equation \ref{3-momentum-eq}. Initially it was included in the `toy' model detailed in Chapter 3 where the effect it had on the results was found to be small. This is in line with research on the strength of its effect in GTAW \cite{weldDrive}. With the `toy' model being two phase, it was possible to add the Lorentz force as a boundary condition that moved with the weld however with three phases the liquid weld is not guaranteed to be on the boundary thus the same boundary condition implementation could not be used. Given how its effect was small, it was judged that the required additional development time to create a three phase implementation would not bring enough additional benefit.

    \subsubsection{Buoyancy term}
    \label{3-buoyancy}
    The Boussinesq approximation is employed to simulate the effects of buoyancy - namely density variation. This means the density in all terms in equation \ref{3-IF-momentum-eq} is assumed constant except for the $(g \cdot \hat{x}) \nabla \rho$ term. Buoyancy was judged to only matter in the liquid metal phase therefore the approximation deals only with $\rho_2$. The variations in density due to temperature is implemented as 
    
    \begin{equation}
    \label{3-bouss-eq}
        \rho_2 = \rho_{2, \ ref}(1.0 - \beta(T - T_{ref}))
    \end{equation}
    where $\beta$ is the volumetric coefficient of thermal expansion and $ref$ refers to the reference density and temperature - this is taken to be at the melting temperature of the liquid metal phase. Given the class implementation detailed in Section \ref{3-class-structure}, the density update in the solver is performed within the class \emph{gtawIncompressibleThreePhaseMixture}. However, all the inbuilt heat transfer solvers such as \emph{buoyantBoussinesqPimpleFoam} apply equation \ref{3-bouss-eq} (divided by $\rho_2$) directly into the pressure gradient term on the RHS of equation \ref{3-IF-momentum-eq}. Therefore, for clarity a flag is included in the solver to revert back to the methodology of the inbuilt solvers with the $\beta$ multiplied by $\alpha_2$ to maintain the effect only on the liquid fraction. Both methodologies achieve the same results. In equation \ref{3-momentum-eq} the implementation of the effects of buoyancy is illustrated through a source term $S_{buoyancy}$.     
    \subsubsection{Darcy source term}
    A standard solid phase 'Darcy' source term of the form proposed by Voller and Prakash \cite{voller1987darcy} is employed. Briefly this term is derived through assuming the solid region is a porous medium and therefore Darcy's law 
    
    \begin{equation}
        \boldsymbol{U} = \frac{K}{\mu} \nabla p
    \end{equation}

    applies. Here $\mu$ is the viscosity, $p$ is pressure, $\boldsymbol{U}$ is the mass flux and $K$ is a permeability factor which for two phase solid-liquid flow corresponds to $1 - \alpha_{solid}$. Darcy's law can be used to formulate the Carman-Kozeny equation \cite{Kruczek2015CarmanKozeny} - this implies a form for a source term for two phase solid-liquid flow in a porous medium as 
    
    \begin{equation}
        S_{darcy} = - C_d \frac{(1 - \alpha_{liquid})^2}{\alpha_{liquid}^3}
    \end{equation}
    
    with a computational factor $C_d$. Given three phases are employed, it is written in terms of the solid fraction $\alpha_3$ as 
    
    \begin{equation}
    \label{4-eq-cd}
        S_{d} = - C_d \cdot \frac{\alpha_3^2}{(1.0 - \alpha_3)^3 + 10^{-3}}        
    \end{equation}
    
    with a small $10^{-3}$ term added to the denominator to stop division by zero. The computational factor, $C_d$, is typically $\mathcal{O}(10^{6})$ - $\mathcal{O}(10^{12})$ depending on the application.

    \subsubsection{Marangoni term}
    The Marangoni effect is implemented using the form suggested by Saldi \cite{saldi2012marangoni} in a thesis on the topic. Here, a tangential component of surface tension, $\nabla_t \sigma$, is added to the CSF model as
    \begin{equation}
        F_{ST} = (\sigma \kappa + \nabla_t \sigma)\nabla \alpha_{i}. 
    \end{equation}
    
    In a weld pool the surface tension to temperature gradient, $\frac{\partial \sigma}{\partial T}$, is created through the large temperature difference across the weld pool. Therefore, the tangential component is expressed as  

    \begin{equation}
        \nabla_{t} \sigma = \frac{\partial \sigma}{\partial T} \left(\nabla T - \boldsymbol{n} (\boldsymbol{n} \cdot \nabla T) \right).
    \end{equation}
    
    Given the large density ratio - $\mathcal{O}(10^{3})$ - between the gas and metal phases, spurious currents can be generated in the interface region where the Marangoni term is applied. Thus, a redistributive term is applied to shift the term into the metal. In fact, this is a well known problem with \IFM and the VoF method in general - see \cite{interFoamEvaluation} for details. Typically, this redistributive term is written as $\frac{2 \rho}{\rho_1 + \rho_2}$ and uses the fact that as compared to the gas phase, $\rho_2 \approx \rho_3$. Thus the redistributive term is $\approx 1$ where the density field $\rho \approx \rho_2$ and $\approx 0$ where $\rho \approx \rho_1$. However, the \vsf object can be exploited to use the temperature dependent $\rho_2$ covered previously for an alternative distributive term: $\frac{\rho_2 (T)}{\rho_2}$. Here, as the phase fraction $\alpha_2$ is used in the calculation of $\rho_2 (T)$ the redistributive term will ensure only the liquid metal phase will be subject to the Marangoni term. The Marangoni term is thus expressed as
    \begin{equation}
    \label{3-marG-eq}
        S_{m} = \frac{\partial \sigma}{\partial T} \left(\nabla T - \boldsymbol{n} (\boldsymbol{n} \cdot \nabla T) \right) \nabla \alpha_{2} \cdot \frac{\rho_2 (T)}{\rho_2}.
    \end{equation}
    %mag(fvc::grad(alpha2))*(2*rho/(rho1 + rho2))
    %\subsubsection{Continuity Error}
%\subsection{Pressure Correction}    
\subsection{Energy equation}
\subsubsection{Overview}
\label{3-energy-section}
This section details the energy equation used and its implementation. The energy is modelled using the enthalpy method \cite{meltingReview2017}. The method involves solving equation \ref{3-enthalpy-equation}

\begin{equation}
\label{3-enthalpy-equation}
\rho \left(\partial_{t}H + \nabla \cdot \boldsymbol{U} H \right) = \nabla \cdot (k\nabla T)
\end{equation}

where enthalpy, $H$, obeys

\begin{subequations}
\label{3-enthalpy-int}
\begin{eqnarray}
&& H = \int c_p dT \\
&& H = \begin{cases}
      c_{p,\ gas} \cdot T  & \text{gas region}\\
      c_{p,\ solid} \cdot T  & \text{metal region where}\ T \leq T_{melt}\\
      c_{p,\ liquid} \cdot T + \alpha_2 L & \text{metal region where}\ T > T_{melt}
    \end{cases}
\end{eqnarray}
\end{subequations}

for a latent heat $L$. It is possible to split $T_{melt}$ into $T_{liquidus}$ and $T_{solidus}$ for alloys. However, an abrupt isothermal transition between phases with a two-state transition was found to integrate easier with the rest of the solver. For instance, a clear melt front allows easier assessment of benchmarks for the solver. Additionally, the ultra-thin-walled tube welding in Chapter 6 requires clear solid and liquid regions and mushy transition region between the phases causes issues with this.

Enthalpy can be solved for directly - this is method used in \ITPCF and was explored for the present work. It was found that solving for $T$ with the substitution $H = c_p T$ was faster and simpler. For instance, solving for $H$ can require multiple loops - this isn't the case when solving for $T$. Further, for the present application, solving for $H$ achieved the same results as solving for $T$ but with longer computation times. Given the long simulations and large $\mathcal{O}(10^{2})$ batch jobs covered in Chapters 4 and 5, this becomes a significant disadvantage. Therefore, due to the advantages of the substitution $H = c_p T$, it was used. To implement it, two additional variables $\rho c_p$ and $\boldsymbol{\rho \hat{U} c_p}$ are introduced and are defined as:

\begin{equation}
\label{3-rhoCp-sum}
\rho c_p = \sum_{i=1}^3 \alpha_i \rho_i c_{p,i},
\end{equation}
\begin{equation}
\label{3-rhoPhiCp-sum}
\boldsymbol{\rho \hat{U} c_p} = \sum_{i=1}^3 \boldsymbol{\alpha_{i} U_{\alpha}} \cdot \rho_i c_{p,i}. 
\end{equation}

Notice the $\alpha_i$ fraction and corresponding $\boldsymbol{\alpha_{i} U_{\alpha}}$ field are pre-packed within these two additional terms. This allows for one equation covering the whole domain with a reduced number of variables. For variations in density and specific heat capacity within a phase, a \vsf is required. Due to this, $\boldsymbol{\rho \hat{U} c_p} \neq \boldsymbol{U} \cdot \rho c_p$. This is because $\boldsymbol{\rho \hat{U} c_p}$ is a \ssf which cannot be multiplied by a \vsf directly. Although, this is technically possible to do and there are tools in \of such as \emph{interpolate} and \emph{reconstruct} that deal with this. However, an issue comes from ensuring consistency with including the compression fluxes when constructing $\boldsymbol{\alpha_{i} U_{\alpha}}$ that are constructed using the $\alpha_i$ object and $\boldsymbol{\rho \hat{U} c_p}$ which would need to be constructed with a separate $\rho c_p \alpha_i$ object. To do this, a parallel $\rho c_p \alpha_i$ equation would be required hence you cannot directly multiply the objects. Given this, $\boldsymbol{\rho \hat{U} c_p}$ is created from a \ssf ($\boldsymbol{\alpha_{i} U_{\alpha}}$) and two \ds variables ($\rho$ and $c_p$). Whereas, $\rho c_p$ is \vsf already due to its $\alpha_i$ constituent. Thus, $\rho c_p$ can use a \vsf for $\rho$ and $c_p$ enabling variations of density and specific heat capacity within the phase. Due to this, the temporal change term and the convective change term for the material derivative of $\rho c_p T$ use different values for $\rho$ and $c_p$. This was judged to be acceptable given the advantages of integrating variations in density and specific heat capacity. However, a boolean switch is provided to revert to \ds for $\rho_i$ and $c_{p,i}$.

Finally, to complete the enthalpy formulation, sources terms are required for latent heat ($S_{Latent}$) and GTAW ($Q_{GTAW}$). Here, the term GTAW is used in this context for the source term that mimics the heat inputted to the work piece during Gas Tungsten Arc Welding. This is because the arc physics are not simulated and therefore neither is the main heat transfer mechanism of current flow into the work piece \cite{murphy2015perspective}. Thus terms such as \emph{arc} or \emph{welding heat input} are misleading and \emph{heat source} isn't specific. These two terms are covered separately in sections \ref{3-latent-heat-term} and \ref{3-GTAW-source-term} respectively. Through incorporating these source terms with equations \ref{3-enthalpy-int}, \ref{3-rhoCp-sum} and \ref{3-rhoPhiCp-sum}, equation \ref{3-enthalpy-equation} can be rewritten as:  

\begin{equation}
\label{3-energy-equation}
\rho c_p \frac{\partial T}{\partial t} + \nabla \cdot (\boldsymbol{\rho \hat{U} c_p} T) - \nabla \cdot (k\nabla T) = - S_{Latent} + Q_{GTAW}
\end{equation}

where the non-linear terms are collected on the right has side as source terms. The implementation of equation in \of is shown in Listing \ref{3-code-T-eqn} in Appendix \ref{ap-listings}.

% effective diffusion implicit term subtracted for stability
% convective term div is zero in the solid region as no convective flow

    % One solve sufficient
    \subsubsection{Phase change}
    \label{3-alpha phase-change}
    \numParagraph{Overview}
    Given phase change can physically only occur within the metal region, a geometric field is used to isolate the metal from the gas. To recap, the term `geometric field' is used in this thesis to describe code features written to select specific regions of the mesh for operations. It is not related to a specific \of object or class and is used purely to describe the implementation. In addition to isolating the metal region, the geometric field, $g(\alpha_1, \alpha_2, \alpha_3)$, is defined to avoid spurious interface effects in regions where the phase fraction sum may (unphysically) exceed 1 due to phase changes. This issue is exacerbated in cells where three phases are present. An effective field was found to be regions with less than 25\% gas phase and at least 50\% liquid or solid metal. This can be expressed mathematically as
        \begin{equation}
        \label{3-geo-region-dmdt}
        g(\alpha_1, \alpha_2, \alpha_3) = \begin{cases}
                    1, & \text{if}\ \alpha_1 < 0.25 \land (\alpha_2 \geq 0.5 \lor \alpha_3 \geq 0.5) \\
                    0, & \text{otherwise}        
                    \end{cases}.
        \end{equation}    
        
    With the phase change candidate region isolated, a method to change the phase is required. In a popular method proposed by Brent \emph{et al.} \cite{brent1988enthalpy}, equation \ref{3-energy-equation} (with $Q_{GTAW}$ omitted) is written using notation style from \cite{patankarBook} as 
    
    \begin{equation}
    \label{3-patankar}
    a_c H_c = \sum a_{nc} H_{nc} + a_c^0 H_c^0 + d 
    \end{equation}
    % n is neighbouring cells
    % explain what p is etc ...
    Where the subscript $c$ denotes a cell with centre point $c$ and $n$ the neighbouring cells - in two dimensions these would be north south east and west.
    This enthalpy equation can be updated as  
    \begin{equation}
    \label{3-brent-discrit}
    [\Delta H_c]_{k+1} = [\Delta H_c]_k + \frac{a_c}{a_c^0} [\{H_c\}_k - c_p F^{-1} \{\Delta H_c\}_k] 
    \end{equation}
    where $\Delta H_c$ is the latent content of a cell with centre c and $H_c$ is its enthalpy content, $a_c$ and $a_c^0$ are the coefficients of finite volume discretization, $k$ is the iteration level and $c_p$ is the specific heat capacity. The variable $F^{-1}$ is the inverse of the latent heat function which in this case is $= T_{melt}$ (this term depends on the type of model used see \cite{chakraborty2001generalized} for other options). From \cite{brent1988enthalpy}, equation \ref{3-brent-discrit} is rewritten for the isothermal phase change in the present application as 
    % Ask online
    % through explicit discretization \cite{fvOpsolidificationmeltingsource}. Here for a small time $\Delta t$ there is negligible spacial changes are assume and thus the divergence terms go to zero. 
    \begin{equation}
    \label{3-brent-discrit-subt}
    [\Delta H_c]_{k+1} = [\Delta H_c]_k + \frac{a_c}{a_c^0} c_p \{ T - T_{Melt} \}. 
    \end{equation}
    
    For a cell transforming from solid to liquid, the enthalpy changes by approximately $\Delta H \approx \alpha_2 L$ (where $c_{p,\ liquid} \approx c_{p,\ solid} \approx c_p$). Subbing these values into equation \ref{3-brent-discrit} (and neglecting $\frac{a_c}{a_c^0}$) gives

    \begin{equation}
    \label{3-brent-discrit-final}
    \alpha_2 L = 0 + c_p[T - T_{melt}]. 
    \end{equation}
    % \cite{fvOpsolidificationmeltingsource} 
    
    This is then used to update the liquid fraction as 
    \begin{subequations}
    \label{3-brent-alpha-update}
    \begin{eqnarray}
    \alpha_2^{k+1} = \alpha_2^{k} + \gamma \frac{c_p}{L}(T - T_{melt}) \\   
    \alpha_2^{k+1} = max(0, min(1, \alpha_2^{k+1}))
    \end{eqnarray}
    \end{subequations}    
    % bulk out these references from web of science
    where $\alpha_2^{k+1}$ is the new cell value for $\alpha_2$ constructed from its origional ($\alpha_2^{k}$) and an update term with relaxation factor $\gamma$. This method works very effectively in both general CFD codes \cite{brentUpdateCFD1, brentUpdateCFD2} and specifically in \of \cite{fadenOfEnthalpy2018, fvOpsolidificationmeltingsource}. This methodology can even be taken a step further to simply write a variable for the liquid fraction, $f_{liquid}$, with some linear function \cite{hummel2020brentStep, rosler2011brentStep} as 

    \begin{equation}
    \label{3-eq phase-change}
    f_{liquid} = \begin{cases}
                    1, & T > T_{melt}\\
                    0, & T < T_{melt}        
                    \end{cases}
    \ \text{(for isothermal)} \ 
    f_{liquid} = \begin{cases}
                    1, & T > T_s\\
                    \frac{T - T_s}{T_l - T_s}, & T_s > T > T_l \\
                    1, & T < T_l\\        
                    \end{cases}
    \ \text{(for alloys)}
    \end{equation}       
    
    where $T_s$ is the solidus temperature and $T_l$ the liquidus temperature of an alloy. As such it is the recommendation of the author to use it unless there is a specific application that requires a separate $\alpha_2$ equation; this is the case for the present work.  
    
    In order to simulate ultra-thin-walled tube welding, it was judged key to understand the phase evolution of solid-liquid-gas necessitating a VoF approach. Note, another advantage of VoF is additional source terms for additive manufacturing or MIG welding can be added to the $\alpha$ equation. To integrate phase change into a VoF approach a source term in equation \ref{3-full-alphas-equation} is required. This is not a trivial change and so a computational constant, $C_{pc}$, is introduced. Combing the computational constant and geometric field with equation \ref{3-brent-discrit-final} gives a full source term of
    % What about 
    \begin{equation}
    \label{4-eq-cpc}
    S_{\alpha_2} = C_{pc}\left(\frac{c_p(T - T_{melt})}{L}\right) \cdot g(\alpha_1, \alpha_2, \alpha_3)
    \end{equation}
    
    where $c_p$ is a \vsf and therefore automatically uses the specific heat capacity that corresponds to the phase fraction of the cell. This is due to how $\rho c_p$ is constructed in equation \ref{3-rhoCp-sum}, it can be divided by $\rho \equiv \sum_{i=1}^3 \rho_i$ to extract a phase weighted specific heat capacity $\alpha_i c_p$. Finally, $S_{\alpha_2}$ is then added as an explicit source term to the $\alpha_2$ equation (and subtracted from the $\alpha_3$ equation) discussed in Section \ref{3-alpha-section}. %Su Sp etc...
    %\footnote{As detailed in equation $\rho c_p$ is one variable}

\numParagraph{Phase change constant $C_{pc}$ and Darcy Constant $C_{d}$}
 The Darcy computational constant, $C_{d}$ introduced in equation \ref{4-eq-cd} and the phase change computational constant, $C_{pc}$ introduced in equation \ref{4-eq-cpc} have a large effect on the phase fraction evolution. $C_{d}$ is a scaling factor for the momentum sink term. When $C_{d} \geq \mathcal{O}(10^{3})$ it is usually sufficient to stop all flow where the solid phase fraction $\alpha_3 = 1$ yet it is typically $\mathcal{O}(10^{6})$ - $\mathcal{O}(10^{12})$. This is because the absolute value of $C_{d}$ will affect the gradient of the interface between where $\alpha_3 = 0$ and $\alpha_3 = 1$. Therefore the physical interpretation of $C_{d}$ is how `solid' the interface at the edge of the weld pool is.

$C_{pc}$ is a scaling factor for the phase change source term introduced in the previous section. In principle this could be assumed to equal one as all the melted material should come from prior solid material. However, most likely due to the numerical diffusion of VoF based solvers, some phase is `lost' during phase change thus the added liquid fraction that replaces the solid fraction needs to be scaled by $> 100\%$. The physical interpretation of $C_{pc}$ is that of preventing unphysical numerical diffusion. The situation is complicated by the fact $C_{d}$ affects how `solid' the interface is and thus how much solid is available to be turned into liquid. Naively it could be expected that there is a straightforward mathematical relationship between $C_{pc}$ and $C_{d}$ however this was found to be empirically incorrect. In fact their interaction is quite complex and a robust benchmarking process is undertaken in Chapter 5 to optimize their values.

    \subsubsection{Latent heat}
    \label{3-latent-heat-term}
    Through subbing equation \ref{3-enthalpy-int} into equation \ref{3-enthalpy-equation} a latent heat term, $S_{Latent}$ can be formulated as  
    \begin{equation}
    \label{3-latent-heat-eq}
    S_{Latent} = \rho L \frac{\partial \alpha_2}{\partial t} + \alpha_2 L(\nabla \cdot \boldsymbol{\rho U}).
    \end{equation}

% REMEMBER TO CHECK REFS HERE    
    Using the same notation outlined in equations \ref{3-patankar} to \ref{3-brent-alpha-update}, this term is oft linearized \cite{vollerSourceBasedLinear} as

    \begin{subequations}
    \begin{gather}
    S_{Latent} = S_C T + S_U \\
    \text{where} \ S_C =  [\Delta H_c]_k \frac{\partial F}{\partial T} \ \text{and} \ S_U = - S_C F^{-1} + [\Delta H_c]_k \left(\alpha_2^{k-1} - \alpha_2^{k}\right)
    \end{gather}
    \end{subequations}

    However, with \of equation \ref{3-latent-heat-eq} can be added directly in an \emph{fvScalarMatrix} as an explicit source term. Further, the convective (second) term is technically $= 0$ for isothermal phase change so it does not actually need to be implemented and is thus dropped. Finally, to smooth the implementation at metal-gas interface cells where there is a large difference in density, the global $\rho$ term in equation \ref{3-latent-heat-eq} is switched for a \vsf of $\rho_2$ with values only $\neq 0$ in the liquid region - this helps to prevent spurious currents. Thus, the final implementation for the latent heat term is
    \begin{equation}
    \label{3-latent-heat-eq-final}
    S_{Latent} = \rho_2 L \frac{\partial \alpha_2}{\partial t}.
    \end{equation}
    % separate linear term e.g. no convective term in OPENFOAM SIMULATIONS OF ISOTHERMAL PHASE-CHANGE IN THE ABSENCE AND PRESENCE OF SHRINKAGE
    % div term in fvOptions Solidification melting source

    \subsubsection{GTAW source term}
    \label{3-GTAW-source-term}
    \numParagraph{Overview}    
    The core scope of this work is to predict ultra-thin-walled titanium tube welding and not to explore the physics of arc welding. In this work, formulating the value and position of the inputted heat is vital. However, the route to formulation is of secondary importance; neglecting the physics calculations in favour of a calculable source term achieves the core scope in a far simpler route. Thus, heat input modes inseparable from the arc (joule heating and conduction from the arc) are reformulated into one volumetric heat source term of known value, $Q_{GTAW}$, that is added to the energy equation. %The derivation and implementation of this heat source term is covered in the next two sections. 
    % Goldak etc... a bit of history
    % move Q = VI eta

    Another requirement for the present application is being able to benchmark the heat source. Detailed further in Chapter 6, there are ample trial weld results available from research and development of ultra-thin-walled tube welding but no systematic or real time measurements deliberately designed to benchmark a multiphysics solver.  Due to their popularity, a Goldak-type source would provide candidate studies to benchmark against and as detailed in Chapter 2 they are apt for a volumetric heat source term and thus they are chosen for the present application. This implementation is detailed in the following sections. 

    % why not surface model? why volumetric
        \numParagraph{Derivation}
        In the present work, an `elliptic paraboloid' heat source model of the type detailed in \cite{paraboloid-paper} is used. This `elliptic paraboloid' model is a simplification of the general Goldak-type volumetric heat source that reduces the required parameters to the weld power, and weld width and depth (plus some computational factors). As discussed in Section \ref{2-sec-hs-bv}, whilst there are many Goldak-like heat source implementations for GTAW \cite{lindgren2011metal, Lundback2012} many require additional parameters which are hard to quantify for the present application where the full heat source is enclosed inside an orbital weld head and thus cannot be seen. The another key advantage here is that the weld power and weld width and depth are widely reported in both computational and experimental welding research - thus there is considerably flexibility in the validation of this model. 
%%%%% New comment
% put in others from table in lit review

The model in \cite{paraboloid-paper} is for a two phase ANSYS-FLUENT implementation thus it has been adapted to a three phase free surface in \of. Further there are a few flaws in \cite{paraboloid-paper} that are covered in detail in Chapter 5. Essentially, the heat source term works through specifying a geometric region in the shape of an elliptic paraboloid and applying a Gaussian heat distribution in that region; only a singular region is defined rather than a two or more. An elliptic paraboloid has a general shape governed by the equation
        
        \begin{equation}
        \label{3-paraboloid-eq}
        z = \frac{x^2}{a^2} + \frac{y^2}{b^2}.
        \end{equation}
    
\begin{figure}[!htb]
%            \centering
            \includegraphics[width=15cm]{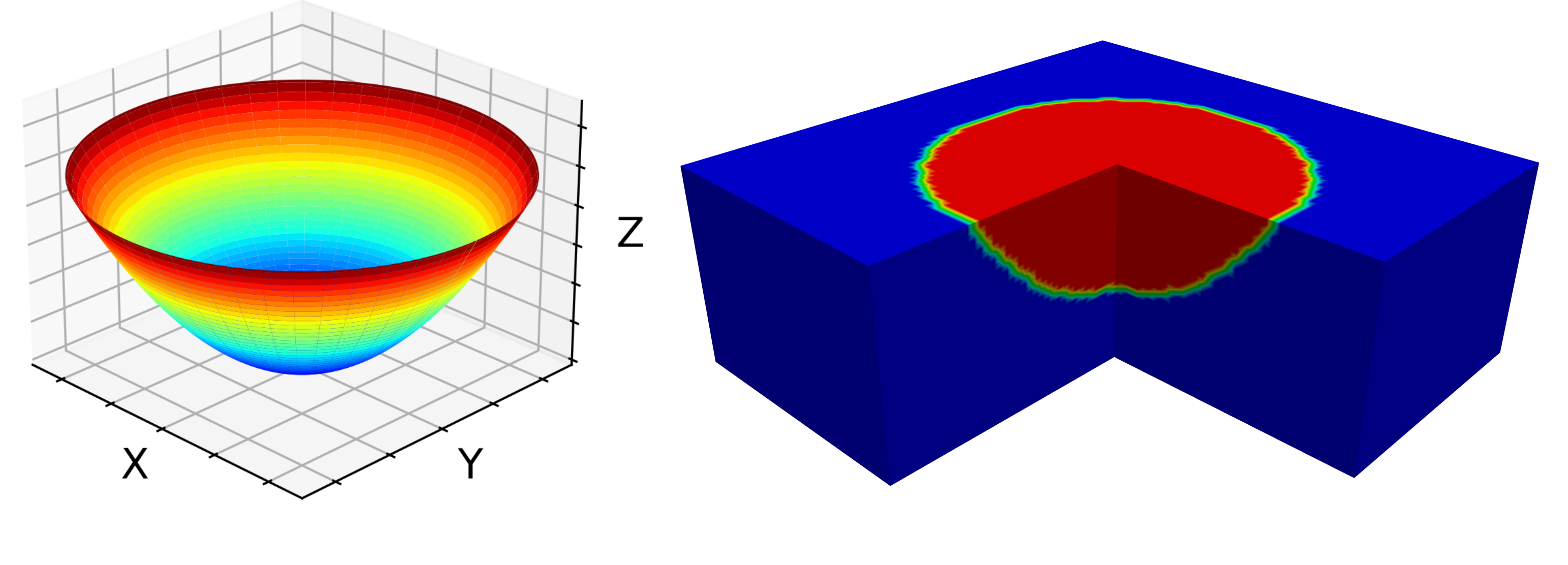}
            \caption{Elliptic paraboloid surface of the form of equation \ref{3-paraboloid-eq} (left) and an \of mesh with a geometric region defined an \of implementation of equation \ref{3-paraboloid-eq} (right). Colour is added to both images purely to aid visualization.}
        \label{3-fig-paraboloid-plt-of}
        \end{figure}

        This quadratic surface can be used to define a region of an \of mesh where the heat source is applied. Through altering parameters $a$ and $b$, the shape can be adjusted to match different sized heat sources e.g. for larger and smaller heat inputs. To illustrate this, Figure \ref{3-fig-paraboloid-plt-of} shows a plot of equation \ref{3-paraboloid-eq} next to an \of mesh with a geometric region defined by an \of implementation of equation \ref{3-paraboloid-eq}.
    
        With a geometric region defined, a heat source can be applied within the region. Starting through assuming the energy input has a Gaussian shape and an energy equal to the standard voltage multiplied by current and efficiency gives
    
        \begin{equation}
        \label{3-heat-source-IV}
        Q_{Source}(x_{dir}) = \eta V I = \int_{\infty}^{\infty} q(x = 0) e^{-Ax^2} dx = q(x = 0) \cdot \sqrt{\frac{\pi}{A}}        
        \end{equation}    
        
        in the $x$-direction, where $q(0)$ is the heat source intensity at the origin and $A$ is a constant. The same procedure applies in the $y$-direction with constant $B$. In the $z$-direction however, instead of a Gaussian shape, exponential decay is assumed:     
        
        \begin{equation}
        \label{3-heat-source-z-dir}
        Q_{Source}(z_{dir}) = \int_{\infty}^{\infty} q(z = 0) e^{-\sqrt{C} z} dz = q(z = 0) \cdot \sqrt{\frac{1}{C}}.        
        \end{equation}       
    
        To make the derivation clearer, a $D$ is defined as $D = C \pi$. Thus, $Q_{source}$ can be written as
        
        \begin{equation}
        \label{3-heat-source-const}
            Q_{Source} = q(0, 0, 0) \sqrt{\frac{\pi ^ 3}{ABD}}.
        \end{equation}
        
        Using the standard assumption for Goldak-type heat sources \cite{fariasEff, malik2008analysis, asaidOrbital} of the paraboloids accounting for 95\% of the heat input, ABD are estimated in \cite{paraboloid-paper} through taking 95\% of the heat source. Thus for the $x$-direction taking $q$ as 5\% of its maximum to be at $x = a$, mathematically equivalent to $q(a, 0, 0) = 0.05q(0, 0, 0)$, results in
        
        \begin{equation}
            0.05 q(0) = q(0) e^{-Aa^2}        
        \end{equation}
        which can be rearranged into
        \begin{equation}
        \label{3-heat-source-const-A}
            A = \frac{-\ln{0.05}}{a^2} \approx \frac{3}{a^2}
        \end{equation}
        with the same procedure applying to the $y$-direction/$B$. For $C$ however the $z$-direction term is linear. Thus $q(0, 0, c) = 0.05q(0, 0, 0)$ results in
    
        \begin{equation}
            0.05 q(0) = q(0) e^{-\sqrt{C} c}        
        \end{equation}
    
        which rearranged for $C$ gives
        
        \begin{equation}
            \sqrt{C} = \frac{-\ln{0.05}}{c} \approx \frac{3}{c}.
        \end{equation}
        
        Introducing a new variable, $l$ gives
        
        \begin{equation}
        \label{3-heat-source-const-C}
            D = \frac{-\ln{0.05}}{l} \approx \frac{3}{l} \text{ where } l = - \frac{c^2}{\ln{0.05} \pi}.
        \end{equation}
        
        Subbing these constants back into equation \ref{3-heat-source-const} gives
        \begin{equation}
        \label{3-heat-source-const-ABC}
            q(0, 0, 0) = Q_{Source} \sqrt{\frac{(\frac{3}{a^2})(\frac{3}{b^2})(\frac{3}{l})}{\pi ^ 3}} =  3\sqrt{3} \left( \frac{Q_{Source}}{a b \pi \sqrt{\pi l}} \right) =  C_{N} \left( \frac{Q_{Source}}{a b \sqrt{l}} \right)
        \end{equation}
        where $C_{N}$ is a constant equal to the gathered numerical factors. For implementation purposes it is prudent to add a computational constant, $C_{C}$, which $C_{N}$ is incorporated into. This term replaces $f_h$ in \cite{paraboloid-paper}. With axisymmetry in the x-y plane $a^2 = b^2 = ab = \omega^2$. Thus, through combining equations \ref{3-heat-source-IV}, \ref{3-heat-source-z-dir} and \ref{3-heat-source-const-ABC}, the final equation for the heat source is  
    
        \begin{equation}
        \label{3-heat-source-equation}
        q(x,y,z)  = C_{C} \left( \frac{Q_{Source}}{\omega^2 \sqrt{l}} \right) \cdot \exp\left(\ln{0.05}\left(\frac{(x^2 + y^2)}{\omega^2} + \frac{z}{l}\right)\right).
        \end{equation}        
    
        \numParagraph{Implementation}
        As terms such as $Q_{Source}$, $C_C$, $\omega$ and $l$ in equation \ref{3-heat-source-equation} are constants, the core implementation is applying the exponential term to a geometric mesh. Simply applying equation \ref{3-heat-source-equation} to a mesh in \of causes the heat source to extend to $\infty$ in all directions leading to floating point exceptions where $|q(x, y, z)| < 10^{-300}$. Further, small terms of the order of $<10^{-10}$ are computationally intensive and provide little (or negative) improvements in simulation accuracy. Thus, to implement equation \ref{3-heat-source-equation} in OpenFOAM, a geometric region where the heat source applies is employed.   
        
        Inspecting equation \ref{3-heat-source-equation} reveals that at $\omega^2 = x^2 = y^2$ and $z = l$ the term inside the exponential resolves to $\ln{0.05}$ in each direction. The exponential and natural log then cancel out leaving just the 0.05 value. Hence, the exponential term will resolve to $\geq 0.05$ when $x^2, y^2 \leq \omega^2$ and $z \leq l$ and will resolve to $< 0.05$ when $x^2, y^2 > \omega^2$ and $z > l$. Through temporarily only considering the exponential term ($q_{exp}$) in equation \ref{3-heat-source-equation} and writing 0.05 as $C_{cut}$, a geometric region $g(x, y, z)$ for the GTAW source can be defined as
        
        \begin{equation}
        \label{3-geo-region}
        g(x, y, z) = \begin{cases}
                    1, & \text{if}\ q_{exp}(x, y, z) \geq C_{cut} \\
                    0, & \text{otherwise}        
                    \end{cases}.
        \end{equation}
        
        Using equations \ref{3-heat-source-equation} and \ref{3-geo-region}, parameter $l$ matches the z-direction distance from the origin to the edge of the geometric region and parameter $\omega$ the $x$ and $y$-direction distances from the origin to the edge of the geometric region. Hence, by using $\omega$ and $l$ a geometric region can be defined dimensionally e.g. a \SI{5}{\milli\meter} radius and a \SI{3}{\milli\meter} depth. The value of the GTAW source term at its edge was chosen (arbitrarily) in \cite{paraboloid-paper} as 5\% of the peak. However, through switching 5\% to a generalized $C_{cut}$ another tuneable parameter is created. $C_{cut}$ is thus defined 
        
        \begin{equation}
            C_{cut} = \frac{q(\omega, \omega, l)}{q(0, 0, 0)}.           
        \end{equation}
        
        Further, given the constants $ABD$ in equation \ref{3-heat-source-const} are integrated into $C_C$ anyway, the values of $C_C$, $C_{cut}$, $\omega$ and $l$ can be matched to experimental results for weld sizes. In the ANSYS-FLUENT implementation used in \cite{paraboloid-paper}, with a fixed $C_{cut} = 5\%$, $l$ is identified as the weld pool depth and $\omega$ as one third of the fusion zone width. The identification and optimization of these parameters in the present work shall be explored in Chapter 5.  

        The geometric region is defined in \of by first creating a vector field of the coordinates for the centre of each cell in the mesh. This cell centre field is split into its component $x$, $y$ and $z$ vectors so that each can be handled individually. A \emph{position0} vector is then subtracted from each component to define an origin position for the source. This defaults to (0, 0, 0) but depending on the mesh the heat source may need centring elsewhere. In the $x$ and $y$ directions, a $velocity \cdot time$ term is subtracted to move the centre of the source around at a specified velocity. As GTAW heat sources typically require some ramp-up time as the weld pool forms, a pause time - $t_{pause}$ - value is subtracted from the current simulation time - $t$ - to keep the source stationary for a specified time. Whilst $t < t_{pause}$ a different equation is used with the velocity term omitted. This vector field in then substituted into $x$ in equation \ref{3-heat-source-equation}. Separating the exponential part of this equation in the $x$-direction creates equation \ref{3-expX-eq}. A code snippet with the implementation of equation \ref{3-expX-eq} is shown in Listing \ref{3-code-expX} in Appendix \ref{ap-listings}.       
        
        \begin{equation}
        \label{3-expX-eq}
            f(x) = exp\left(\ln{C_{cut}}\frac{(x_{cell centre} - x_{position 0} - x_{velocity} \cdot (t - t_{pause}))^2}{\omega^2}\right).
        \end{equation}

        As equation \ref{3-expX-eq} will return $C_{cut}$ at $x = \omega$, a simple conditional operation can be used to create the geometric region. Looping through each cell, those with a value greater than $C_{cut}$ are assigned a value of 1 and those with less than $C_{cut}$ are assigned a value of 0. This creates a region defined by equation \ref{3-geo-region}.  
        
\begin{figure}[!htb]
            \centering
            \includegraphics[width=15cm]{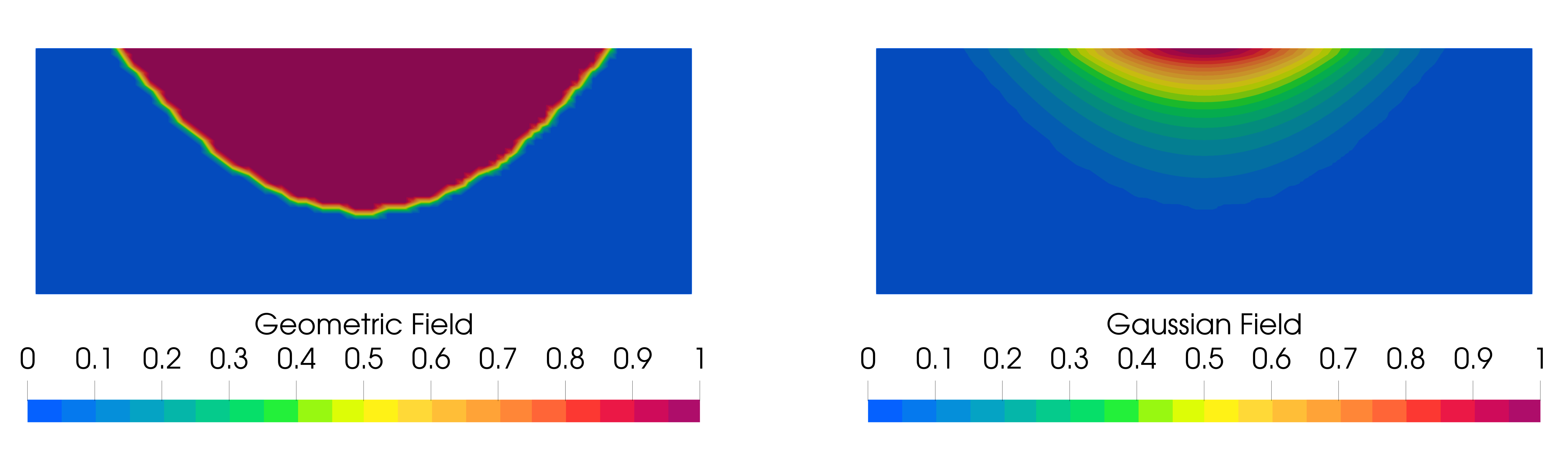}
            \caption{Elliptic paraboloid geometric field defined by equation \ref{3-geo-region} on an \of mesh (left) and a normalized Gaussian output field with $C_{cut} = 0.05$ generated from the geometric field and equation \ref{3-heat-source-equation} (right).}
        \label{3-fig-paraboloid-comp}
        \end{figure}

        With this geometric region defined, it is multiplied by the (normalized) exponential term defined in equation \ref{3-heat-source-equation}. Where the geometric field is 0, the exponential term is zero and where the geometric field is 1 the exponential term is unaffected. Figure \ref{3-fig-paraboloid-comp} shows the geometric field and the outputted field post multiplication.     
        % mesh comparisons for different cut offs

        Given the mesh is three dimensional, the equations presented thus far actually create a full ellipsoid constituted by an elliptic paraboloid reflected in the z direction at $position0$. This is not an issue for two phase flow where the source can be centred at a domain boundary but for three phase flow the source should only act on the solid and liquid phases. This is addressed through introducing a composite field $\alpha_{metal} = \alpha_2 + \alpha_3$ normalized between 0 and 1. Through multiplying equation \ref{3-geo-region} by $\alpha_{metal}$ the geometric field will be reduced inversely proportionally to its $\alpha_{metal}$ fraction e.g. where $\alpha_{metal} = 0,\ g(x, y, z) = 0$ (the source still needs to be positioned at the edge of $\alpha_{metal}$ in the same manner as experimental GTAW). Therefore the final implementation equation for the source term is written

        \begin{equation}
        \label{3-heat-source-equation-final}
        Q_{GTAW}(x,y,z)  = C_{C} \left( \frac{Q_{Source}}{\omega^2 \sqrt{l}} \right) \cdot \exp\left(\ln{0.05}\left(\frac{(x^2 + y^2)}{\omega^2} + \frac{z}{l}\right)\right) \cdot g(x, y, z) \cdot \alpha_{metal}.
        \end{equation}   

        Finally, it should be noted that a class based implementation was chosen over an \fvOp based implementation. \fvOp is a runtime selectable module that allows custom sources or constraints to be applied to the solver's governing equations. During development on the solver covered in this chapter, a class-based implementation was found to be superior. The class-based implementation involves creating a new class that solves the relevant equations and produces the source term $Q_{GTAW}$. This change was due to better computational performance and simpler interoperability with dynamic meshing. 
        
    \subsubsection{Temperature dependent thermophysical properties}
\begin{figure}[!htb]
            \centering
            \includegraphics[width=12cm]{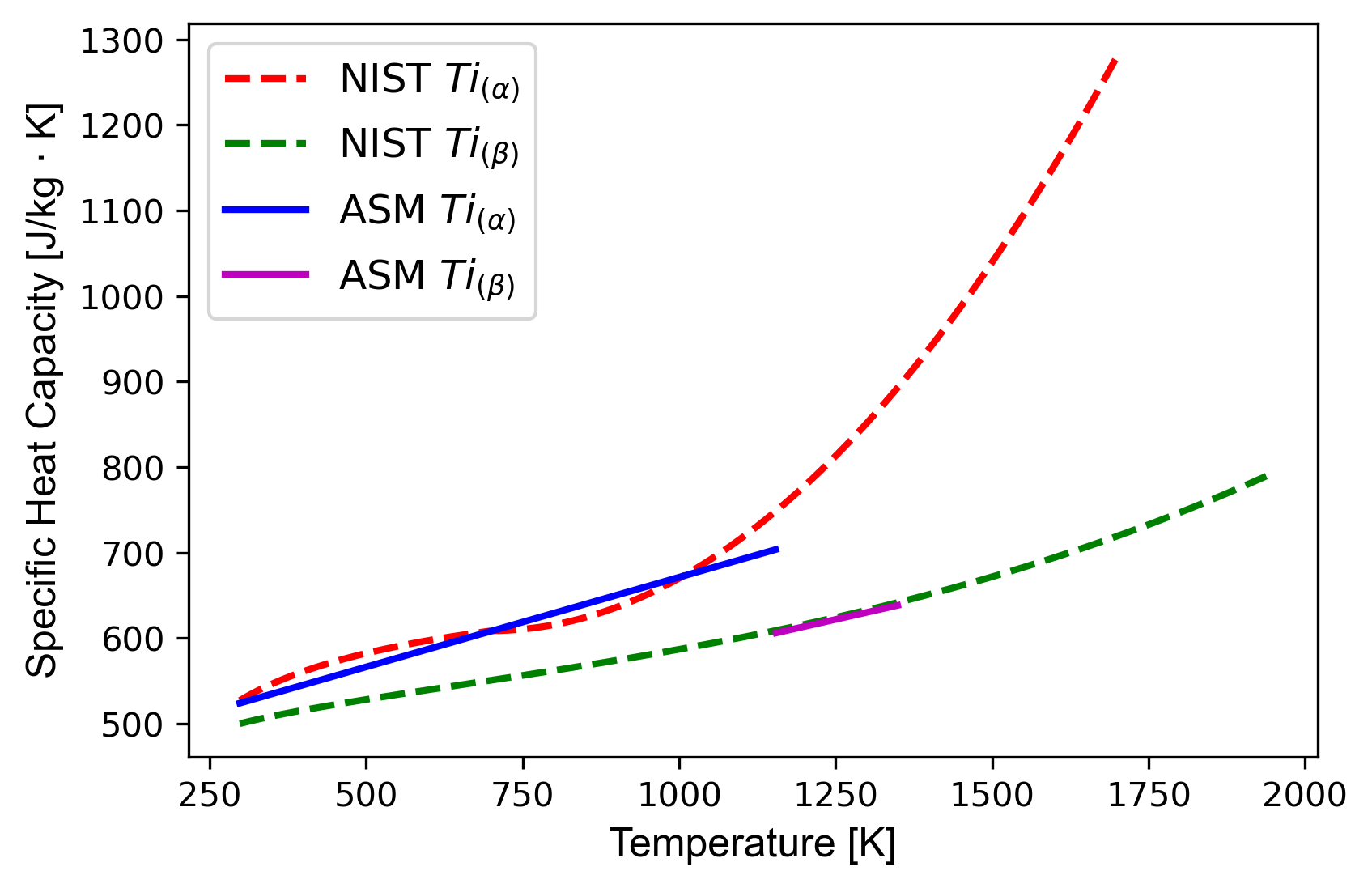}
            \caption{Comparison of the temperature dependence of specific heat capacity for $\alpha$ and $\beta$ titanium between the ASM forumation (equation \ref{3-temp-dep-cp-ASM-eq}) and the NIST formulation (equation \ref{3-temp-dep-cp-NIST-eq}). Note data for the ASM formulation stops at \SI{1350}{\kelvin}, below the melting point of titanium. Also, note the large change is $c_p$ for $Ti_{(\alpha)}$ between 298 and \SI{1700}{\kelvin} showing the potential large effect of temperature dependence on specific heat capacity.}
        \label{3-fig-cp-nist-asm}
        \end{figure}

    \numParagraph{Overview}
    \label{3-temp-dep}
    To improve the accuracy of the solver the temperature dependence on thermophysical properties is considered. A key advantage of the class structure detailed in Section \ref{3-class-structure} is the temperature inheritance from \emph{gtawThreePhaseMixture} enable many properties of the phases to be temperature dependent. Temperature dependence has been implemented for specific heat capacity, viscosity and density. Given the inherent impact on performance additional calculations have, these features can all be switched on or off.
        \numParagraph{Temperature dependent specific heat capacity}
        Given the temperature evolution in the gas phase is not of interest, the temperature dependence on specific heat capacity is implemented only for the metal phases. By defintion, at constant pressure a change in enthalpy between temperatures $T_1$ and $T_2$ can be defined as
        \begin{equation}
        \label{3-cp-dt}
            \Delta H = \int_{T_1}^{T_2} c_p(T) dT.
        \end{equation}
        This equation can be approximated as a power series. In their handbook on casting \cite{thermoPhysProp}, ASM uses the formulation:
        \begin{equation}
        \label{3-temp-dep-cp-ASM-eq}
            c_p(T) = a + bT + cT^{-2}
        \end{equation}
        and provides the values of $a$, $b$ and $c$ for a set of commonly used metals and their alloys. However, these values are limited and, for instance, do not include gallium. Conversely, the National Institute of Science and Technology (NIST) thermochemical tables \cite{nistJanaf} comprehensibly covers all elements in the periodic table and use an alternative formulation - the Shomate Equation:         
        \begin{equation}
        \label{3-temp-dep-cp-NIST-eq}
            c_p(T) = a + bt + ct^{2} + dt^{3} + et^{-2}
        \end{equation}
        % check
        where $1000t = T$. The purpose of having both ASM and NIST formulations of $c_p(T)$ implemented is that the ASM values are not a subset of the NIST values as the ASM include metal alloys of commercial interest. Thus, the solver has more flexibility with both included. Additionally, as shown in Figure \ref{3-fig-cp-nist-asm}, the difference between these formulations is non-trivial so a qualitative choice for the most apt in each situation must be made.  

        There are three elements to the implementation of $c_p(T)$: the \emph{mixture} object must update its $c_p$ member variable for the phase; the constants used to calculate $c_p(T)$ should be modifiable for different metals; and the prior two elements should work for both equations \ref{3-temp-dep-cp-ASM-eq} and  \ref{3-temp-dep-cp-NIST-eq}. This is achieved through, should a Boolean flag be set to \emph{true}, overwriting all calls to the \ds specific heat capacity of the phase with a function that returns a \vsf with the modified specific heat capacities. This is because the \ds is always ultimately multiplied by a \vsf through modifying the various calls $c_p(T)$ can be switched on and off. This member function has its own flag depending on whether equation \ref{3-temp-dep-cp-ASM-eq} or \ref{3-temp-dep-cp-NIST-eq} is used and is shown in Listing \ref{3-code-cpT-memberfunction} in Appendix \ref{ap-listings}.
        % ideally you want a look-up table implemented for ease of use
        % look up table works + want up to three sets of constants 0, 1, 2 for different regions
        No temperature argument is required as \emph{gtawIncompressibleThreePhaseMixture} has a \emph{gtawThreePhaseMixture} and access the same temperature object. At compile time the values for $a$ to $c$ in equations \ref{3-temp-dep-cp-ASM-eq} and \ref{3-temp-dep-cp-NIST-eq} are defaulted to one and for $e$ and $d$ in equation \ref{3-temp-dep-cp-NIST-eq} are defaulted to zero. This allows either formulation to be used to overwrite these values with an \of dictionary at run time. When this temperature object is raised to a power it can lead to floating point exceptions. To prevent this, the maximum value of either \SI{1}{\micro\kelvin} (named smallT\_) or the temperature cell value is taken. 
%        \rule{\textwidth}{0.4pt}
%        \rule{\textwidth}{0.4pt}    

        % for the energy equation!?!?
        As $c_p(T)$ is returned as a \vsf it has values on the entire mesh - both where the phase is and is not present. Thus, the cell average for the heat capacity of the phase across the mesh will not be equivalent to average heat capacity of the phase. This therefore obscures the results for average heat capacity when analysing solver results. Additionally, this change from \ds to \vsf is problematic for the energy equation that requires a \ds for $c_p$. However, this discrepancy does not affect the simulation too strongly so is judged acceptable. This issue is probably solvable but ultimately will require a time investment that is larger than the possible benefit. This is because the inbuilt multiphase features do not include $c_p$ so multiple modifications would be required.

        \numParagraph{Temperature dependent viscosity}
        The base \IFM solver handles kinematic viscosity, $\nu$, in a separate class from which \emph{incompressibleTwoPhaseMixture} is constructed from; \gIF uses the same structure. \of comes with several inbuilt viscosity models (e.g. Newtonian, power law etc.) that return a value for $\nu$ depending on the model chosen. Thus to implement temperature dependent viscosity a new derived class from the base viscosity model with an apt equation is required. In their handbook on casting, ASM suggest the following formula for the dynamic viscosity of liquid metals \cite{thermoPhysProp}
        
        \begin{subequations}
        \label{3-nu-Tdep}
        \begin{gather}
        \eta(T) = \eta_0 \cdot exp\left(\frac{2.65T^{1.27}}{RT}\right) \\
        \text{where} \ \eta_0 = \eta_{melt} \cdot exp\left(\frac{2.65T_{melt}^{1.27}}{RT_{melt}}\right) 
        \end{gather}
        \end{subequations}
        where $R$ is the ideal gas constant, $T_{melt}$ the melting temperature of the metal and $\eta_{melt}$ the dynamic viscosity at the melting temperature. Given the solid phase does not move and the viscosity of the gas phase does not evolve in a manner described by equation \ref{3-nu-Tdep}, only the liquid phase uses equation \ref{3-nu-Tdep}. However, as the solver requires the value for kinematic viscosity so equation \ref{3-nu-Tdep} is divided by the appropriate density ($\rho_2$) when implemented.  
        \numParagraph{Temperature dependent density}
        This section only considers the density of the solid phase. The temperature dependence on density for the liquid phase is covered in Section \ref{3-buoyancy} and the dependence for the gas phase is not of interest.   %The impact of this feature is minimal yet the implementation is trivial so it was added.        
        
        If free to thermally expand, a material of length $L_{0}$ at a temperature of $T_{0}$ can be estimated to expand to length $L_{0} + \Delta L$ at temperature $T_{0} + \Delta T = T$. Thus the length at temperature $T$ will be the original length the plus a proportional increase. This proportional increase is the coefficient of linear thermal expansion ($CLTE$):  
        \begin{equation}
        \label{3-CLTE}            
        L(T) = L_{0}(1 + CLTE). 
        \end{equation}

        For the three dimensional work piece, equation \ref{3-CLTE} will be extended to length width and height giving a volumetric change. However, the phase evolution, and therefore volume, is calculated using equation \ref{3-full-alphas-equation}. Thus, to implemented thermal expansion in \emph{gtawFoam}, the density of the solid phase is reduced proportionally to a volumetric change but the volume remains constant:  
        \begin{equation}
        \label{3-thermal-expansion-equation}
            \rho(T) = \frac{\rho(T_0)}{(1 + CLTE)^3}.
        \end{equation}

        \numParagraph{Temperature dependent thermal conductivity}
        The temperature dependence of thermal conductivity is also included. Similar to specific heat capacity, this is handled as essentially a member variable for a particular phase. Because of this, it is implemented (using the same member function) for all phases. The formula used is a modified version of the one provided by \cite{thermoPhysProp} to account for units of Kelvin and is presented in equation \ref{3-eq-t-dep-k}.

        \begin{equation}
        \label{3-eq-t-dep-k}
        \kappa(T) = a + b\left(\frac{T}{273.15}\right) + c\left(\frac{T^2}{273.15}\right)
        \end{equation}

    \subsubsection{Convection-radiation boundary condition}
    During welding, the work piece will lose heat to it surroundings. To simulate this, boundary conditions that match heat transfer processes are required. With a known temperature at the boundary, a Dirichlet boundary condition with a constant value is apt (\emph{fixedValue} in \emph{OpenFOAM}). Whereas with a known heat flux at the boundary, a Neumann boundary condition with a constant gradient is appropriate (\emph{fixedGradient} in \emph{OpenFOAM}). However, in order to simulate convective and radiative heat loss at the boundaries a Robin boundary condition is required \cite{vilums2011implementation}. Robin boundary conditions are linear combination of both Dirichlet and Neumann boundaries and can be expressed for a temperature field as   

    \begin{equation}
    \label{3-robin-equation}
        T_{Boundary} = f \cdot T_{\infty} + (1 - f) \cdot T_{Centre}
    \end{equation}

    \begin{figure}[!htb]
    \centering
    \includegraphics[width=10cm]{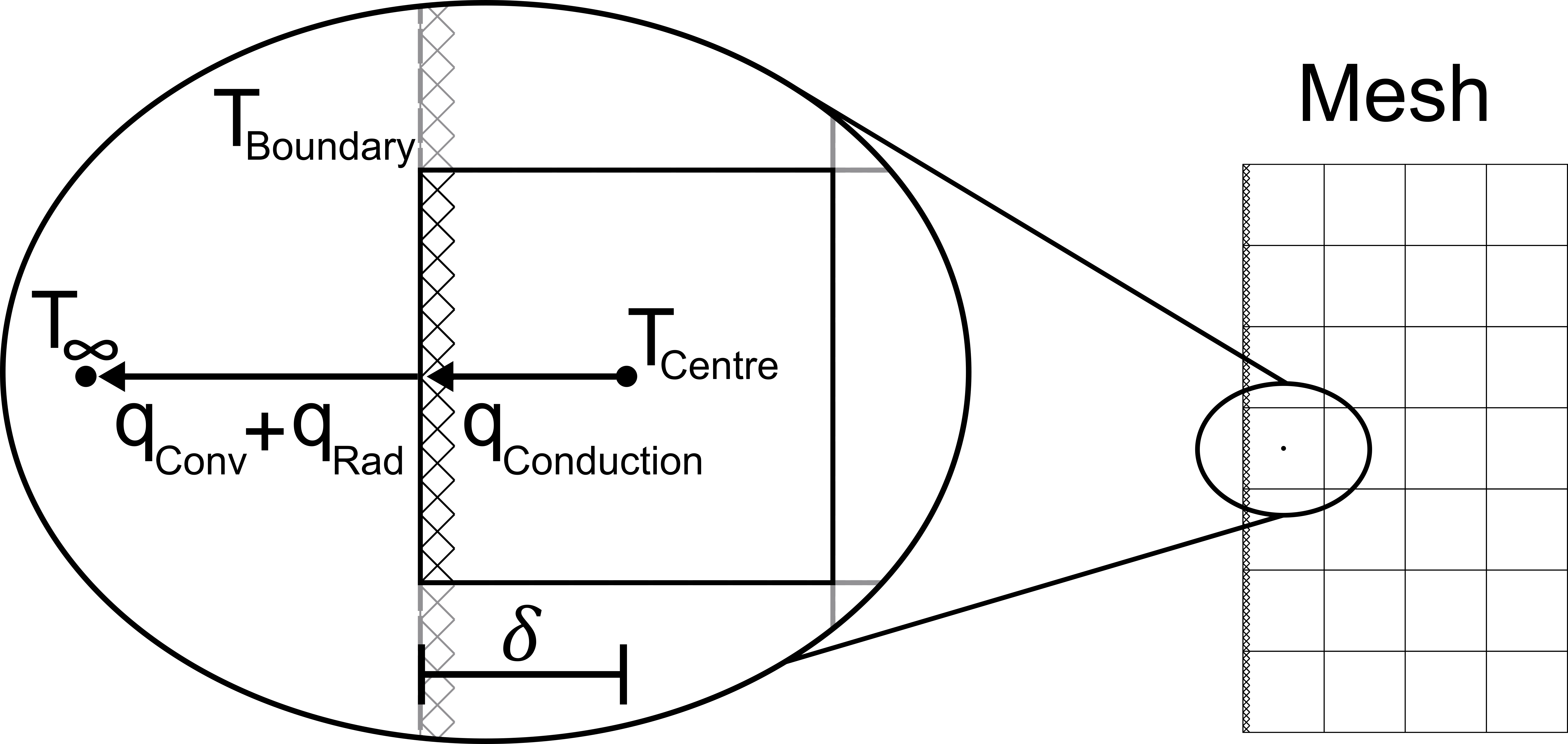}
    \caption{Diagram of boundary cell within a mesh for a temperature field showing the key values for the custom Robin boundary condition.}
    \label{3-fig-robin-bc}
    \end{figure}   

    Figure \ref{3-fig-robin-bc} shows these values on a CFD mesh. To implement this in OpenFOAM, a custom boundary condition library \gBC of an \of extension project \sFF \cite{swakGschaider2013} is employed. This is because the inbuilt \emph{mixed} boundary conditions are insufficient for handling Robin Boundary conditions. Note, in the latest versions of \of the \emph{mixed} condition is more robust and can better handle Robin boundary conditions. The implementation is separate to that of the rest of the solver as the boundary conditions are set in the individual cases and therefore are not part of the compiled solver executable. \gBC allows \emph{mixed} boundaries to be expressed as arbitrary functions - in the present case, this is variable $f$ in equation \ref{3-robin-equation}. To find $f$, the boundary is first expressed as a balance of internal conduction to convective and radiative heat loss: 
    
    \begin{equation}
    \label{3-boundary-eq-q}
    q_{Conduction} = q_{Convection} + q_{radiation}. 
    \end{equation}

    Substituting in the equations for conduction, convection and radiation into equation \ref{3-boundary-eq-q} gives  

    \begin{equation}
    \label{3-cond-conv-rad-equation}
    k \frac{\partial T}{\partial n} = h_{Conv} (T_{\infty} - T_{Boundary}) + \epsilon \sigma (T_{\infty}^4 - T_{Boundary}^4)
    \end{equation}

    where $h_{Conv}$ is the convective heat transfer coefficient, $\epsilon$ is the emissivity factor and $\sigma$ is the Stefan-Boltzmann constant. The thermal conductivity $k$ has to be set to match the same value as the solver. This convective-radiative boundary condition is well known in GTAW simulation research. However, for implementation in \of the boundary has to be expressed in the form $T_{Boundary} = X$. To achieve this, the following two substitutions are used: 
        
    \begin{equation}
    \label{3-boundary-eq-k}
    k \frac{\partial T}{\partial n} = k \frac{(T_{Boundary} - T_{Centre})}{\delta}
    \end{equation}

    \begin{equation}
    \label{3-boundary-eq-rad}
    \begin{split}
    \epsilon \sigma (T_{\infty}^4 - T_{Boundary}^4) & = \epsilon \sigma (T_{\infty}^2 + T_{Boundary}^2)(T_{\infty} + T_{Boundary})(T_{\infty} - T_{Boundary}) \\
    & = h_{Rad}(T_{\infty} - T_{Boundary}).
    \end{split}
    \end{equation}
        
    Subbing in equations \ref{3-boundary-eq-k} and \ref{3-boundary-eq-rad} into equation \ref{3-cond-conv-rad-equation} results in
        
    \begin{equation}
    k \frac{(T_{Boundary} - T_{Centre})}{\delta} = h_{Conv} (T_{\infty} - T_{Boundary}) + h_{Rad}(T_{\infty} - T_{Boundary})
    \end{equation}
        
    and then gathering the terms produces
        
    \begin{equation}
    \left( \frac{k}{\delta} + (h_{Conv} + h_{Rad})\right) T_{Boundary} = (h_{Conv} + h_{Rad}) T_{\infty} + \left( \frac{k}{\delta} \right) T_{Centre}.             
    \end{equation}
        
    Rearranging for the form $T_{Boundary} = X$ results in
    \begin{equation}
    T_{Boundary} = \frac{(h_{Conv} + h_{Rad})}{\left( \frac{k}{\delta} + (h_{Conv} + h_{Rad})\right)} \cdot T_{\infty} +  \frac{\frac{k}{\delta}}{\left( \frac{k}{\delta} + (h_{Conv} + h_{Rad})\right)} \cdot T_{Centre}.
    \end{equation}
    
    Thus a Robin boundary condition in the form of equation \ref{3-robin-equation} is recovered:
    \begin{equation}
    \begin{gathered}
    \label{3-boundary-eq-robin}
    T_{Boundary} = f \cdot T_{\infty} + (1 - f) \cdot T_{Centre}, \;\;\;\; f  = \frac{1}{\left( 1 + \frac{k}{\delta (h_{Conv} + h_{Rad})} \right)} \\
    \text{where} \ h_{Rad} = \epsilon \sigma (T_{\infty}^2 + T_{Boundary}^2)(T_{\infty} + T_{Boundary}). 
    \end{gathered}
    \end{equation}
    
    Implementation wise, $h_{Rad}$ is tuned via the emissivity ($\epsilon$) as the other parameters are fixed. Whereas, $h_{Conv}$ is tuned directly.

\section{Chapter summary}
The mathematical details of a multiphysics model for GTAW simulation has been described. This contributes to achieving the second objective detailed in Chapter 1. Combining all of the equations covered in this chapter gives a final multiphysics model expressed as
\begin{equation}
\begin{gathered}
\partial_{t}\alpha_{1} + \nabla \cdot \boldsymbol{\alpha_{1} U} + \nabla \cdot (\alpha_{1} \alpha_{2} \boldsymbol{U_{c_{12}}}) + \nabla \cdot (\alpha_{1} \alpha_{3} \boldsymbol{U_{c_{13}}})  = 0 \\ 
\partial_{t}\alpha_{2} + \nabla \cdot \boldsymbol{\alpha_{2} U}  + \nabla \cdot (\alpha_{2} \alpha_{3} \boldsymbol{U_{c_{23}}}) = S_{\alpha_{2}} \\
\partial_{t}\alpha_{3} + \nabla \cdot \alpha_{3} \boldsymbol{U}  = S_{\alpha_{3}} \\
\partial_t \rho \hat{U} + \nabla \cdot \boldsymbol{\rho U} \hat{U} + S_{m}  = -\nabla p_{rgh} + (g \cdot \hat{x}) \nabla \rho  + F_{ST} + S_{d} + S_{buoyancy} + \mu \nabla^2 \hat{U} \\
\rho c_p \partial_t T + \nabla \cdot (\boldsymbol{\rho \hat{U} c_p} T) - \nabla \cdot (k\nabla T) = - S_{Latent} + Q_{GTAW}.
\end{gathered}
\end{equation}

Following the procedure outlined in Section \ref{3-solver-algorithm}, these equations can be solved for a given case to predict the results of the GTAW process. However, to have confidence in the predictions of this solver it needs to be benchmarked against known results; this is the subject of Chapter 5.

%% file: chapters/thesis-chapter-5.tex
\lhead{Chapter 5: Benchmark cases}
\section{Introduction}
This chapter details the performance of the \gIF solver compared to various benchmark cases. Each section in the chapter addresses a particular element of the solver. Table \ref{5-tab-benchmarks} details the materials used for the benchmarking showing the variety in their thermophysical properties. Unless otherwise stated, these thermophysical properties are from \cite{thermoPhysProp}. For relevance, the popular benchmark cases in the literature are covered with additional cases included to validate specific features. The benchmarks are a combination of experimental and computational works. For the experimental benchmarks, there is an inherent uncertainty in the melt front position due to the experimental practice. Therefore, a reasonable margin for error is assumed so that a simulation that visually matches experimental results closely was judged to be `perfectly matched'. The melt fronts - defined as the region where $0.4 \leq \alpha_2 \leq 0.6$ - are all extracted using the method detailed in Appendix C. As with all \gIF examples in this thesis, $\alpha_1$ refers to the gas phase fraction, $\alpha_2$ to the liquid phase fraction and $\alpha_3$ to the solid phase fraction. Table \ref{4-tab-bc} lists the abbreviations used for various boundary conditions used in the chapter. The benchmark cases are available on GitHub at \url{https://github.com/WillYeadon/thesis} in the cases directory.  

\begin{table}[h]
\caption{\label{5-tab-benchmarks}Summary of the thermophysical properties of the benchmarking cases covered in this chapter.} 
\centering
\begin{tabular}{c c c c c}
\toprule
Material & $T_{melt}$ & $\rho$ & $c_p$ & $k$ \\
\hline
Gallium\rule{0pt}{2.6ex} & Low & Medium & Low & Medium \\
Tin & Medium & Medium & Low & Medium \\
Water & Low & Low & High & Low \\
Aluminium & Medium & Low & Medium & High\\
Bismuth & Medium & High & Low & Low \\
304 Stainless Steel & High & High & Medium & Medium \\
316 Stainless Steel & High & High & Medium & Medium \\
Titanium & High & Medium & Medium & Medium \\
\bottomrule
\end{tabular}
\end{table}

\begin{table}[ht]
\setlength{\tabcolsep}{10pt}
    \centering
    \caption{\label{4-tab-bc} Boundary condition abbreviations}
    \begin{tabular}{l l}
    \toprule
    Boundary condition & Abbreviation \\
    \hline
    \emph{zeroGradient}\rule{0pt}{2.6ex} & $\partial_n = 0$ \\
    \emph{pressureInletOutletVelocity} & \emph{PIOV} \\
    \emph{fixedFluxPressure} & \emph{FFP} \\
    \emph{totalPressure} & \emph{TP} \\ 
    \emph{inletOutlet} & \emph{IO} \\
    \emph{noSlip} & \emph{NS} \\
    \emph{fixedValue} & value used e.g. $T_{hot}$ \\    
    \bottomrule
    \end{tabular}
\vspace{-0.5cm}
\end{table}

\section{Melting benchmark cases}
\label{4-melting-section}
\subsection{Gallium melting}
\label{4-melting-section-ga}
\subsubsection{Optimization of gallium melting}
\label{4-ga-opt}
In order to benchmark \gIF the popular benchmark case by Gau and Viskanta \cite{GauGaMeltExp} used for the `toy' model in Chapter 3 can be reused. The case outline is identical to Chapter 3 except three phase thermophysical properties are used. These properties are shown in Table \ref{4-tab-Ga-TPP}. The values for the computational factor for the phase change source term, $C_{pc}$, and the computational factor for the Darcy source term, $C_{d}$ have a large effect on the efficacy of the simulation. The `best' values for these factors were judged to be the ones that match the experimental data most accurately whilst also being robust in the face of changes to the case and solution method. The aim of using this combination of robustness and accuracy was to overcome possible `over-fitting' of the gallium melting case. In this context, over-fitting refers to a pair of computational factors that very accurately match one simulation case whilst performing poorly with others. To achieve the required predictability for general-case ultra-thin-walled tube welding and to create a robust solver for any GTAW procedure requires a solver that matches well with many cases.     
%\begin{center}
\begin{table}
\caption{\label{4-tab-Ga-TPP}Thermophysical properties of gallium used for benchmark case. Temperature dependent $\rho_3$, $c_p$ and $\nu$ were not used.}
\begin{tabular}{l l l l}
\toprule
Phase & Property & Value & Units\\
\hline
$\alpha_1$ & Density, $\rho_1$ & 1 & \si{\kilogram\per\metre\cubed} \\
 & Specific Heat Capacity, $c_{p,1}$ & 1000 & \si{\metre\squared\per\second\squared\per\kelvin} \\
 & Thermal Conductivity, $k_1$ & 0.02 & \si{\kilogram\metre\per\second\cubed\per\kelvin} \\
 & Kinematic Viscosity, $\nu_1$ & \num{1.48e-5} & \si{\metre\squared\per\second} \\
\hline
$\alpha_2$ & Density, $\rho_2$ & 6093 & \si{\kilogram\per\metre\cubed} \\
 & Specific Heat Capacity, $c_{p,2}$ & 381 & \si{\metre\squared\per\second\squared\per\kelvin} \\
 & Thermal Conductivity, $k_2$ & 32 & \si{\kilogram\metre\per\second\cubed\per\kelvin} \\
 & Melting Point, $T_{melt}$ & 302.93 & \si{\kelvin} \\
 & Reference Temperature, $T_{ref}$ & 302.93 & \si{\kelvin} \\
 & Latent Heat of Fusion, $L_f$ & 80160 & \si{\metre\squared\per\second\squared} \\
 & Volumetric Thermal Expansion Coefficient, $\beta$ & \num{1.2e-4} & \si{\per\kelvin} \\
 & Kinematic Viscosity, $\nu_2$ & \num{2.97e-7} & \si{\metre\squared\per\second} \\
\hline
$\alpha_3$ & Density, $\rho_3$ & 5910 & \si{\kilogram\per\metre\cubed} \\
 & Specific Heat Capacity, $c_{p,3}$ & 385 & \si{\metre\squared\per\second\squared\per\kelvin} \\
 & Thermal Conductivity, $k_3$ & 30 & \si{\kilogram\metre\per\second\cubed\per\kelvin} \\
 & Kinematic Viscosity, $\nu_3$ & \num{3.06e-7} & \si{\metre\squared\per\second} \\
%\hline
%All & Specific Heat Capacity & 381.5 & \si{\kilogram\per\metre\cubed} \\
\bottomrule
\end{tabular}    
\end{table}
%\end{center}
% what about the cases that you do not do
%\vspace{-0.5cm}

\begin{figure}[ht]
\centering
\includegraphics[width=10.5cm]{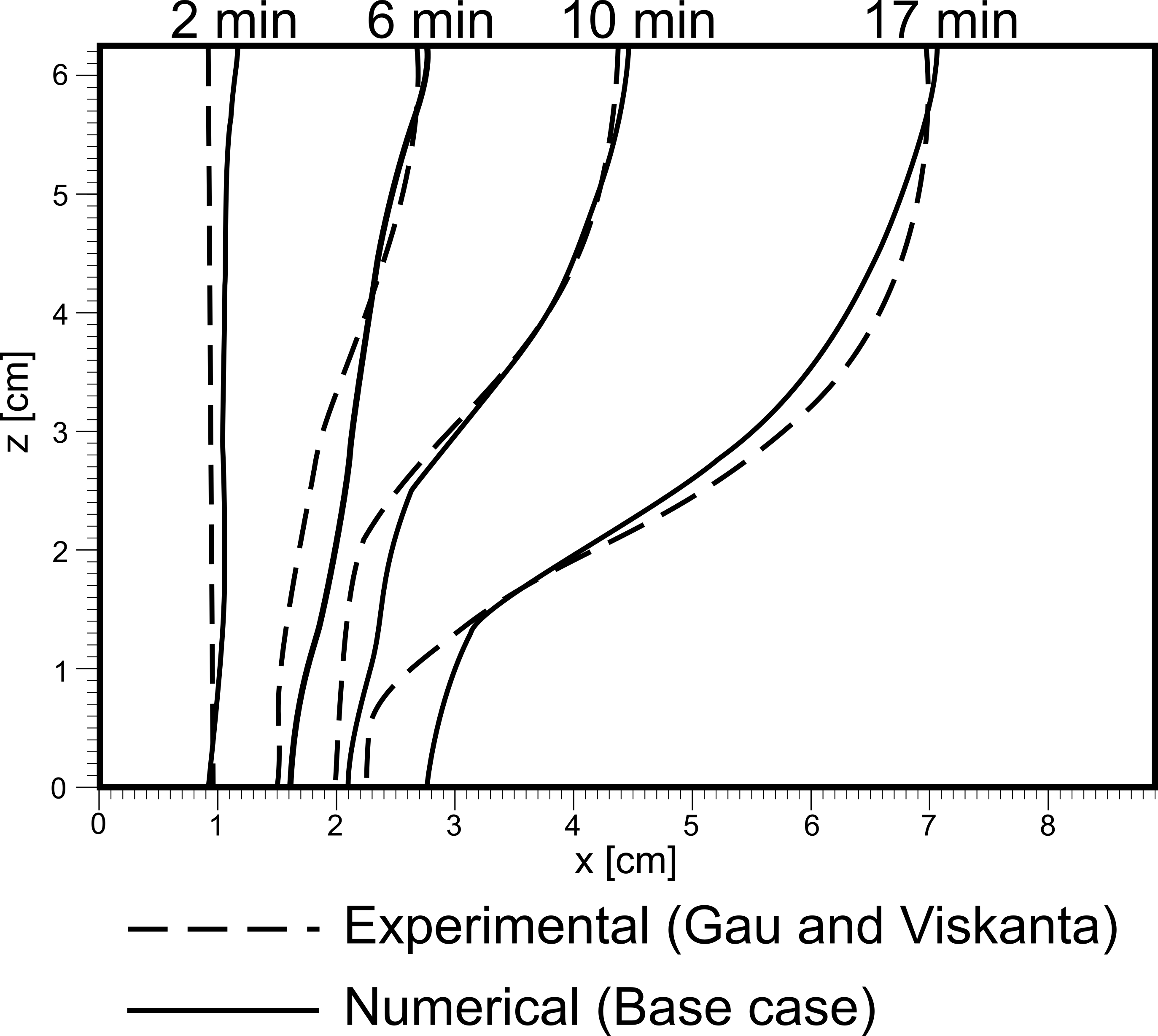}
\caption{Melt front position of the melting of gallium case \cite{GauGaMeltExp}. Dashed line shows experimental results and solid line shows numerical results for the base case. From left the right the times are \SI{120}{\second}, \SI{360}{\second}, \SI{600}{\second} and \SI{1020}{\second}.}
\label{4-fig-ga-melt-base}
\end{figure}

To achieve this, initially a `modified' simulation was run that included extra factors to match the experimental data closely; this `modified' case is termed the base case. A comparison between the base case and experimental results in shown in Figure \ref{4-fig-ga-melt-base}. In line with other research on gallium melting simulations \cite{brent1988enthalpy, rosler2011brentStep, brentUpdateCFD1}, comparison times of 120, 360, 600 and \SI{1020}{\second} were used. This base case was then rerun with \gIF to create a test case for assessing the computational factors. 

The phase space investigated for the computational factors was $10^{0} - 10^{2}$ for $C_{pc}$ and $10^{7} - 10^{11}$ for $C_{d}$. The step in both factors were the inclusive integers 1 to 10 multiplied by $10^n$. Thus for $C_{pc}$ the series is 1, 2, 3, 4, 5, 6, 7, 8, 9, 10, 20, 30, 40, 50, 60, 70, 80, 90, 100 and the next value would be 200. This step was chosen as it allows for the correct order of magnitude to be identified with some level of granularity in between each order of magnitude. Investigating such a broad parameter space necessitates a loss of precision to keep the overall cases investigated low. However, the observation of the broad range of constants used (for $C_{d}$) in the literature led to the conclusion that it was important to cover the largest possible range rather than the most granular. With a computational factor pair identified, to assess the quality of the match between the base and a test case for a given $C_{pc}$ and $C_{d}$ pair the following procedure was used:

\begin{enumerate}
\label{5-enum-1}
    \item \label{a} Bin values for $\alpha_2$ value in each ordered\footnote{All cases use an identical mesh so there is a 1:1 match between the cells. Thus cell \#95 in the base case will be in the same position as cell \#95 in the test case and the two can be directly compared.} cell into 11 discrete bins between $-0.05$ and $1.05$ with a bin width of 0.1.
    \item \label{b} Assign each ordered cell the value of the corresponding bins centre. For example, a cell with a $\alpha_2$ value in the $0.15-0.25$ bin would be assigned the value of $0.2$. 
    \item \label{c} Calculate the percentage of matching values in the base and test case for each cell. So if the value for $\alpha_2$ in a given cell is assigned 0.3 in the base case and 0.3 in the test case they would qualify as a match but values of 0.3 and 0.4 would not qualify as a match.
    \item \label{d} Repeat steps \ref{a} to \ref{c} for all four comparison steps (120, 360, 600 and \SI{1020}{\second}) and find the average of their percentage scores.
\end{enumerate}

\begin{figure}[htp]
\centering
\includegraphics[width=15cm]{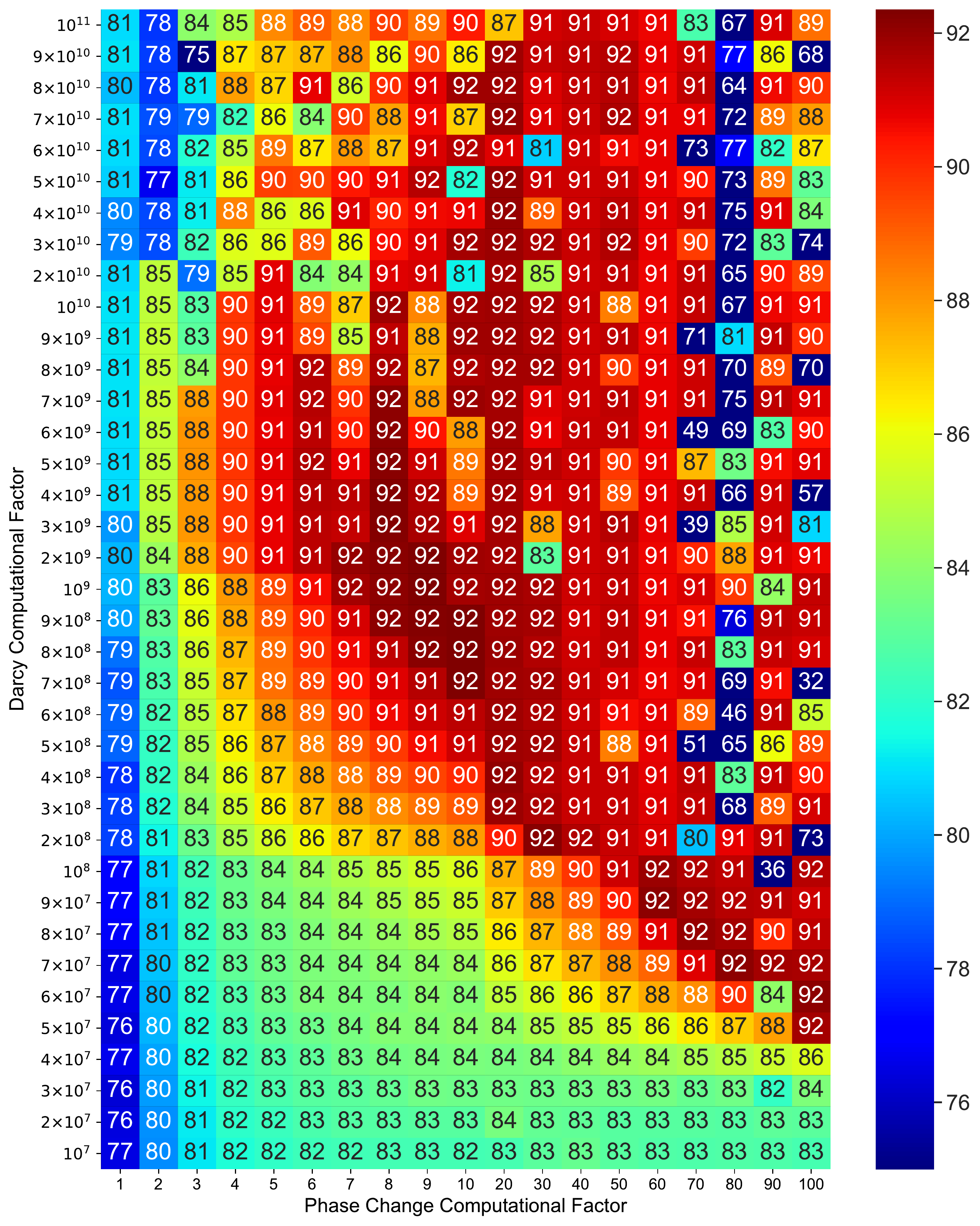}
\caption{Percent score for \gIF on gallium melting test cases calculated through steps 1 to 4 comparing to the base case shown in Figure \ref{4-fig-ga-melt-base}. To aid visualization, percent score is colourized between 75\% and 93\% as most values sit in this range. Phase space investigated is between $10^{0} - 10^{2}$ for $C_{pc}$ and $10^{7} - 10^{11}$ for $C_{d}$. Below this range, the percent score decreases and above it, instability and computation time becomes excessive. Instability refers to large score variation in adjacent cells. Despite the presence of high scores, unstable regions should be avoided as small changes in the case and solution method could dramatically affect the score.}
\label{4-fig-ga-melt-batch}
\end{figure}

Steps 1 to 4 were performed for all $C_{pc}$ and $C_{d}$ combinations in the phase space, shown in Figure \ref{4-fig-ga-melt-batch}. This averaging serves to allow better delimitation between computational factor pairs; given the size of the mesh this is important. An advantage of using this procedure is that is allows the melt front match to be accessed. Understanding the melt front is important in welding simulations as it will determine the shape of the weld and hence be important for weld integrity. For example, a wide and shallow weld could have the same melt fraction in the domain as a deep and narrow weld but they would have different melt fronts. Given the reasonably high volumetric expansion coefficient of gallium $(\beta = \num{1.2e-4})$ it is a suitable material to assess this.   

% Conversely, the bottom left region shows a reasonably smooth transition into higher scores and is therefore more stable and resistant to large variations with different cases and solution methods

Figure \ref{4-fig-ga-melt-batch} shows several computational factor pairs with similar high scores of $\approx 92\%$. However, the adjacent, second adjacent and third adjacent values for some of these high scores are dramatically smaller. This change (or lack thereof) in scores in adjacent values is termed `(in)stability' where a `stable' region is where adjacent scores are reasonably similar. In general, the instabilities occur at higher values of $C_{pc}$ and $C_{d}$. These variations combined with the possible variations in results from using different discretization methodologies were judged to make the solver less robust. Ideally, optimal values for the computational factor pair should work effectively for many different cases and a pair subject to large changes in results with slight variations should be avoided. Further, given the optimization is matching a `modified' simulation rather than experimental result assessing the stability is actually a key result from the optimization. Therefore, the stability of each cell was found through calculating the gradient in the x and y direction between adjacent cells in Figure \ref{4-fig-ga-melt-batch} and taking the absolute product of the two.

\begin{figure}[htp]
\centering
\includegraphics[width=15cm]{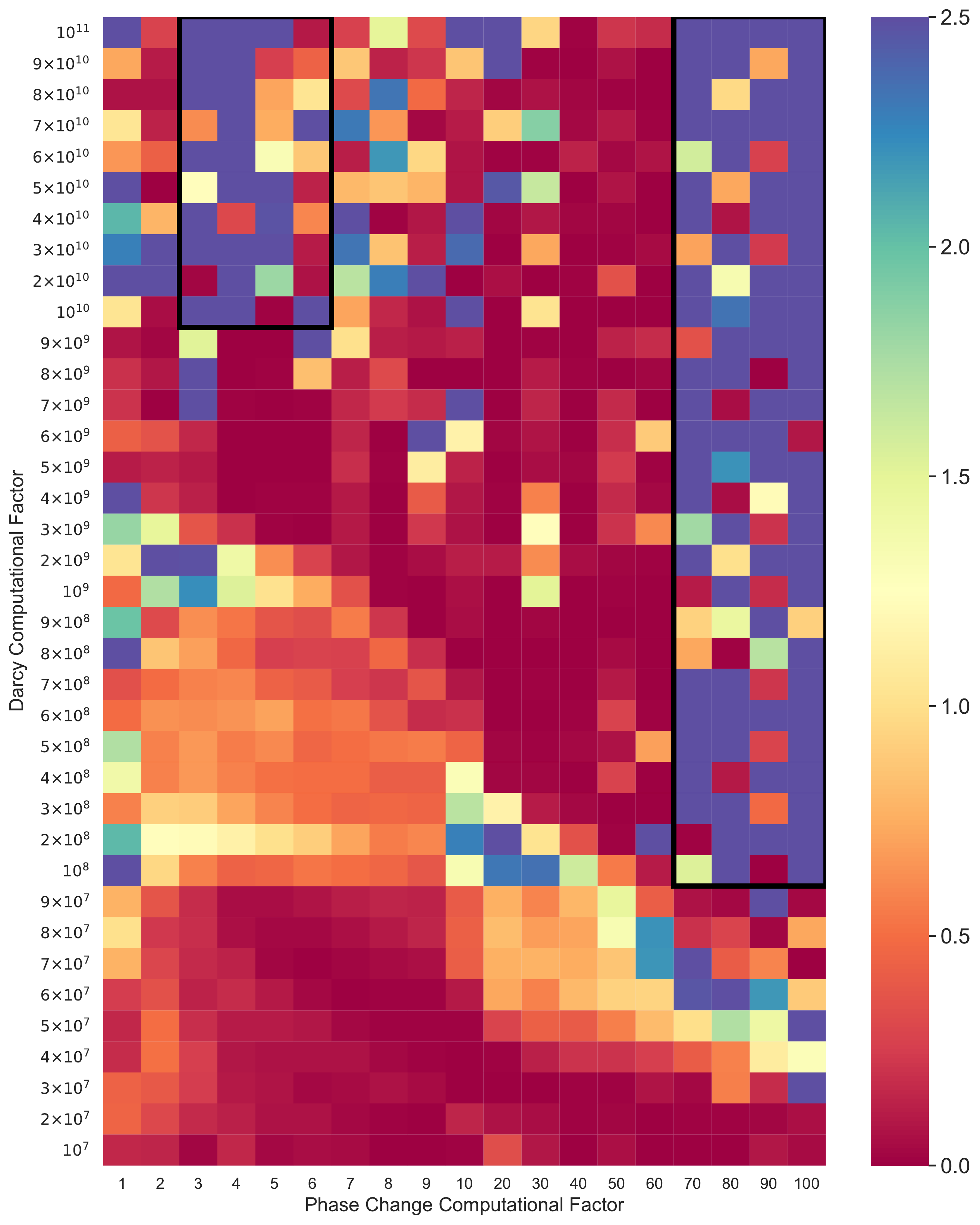}
\caption{Absolute product of $x$ and $y$ gradient of adjacent values in Figure \ref{4-fig-ga-melt-batch} for phase space between $10^{0} - 10^{2}$ for $C_{pc}$ and $10^{7} - 10^{11}$ for $C_{d}$. The gradients are found using the $numpy.gradient$ function in the numpy library for the $37 \times 19$ matrix of percent scores. The absolute gradient product allows for an assessment of stability as regions with a steady change in values have a lower gradient between cells and hence more stability. To aid visualization, the values are colourized between 0 and 2.5. The colour map is capped at 2.5 as values within the unstable regions bordered by the black boxes are significantly larger. The large values are caused by the corresponding large differences in adjacent cell values due to the instabilities at large computational constants. Figure \ref{4-fig-ga-melt-batch-gradZoom} shows natural logarithm of these boxed values.}
\label{4-fig-ga-melt-batch-grad}
\end{figure}

\begin{figure}[ht]
\centering
\includegraphics[width=10.5cm]{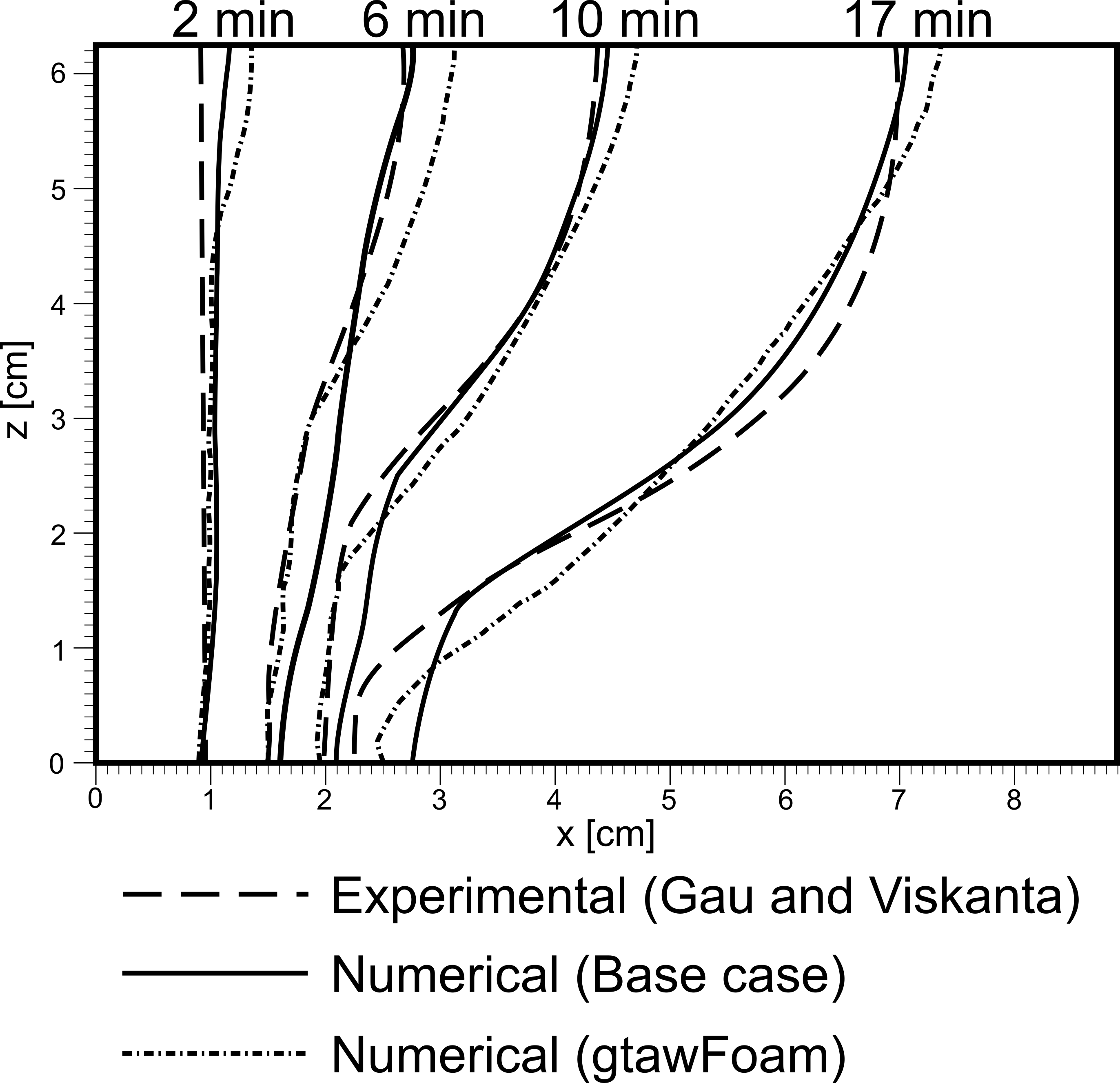}
\caption{Melt front position of the optimized simulation output from \gIF for the melting of gallium using $C_{pc} = 10$ and $C_{d} = \num{8e8}$ overlaid onto Figure \ref{4-fig-ga-melt-base}.}
\label{4-fig-ga-melt-opt}
\end{figure}

To find a final score, each cell in Figure \ref{4-fig-ga-melt-batch-grad} was multiplied by 10 and subtracted element wise from Figure \ref{4-fig-ga-melt-batch}. The $10 \times$ multiplication is used weight the gradient more heavily so that stable regions would be less affected. Figure \ref{4-fig-ga-melt-batch-weight} shows the result of this computation and indicates $C_{pc} = 10$ and $C_{d} = \num{8e8}$ is the optimum pair. With the $C_{pc}$ factor in this pair being on the edge of a transition (from 1 to 10 increment), confidence in the stability is high. The results from this pair is CLTE on Figure \ref{4-fig-ga-melt-base} in Figure \ref{4-fig-ga-melt-opt}.  

\FloatBarrier

\begin{figure}[htp]
\centering
\includegraphics[width=15cm]{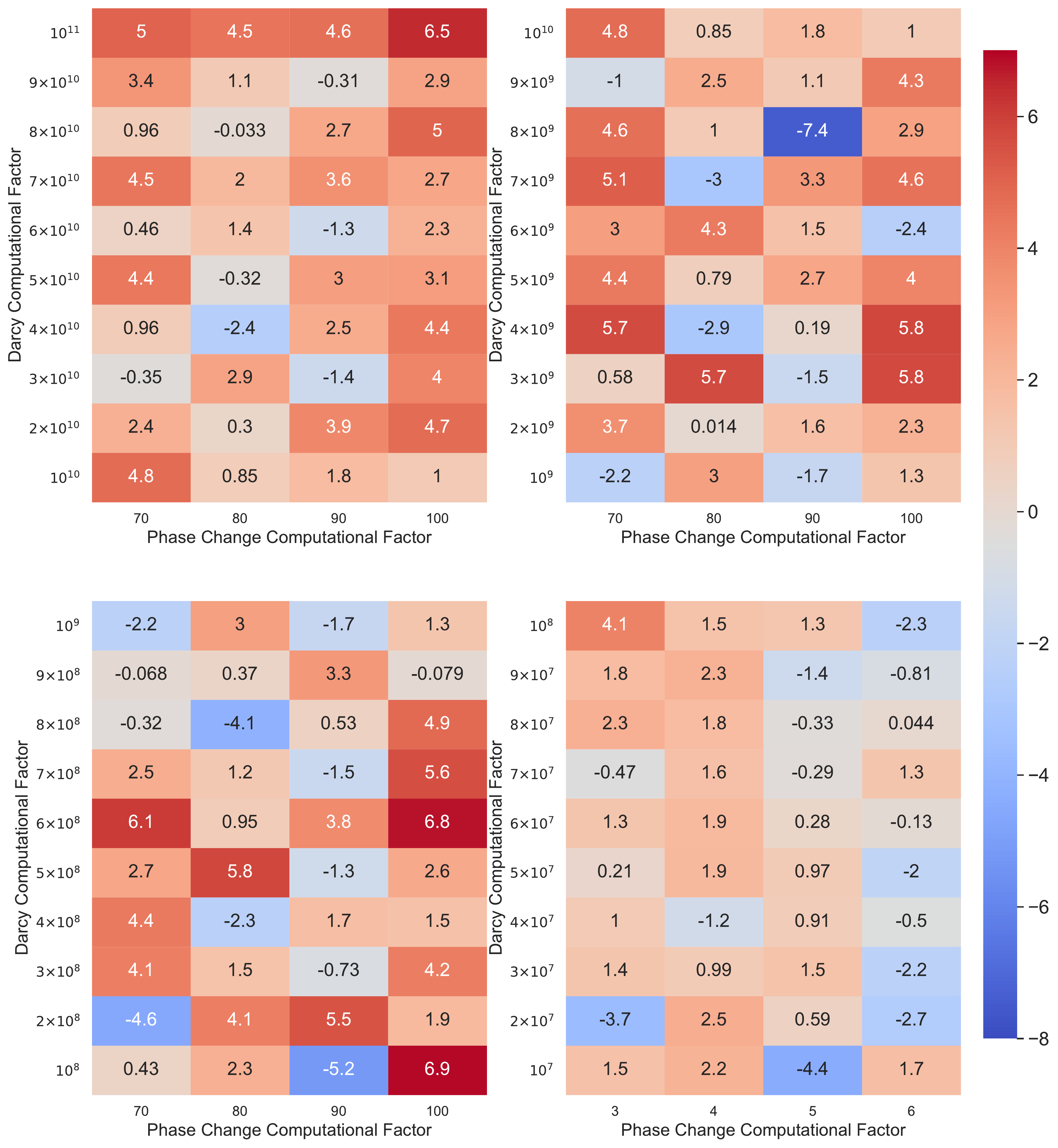}
\caption{Natural logarithm of absolute product of $x$ and $y$ gradient in regions highlighted in Figure \ref{4-fig-ga-melt-batch-grad}. for phase space between $10^{0} - 10^{2}$ for $C_{pc}$ and $10^{7} - 10^{11}$ for $C_{d}$. To aid visualization, the values are colourized between -8 and 7. Note how within these 40 cell subregions the values are often highly positive and highly negative compared to $ln(1) = 0$ which is a typical result for the natural logarithm of cells in Figure \ref{4-fig-ga-melt-batch}. The negative values are due to `islands' of stability between two adjacent cells that have very similar values that therefore have a gradient that is $< 1$ and hence a $ln(\text{gradient})$ that is negative. Whereas the positive values are due to steep changes between adjacent cells and therefore large gradients. Further, notice that the adjacent, second adjacent and third adjacent values for gradient vary greatly for a given cell. Therefore when viewing the subregions in their entirety the large positive values are not outliers but an integral constituent part of the subregion.}
\label{4-fig-ga-melt-batch-gradZoom}
\end{figure}

\begin{figure}[htp]
\centering
\includegraphics[width=15cm]{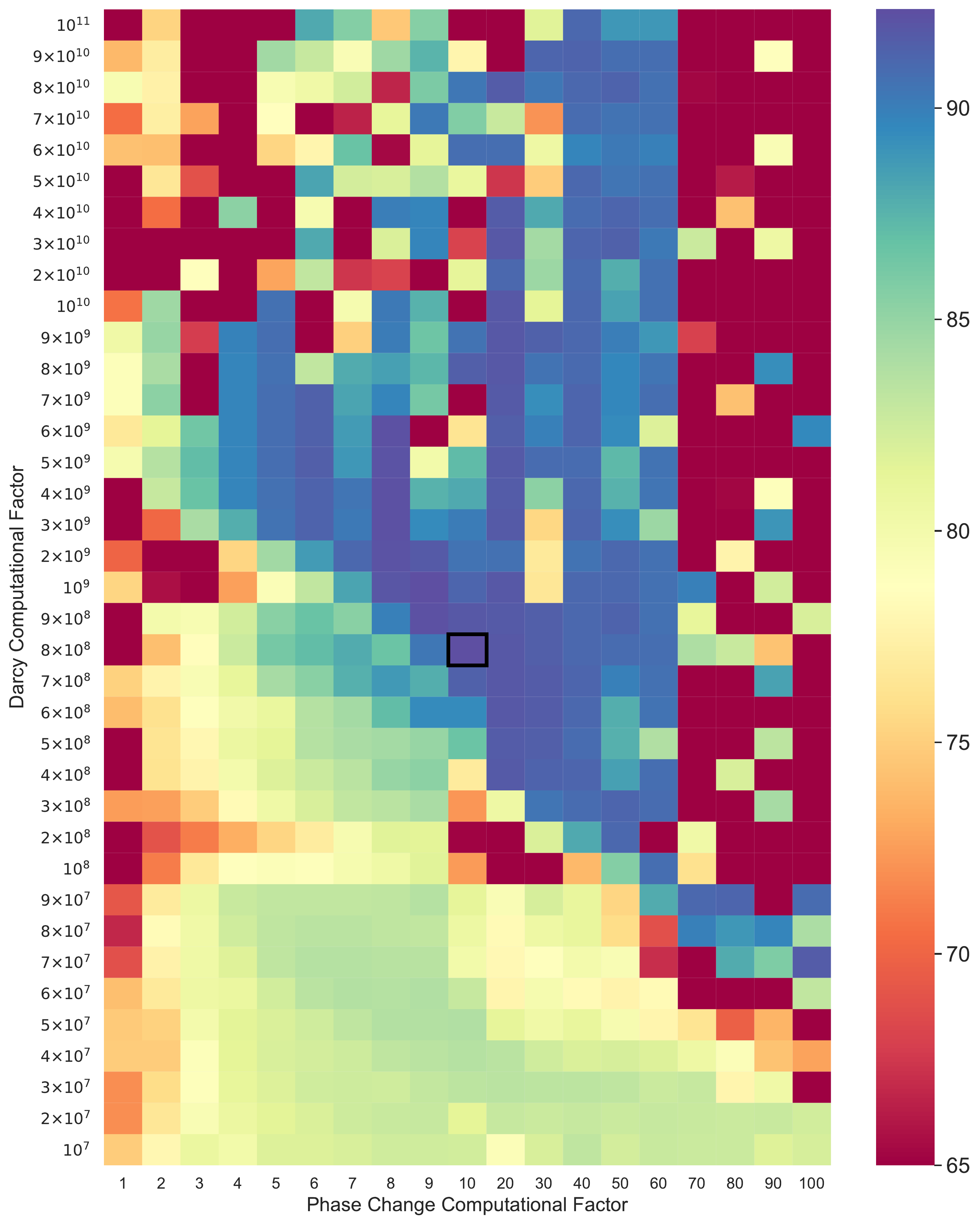}
\caption{Final weighed score of \gIF performance for gallium melting simulation between $10^{0} - 10^{2}$ for $C_{pc}$ and $10^{7} - 10^{11}$ for $C_{d}$. Here, the absolute gradient product values in Figure \ref{4-fig-ga-melt-batch-grad} are subtracted ten times from the percent score in Figure \ref{4-fig-ga-melt-batch}. This process effects unstable regions with high gradients considerably whereas stable regions with gradients close to zero are largely unaffected. Therefore the computational constant pair with a combination of a high percent score and a stable region can be identified; this pair is highlighted by a black box at $C_{pc} = 10$ and $C_{d} = \num{8e8}$. To aid visualization, the values are colourized between 65 and 93. The colour map is limited to $\geq 65$ as values in unstable areas such as those accentuated in Figure \ref{4-fig-ga-melt-batch-gradZoom} are considerably $\mathcal{O}(10^{3})$ negative.}
\label{4-fig-ga-melt-batch-weight}
\end{figure}

\FloatBarrier
\subsubsection{Analysis of the computational factors} 
\begin{figure}[ht]
\centering
\includegraphics[width=14cm]{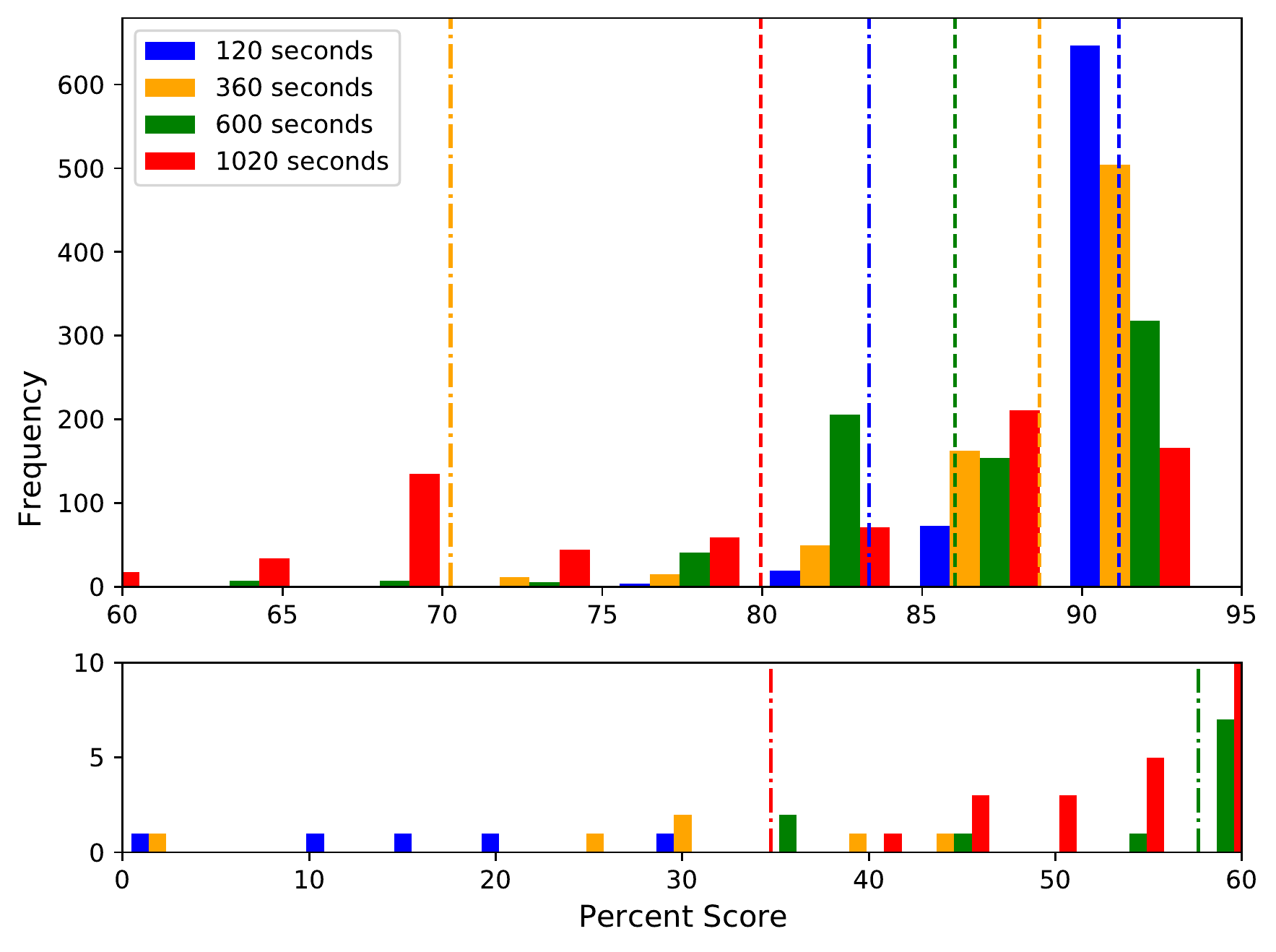}
\caption{Histogram of simulation outputs for individual times used to create Figure \ref{4-fig-ga-melt-batch}. Dashed lines show averages of the corresponding time. Dotted dash lines show the scores for the `zero case' where there is no phase change ($\alpha_2 = 0$ for all cells) for each time. Top histogram shows frequency of scores above 60 and bottom histogram shows scores below 60.}
\label{4-fig-ga-melt-hist}
\end{figure}

Qualitatively observing the initial percentage scores in Figure \ref{4-fig-ga-melt-batch} shows a gentle change from the lower value pairs in the bottom left to higher values in the centre peaking at $\approx 92\%$. Beyond this point, the peak values plateau and instabilities - some drastic - appear. As these scores are found from the average of the four sample times, they can be investigated through looking at the distribution of the constituent scores. Figure \ref{4-fig-ga-melt-hist} is a histogram of the scores at each sample time, here most configurations initially perform very well ($>90\%$) with a high average score however some perform very poorly. The available phase space for high scores reduces over time. Specifically, 80.27\% of the computational pairs tested have a score $>90\%$ at \SI{120}{\second}. This drops to 54.8\% at \SI{360}{\second}, 38.27\% at \SI{600}{\second} and finally to 14.13\% at \SI{1020}{\second}. 

The reduction in average performance at later times shows that in cases with little liquid flow (or cases where the liquid flow lacks the time to fully develop) there are a wide variety of acceptable pairs of computational factors. However, as the flow develops (at later times) the range of acceptable pairs reduces. This phenomenon can be described as either conduction dominated flow for earlier times or convection dominated flow at later times. The deterioration of scores at later times shows the importance of the treatment of the convection dominated flow - the main mechanism for this in the present case is the density variation driving the flow. At later times the `zero case' score - where phase change is disabled - reduces. This is expected as more of the domain is occupied by the melt fraction and therefore in many cells $\alpha_2 \neq 0$. Cases scoring below the `zero case' indicates unphysical melting as they include melting which is not present in the base case.      

%Regardless of this suggests that the solver should perform well in benchmark cases with lower flow? Link back to this later.     

\begin{figure}[ht]
\centering
\includegraphics[width=12.5cm]{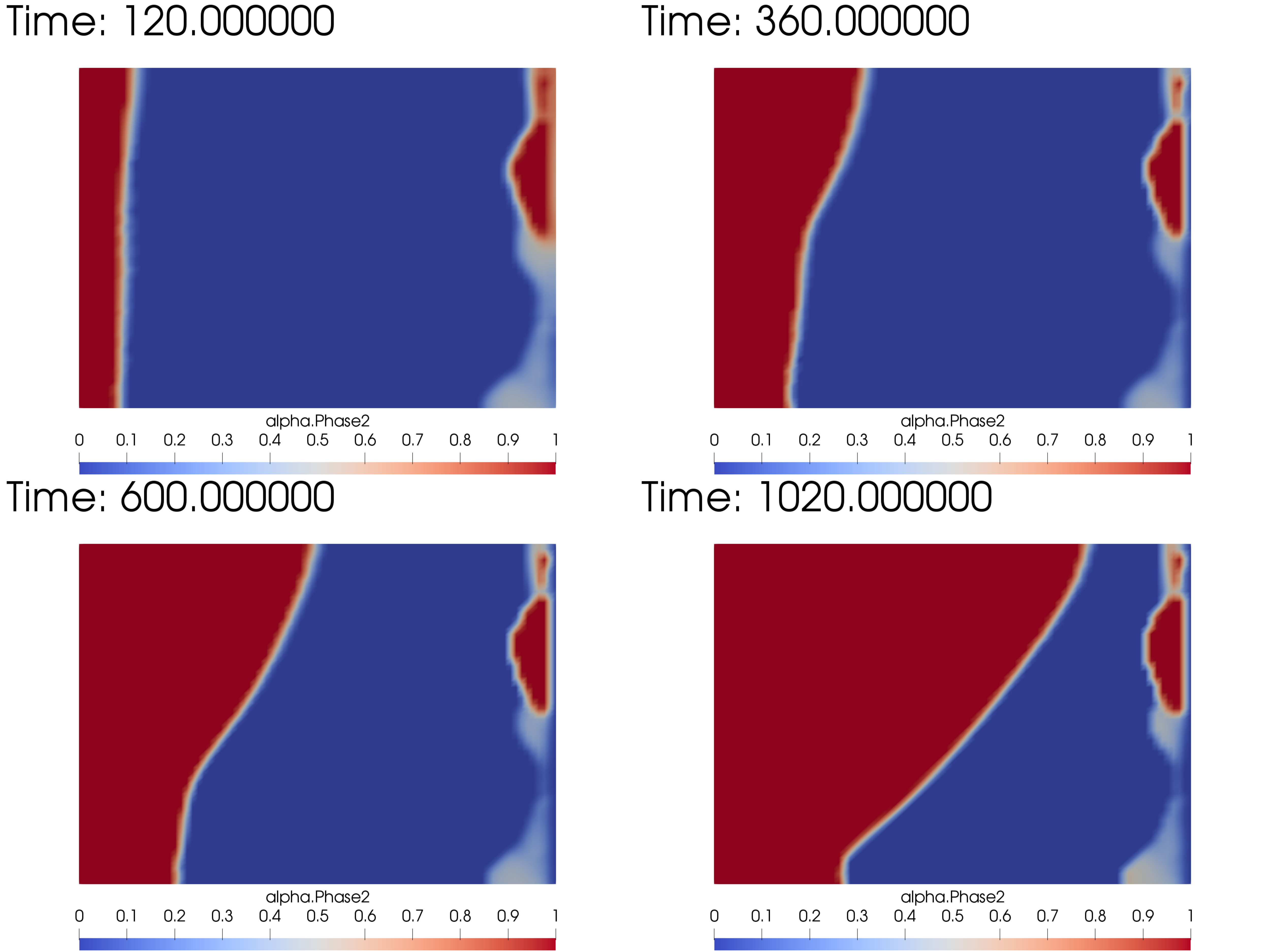}
\caption{Melt fraction simulation output from \gIF for gallium melting with $C_{pc} = 7$ and $C_{d} = \num{2e10}$. Figure shows melt fraction ranging from pure solid ($\alpha_2 = 0$) in blue to pure liquid ($\alpha_2 = 1$) in red. This simulation is an 84\% match with the base case and yet is clearly unphysical.}
\label{4-fig-ga-melt-odd}
\end{figure}

Inspecting the few cases in the 120 second group that score below 85 show that this is sometimes due to the melting taking a while to begin (the zero case score is 83.3\%). If melting starts 100 seconds behind the base case but melts identically then all subsequent scores will be affected despite a nominal match in melt front. This is a feature and not a bug of the scoring system as an accurate time match is critical with respect to matching the heat flux. Conversely, some reasonable scores correspond to unphysical results. For instance Figure \ref{4-fig-ga-melt-odd} shows the appearance of an unphysical melted region on the right hand side of the domain - far from the hot wall on the left hand side. Whilst this pair ($C_{pc} = 7$ and $C_{d} = \num{2e10}$) scores 84\% with the procedure outlined in Section \ref{4-ga-opt}, the result should be heavily penalized for the unphysicality. This is ultimately done through human judgment with unphysical solutions avoided. From experience, these unphysical solutions occur in unstable regions which is another reason to avoid them. The below `zero case' score in this instance however is another reason why simple melt fraction percentage is insufficient to assess computational pair quality; accurate prediction of the location of the melt front is vital for a useful solver. An option to solve this computationally would be to remake the scoring system to only assess possible candidate melt front cells - such as those with $0.1 \leq \alpha_2 \leq 0.9$. However, whilst the melt front is important it doesn't follow that the melt fraction is not also important and therefore both are used in the scoring system. 

The cases that score very poorly ($<35\%$) correspond to similar unphysical solutions with bizarre melt fronts that are worse than the `zero case'. Often these unphysical solutions correspond to very long ($10 \times$ typical) computation times. From a computational perspective, this is due to extremely large gradients between the cells on the discretized velocity \emph{volVectorField} or discretized $\alpha_2$ \vsf resulting in a large face fluxes. The solver resolves by dropping the time step to $\leq 10^{-10}$ (resulting in long computation times) and altering other cell values in the mesh (unphysical solutions).      

Typically, a change in regime in CFD results may coincide with a dimensionless number threshold. However, Figure \ref{4-fig-ga-melt-batch} shows scores continuing at high ($>90 \%$) levels well into the top right of the figure. This is indicating that there is not a smooth function to explain the shape. This is compounded by the fact that the simulations are all completely deterministic - producing the same results every time - and precise to $\approx 10^{-300}$ so a measurement error argument is invalid. Further, as prior mentioned, there is a danger in local over-fitting for this particular case if a mathematical fit is applied to the results and taken as an interpretation. In fact, given that there are only two factors that change both of which are purely computational, fitting the scores to a function would not reveal any relevant physical information. This is in contrast to something like the Prandtl number that actually reveals information about the flow in the case. Finally, whilst one pair of numbers for $C_{pc}$ and $C_{d}$ are required to run a case, a \emph{unique} set is not required and the pair needs only to perform as well as other pairs and have good stability. Therefore, instead of fitting the scores the way to validate the computational factors is to assess their performance in different cases. 
% spurious currents interfoam
%\FloatBarrier
\subsubsection{Extension to three phases} 
\begin{figure}[!htbp]
\centering
\includegraphics[width=10.5cm]{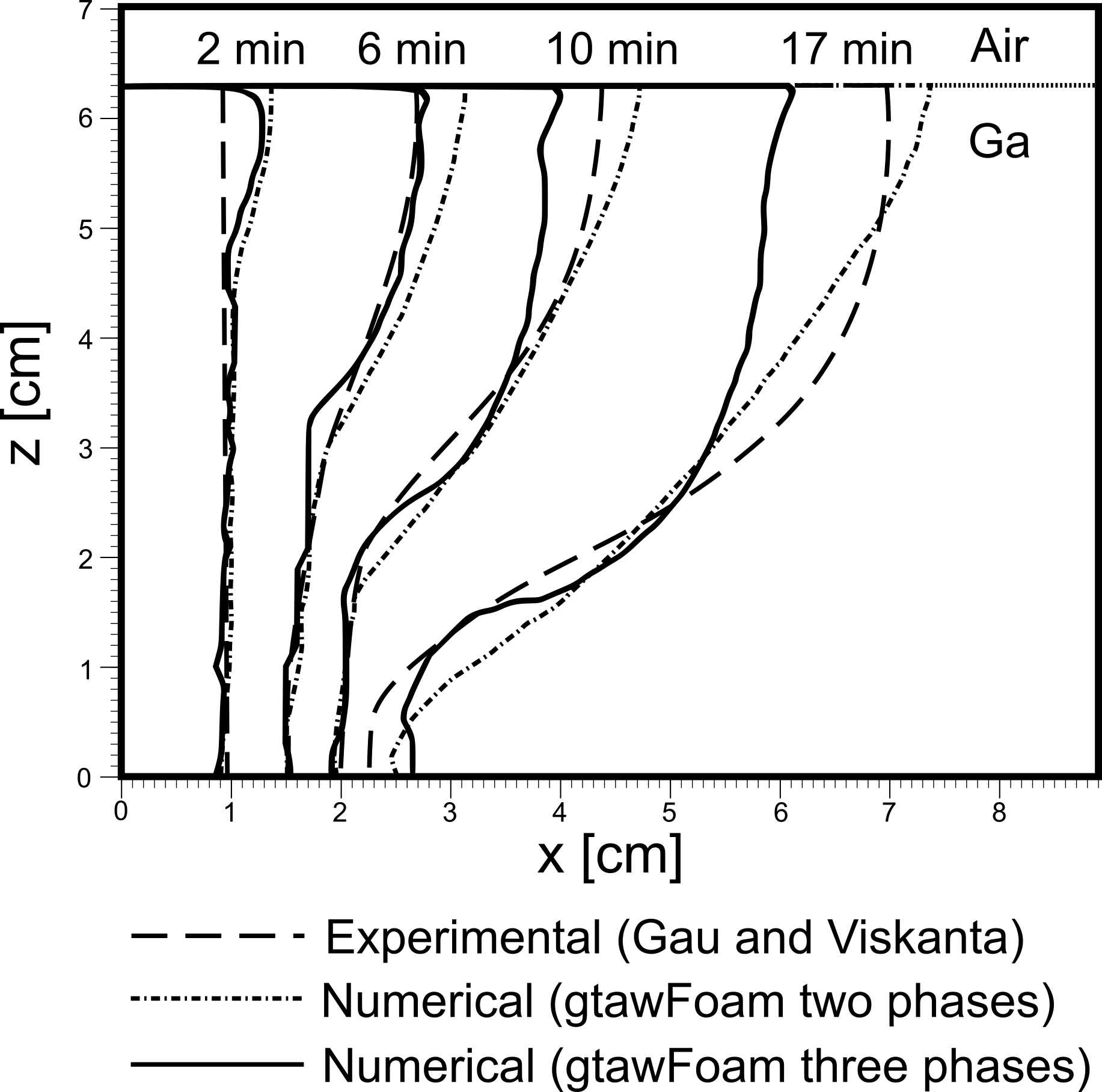}
\caption{Melt front positions calculated with \gIF of a hypothetical gallium melting situation with three phases where an enclosure of air is present above the melting gallium. Some results from Figure \ref{4-fig-ga-melt-opt} are overlaid to illustrate the changes when adding a third phase although no match would be expected given different situations are being simulated. Both simulations use $C_{pc} = 10$ and $C_{d} = \num{8e8}$.}
\label{4-fig-ga-three-mf}
\end{figure}

\begin{figure}[!htbp]
\centering
\includegraphics[width=10cm]{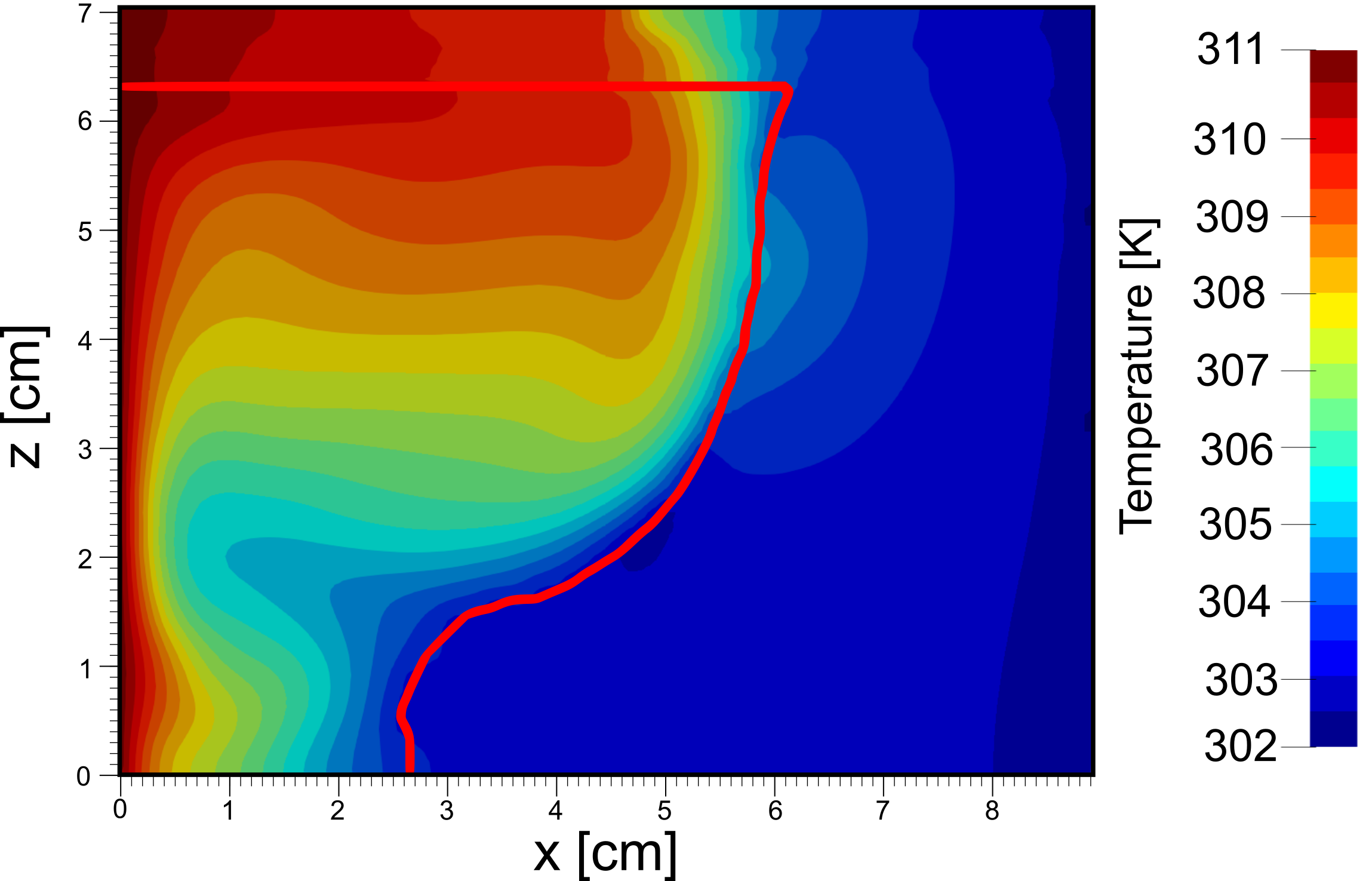}
\caption{Temperature profile for the \gIF simulation output for the hypothetical three phase gallium melting at \SI{1020}{\second}. The red line shows the melt front from Figure \ref{4-fig-ga-three-mf} at \SI{1020}{\second}.}
\label{4-fig-ga-three-T}
\end{figure}

A key feature of \gIF is its ability to handle three phase solid-liquid-gas situations. To demonstrate this, the melt front for a hypothetical three phase gallium melting case with an air pocket above the melting gallium is shown in Figure \ref{4-fig-ga-three-mf}. All surface tension coefficients are set to zero with a \SI{90}{\degree} contact angle on the walls. Apart from the third phase, the case is identical to the two phase case - for brevity, the full case details are presented in Appendix B.3. A match is not expected as the experiment by Gau and Viskanta did not include a free surface like the hypothetical case in Figure \ref{4-fig-ga-three-mf}. However, the match between the three phase case and the two phase simulation and experimental results is still reasonable. One possible reason for the discrepancy is the fact the air phase above the gallium conducts heat (as opposed to the adiabatic top boundary for the two phase case), Figure \ref{4-fig-ga-three-T} illustrates this.  

\subsection{Tin melting}
\label{4-tin-melting}
To evaluate the results from the melting of gallium benchmark, the solver is benchmarked against the melting of tin. Here, the peak score values of $C_{pc} = 10$ and $C_{d} = \num{8e8}$ are used. Wolff and Viskanta performed an experiment for the melting of tin in a rectangular cavity \cite{WolffTinMelt}. A solidification experiment, covered in Section \ref{4-tin-solid}, was also performed using the same experimental set up. Compared to the melting of gallium, the time covered is considerably longer - of the order of hours rather than minutes. The experimental time steps were uneven so the simulation was done to the nearest minute and then copy/pasted into a separate case to match the time exactly. Therefore a reduction in the match between experiment and simulation is expected at later times. Additionally, the melting of tin case can be used to showcase the implementation of a temperature dependent thermal conductivity given in equation \ref{4-eq-tin-kT} for liquid tin with values from \cite{peralta2001TinK}.  

\begin{equation}
\label{4-eq-tin-kT}
    k(T) = 10.204 + 32.063 \cdot \frac{T}{273.15} + 5.686 \cdot \left(\frac{T}{273.15}\right)^2
\end{equation}

\begin{figure}[ht]
\centering
\includegraphics[width=8cm]{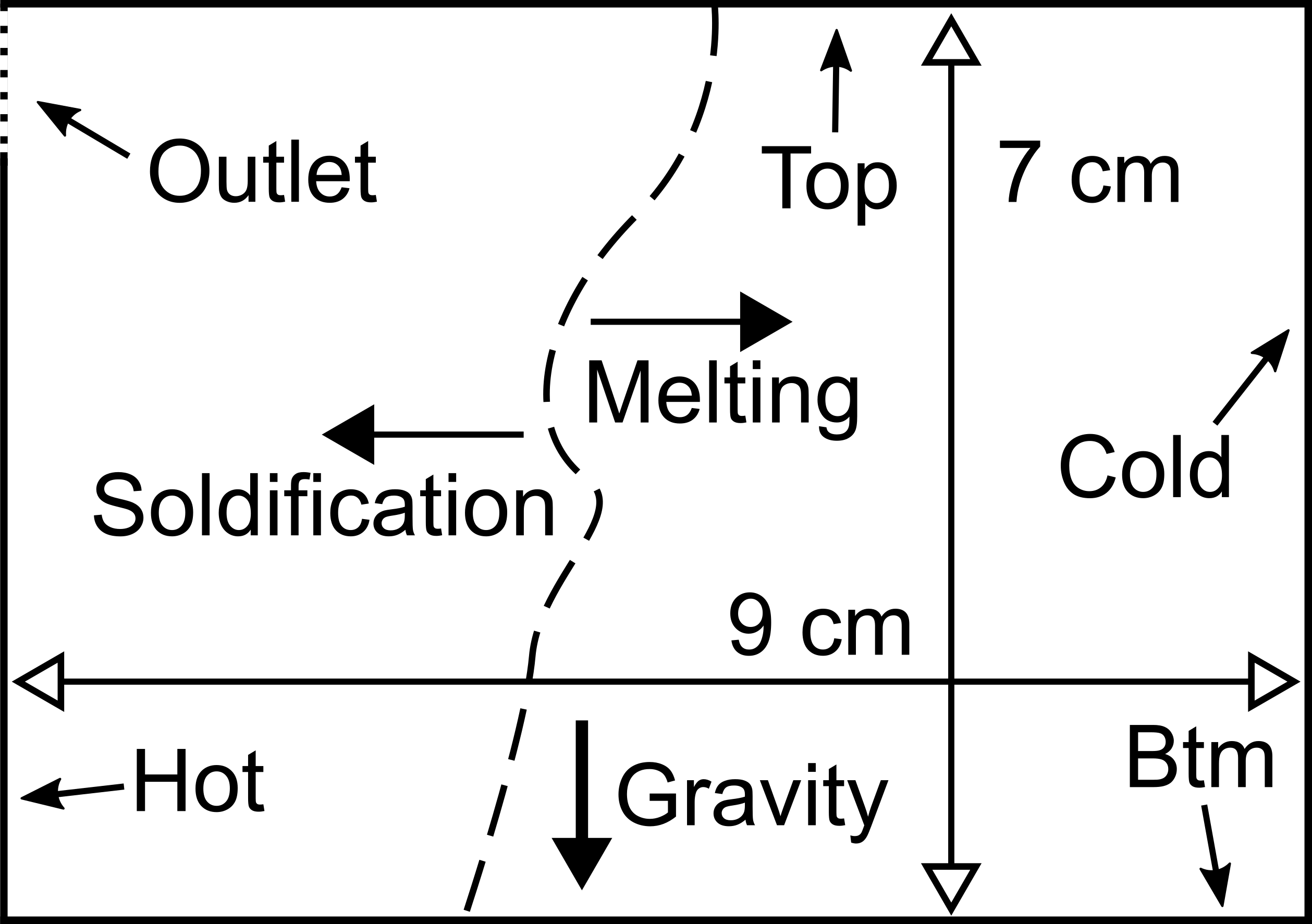}
\caption{2D computational domain used for tin melting and solidification simulations. The dashed line shows a hypothetical melt front during either melting or solidification. Note the addition of the outlet on the hot wall compared to the gallium melting case.}
\label{4-fig-tin-melt-domain}
\end{figure}

This formula is used for the thermal conductivity of the liquid tin phase. For brevity, the other thermophysical properties are presented in Appendix B.4. Further `tools' are also employed in the case; given the density change from solid to liquid tin, an outlet of the form proposed by \cite{FadennOctOutlet} is used. This change is also present in gallium and many other metals and thus the outlet technique could also be used for them. Similar to temperature dependent thermophysical properties, the apt additional `tools' employed for each case are judged on a case by case basis for their usefulness. Shown in Figure \ref{4-fig-tin-melt-domain}, a small outlet is present in on of the walls to create two separate boundary conditions and allow outflow from the domain.  

\begin{table}[ht]
\caption{\label{4-tab-tin-case-bc}Boundary conditions for tin melting benchmark case. (*) Cold is technically two boundaries to match the outlet on the hot wall but both boundaries are set identically.} 
\centering
\begin{tabular}{l l l l l l l}
\toprule
Boundary & Velocity & Pressure & Temperature & $\alpha_1$ & $\alpha_2$ & $\alpha_3$ \\
\hline
Top\rule{0pt}{2.6ex} & \emph{NS} & \emph{FFP} & $\partial_n = 0$ & $\partial_n = 0$ & $\partial_n = 0$ & $\partial_n = 0$ \\
Outlet & \emph{PIOV} & \emph{TP} & $T_{hot}$ & $\partial_n = 0$ & \emph{IO} & $\partial_n = 0$ \\
Hot & \emph{NS} & \emph{FFP} & $T_{hot}$ & $\partial_n = 0$ & \emph{IO} & $\partial_n = 0$ \\
Cold* & \emph{NS} & \emph{FFP} & $T_{cold} = \SI{505}{\kelvin}$ & $\partial_n = 0$ & $\partial_n = 0$ & $\partial_n = 0$ \\
Btm & \emph{NS} & \emph{FFP} & $\partial_n = 0$ & $\partial_n = 0$ & $\partial_n = 0$ & $\partial_n = 0$ \\
\bottomrule
\end{tabular}
\end{table}

Two cases are investigated where both have a cold wall and an initial temperature set to \SI{505}{\kelvin} - \SI{0.1}{\kelvin} below the melting of point of tin (\SI{505.1}{\kelvin}). Whereas, the hot wall temperatures are set to \SI{507.05}{\kelvin} in one case and \SI{507.65}{\kelvin} in the other. These temperature differences are $\approx 20\%$ of the gallium case and the temperature of the hot walls are only slightly higher than the melting point of tin creating the long melt times. The experimental results are compared against simulation output using the optimum computational constants for gallium ($C_{pc} = 10$ and $C_{d} = \num{8e8}$) and with locally optimized constants. The second comparison is to illustrate the improved results from using different constants. As covered previously, in this thesis this is termed local optimization and it reduces the predictability of a simulation. As is demonstrated in the base case of the gallium melting simulation, it is possible to add extra computational factors until the simulation matches experimental results. However, it is the view of the author that computational constants should be kept to a minimum and the better solution to improved results is to include extra tools such as temperature dependent thermophysical properties. 

\begin{figure}[ht]
\centering
\includegraphics[width=15cm]{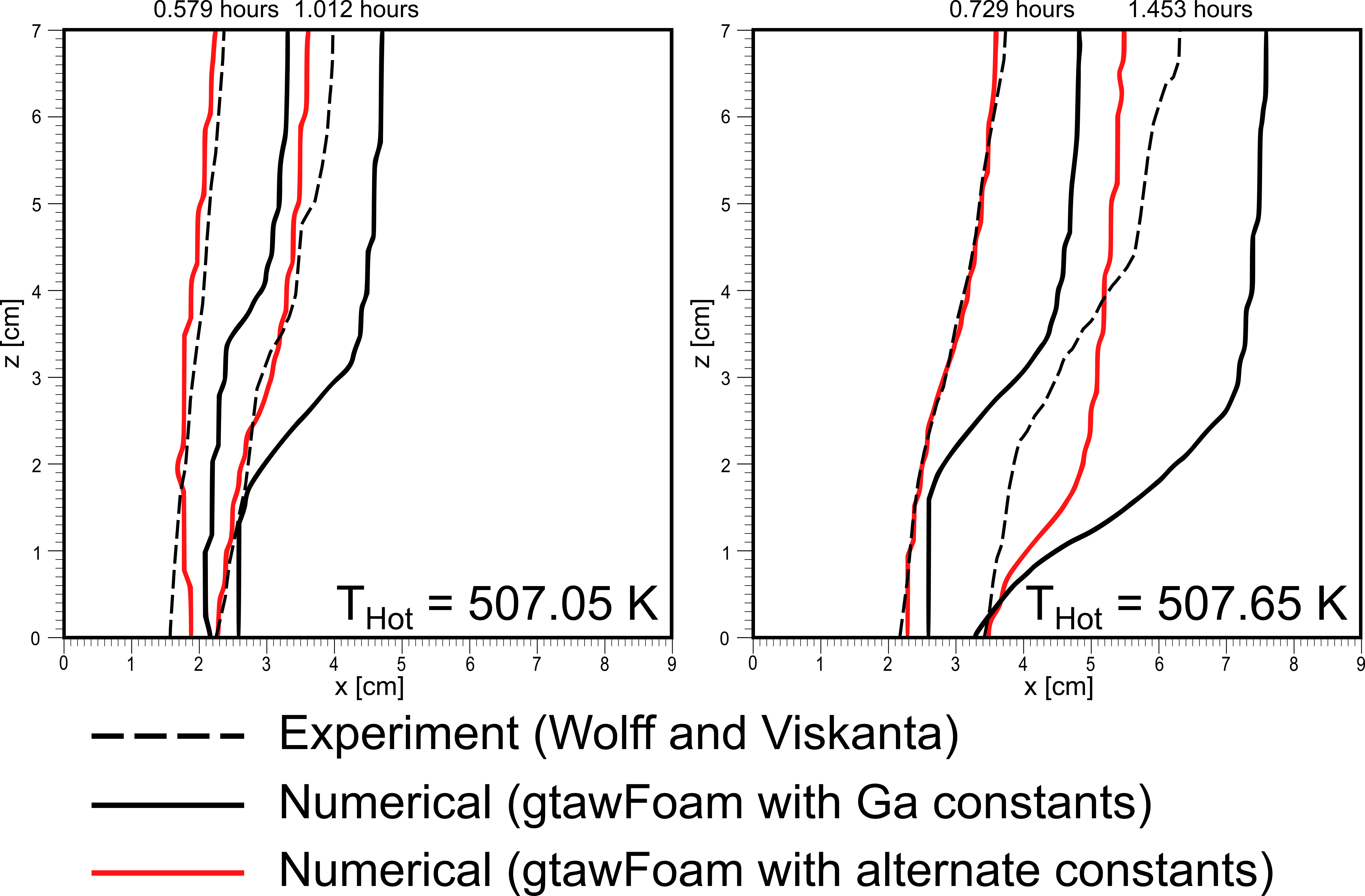}
\caption{Melt front positions from the tin melting simulation with constants from gallium optimization (solid black line) vs experiment (dashed black line). To illustrate local optimization, the red lines show simulations with $C_{d} = \num{5e9}$ and $C_{pc} = 2.5$ for the first time and $C_{pc} = 3.5$ for the second time for \SI{507.05}{\kelvin} (left image). For the right image, $C_{d} = \num{7.5e9}$ and $C_{pc} = 3.5$ are used. The left hand side figure shows results (from left to right) at 0.579 hours and 1.012 hours. The right hand side figure shows results (from left to right) at 0.729 hours and 1.453 hours.}
\label{4-fig-tin-melt-results}
\end{figure}

The simulation output from melting of tin benchmark case are shown in Figure \ref{4-fig-tin-melt-results}. In general, they show an aggressive melt front prediction for the simulation results with the optimized gallium case computational constants. This over-prediction exacerbates at later times as the flow develops further. It is incorrect to conclude that this shows \gIF cannot handle convection driven flow as the results using different computational factors match the experiment a lot better despite using the same equations. Instead, the correct conclusion is that with the optimum computational constants from the gallium case the performance of \gIF deteriorates at later simulation times. Given weld pools are in general liquid for seconds rather than minutes, the deterioration at later simulation times is not too deleterious. The second conclusions is that for specific scenarios it is apt to change the computational constants given their ability to improve the performance of the solver. Finally, for both of these conclusions it is worth considering that the experiments are the results chosen for publication by a third party and thus there will be some discrepancy between experiment and simulation.   

% Check this is 0.05 and not 0.15

\section{Solidification benchmark cases}
\label{4-solidification-section}
\subsection{Tin solidification}
\label{4-tin-solid}
\subsubsection{Solidification front evolution}
Using the same experimental set up as with the melting of tin (detailed in Section \ref{4-tin-melting}), Wolff and Viskanta also conducted an experiment into the solidification of tin \cite{WolffTinSolid}. To simulate this, the computational case outlined in Figure \ref{4-fig-tin-melt-domain} and Table \ref{4-tab-tin-case-bc} can be reused with an initial liquid tin $\alpha_2$ field used and different wall temperatures. In the two cases investigated, a cold wall temperature of \SI{499.15}{\kelvin} was used with one case having a hot wall and initial temperature set to \SI{506.15}{\kelvin} and the other set to \SI{507.15}{\kelvin}. 

\begin{figure}[ht]
\centering
\includegraphics[width=15cm]{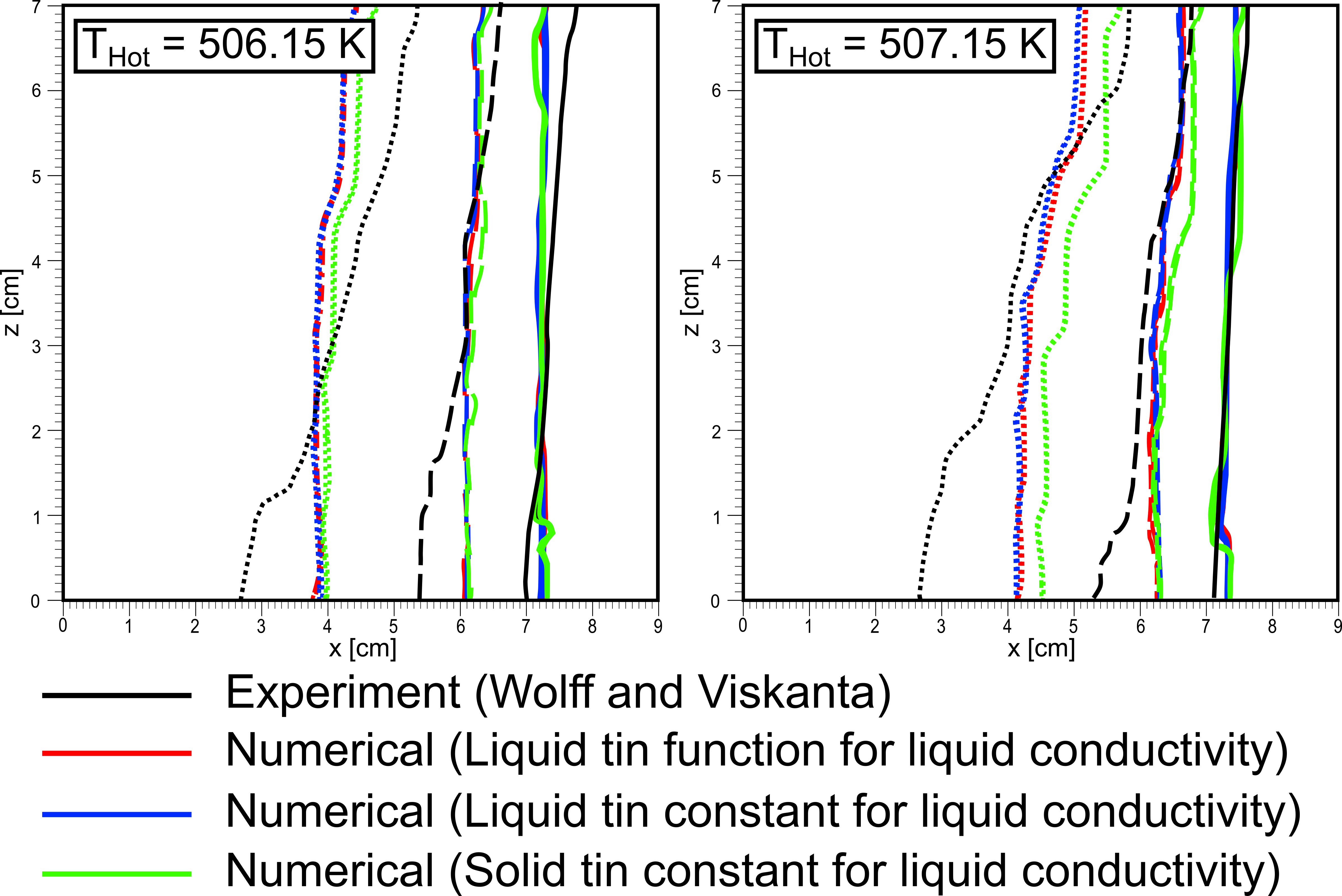}
\caption{Melt front positions from tin solidifying simulation with computational constants from gallium optimization vs experiment (black line). The red lines show the conductivity of liquid tin, $k_2$, determined by equation \ref{4-eq-tin-kT}. The blue lines show $k_2 = \SI{27}{\watt\per\meter\per\kelvin}$ and the green lines show $k_2 = \SI{59.5}{\watt\per\meter\per\kelvin}$. The solid lines are for the melt front after \SI{227.2}{\second}, the dashed lines show it after \SI{594}{\second} and dotted lines show it after \SI{1904.4}{\second}. Note, unexpectedly there is little difference between the solidification fronts, the red, blue and green lines are all very similar despite different values of $k_2$. This suggests convection dominance in the liquid tin during solidification.}
\label{4-fig-tin-solid-results}
\end{figure}

As shown in the previous sections, the choice of computational constants in \gIF have a substantial effect on the results. However, the thermophysical properties used also have a large effect. In the experimental paper Wolff and Viskanta cite multiple sources for the thermophysical properties including an ASM handbook on non-ferrous materials. The 1990 version of this book quotes the thermal conductivity of solid tin at \SI{56.5}{\watt\per\meter\per\kelvin}. The ASM handbook on casting used for the temperature dependent specific heat capacity feature detailed in Chapter 3 quotes solid tin at \SI{59.5}{\watt\per\meter\per\kelvin} and liquid tin at \SI{27}{\watt\per\meter\per\kelvin} \cite{thermoPhysProp}. The conductivity of liquid tin can also be described by equation \ref{4-eq-tin-kT} for with values from \cite{peralta2001TinK}. Notice these values are all slightly different. 

Mathematically the thermal conductivity will affect the heat flux (coming from the boundaries) and hence the phase front as the tin solidifies. With a material such as tin where there is a substantial change in thermal conductivity between solid and liquid the use of different thermal conductivities in simulations is expected to have a large effect. Shown in Figure \ref{4-fig-tin-solid-results}, the solidification of tin is investigated using three different thermal conductivities for liquid tin: the function given in equation \ref{4-eq-tin-kT}, the ASM handbook value for liquid tin (\SI{27}{\watt\per\meter\per\kelvin}) and the ASM handbook value for solid tin (\SI{59.5}{\watt\per\meter\per\kelvin}). All cases use \SI{59.5}{\watt\per\meter\per\kelvin} for solid tin. Interestingly, the thermal conductivity appears to have little effect - only becoming pronounced at later times whereby the simulations with the lower conductivities ($k_2 = \SI{27}{\watt\per\meter\per\kelvin}$ and when $k_2$ is determined by equation \ref{4-eq-tin-kT}) match the experimental results better. It can therefore be concluded empirically that in this solidification case the thermal conductivity of the liquid metal has little effect. 

As a general overview, the results are a good match with experiment. Similar to the melting of tin, the results show an increasing mismatch between the simulation and the experiment as the time progresses yet this mismatch is much reduced. Further, instead of the simulation over-predicting the amount of melting it under-predicts the amount of solidification. Essentially it the solver is `bullish' in its melting predictions and `bearish' in its solidification predictions. As with the melting of tin it is possible to change the constants to create a better fit to the data but this comes with the same prior mentioned issues of local optimization. Additionally, the close match actually validates the choice of the computational constants from the gallium case.  

Applying the conclusion for the mismatch in melting fronts to the solidification fronts suggests that the convective flow is completely dominant at the start of solidification so the longer the simulation times the larger the cumulative effect it has on the solidification front. At the earlier times in the gallium melting the flow is not fully formed and the melting conduction dependent. Whereas, with solidification the flow essentially starts at the most convective dominant and works backward. The issue with this is that the solidification actually performs quite well compared to the melting. Solidification is not the reverse of melting and therefore an exact performance match would be unexpected however, there is still a noticeable difference. A method to probe this is to see what causes the solidification performance to deteriorate. 

\begin{figure}[h]
\centering
\includegraphics[width=15cm]{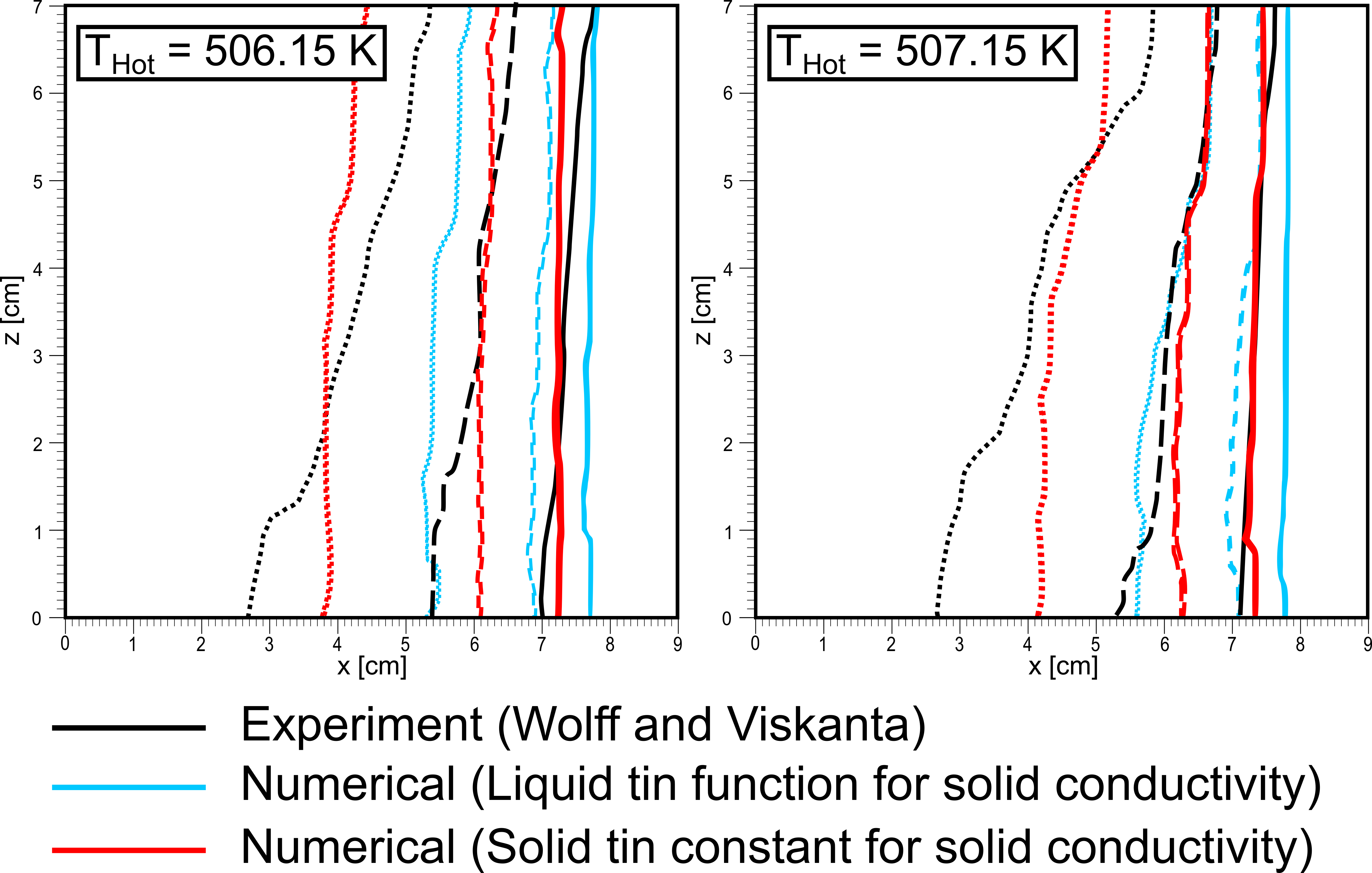}
\caption{Melt front positions from tin solidifying simulation with computational constants from gallium optimization vs experiment (black line). The red lines show the conductivity of solid tin set to $k_3 = \SI{59.5}{\watt\per\meter\per\kelvin}$, this is the same as the red line in Figure \ref{4-fig-tin-solid-results}. The light blue lines show $k_3$ determined by equation \ref{4-eq-tin-kT}. In both numerical cases the liquid conductivity is determined by equation \ref{4-eq-tin-kT}. The solid lines are for the melt front after \SI{227.2}{\second}, the dashed lines show it after \SI{594}{\second} and dotted lines show it after \SI{1904.4}{\second}.}
\label{4-fig-tin-solid-k-comp}
\end{figure}

A test for this is to investigate is whether the reverse situation of thermal conductivity is true: does changing the solid conductivity have much of an effect? Shown in Figure \ref{4-fig-tin-solid-k-comp}, when the function for liquid tin conductivity (equation \ref{4-eq-tin-kT}) is used for the conductivity of the solid tin and the performance deterioration is quite noticeable. Here, a reduction in the thermal conductivity of the solid phase - which has a velocity of zero - changes the melt front drastically. This result somewhat refutes the conclusion from the melting cases and suggests further investigation is required. For instance, dimensionless numbers can be used to identify the regime.

\FloatBarrier
\subsubsection{Rayleigh number and melt fraction}
\label{4-rayleigh-number-section}
\begin{figure}[!h]
\centering
\includegraphics[width=13.5cm]{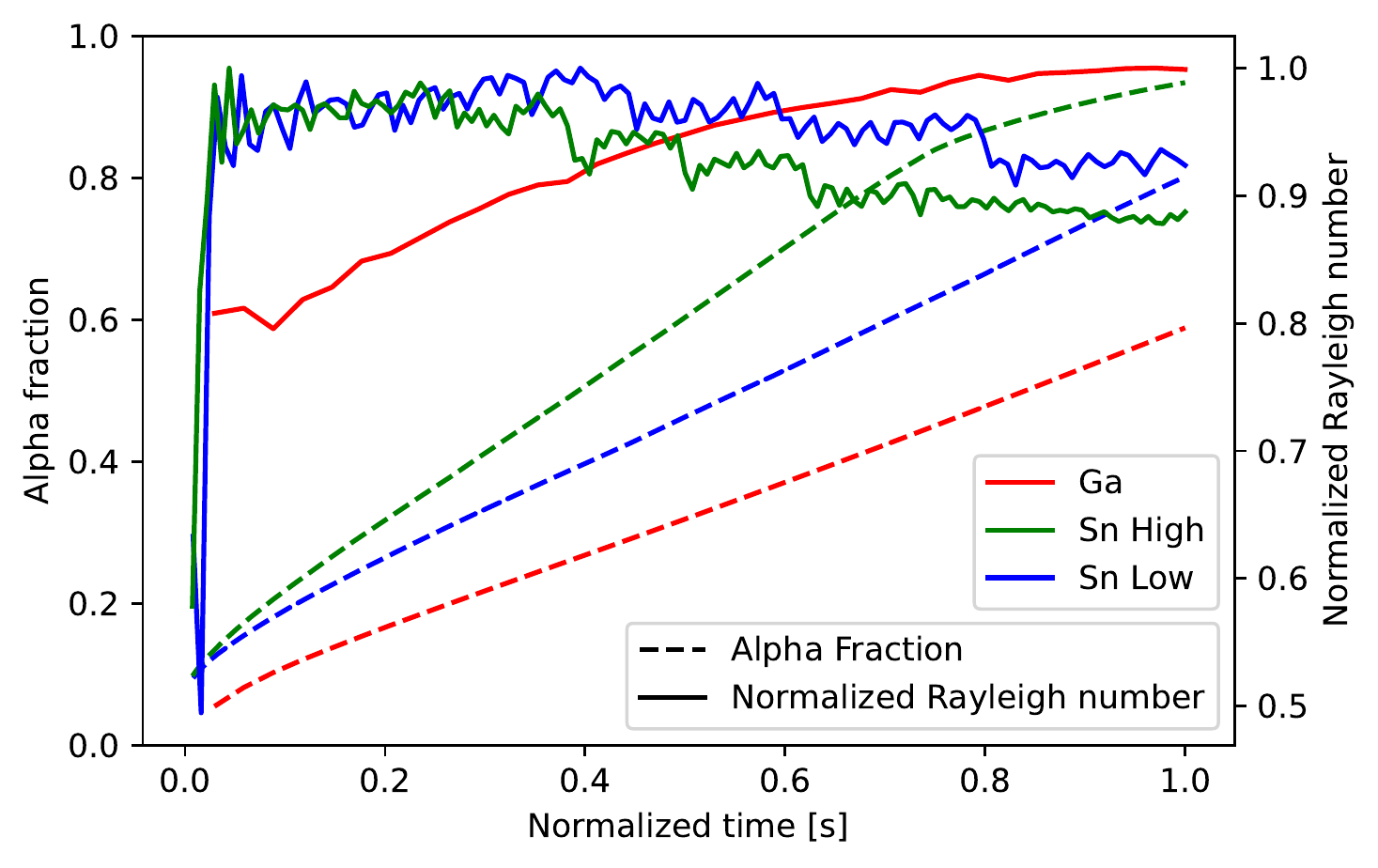}
\caption{$\alpha_2$ fraction and normalized Rayleigh number as a function of normalized time for melting cases. Red line shows gallium case, blue line shows the melting of tin with a hot wall temperature of \SI{507.05}{\kelvin} and the green line shows the melting of tin with a hot wall temperature of \SI{507.65}{\kelvin}.}
\label{4-fig-ray-melt}
\end{figure}

\begin{figure}[ht]
\centering
\includegraphics[width=13.5cm]{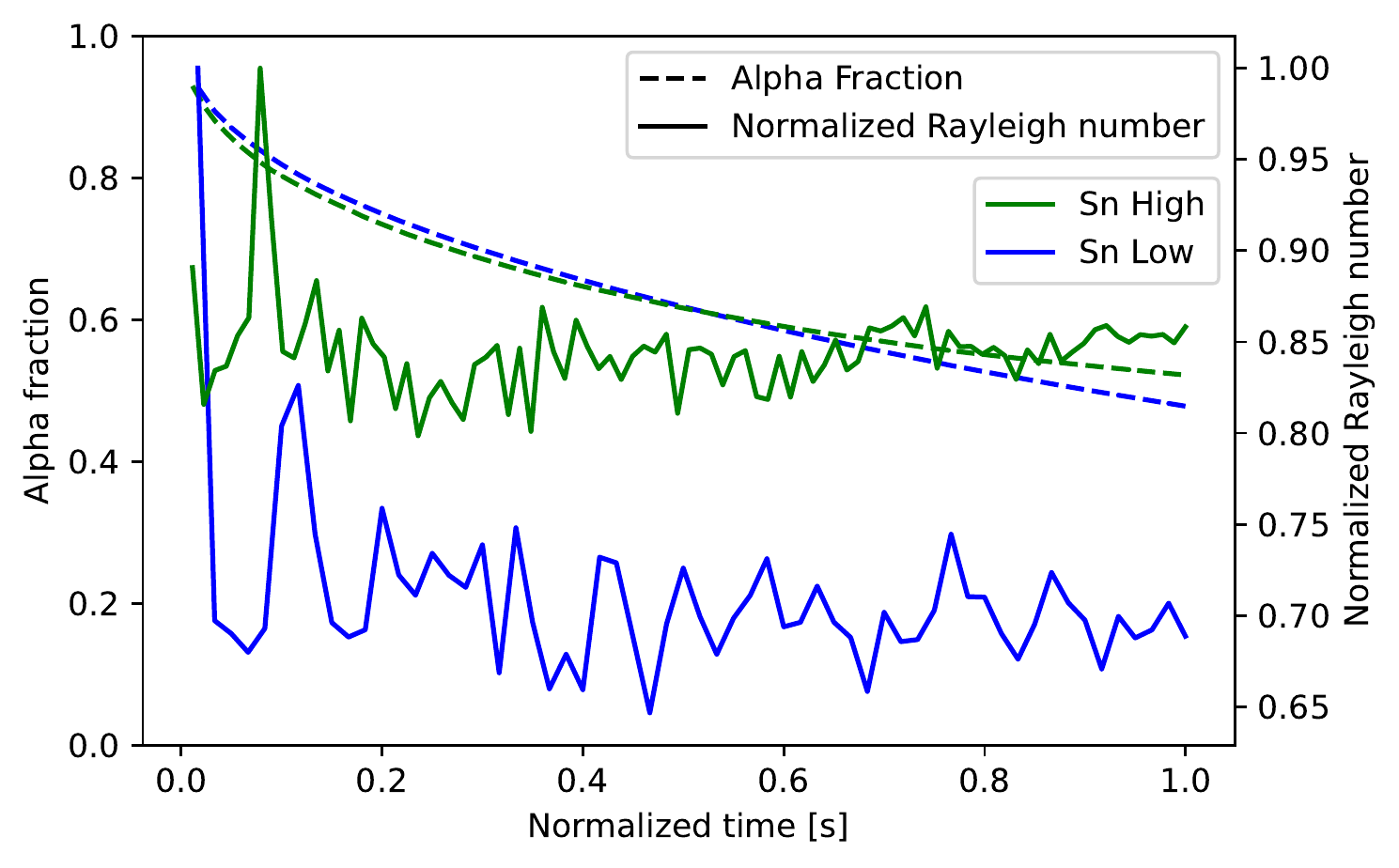}
\caption{$\alpha_2$ fraction and normalized Rayleigh number as a function of normalized time for tin solidification cases. Blue line shows the solidification of tin with a hot wall temperature of \SI{506.05}{\kelvin} and the green line shows the melting of tin with a hot wall temperature of \SI{507.05}{\kelvin}.}
\label{4-fig-ray-sold}
\end{figure}

To investigate the solver performance further, the normalized Rayleigh number and $\alpha_2$ fraction is plotted against normalized time for the cases detailed so far in this chapter. The Rayleigh number is calculated just for the $\alpha_2$ fraction as

\begin{equation}
    Ra = \frac{\rho c_p \beta \Delta T l g}{\nu k}
\end{equation}

for a temperature difference $\Delta T$ over a length $l$ where $g$ is acceleration due to gravity, $\nu$ is the kinematic viscosity, $\rho$ is density, $k$ is thermal conductivity and $c_p$ is specific heat capacity. Figure \ref{4-fig-ray-melt} plots the melting cases and Figure \ref{4-fig-ray-sold} plots the solidification cases. For gallium there is a gradual rise in normalized Rayleigh number with a rise in $\alpha_2$ fraction. This is expected as a gradual transition from conduction to convection. However, for the melting and solidification of the tin cases (neglecting the initial sharp changes at the start of the simulation) the Rayleigh number stays reasonably constant. This is due to how the temperature field is specified - only boundary conditions and an initial domain temperature are given so the first few times have large changes in the temperature distribution within the domain. The lack of a trend with tin suggests there isn't a regime change from conduction to convection and the simulation is only ever convection dominated. 

\FloatBarrier
\subsection{Ice solidification}
\label{4-ice-solid}
The previous three benchmarks demonstrate the importance of the liquid metal flow in successfully simulating the melt front. The principal driver of the flow in these cases is differences in density with the liquid metal phase. These benchmarks all use the Boussinesq approximation (shown in equation \ref{2-eq-bouss}) to calculate this and it is therefore prudent to investigate the performance with a different formulation of density. 

\begin{equation}
\label{2-eq-bouss}
    \rho \cdot (1.0 - \alpha_2\beta(T - T_0)) \equiv \rho - \rho(T)\alpha_2.
\end{equation}

A good test case for this is the density of water, which was investigate by Kowalewski and Rebow \cite{waterFreezeKowa1999}. Water has a peak density at $\approx$ \SI{277.138}{\kelvin} - the Boussinesq will miss this detail. In \cite{waterFreezeKowa1999} the density of water is approximated with a function as  

\begin{equation}
\label{4-eq-rhoT}
\begin{split}
\rho(T) = & \ A + BT + CT^2 + DT^3 + ET^4 \\     
= & \ 999.840281167108 + 0.0673268037314653 \cdot T \\ & - 0.00894484552601798  \cdot T^2 + \num{8.7846286650041e-5}  \cdot T^3 \\ & - \num{6.6213979262754e-7}  \cdot T^4.
\end{split}
\end{equation}

\begin{figure}[!h]
\centering
\includegraphics[width=13.5cm]{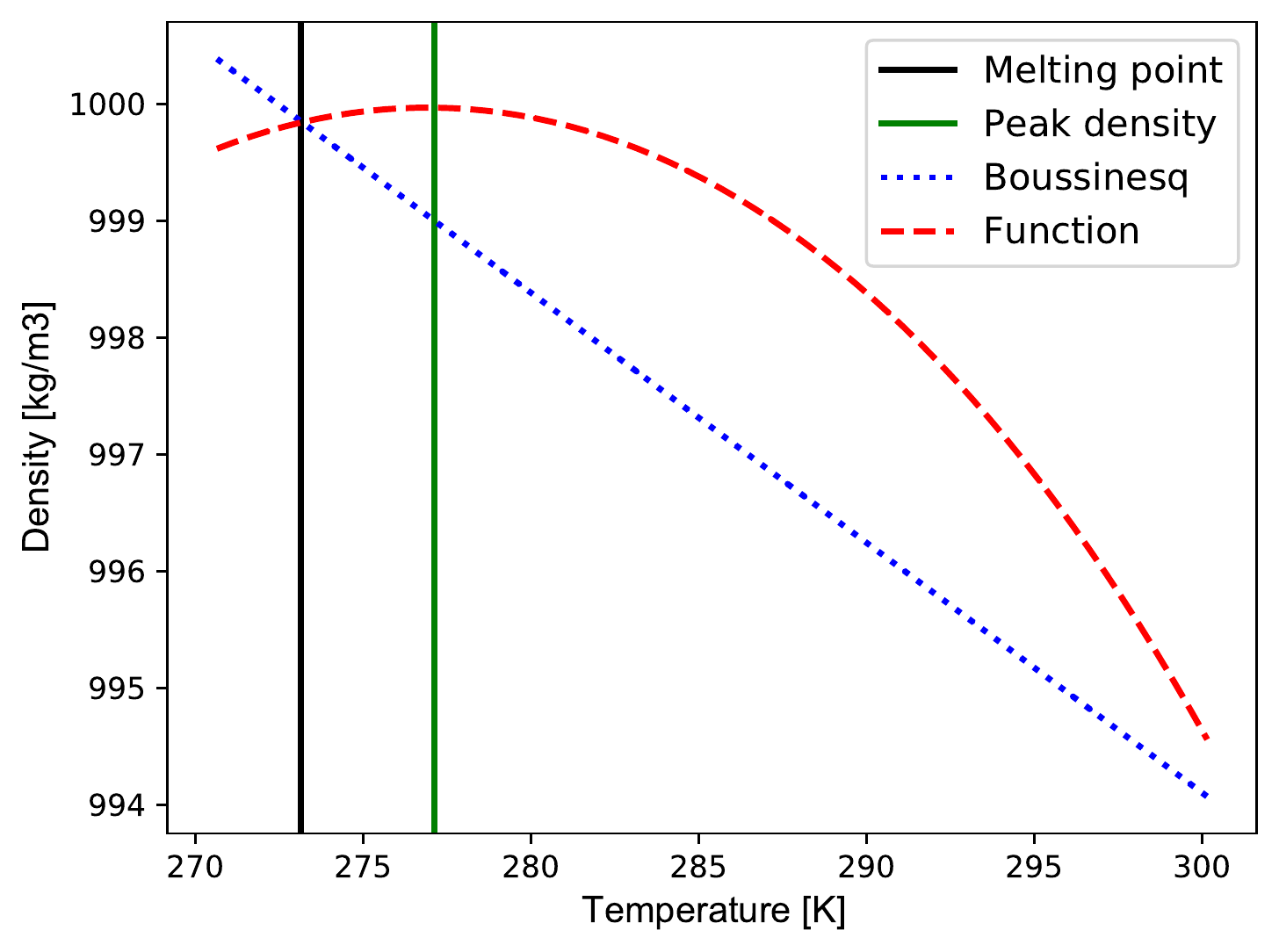}
\caption{Graph of temperature dependence of density, $\rho$, for water. Solid blue line shows \SI{0}{\celsius}, solid green line shows peak density at $\approx$ \SI{277.138}{\kelvin} and dashed red line shows the density of water at different temperatures. Note the peak density is at a few degrees higher than \SI{0}{\celsius} leading to an inflection in the gradient before freezing.}
\label{4-fig-rhoT}
\end{figure}

As shown in Figure \ref{4-fig-rhoT}, these two formulations for $\rho(T)$ have different functional forms. Critically, equation \ref{4-eq-rhoT} produces an inflection point at $\approx$ \SI{277.138}{\kelvin} that is separated from the freezing temperature of water that is set to \SI{273.15}{\kelvin}. 

For flow that is exclusively density driven, this will produce two different flows and hence two different melt/solidification fronts. To investigate this, a 2D \SI{38}{\milli\meter} $\times$ \SI{38}{\milli\meter} domain similar to the one used for the melting of gallium was used to simulate the investigation carried out in \cite{waterFreezeKowa1999}. Here a cold wall is held at temperature \SI{263.15}{\kelvin} and three different hot wall temperatures are used: \SI{278.15}{\kelvin}, \SI{280.65}{\kelvin} and \SI{283.15}{\kelvin} (e.g. \SI{5}{\celsius}, \SI{7.5}{\celsius} and \SI{10}{\celsius}). For brevity, the full case details are given in Appendix B.5. 

\begin{figure}[!h]
\centering
\includegraphics[width=15cm]{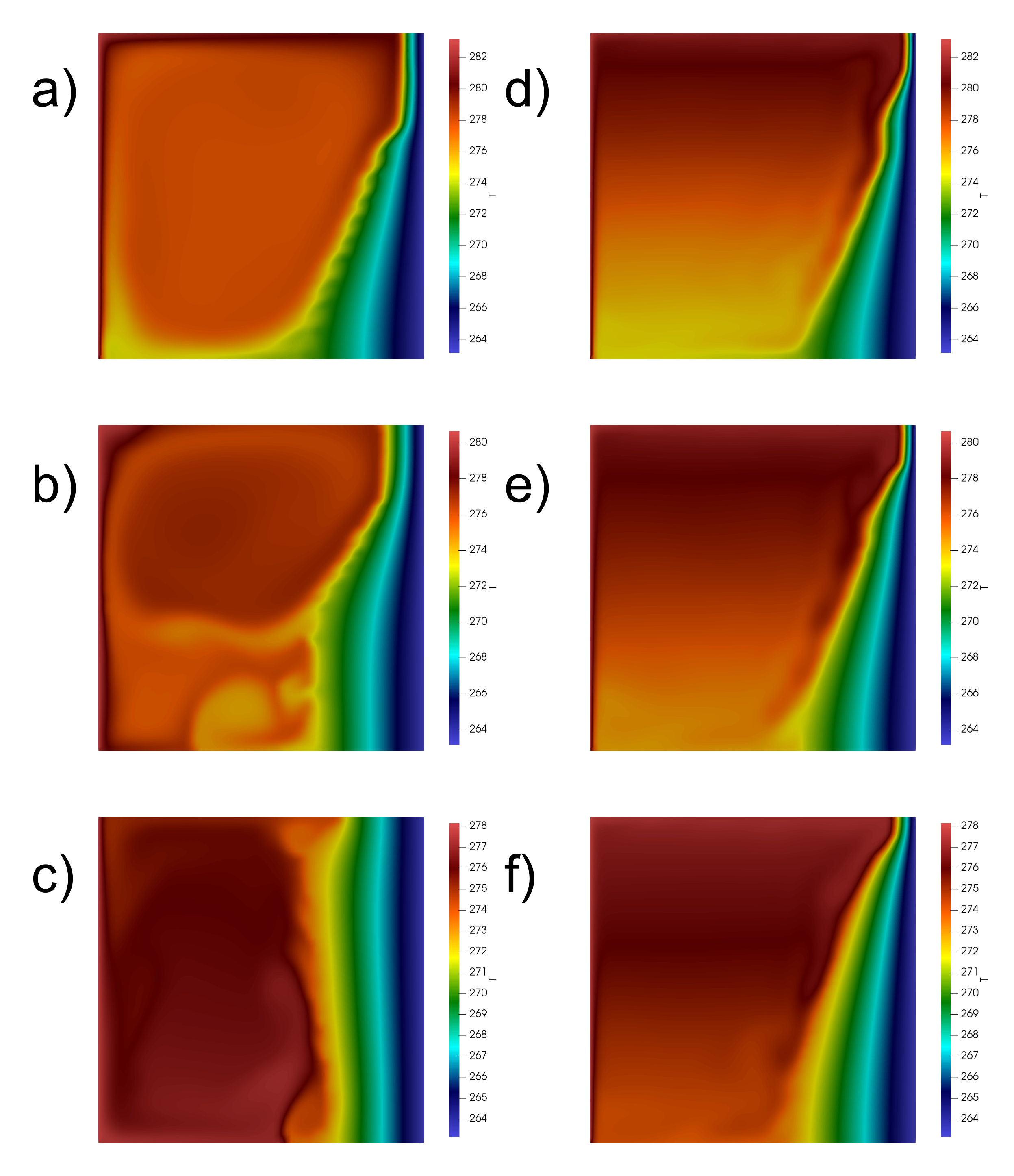}
\caption{Comparison of the temperature fields for water freezing in a 2D square cavity at \SI{2500}{\second} using the computational constants optimized for gallium. Equation \ref{4-eq-rhoT} is used to calculate a), b) and c) for a hot wall temperature of \SI{283.15}{\kelvin}, \SI{280.65}{\kelvin} and \SI{278.15}{\kelvin} respectively. Whereas the Boussinesq approximation is used in d), e) and f) for a hot wall temperature of \SI{283.15}{\kelvin}, \SI{280.65}{\kelvin} and \SI{278.15}{\kelvin} respectively. All cases have the same cold wall temperature of \SI{263.15}{\kelvin} (\SI{-10}{\celsius}). Note how the images d), e) and f) have roughly the same distribution of temperature but different absolute values as a result of the linear Boussinesq approximation. Conversely, a), b) and c) have varying distributions as the density calculated by equation \ref{4-eq-rhoT} can go up and down with rising temperature.}
\label{4-fig-rhoT-Bouss-Temp}
\end{figure}

\begin{figure}[!h]
\centering
\includegraphics[width=15cm]{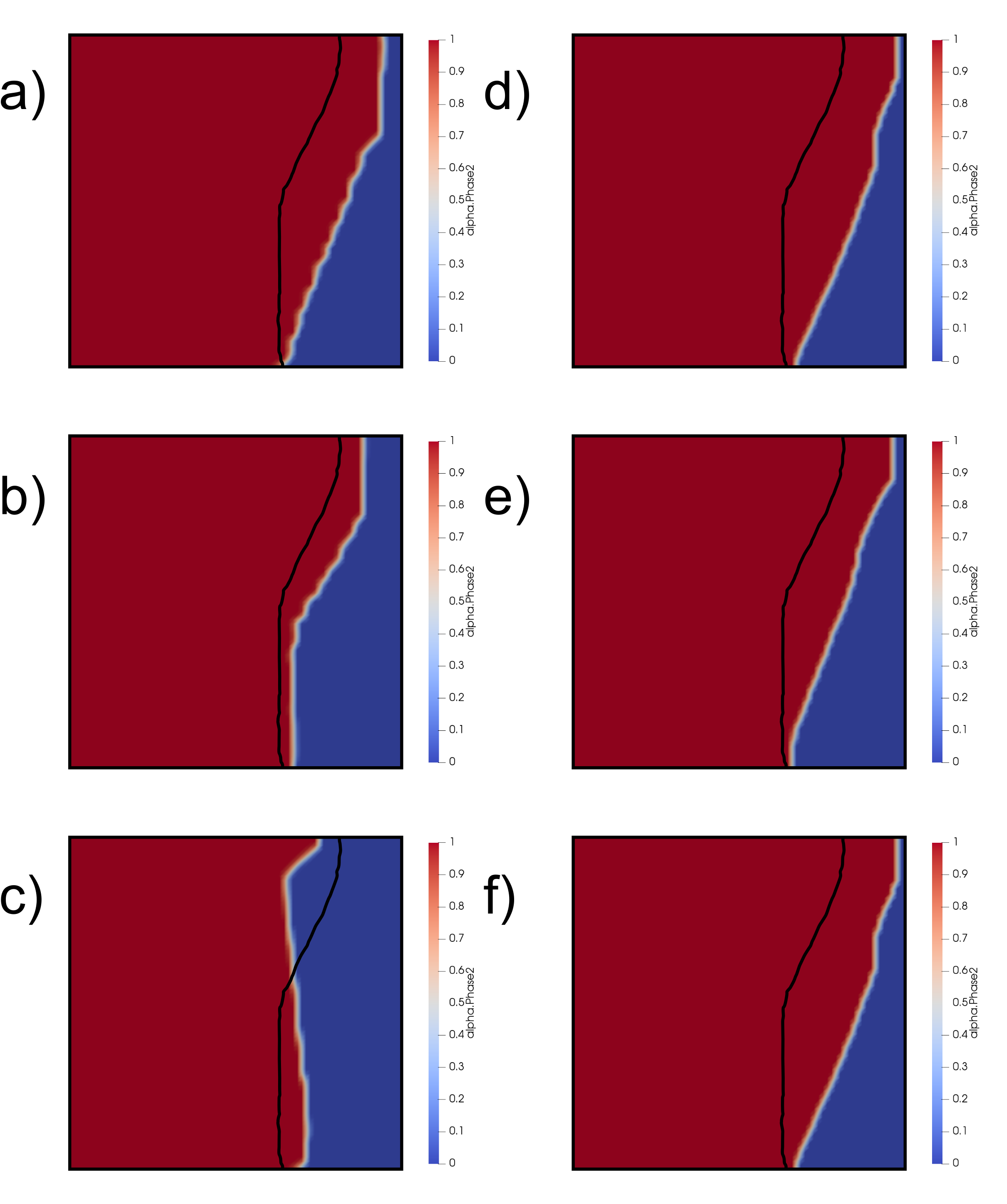}
\caption{Melt fraction resulting from the temperature fields in Figure \ref{4-fig-rhoT-Bouss-Temp} at \SI{2500}{\second}. Equation \ref{4-eq-rhoT} is used to calculate a), b) and c) for a hot wall temperature of \SI{283.15}{\kelvin}, \SI{280.65}{\kelvin} and \SI{278.15}{\kelvin} respectively. Whereas the Boussinesq approximation is used in d), e) and f) for a hot wall temperature of \SI{283.15}{\kelvin}, \SI{280.65}{\kelvin} and \SI{278.15}{\kelvin} respectively. All cases have the same cold wall temperature of \SI{263.15}{\kelvin} (\SI{-10}{\celsius}). The black outline shows the experimental results at \SI{2500}{\second} for \SI{283.15}{\kelvin} from \cite{waterFreezeKowa1999}. Despite a) being set identically to the experiment, b) - which has a hot wall temperature \SI{2.5}{\kelvin} lower - has the closest match. As with the temperature fields, the melt fractions in d), e) and f) all have roughly the same shape and in fact have a similar shape to a). This uniformity is due to the aforementioned treatment of density and further shows how the melt front is dependent on the liquid phase flow.} 
\label{4-fig-rhoT-Bouss-Alpha}
\end{figure}

Figure \ref{4-fig-rhoT-Bouss-Temp} compares temperature for the Boussinesq and the function formulation of $\rho(T)$ whereas Figure \ref{4-fig-rhoT-Bouss-Alpha} compares melt fraction. The Boussinesq driven flows all look very similar except with different temperature ranges. This is expected due to the constant gradient of the linear variation of density with temperature causing the flow to be driven in the same way regardless of temperature difference. For the functional form however, the density can both increase and decrease with temperature thus causing the density driven flow to vary. The contrasting temperature distributions result in different solidification fronts.

% Given the computational constants are optimized for gallium the clear melt front change due to the $\rho(T)$ formulation demonstrate the strong effect it has.

\FloatBarrier
\section{Liquid metal flow benchmark cases}
\subsection{Aluminium Marangoni driven flow}
\label{4-sec-alum-marg-flow}
In addition to the density driven flow covered in Section \ref{4-ice-solid}, there is also flow governed by the Marangoni effect to consider. The implementation of the Marangoni effect is based off of Z.S. Saldi's thesis on the topic \cite{saldi2012marangoni}. Thus to validate the implementation, the same tests of Marangoni driven flow are used. Unlike the prior benchmarks, these are simulation benchmarks and so only qualitative properties are of interest. The first test case is based off of work by Bergman and Keller involving the flow of liquid aluminium in a differentially heated 2D rectangular cavity \cite{AlumMargFlow}. The interesting element of this work is that the surface tension to temperature gradient, $\frac{\partial \sigma}{\partial T}$ is negative for pure aluminum (\SI{-3.5e-4}{\newton\per\meter\per\kelvin}) and positive for an aluminium-tin alloy (\SI{2.0e-4}{\newton\per\meter\per\kelvin}). Therefore, for the negative $\frac{\partial \sigma}{\partial T}$ value the Marangoni effect will `work with' the natural convection as they both push the fluid the same way. However, for the positive $\frac{\partial \sigma}{\partial T}$ value the natural convection the Marangoni effect will `work against' the natural convection. Thus, the prediction is that in cases where $\frac{\partial \sigma}{\partial T}$ is negative or equal to zero the flow should have the same form but this should be altered for a positive $\frac{\partial \sigma}{\partial T}$.

The domain used to investigate this is similar to the one used for the melting of gallium outlined in Figure \ref{4-fig-ga-melt-domain} with a hot and cold wall resulting in a temperature differential, $\Delta T$. This $\Delta T$ is stated as \SI{100}{\kelvin} in Bergman and Keller's research but in a possible typo is listed as \SI{10}{\kelvin} in Saldi's thesis. In the present work, \SI{100}{\kelvin} is used in a domain of size $\SI{20}{\milli\meter} \times \SI{20}{\milli\meter}$. The benchmark cases are for steady state, given \gIF only works for transient simulation a judgement for when the simulation stopped altering was required. All simulations settled into a steady state after \SI{10}{\second} of simulated time however, for robustness, the values were taken after \SI{100}{\second}. In all cases, the thermophysical properties are assumed to be identical to those of pure aluminium. For brevity the full case details are given in Appendix B.6.

Figure \ref{4-fig-alum-marg} shows the results for the streamlines and isotherms for $\frac{\partial \sigma}{\partial T}$ equal to zero, $\SI{-3.5e-4}{\newton\per\meter\per\kelvin}$ and $\SI{2e-4}{\newton\per\meter\per\kelvin}$ for this case. Here the direction of the flow is reversed for the positive $\frac{\partial \sigma}{\partial T}$ value as the Marangoni effects `works against' the buoyancy. Note using $\SI{1.95e-4}{\newton\per\meter\per\kelvin}$ instead of $\SI{2.0e-4}{\newton\per\meter\per\kelvin}$ matches the results in \cite{saldi2012marangoni} and \cite{AlumMargFlow} better. When $\SI{2.0e-4}{\newton\per\meter\per\kelvin}$ is used the Marangoni effect is overwhelms the buoyancy driven flow and reverses it completely making the different streamlines harder to see. This discrepancy is due to the specific implementation and slightly different values throughout the solver - as usual it can be corrected with computational factors. These varying streamlines validate the implementation of Marangoni flow for two phases (top boundary is set to gas phase) however for three phases an additional test case is needed. 

\begin{figure}[!h]
\centering
\includegraphics[width=15cm]{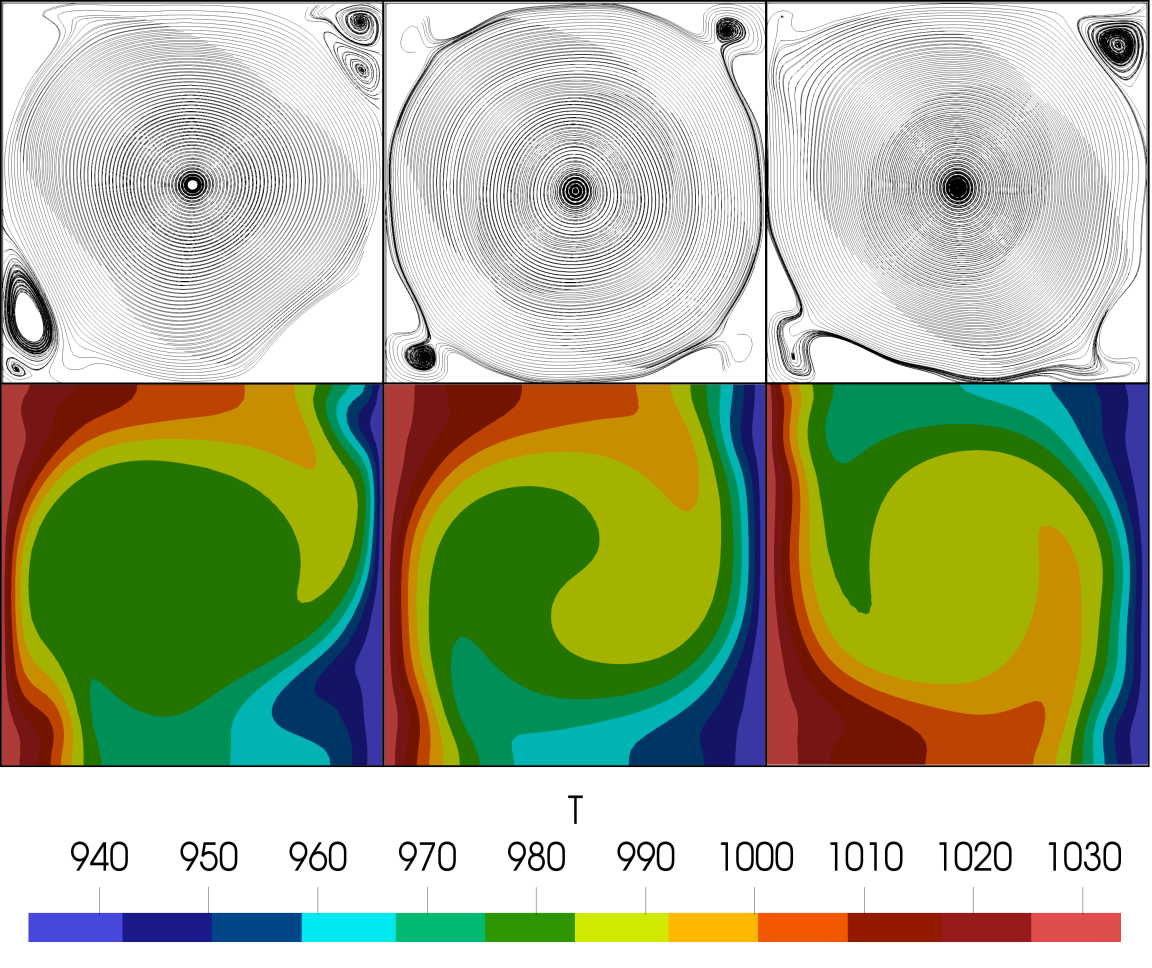}
\caption{Streamlines (top) and isotherms (bottom) for Marangoni driven flow of Aluminium for a 2D differentially heated cavity with $\Delta T = \SI{100}{\kelvin}$ between hot and cold walls. From left to right for both top and bottom $\frac{\partial \sigma}{\partial T}$ is equal to $\SI{-3.5e-4}{\newton\per\meter\per\kelvin}$, zero, and $\SI{1.95e-4}{\newton\per\meter\per\kelvin}$. Note the reversed isotherms for $\SI{1.95e-4}{\newton\per\meter\per\kelvin}$ as the Marangoni effects works against buoyancy.}
\label{4-fig-alum-marg}
\end{figure}

\FloatBarrier
\subsection{Bismuth Marangoni driven flow}
\subsubsection{Overview}
\label{4-bis-marg}
\begin{figure}[!h]
\centering
\includegraphics[width=12.5cm]{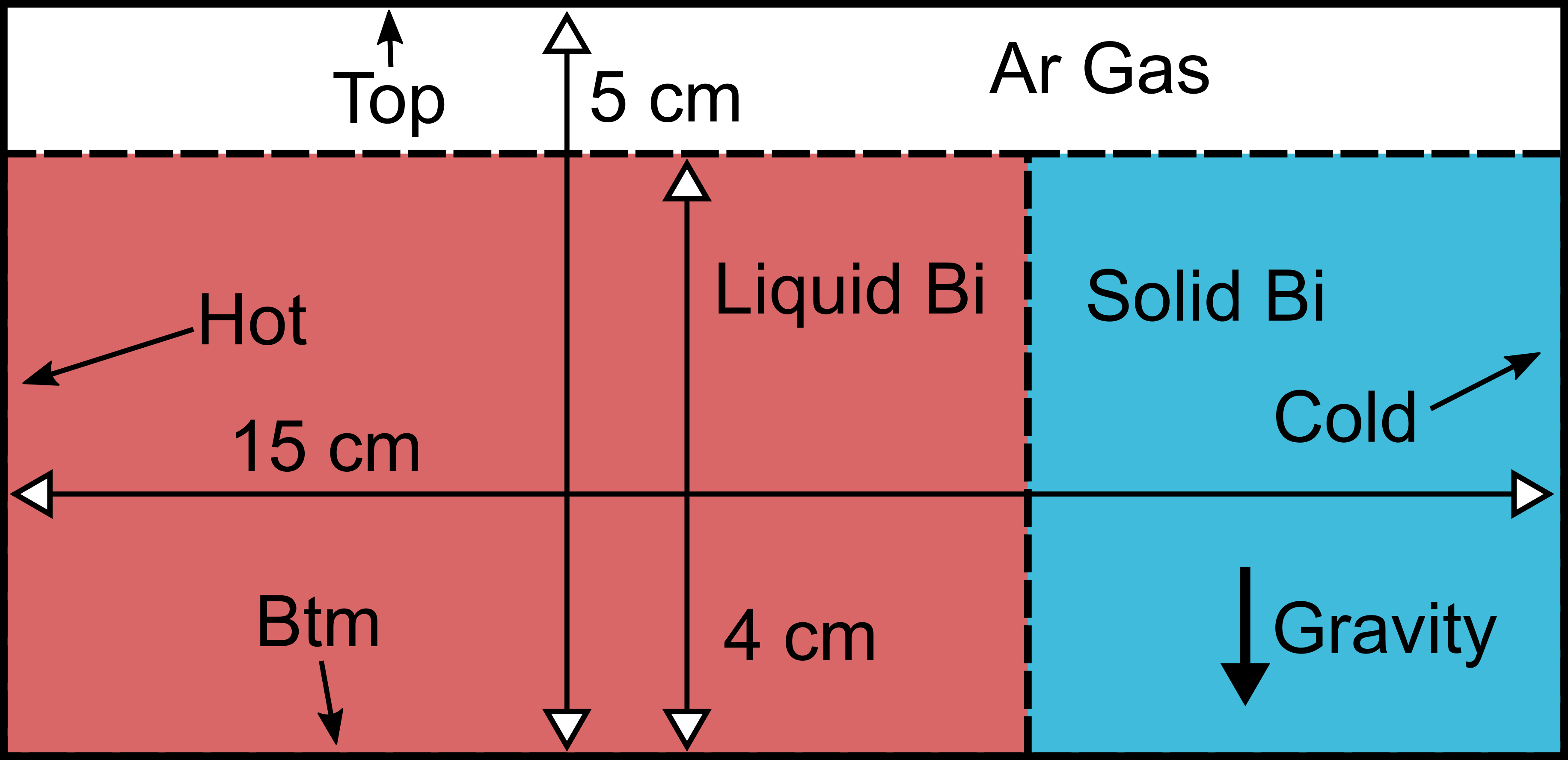}
\caption{2D domain used for the Marangoni driven melting of bismuth benchmark with centimetres as opposed to millimetres as performed by Tan et al. \cite{BiTan2005} and matched by Saldi \cite{saldi2012marangoni}. As detailed in Section \ref{4-bis-marg}, with low gravity and a \SI{15}{\milli\meter} width (as stated in both \cite{BiTan2005} and \cite{saldi2012marangoni}) the surface tension force is overwhelmingly dominant. The solution used to resolve this is to scale the domain to \SI{15}{\centi\meter} width.}
\label{4-fig-bismuth-domain}
\end{figure}

Another benchmark used in Saldi's thesis is for the validation of the solver for three phase Marangoni driven flow. The benchmark, performed by Tan et al. \cite{BiTan2005}, is an ANSYS FLUENT simulation of liquid bismuth, solid bismuth and argon gas in a differentially heated rectangular domain. Their simulation involved finding the steady state melt front for various linear temperature differences between the hot and cold walls in a microgravity environment. As shown in Figure \ref{4-fig-bismuth-domain}, the liquid bismuth covers two thirds of the width of the domain. The linear temperature difference is set so that at the liquid-solid interface, the temperature is the melting point of bismuth (\SI{544.45}{\kelvin}). Thus for $\Delta T = \SI{12}{\kelvin}$ - which is the case covered in Saldi's thesis - the hot wall is set to \SI{552.45}{\kelvin} and the cold wall is set to \SI{540.45}{\kelvin}. In this case, the Marangoni number, Ma, is set to 244. The Marangoni number is defined in equation \ref{4-eq-Ma-No} where $h$ is the height of the free surface, $c_p$ the specific heat capacity, $\nu$ the kinematic viscosity and k the thermal conductivity of a material with a surface tension to temperature gradient and a temperature difference over a free surface of $\Delta T$. For brevity, the full case details are covered in Appendix B.7.

\begin{equation}
\label{4-eq-Ma-No}
Ma = - \frac{\partial \sigma}{\partial T} \frac{h c_p}{\nu k} \Delta T
\end{equation}

In the benchmark case a microgravity environment where $g = \SI{4.4145e-4}{\meter\per\second\squared}$ is used. An issue with this is that at the \SI{15}{\milli\meter} domain width and thus a \SI{10}{\milli\meter} liquid fraction width as stated in \cite{BiTan2005, saldi2012marangoni} the surface tension force of bismuth (\SI{0.0378}{\newton\per\meter}) \cite{thermoPhysProp} is overwhelmingly strong and the phase fraction drags up the walls through essentially capillary action. This results in very different results as the phase fractions behave like small water drops and nothing like the results in \cite{BiTan2005, saldi2012marangoni}. This can be understood intuitively with what happens to the meniscus of water in a tube \SI{10}{\milli\meter} wide and a tube \SI{10}{\centi\meter} wide, and this is in a case where $g = \SI{9.81}{\meter\per\second\squared}$ if $g = \SI{4.4145e-4}{\meter\per\second\squared}$ the difference would be overwhelming. Thus \gIF is completely unable to match the benchmark as stated in both \cite{BiTan2005} and \cite{saldi2012marangoni}. With both references giving the same value it is unlikely both have made the same typo (or were read incorrectly) but intuitively at \SI{10}{\milli\meter} width the simulation simply cannot work with a reasonable (the capillary action subsides at $\approx \SI{1e-6}{\newton\per\meter}.$) value for bismuth surface tension. 

\begin{figure}[!h]
\centering
\includegraphics[width=12.5cm]{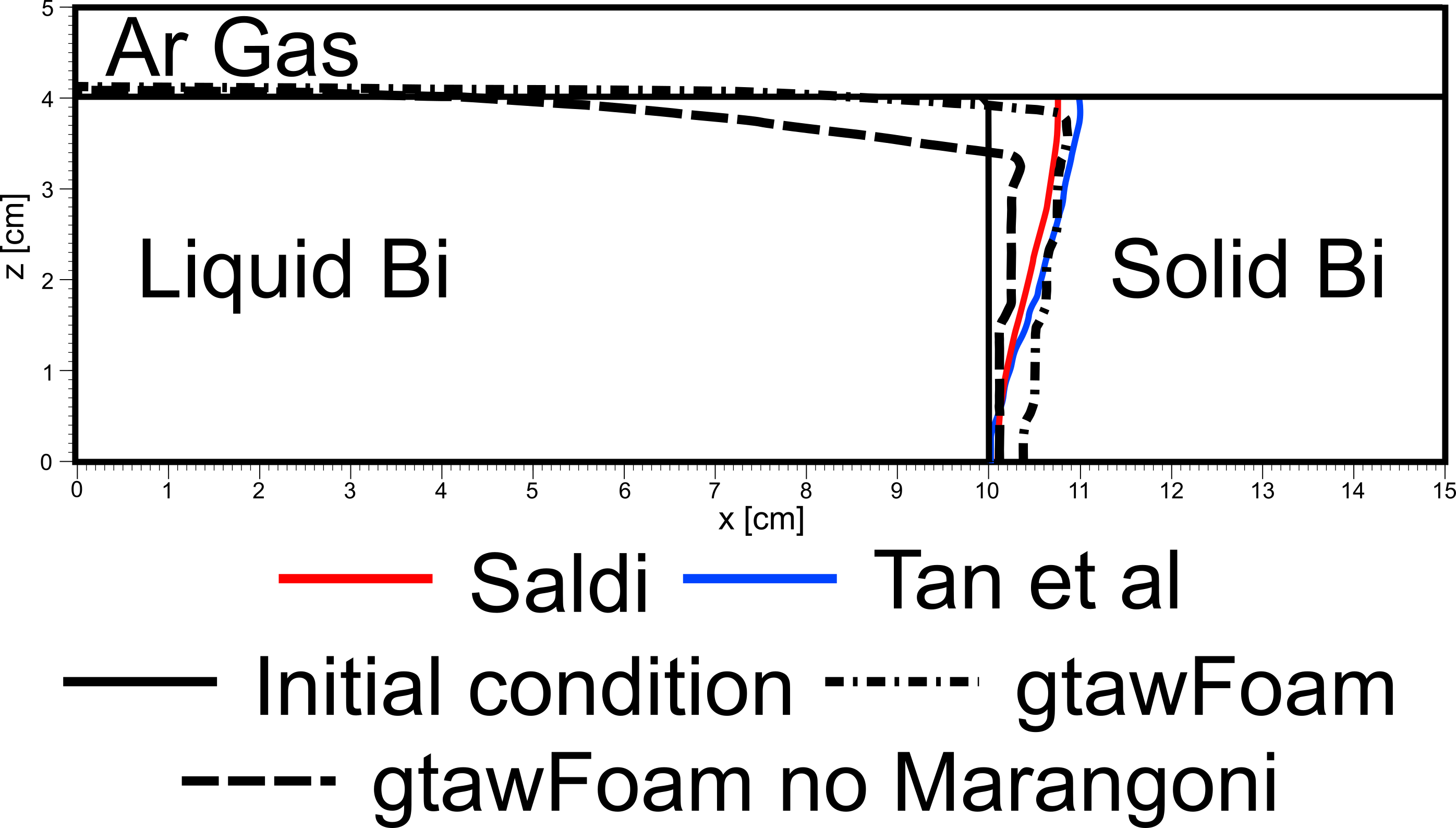}
\caption{Melt front for the initial state, Marangoni driven flow and condition with just surface tension (no Marangoni) for $\Delta T = \SI{12}{\kelvin}$ and $Ma = 244$ compared to the results in \cite{BiTan2005, saldi2012marangoni} scaled by a factor of 10. As the initial condition is set so that the melting temperature isotherm is at the liquid-solid interface, the differences between the initial condition and the other results show that the flow develops due to free surface flow. The difference between where the Marangoni force is turned off and turned on to show its effect on the free surface. Note there is a 0.9\% mass gain with the Marangoni effect enabled and a 5.4\% mass loss with the Marangoni effect disabled.}
\label{4-fig-bismuth-alpha}
\end{figure}

The solution here is to scale the domain by a factor of 10 so that it is \SI{15}{\centi\meter} in width. At these dimensions the results are considerably more similar to those in \cite{BiTan2005, saldi2012marangoni}. Along with the results from Tal et al. and Saldi, Figure \ref{4-fig-bismuth-alpha} shows a comparison between the initial condition (shown in Figure \ref{4-fig-bismuth-domain}), \gIF with $\frac{\partial \sigma}{\partial T} = \SI{4.03e-5}{\newton\per\meter\per\kelvin}$ and \gIF where $\frac{\partial \sigma}{\partial T} = 0$. As the initial condition is set so that the temperature isotherm at the melting temperature of bismuth is at the liquid-solid interface, without surface tension ($\sigma = 0$) the liquid fraction will stay completely still creating a steady state solution from the first time step \footnote{As there is no change in liquid fraction the melt fronts completely overlap at \SI{0}{\second}, \SI{1e3}{\second}, \SI{1e5}{\second} etc... this is hard to show graphically so only the initial condition is shown.} thus any changes in liquid fraction is due to the free surface movement. 

The results in Figure \ref{4-fig-bismuth-alpha} show good match of the Marangoni-driven melt front calculated by \gIF and the simulations of Saldi and Tan et al. Notably the free surface in \gIF deforms due to surface tension whereas in the results from Saldi and Tan the surface remains completely flat (flat surfaces only occur in \gIF when $\sigma = 0$). The melt front for surface deformation due purely to surface tension - where $\frac{\partial \alpha_2}{\partial t} = 0$ - is also included. Here, \gIF begins to fail as the solid phase is treated as a fluid rather an internal boundary and there is no calculation of contact angle so the surface tension force gradually curves the free surface more and more. This effect only becomes pronounced after $\approx \SI{3e3}{\second}$ so it is not a major issue. Whilst this deformation does effect the Marangoni-driven case precisely because there is another driving force (Marangoni) the effect is not as severe.     

\subsubsection{Surface artefacts}
Interestingly due to the implementation of latent heat - shown in equation \ref{4-eq-latent-heat} - any change in the alpha fraction causes the cell elements of $S_{Latent}$ to be non-zero. This means that due to the surface tension force causing the free surface of the liquid metal to move there is artificial latent heat added to the domain as $\frac{\partial \alpha_2}{\partial t} \neq 0$ and thus $S_{Latent} \neq 0$.      

\begin{equation}
\label{4-eq-latent-heat}
S_{Latent} = \rho_2 L \frac{\partial \alpha_2}{\partial t}.
\end{equation}

\begin{figure}[!h]
\centering
\includegraphics[width=15cm]{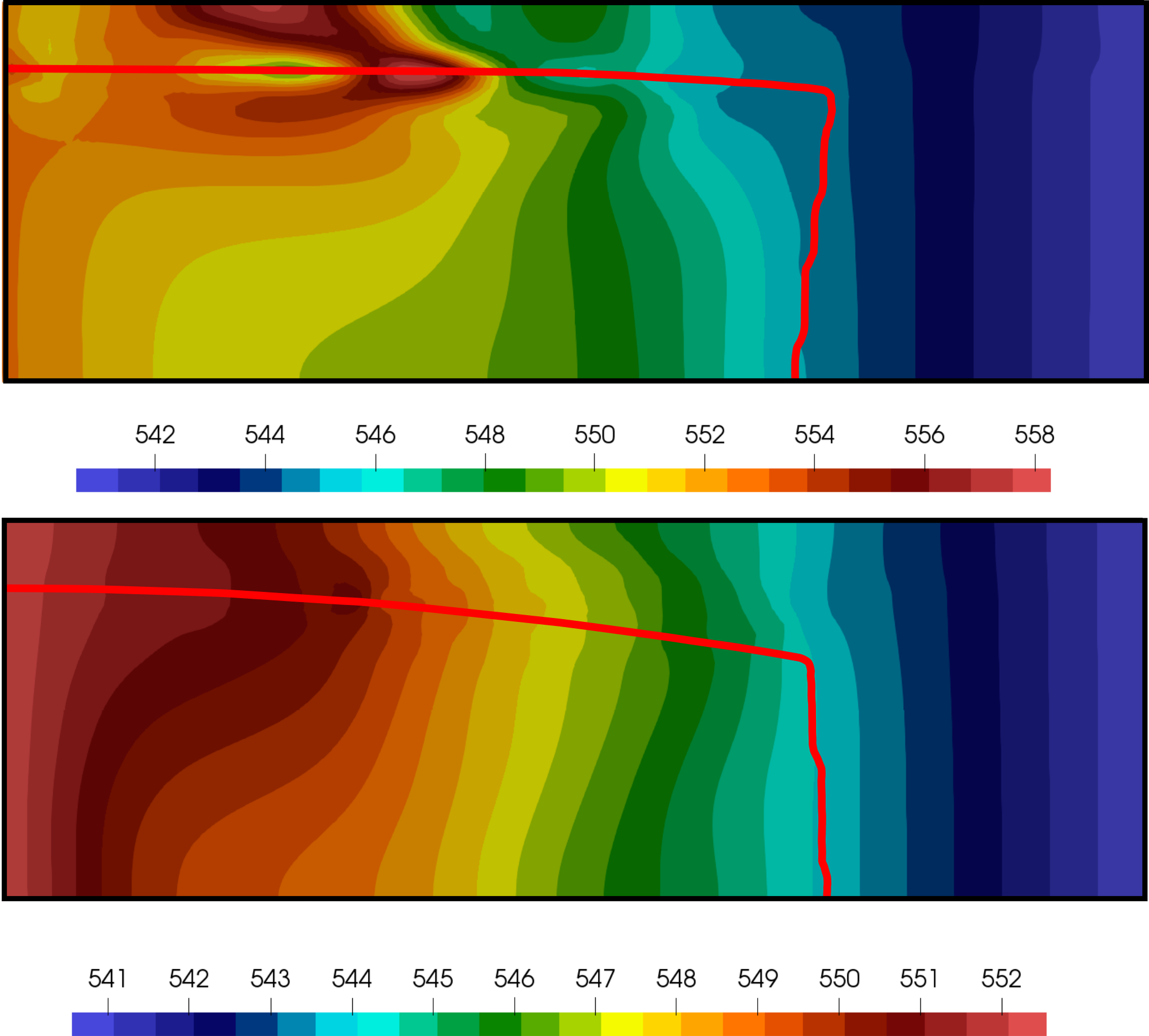}
\caption{Temperature profile (in kelvin) of the default \gIF implementation (top) and with a latent heat term only present where there was metal in the initial conditions (bottom). The corresponding phase fronts for each case is overlaid in red. Note the smooth temperature transition in the bottom image with unphysical equations compared to the temperature variations at the free surface at the top of the top image. Both cases have the same $Ma = 244$.}
\label{4-fig-bismuth-temp}
\end{figure}

There are few methods to address this principally $S_{Latent}$ is multiplied by the total metal phase of the current and previous time step: $\alpha_{metal} \cdot (1.0 - \alpha_1(t - \Delta t))$. This serves to zero out the cells where there isn't a metal fraction and thus there is not the possibility of phase change. It has a side benefit of allowing the creation of new $\alpha_{metal}$ fraction within the domain without causing additional latent heat; this is useful for modelling additive manufacturing. Zeroing the latent heat where there isn't phase change is a sensible physical addition to the solver and is implemented as the default option. However, due to the way \of automatically blends fields there is still spurious latent heat. The most forceful way to address this is to set $\rho_2 L$ as an initial condition to be one value for where $\alpha_{metal} = 1$  and zero where $\alpha_{metal} = 0$ then never update the field. Therefore, even if $\frac{\partial \alpha_2}{\partial t} \neq 0$, as the free surface moves $\rho_2 L$ will be $= 0$ in regions initially occupied by the gas phase. This is the most unphysical solution but it works the best for modelling temperature accurately. As will be covered further later in this thesis, this is an example of the trade-off between phase-fraction accuracy and temperature accuracy. In this thesis, the phase fraction was judged to be the priority. Figure \ref{4-fig-bismuth-temp} shows a comparison between the temperature fields produced without the $S_{Latent}$ `\emph{correction}' and the temperature field with the most unphysical correction. 

\FloatBarrier
\section{Welding benchmark cases}
\subsection{Elliptic paraboloid benchmark}
\label{4-sec-ellip-paraboloid}
\subsubsection{Elliptic paraboloid parameters}
With the individual elements of the solver investigated, they are all combined to test the solver's performance at simulating welding. Given the heat source used is similar to the one proposed in \cite{paraboloid-paper}, the same test case is investigated. As detailed in Chapter 3, the heat source features a computational parameter $C_C$ in addition to the physical parameters of the weld power ($Q_{Source}$), the weld depth ($l$) and one third of the weld pool width ($\omega)$. For convenience, the heat source equation detailed in Chapter 3 is repeated here.

\begin{equation}
        \label{5-heat-source-equation}
        Q_{GTAW}(x,y,z)  = C_{C} \left( \frac{Q_{Source}}{\omega^2 \sqrt{l}} \right) \cdot \exp\left(\ln(C_{cut})\left(\frac{(x^2 + y^2)}{\omega^2} + \frac{z}{l}\right)\right) \cdot g(x, y, z) \cdot \alpha_{metal}.
\end{equation}   

The five parameters ($Q_{Source}$, $l$, $\omega$, $C_{cut}$ and $C_C$) can all be changed to fit a specific results however philosophically the weld power should not be tuned. This means a weld using \SI{10}{\ampere} and \SI{10}{\volt} should always be modelled as \SI{100}{\watt}. Secondly, other variables such as the efficiency of the GTAW arc and any other constants can all be incorporated in a catch all constant $C_C$ which is an obvious candidate for tuning. The $C_{cut}$ variable is set equal to an arbitrary 5\% (0.05) - this determines the reduction at the edge of the heat source compared to the peak. So 5\% means that the heat source is 5\% of its maximum (central) value at its edge. This leaves $l$ and $\omega$ which are identified as the weld depth ($l$) and one third of the weld pool width ($\omega)$ in \cite{paraboloid-paper}. However, as $\omega$ is given a somewhat arbitrary $\frac{1}{3}$ multiplier it is the view of the author that - providing it is consistent - this multiplier should also be tunable; this also applies to the $1 \times$ multiplier on $l$. 

\begin{figure}[!htbp]
\centering
\includegraphics[width=13.5cm]{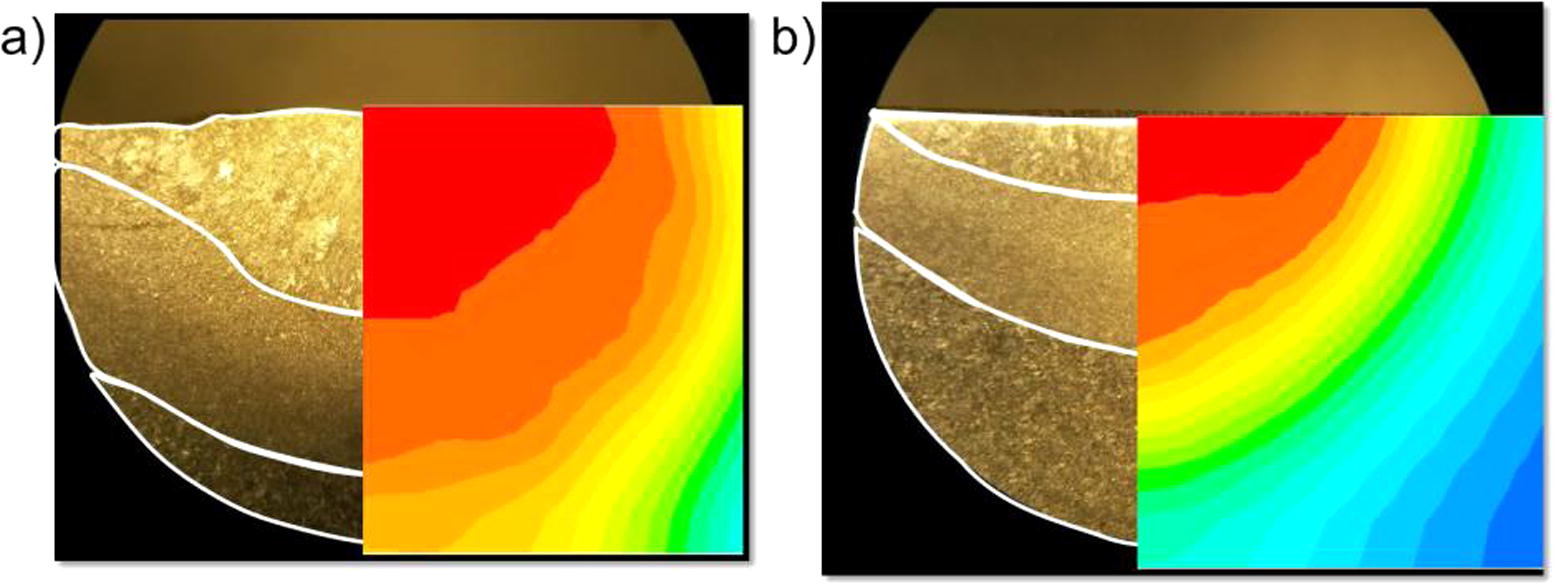}
\caption{Experimental and simulation fusion zone cross-section from Figure 16 in \cite{paraboloid-paper}. Both a) and b) are GTAW with a \SI{130}{\ampere} welding current and a \SI{13.4}{\volt} arc voltage on stainless steel. The travel speed for a) is \SI{1.25}{\milli\meter\per\second} and for b) is \SI{2.5}{\milli\meter\per\second}.}
\label{4-fig-paper-sim-exp}
\end{figure}

As with the weld power, a key point to consider is that to ensure predictability physical causes should be prioritised. That is to say, $C_C$ shouldn't be increased to match a benchmark result better when a better formulation of a physical quantity - such as thermal conductivity as demonstrated in Section \ref{4-tin-solid} - could be used instead. To this end, Figure \ref{4-fig-paper-sim-exp} shows a comparison between the model used and the experiment performed in \cite{paraboloid-paper}. Here, the match is very close yet in their model only a Darcy source term and a buoyancy source term are added to the Navier-Stokes equation. Given gravity only acts in one direction (e.g. on the z axis), the fact density typically decreases with a temperature increase and the Gaussian-esk shape of the temperature distribution presented in \cite{paraboloid-paper} the buoyancy force should generate a wide and shallow weld profile. Whereas, the fact $l$ has a $1 \times$ multiplier but $\omega$ has a $\frac{1}{3} \times$ multiplier combined with the $x^2$ vs $z$ form equation \ref{5-heat-source-equation} means that the heat input drops faster horizontally than vertically. Therefore the model proposed in \cite{paraboloid-paper} can be understood as:

\begin{itemize}
    \item Calculated paraboloid heat input shape ($\approx$ Gaussian shape with low variance) creates weld depth.  
    \item Buoyancy force pushes the weld pool `outwards' which broadens the weld and creates the matching weld shape.
\end{itemize}

%\begin{figure}[!htbp]
%\centering
%\includegraphics[width=10cm]{images/2021-07-04-paper-sim-exp-comp.png}
%\caption{Representation of the experimental weld cross sections (left) and corresponding simulated weld cross sections (right) from the paper the heat source in \gIF is based upon (\cite{paraboloid-paper}). Note the good match despite only the buoyancy force driving the flow.}
%\label{4-fig-paper-sim-exp}
%\end{figure}

\begin{figure}[!htbp]
\centering
\includegraphics[width=13.5cm]{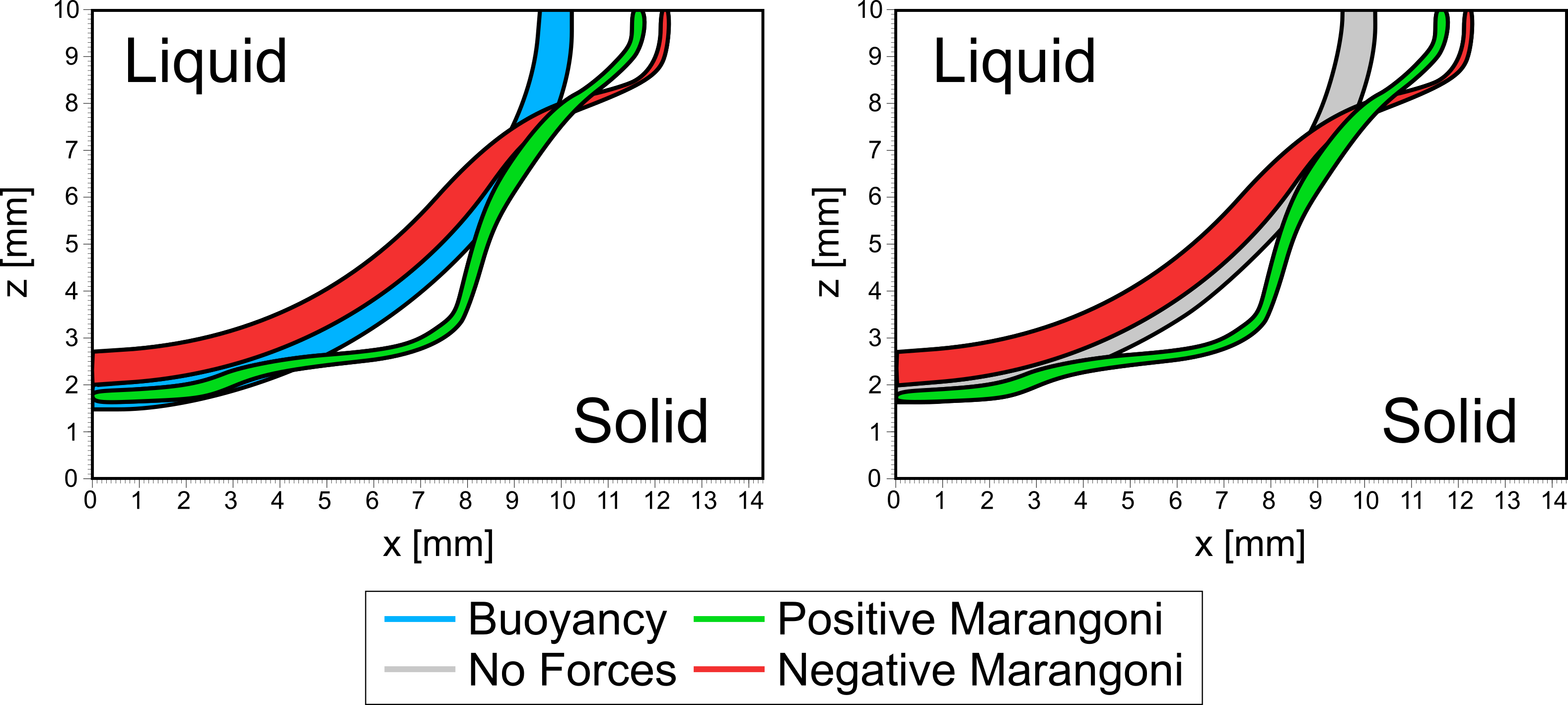}
\caption{Weld pool melt fronts at \SI{2.5}{\second} for a heat source with weld power of \SI{1742}{\watt} and parameters $\omega = \SI{5.2}{\milli\meter}$ and $l = \SI{5.3}{\milli\meter}$ on stainless steel. The melt front in blue has only the Buoyancy force activated. The melt front in green has only the Marangoni effect with a positive $\frac{\partial \sigma}{\partial T}$ value. The melt front in red has only the Marangoni effect with a negative $\frac{\partial \sigma}{\partial T}$ value.  The melt front in grey has no forces switched on. The grey and blue lines are effectively identical so the comparison is split into two images of red-green-blue and red-green-grey. As with other images in this chapter, the melt front widths (defined as the region where $0.4 \leq \alpha_2 \leq 0.6$) vary.}
\label{4-fig-forces}
\end{figure}

The challenge with this methodology is that when implemented in \gIF with the same parameters the buoyancy force alone is not strong enough to create the weld shape. The code from \cite{paraboloid-paper} is not publicly available so it is difficult to assess quite how the results were generated. In fact, for shorter times ($< \SI{3}{\second}$) it does not seem to change the results whether it is on or off. This is because the thermal gradients generated using $C_{cut} = 5\%$, $l =$ weld pool depth and $\omega = \frac{1}{3}$ weld pool width are not large enough. A possible alleviation to this is to incorporate the Marangoni effect. The Marangoni effect will typically drive flows $10 \times$ as strongly as the buoyancy force \cite{weldDrive}. Figure \ref{4-fig-forces} shows this approach with a comparison of the welds produced using either no forces, only the buoyancy force, only the Marangoni effect with a positive $\frac{\partial \sigma}{\partial T}$ or only the Marangoni effect with a negative $\frac{\partial \sigma}{\partial T}$. Here, the incorporation of the Marangoni effect pushes the results from \gIF closer to the weld shape in Figure \ref{4-fig-paper-sim-exp}.

The drawback in turning on the Marangoni source term is that computation times increase dramatically which for large batch simulations can add up. For the batch simulations in Chapter 6 this computation time becomes very important. For the ultra-thin-walled tube welding, the melt fraction is the most important thus it was judged best to benchmark the volumetric heat source in such as way as to match experimental welds irrespective of which forces are used. Therefore, similar to other features in \gIF the Marangoni effect serves as an option rather than a necessity with the welding benchmarks just using the volumetric heat source without any additional forces.

Finally, the optimum values for $C_C$ and $C_{cut}$ and optimum definitions for $\omega$ and $l$ need to be identified. The main quality criteria to assess \gIF by should be the predicted weld. Ideally, the real-time temperature of the weld pool would be predicted but it is the view of the author that this can be sacrificed to achieve better weld prediction capabilities. Through loosening the temperature prediction accuracy aims, the weld pool can be modelled more accurately. To execute this, $C_{cut}$ could be lowered so that for a given $C_C$ value there is a larger contrast between peak and edge values in the source. Further, the definitions (e.g. the multipliers) of $\omega$ and $l$ can be change so - as an example - $\omega$ could be defined as one half of the weld pool width. This would mean that there will be a larger region that the heat source changes temperature over and hence a more pronounced temperature gradient.

\subsubsection{Elliptic paraboloid test case}
\label{4-sec-ellip-para-case-spec}
\begin{figure}[!htbp]
\centering
\includegraphics[width=12.5cm]{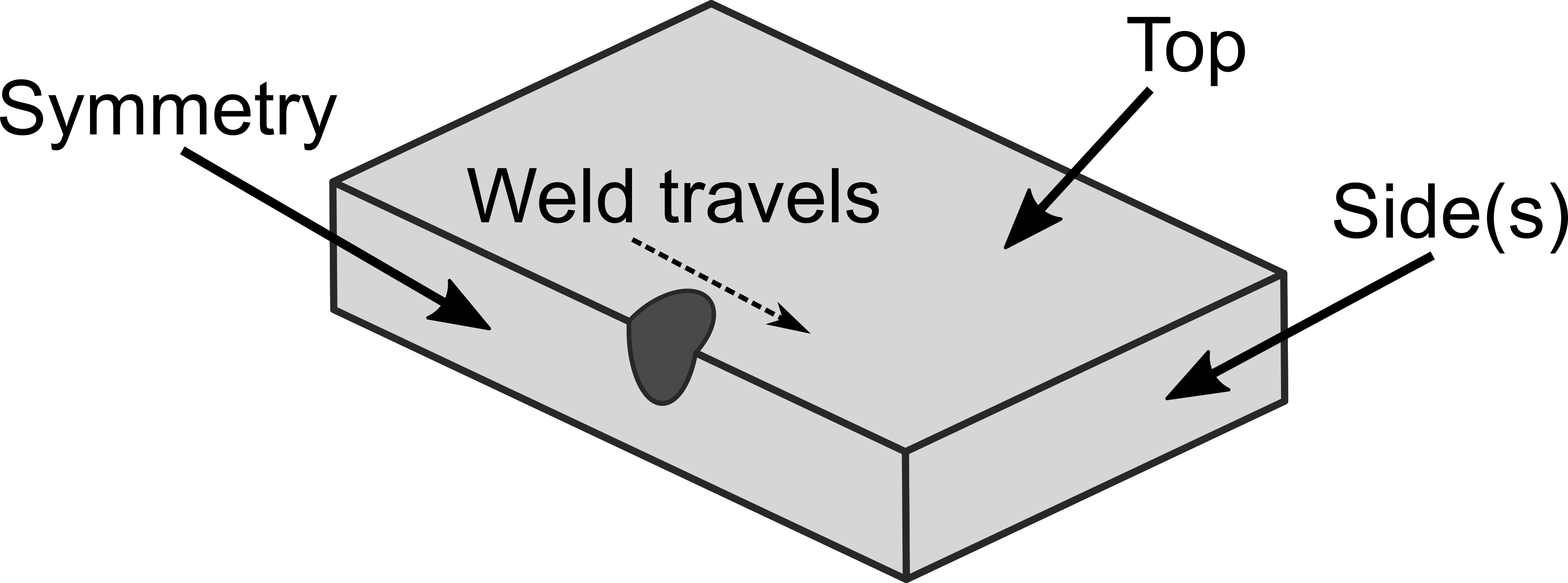}
\caption{3D domain used for the elliptic paraboloid benchmark. The boundary marked symmetry uses the symmetry plane condition which allows the simulation to only have to run half of a full weld. All boundaries not marked are grouped into the `side' boundary.}
\label{4-fig-paraboloid-domain}
\end{figure}

The test case presented in \cite{paraboloid-paper} involves autogenous welding on an AISI CrMo 12-1 stainless steel plate. The test case is in 3D; a sketch of the domain is shown in Figure \ref{4-fig-paraboloid-domain}. To improve computational performance, a symmetry plane boundary condition is used which allows for a domain half the size to be simulated. Compared to the other benchmarks, this one includes the custom boundary conditions for heat loss describes in Chapter 3 - the top of the domain serves as a free surface that is radiative and convective whereas the remaining sides are just convective. To avoid direct reproduction of the figures in \cite{paraboloid-paper} illustrations are used for comparison. For brevity, the full case details are given in Appendix B.8. 

\begin{figure}[!htbp]
\centering
\includegraphics[width=12cm]{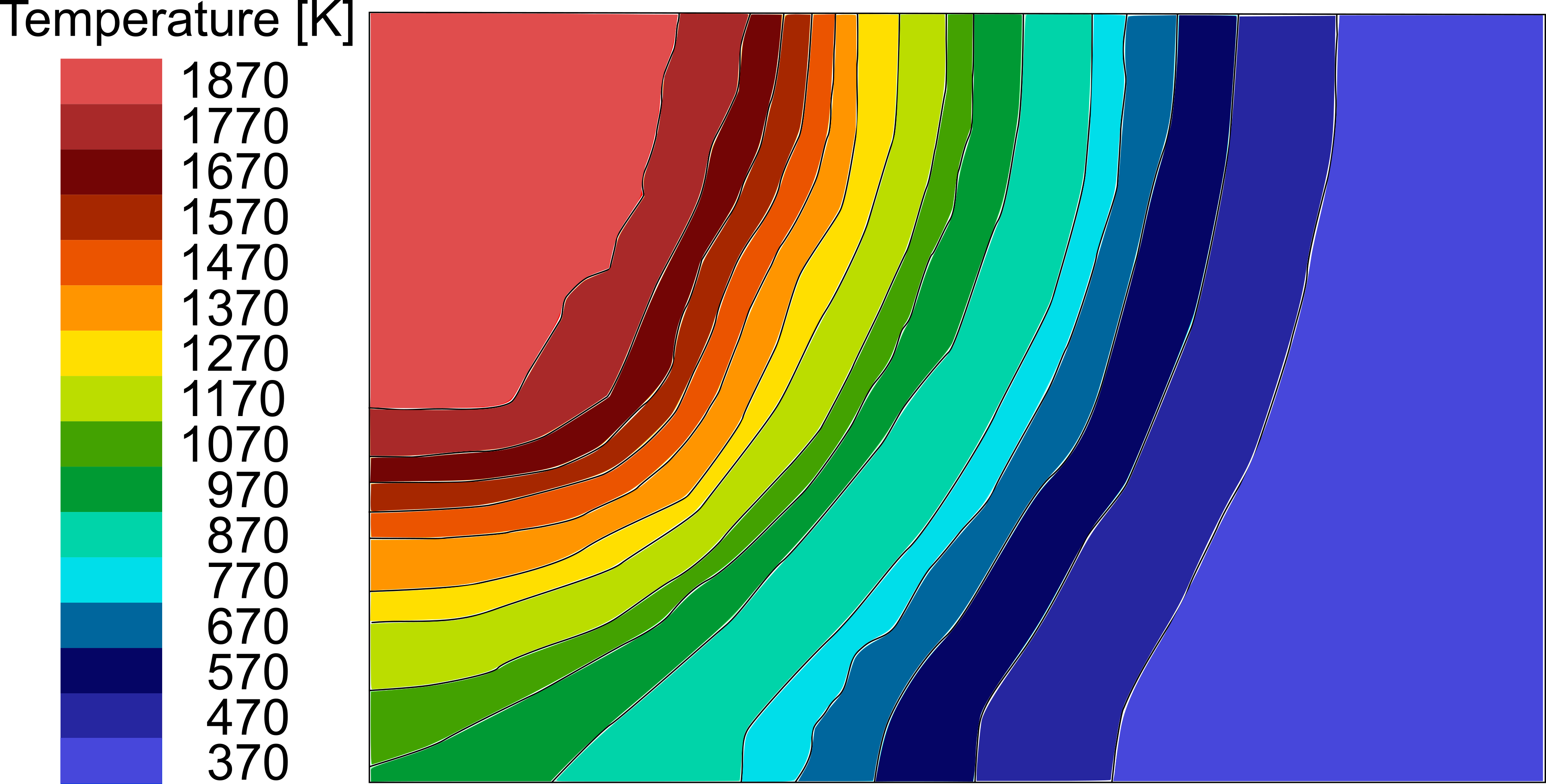}
\caption{Illustration of the Temperature profile presented in the elliptic paraboloid benchmark paper \cite{paraboloid-paper}. Note the binning used.}
\label{4-fig-paraboloid-exp-temp}
\end{figure}

\begin{figure}[!htbp]
\centering
\includegraphics[width=12cm]{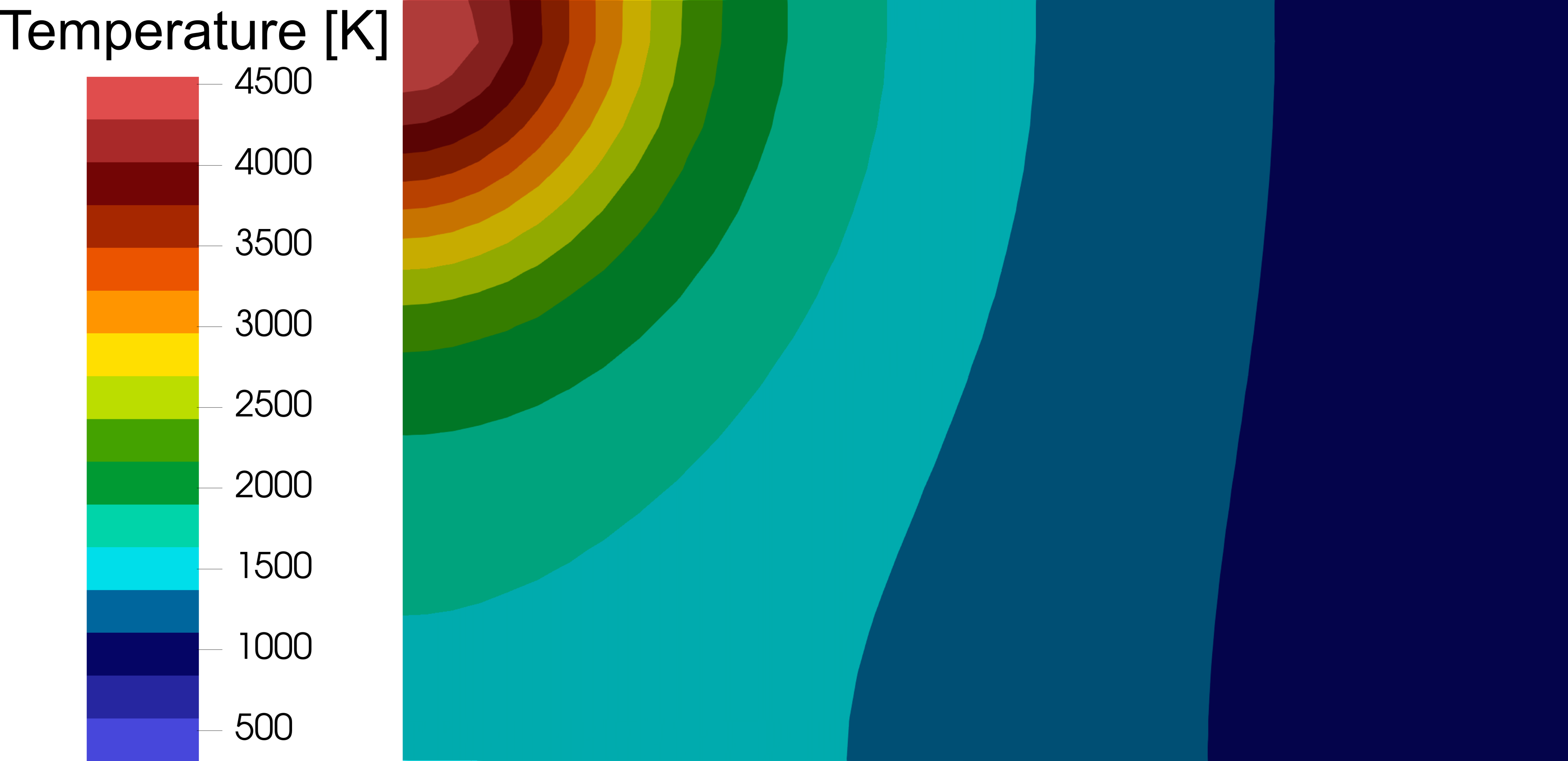}
\caption{Temperature profile produced by \gIF for case 2 in Table \ref{4-tab-para-case}.}
\label{4-fig-paraboloid-sim-temp-high}
\end{figure}

\begin{figure}[!htbp]
\centering
\includegraphics[width=12cm]{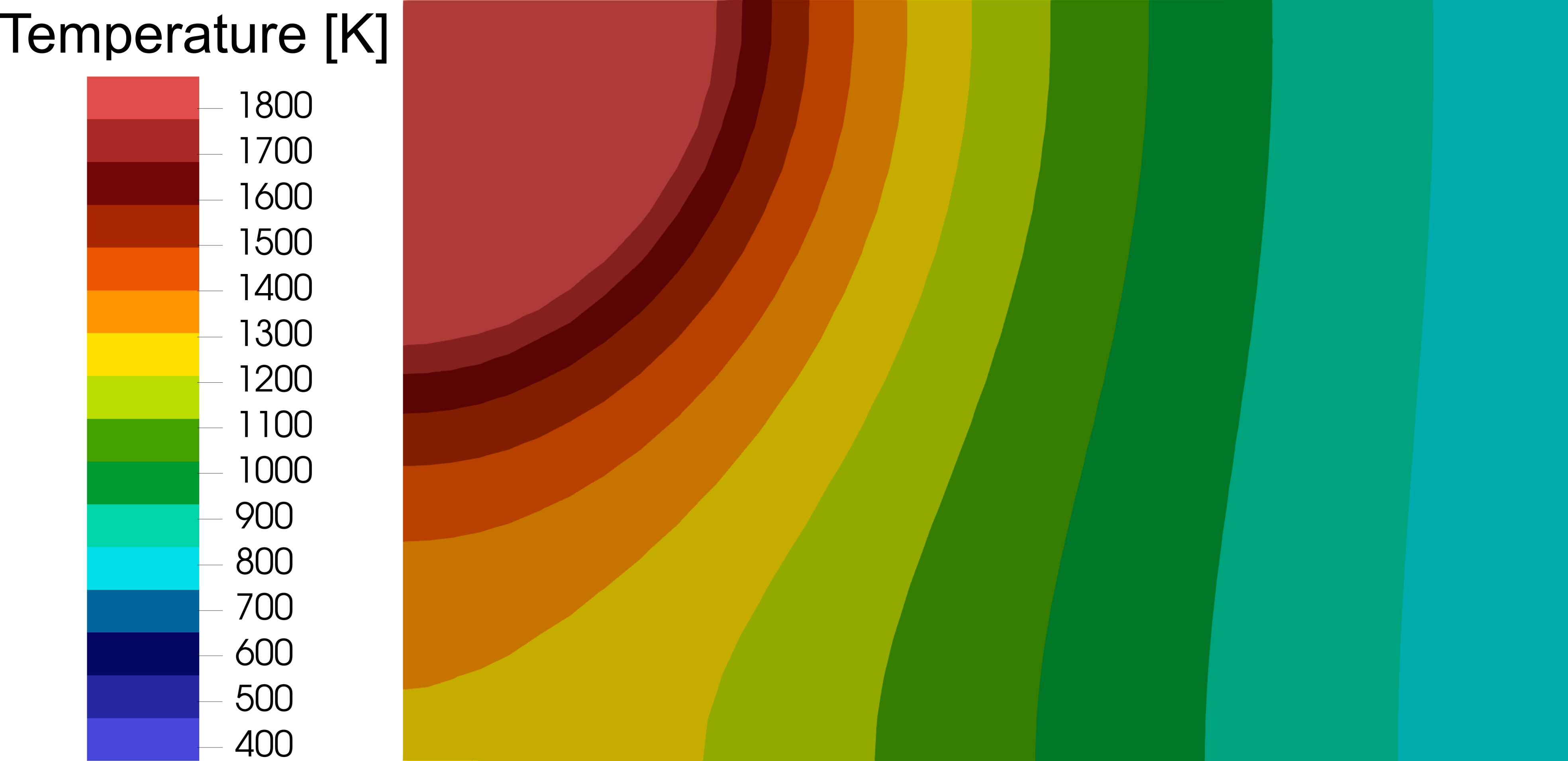}
\caption{Temperature profile in Figure \ref{4-fig-paraboloid-sim-temp-high} re-binned to match the binning in Figure \ref{4-fig-paraboloid-exp-temp}. Note the now reasonable match with Figure \ref{4-fig-paraboloid-exp-temp} for the left hand side of the figure.}
\label{4-fig-paraboloid-sim-temp-low}
\end{figure}

In \cite{paraboloid-paper}, the researchers investigated two weld configurations (weld profiles shown in Figure \ref{4-fig-paper-sim-exp}) specified in Table \ref{4-tab-para-case}. To benchmark \gIF these configurations can be run with the same parameters and the optimum tuning constants, $C_C$ and $C_{cut}$, can be established. Note, $C_{cut}$ is fixed to 0.05 in \cite{paraboloid-paper} and with the other parameters specified in Table \ref{4-tab-para-case}, $C_C = 40$ was found to give the best results. Increasing $C_C$ further leads to an excessively deep weld as the shape of the weld does not change, only the size does. Temperature isotherms were only provided for provided for case 2 - an illustration of which is shown in Figure \ref{4-fig-paraboloid-exp-temp}. The temperature isotherms produced with the same parameters in \gIF is shown in Figure \ref{4-fig-paraboloid-sim-temp-high}. Here, the peak temperature is considerably higher at center of the heat source but further away the results are more similar. Curiously, if the results from \gIF are re-binned to match the binning in \cite{paraboloid-paper} the match is a lot closer; this is shown in Figure \ref{4-fig-paraboloid-sim-temp-low}. Given the code in \cite{paraboloid-paper} isn't publicly available the cause of this discrepancy is difficult to access.      

\begin{table}[ht]
\setlength{\tabcolsep}{10pt}
    \centering
    \caption{\label{4-tab-para-case} Heat source specifications from \cite{paraboloid-paper}. Note is $C_C$ unique to \gIF due to the handling of constants but is added for context.}
    \begin{tabular}{l l l l l l l}
    \toprule
    Case \# & $Q_{Source}$ [\si{\watt}] & $l$ [\si{\milli\meter}] & $\omega$ [\si{\milli\meter}] & Velocity, $v$ [\si{\milli\meter\per\second}] & $C_{cut}$ & $C_C$ \\
    \hline
    1 & 1742 & 2.65 & 2.6$\dot{6}$ & 1.25 & 0.05 & 40 \\
    2 & 1742 & 1.42 & 2.3$\dot{3}$ & 2.5 & 0.05 & 40 \\
    \bottomrule
    \end{tabular}
\end{table}

Looking again at Figure \ref{4-fig-forces} and comparing it to Figures \ref{4-fig-paraboloid-sim-temp-high} and \ref{4-fig-paraboloid-sim-temp-low} reveals that the melt fraction closely matches the temperature isotherm at the welded material's melting temperature. Given the shape of these isotherms, the core issue with \gIF when using the parameters in Table \ref{4-tab-para-case} seems to be that they are not wide enough. Further, Figure \ref{4-fig-forces} shows that the forces within the liquid fraction such as buoyancy and Marangoni have a somewhat muted effect. Thus, in order to better match the melt fractions like in Figure \ref{4-fig-paper-sim-exp} the temperature isotherms need to change. As previously stated, in this thesis the prediction accuracy temperature is secondary to the melt fraction prediction thus there is a reasonable amount of flexibility in changing the heat source parameters. In addition to changing the parameters, there is also an option of adding in an additional heat source in essentially the same way as the Goldak double ellipsoid \cite{goldak1984} heat source works. After testing it was found it does not actually change the shape of the melt front as much as expected and it muddies the direct connection between $\omega$ and $l$ and the weld dimensions. Thus, the better option was judged to be to keep one heat source and modify the parameters within the base paraboloid model. Inspecting equation \ref{5-heat-source-equation}, which defines the heat source, can be used to establish principles to change it:  
\begin{itemize}
    \item Increasing $\omega$ and $l$ causes the overall region defined in $Q_{GTAW}(x,y,z)$ to increase.
    \item Increasing $\omega$ and $l$ causes the peak value of $Q_{GTAW}(x,y,z)$ to decrease and thus any increase in $\omega$ and $l$ needs to be compensated by an increase in $C_C$.  
    \item $C_{cut}$ has essentially the same role as standard deviation does in a Gaussian curve so to broaden the heat source it should be reduced.
\end{itemize}

\begin{figure}[!htbp]
\centering
\includegraphics[width=12.5cm]{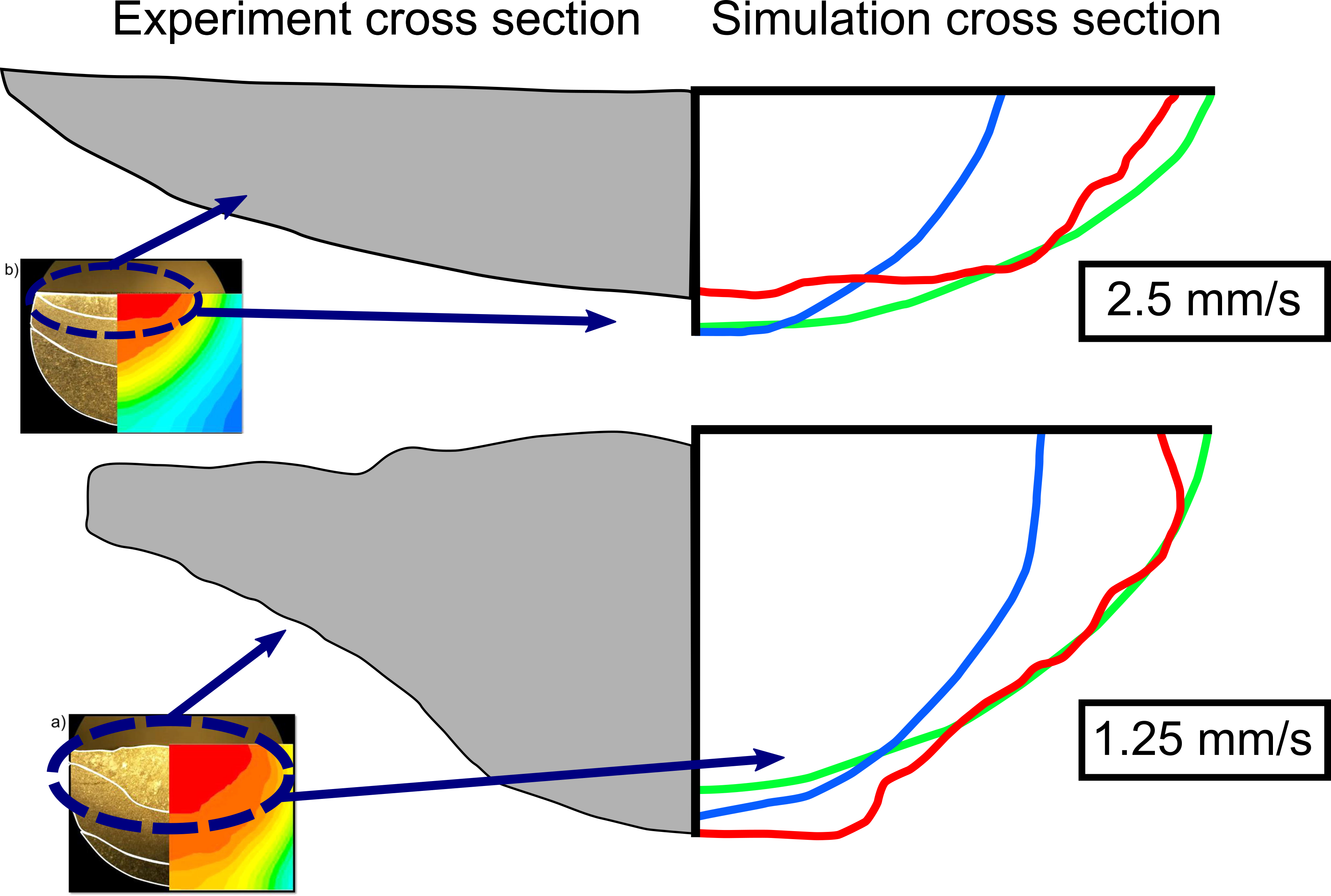}
\caption{Enlarged representations of the fusion zone cross-section experimental results (left) from \cite{paraboloid-paper} compared to simulation results (right) for \cite{paraboloid-paper} in red and simulation output from \gIF in blue and green. The \gIF procedures are in Table \ref{4-tab-para-case} for blue and Table \ref{4-tab-para-case-gIF} for red. Bottom image is for a welding speed of \SI{1.25}{\milli\meter\per\second} and the top image is for \SI{2.5}{\milli\meter\per\second}. The left hand side of the image shows the regions from Figure \ref{4-fig-paper-sim-exp} that the representation is from.}
\label{4-fig-paraboloid-exp-vs-sim}
\end{figure}

Not included in equation \ref{5-heat-source-equation} is the effect of travel speed - although this can be accounted for by changing $C_C$. With these points in mind, Table \ref{4-tab-para-case-gIF} shows new specifications for the heat source. The numbers were picked deliberately so that $\omega$ would be close to the weld width and to be whole numbers so as to avoid over-fitting. Here $\omega$ defines the full weld radius, $l$ the weld pool depth and $v$ the velocity (travel speed). $C_{cut}$ is reduced to 0.01 and $C_C$ increased in accordance with velocity. Figure \ref{4-fig-paraboloid-exp-vs-sim} shows the resultant weld melt fronts for cases 1 - 4. Here, cases 3 and 4 perform better than 1 and 2 and therefore the new parameters are accepted. The final tinkering of the weld pool shape can then be supplied via forces internal to the weld pool such as buoyancy and Marangoni. Given Marangoni was not included in \cite{paraboloid-paper} and buoyancy has little effect (shown in Figure \ref{4-fig-forces}), this final tinkering is omitted. Also, there is the danger of over-fitting to a benchmark case. As was proven in the variety of cases shown in sections \ref{4-melting-section} and \ref{4-solidification-section}, a close match to a benchmark (e.g. gallium melting) does not universally translate to all benchmark cases. Thus, the fit presented in Figure \ref{4-fig-paraboloid-exp-vs-sim} was judged to be an acceptable match.

\begin{table}[ht]
\setlength{\tabcolsep}{10pt}
    \centering
    \caption{\label{4-tab-para-case-gIF} New heat source specifications.}
    \begin{tabular}{l l l l l l l}
    \toprule
    Case \# & $Q_{Source}$ [\si{\watt}] & $l$ [\si{\milli\meter}] & $\omega$ [\si{\milli\meter}] & $v$ [\si{\milli\meter\per\second}] & $C_{cut}$ & $C_C$ \\
    \hline
    3 & 1742 & 2.65 & 8 & 1.25 & 0.01 & 100 \\
    4 & 1742 & 1.42 & 7 & 2.5 & 0.01 & 150 \\
    \bottomrule
    \end{tabular}
\end{table}

%\subsection{Weld melt pool formation}
%Time scales
\subsection{Independent welding benchmark}
The final outstanding issue with \gIF is whether the results from the weld benchmark in Section \ref{4-sec-ellip-paraboloid} translate to other welding scenarios. So far, the benchmarks used for welding are from the paper that first proposed the paraboloid model. Thus, to have confidence in the predictability of \gIF an independent welding benchmark should be investigated. The chosen benchmark is by Jamshidi et al. \cite{independent-paper} investigating autogenous welding on \SI{5}{\milli\meter} thick 304 stainless steel plates. This case is modelled with a similar domain to that shown in Figure \ref{4-fig-paraboloid-domain} except for the domain is set to \SI{12}{\milli\meter} thick. This is because some of the depths on the experimental welds - such as \SI{4.8}{\milli\meter} - very nearly fully penetrate the 304 stainless steel plates. If this is simulated exactly, then the liquid weld gets too close to the bottom domain boundary resulting in computational artefacts that effect the weld predictions. A \SI{12}{\milli\meter} thick domain avoids this. The boundaries are the same as described in Section \ref{4-sec-ellip-para-case-spec} with a radiative and convective free surface top boundary with the other boundaries just convective. For brevity, the full case details are given in Appendix B.9.

The case specified in \cite{independent-paper} is quite different to the cases that will be covered in Chapters 5 and 6. This paper was chosen as it provided six comparison cases with the required information to apply the elliptic paraboloid method that were different to cases 3 and 4 whilst also being roughly similar. As has been demonstrated in this chapter thus far, there is a limit to what `benchmarking' means with respect to over-fitting \gIF to experimental results. By choosing an independent study with an appropriate level of dissimilarity for benchmark cases, a user can have confidence in the results\footnote{Not to labour the point but this is a key design choice when creating materials processing simulation tools such as \emph{gtawFoam}. Realistically, it is possible to get any results the user wants with the inclusion of apt computational constants. However, to have confidence in the results one set of constants needs to perform well for all cases.}. If a case is slightly dissimilar but still produces apt results a user can have confidence the benchmarking worked (or failed if the results are inapt) but with a highly dissimilar case it is difficult to know whether it is the case or the solver. For instance, the melting of gallium and the melting on tin require slightly different case specifications. It would be difficult to determine if the cases did actually require different specifications if there could be an issue with the solver instead. Secondly, the current levels and produced weld widths for the ultra-thin-walled tubing used in Chapter 6 are known exactly. Similarly, the experimental results from Chapter 7 are well known. Therefore, these two chapters can actually serve as test cases to assess the benchmarking.

\begin{figure}[!htbp]
\centering
\includegraphics[width=15cm]{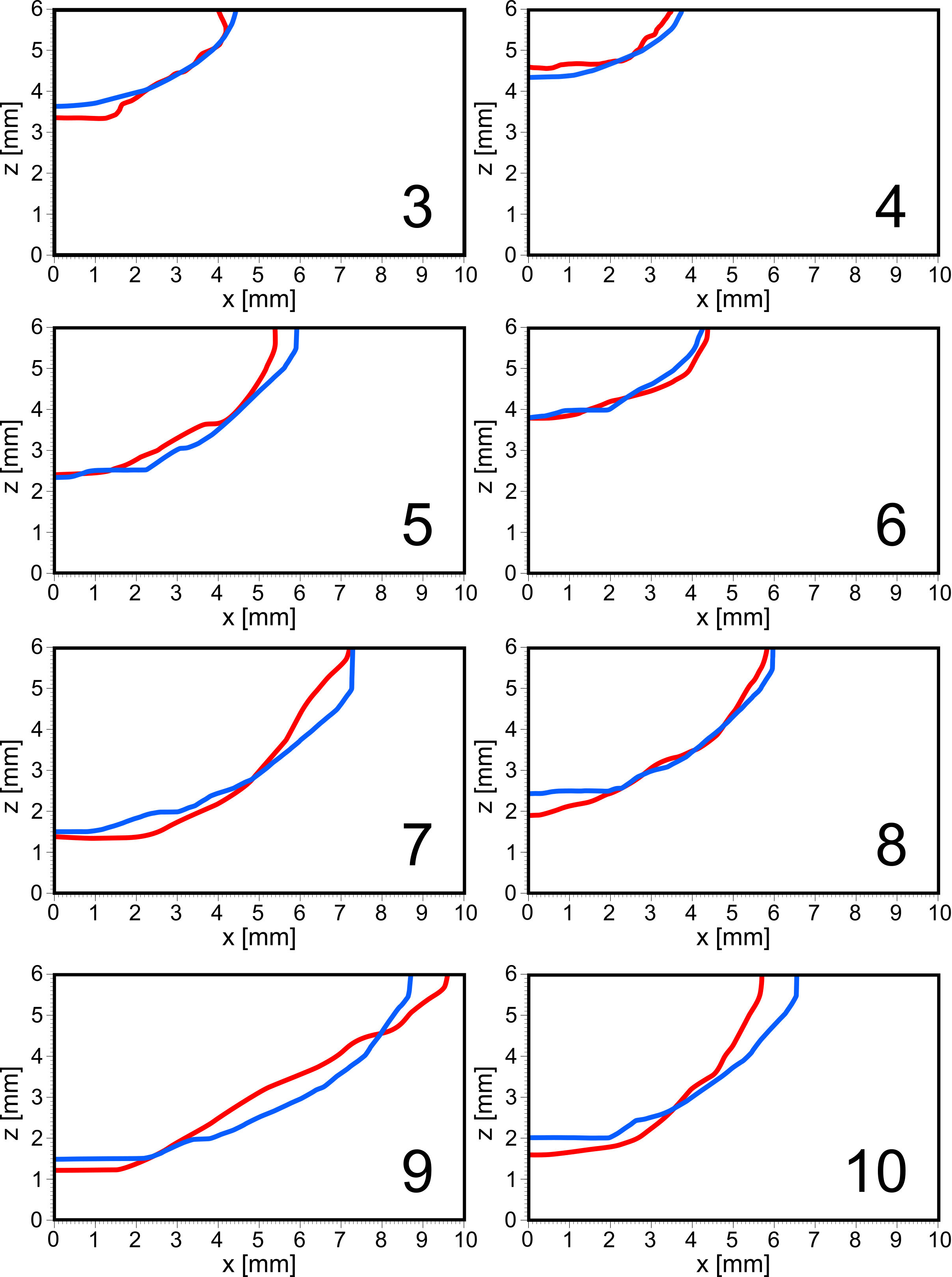}
\caption{Melt fronts for cases outlined in Table \ref{4-tab-para-case-gIF} (3 - 4) and Table \ref{4-tab-304-case} (5 - 10). All melt fronts are illustrated to the same scale where each box represents $\SI{10}{\milli\meter} \times \SI{6}{\milli\meter}$ of the domain. The experimental results from \cite{paraboloid-paper} and \cite{independent-paper} are shown in red and the simulation results are shown in blue. The matches between simulation and experiment are all good.}
\label{4-fig-independent-exp-vs-sim}
\end{figure}

Appropriate values for $C_C$ for the six cases outlined in \cite{independent-paper} are shown in Table \ref{4-tab-304-case} and their produced melt fronts are shown in Figure \ref{4-fig-independent-exp-vs-sim}. Here, $C_{cut}$ is fixed at $0.01$ enabling $C_C$ to be manually tuned through a trial an error approach to an apt value. There is an inherent variability in experimental results so an exact match is not expected thus - as with the optimization of $C_{pc}$ and $C_{d}$ - a uniquely best $C_C$ is unavailable. However, the trial and error values for $C_{cut}$ can then be combined with the results from Table \ref{4-tab-para-case-gIF} and used to find a relationship between the experimental weld specifications (e.g. $Q_{source} = I \times V$, $l =$ weld depth, $\omega = $ weld width and $v$ the velocity (travel speed)) and an appropriate computational constant for simulation, $C_C$. 

\begin{table}[!htbp]
\setlength{\tabcolsep}{10pt}
    \centering
    \caption[Heat source specifications given in \cite{independent-paper} and computation constants $C_C$ and $C_{cut}$ used for simulation that have been found through trial and error. To reduce the computational constants, $C_{cut}$ is set to 0.01 for all cases and $C_C$ is manually optimized.]{\label{4-tab-304-case} Heat source specifications\footnotemark\ given in \cite{independent-paper} and computation constants $C_C$ and $C_{cut}$ used for simulation that have been found through trial and error. To reduce the computational constants, $C_{cut}$ is set to 0.01 for all cases and $C_C$ is manually optimized.}
    \begin{tabular}{l l l l l l l}
    \toprule
    Case \# & $Q_{Source}$ [\si{\watt}] & $l$ [\si{\milli\meter}] & $\omega$ [\si{\milli\meter}] & $v$ [\si{\milli\meter\per\second}] & $C_{cut}$ & $C_C$ \\
    \hline
    5 & 1265 & 3.6 & 10.8 & 1.6$\dot{6}$ \rule{0pt}{2.6ex} & 0.01 & 120 \\
    6 & 1265 & 2.2 & 8.8 & 2.5 & 0.01 & 110 \\
    7 & 1890 & 4.6 & 14.4 & 1.6$\dot{6}$ & 0.01 & 150 \\
    8 & 1890 & 4.1 & 11.6 & 2.5 & 0.01 & 140 \\
    9 & 2760 & 4.8 & 19.2 & 1.6$\dot{6}$ & 0.01 & 180 \\
    10 & 2760 & 4.4 & 11.4 & 2.5 & 0.01 & 160 \\
    \bottomrule
    \end{tabular}
\end{table}
\footnotetext{Note in \cite{independent-paper} the values labeled as `width' are actually radius!}

To calculate an estimating formula for $C_C$ the linear regression function from Scikit-learn \cite{scikitPackage} is used. The relationship calculated by this function is shown in equation \ref{4-eq-fit}. Here, the positive coefficients on $Q_{Source}$, $v$ and $\omega$ show that welds with a large arc power, velocity and weld width need a higher value of $C_C$. Whereas the negative coefficient on $l$ show deeper welds need a lower value of $C_C$. This change in the sign of the coefficients is one reason why finding an apt value for $C_C$ is not immediately intuitive. For instance, in welds on materials with a positive $\frac{\partial \sigma}{\partial T}$ coefficient as $\omega$ increases so will $l$. Therefore, say a value of $C_C \approx 200$ is calculated from equation \ref{4-eq-fit} for a large weld with this (positive $\frac{\partial \sigma}{\partial T}$) material slightly under-predicts the melt pool width it is not trivial just to increase $C_C$ to get a better melt front match. Close inspection of Figure \ref{4-fig-independent-exp-vs-sim} will reveal this; some simulations predict a wider weld than experiment and some predict a deeper weld - it is therefore down to the operator to judge what is apt for their purpose.       

\begin{equation}
\label{4-eq-fit}
\begin{split}
C_C \ [a.u.] &= 0.02425944 \cdot Q_{Source} \ [\si{\watt}] 
\\&+ 16.32170162 \cdot v \ [\si{\milli\meter\per\second}] 
\\&- 6.69688692 \cdot l \ [\si{\milli\meter}] 
\\&+ 5.16127479 \cdot \omega \ [\si{\milli\meter}]
\end{split}
\end{equation}

\begin{figure}[!htbp]
\centering
\includegraphics[width=12.5cm]{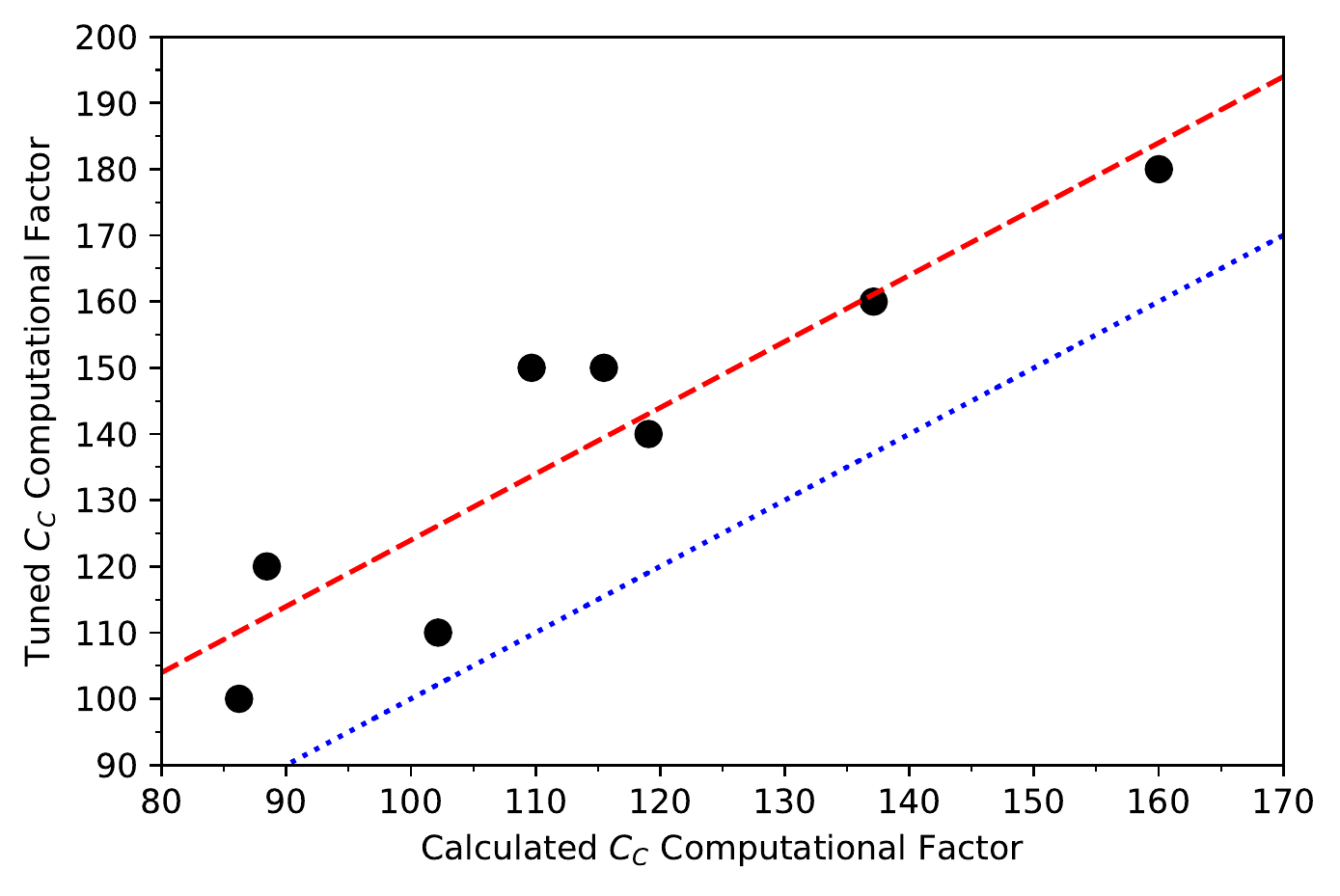}
\caption{Plot of the apt value for $C_C$ calculated from equation \ref{4-eq-fit} against the values found through trial and error shown in Figure \ref{4-fig-independent-exp-vs-sim}. The dashed red line is a linear fit of these points showing a positive correlation and hence that in general larger values for $Q_{Source}$, $v$ and $\omega$ - and smaller values for $l$ - result in larger $C_C$ values. The dotted blue line shows a perfect $y = x$ correlation; all of the trial and error values are above this showing equation \ref{4-eq-fit} under-predicts the value for $C_C$.}
\label{4-fig-CC-plot}
\end{figure}

Only eight data points are used to formulate this equation but as shown in Figure \ref{4-fig-CC-plot} it manages reasonable - although consistently under - predictions for the values of $C_C$ found through trial an error. These predictions are of course only rough estimates and will need to be tuned for each case. Realistically, to find a apt starting value for $C_C$ two or three significant figures should be used for each prefactor in equation \ref{4-eq-fit} and then the value for $C_C$ should be rounded to the nearest 5. Ideally, equation \ref{4-eq-fit} could be refined further through investigating 20 or 30 additional cases over a broad range of procedures although due to the scope of this thesis this was omitted. In fact, this is a potential avenue for further study as a possible weakness in equation \ref{4-eq-fit} is its applicability to a wide range of procedures when extrapolating to large and small values of $C_C$. However, as has been shown with the other benchmarks featured in this chapter, computational constants should be refined to a `practically applicable' level and then further refinement can be performed through altering the case specification.

%Do you need this plot?

\section{Chapter summary}
This chapter details extensive benchmarks for all elements of \gIF against known results. In general, all results are a reasonable match and therefore a user can have confidence in the results produced by \emph{gtawFoam}. A key takeaway in this chapter is the benefits of and the limits to parameter tuning; changing the parameters and the case slightly can produce very different results. The specific values for the purely computational constants ($C_C$, $C_{cut}$, $C_d$, $C_{pc}$) are found enabling the physical parameters to be adapted; for convenience, these values are shown in equation \ref{5-eq-final-comp-const-summary}. However, this does not `solve' parameter tuning as changing the case variables between reasonable values can drastically effect the result. For instance, with the solidification of tin the choice of liquid conductivity (Figure \ref{4-fig-tin-solid-results}) doesn't really affect the results drastically. Whereas, the choice of solid conductivity (Figure \ref{4-fig-tin-solid-k-comp}) is a key parameter. Similarly, in the case of water freezing (Figure \ref{4-fig-rhoT-Bouss-Alpha}) aptly choosing the formulation of density dependence on temperature determines whether an appropriate melt front is calculated or not. Further, there does not seem to be a typical dimensionless number that can be applied to identify a regime that can predict the required formulation. Therefore, ultimately operator experience needs to be applied to use \gIF as effectively as possible. The strategy executed in this chapter addresses the third objective detailed in Chapter 1.

\begin{gather}
\label{5-eq-final-comp-const-summary}
C_{pc} = 10, \quad
C_{d} = \num{8e8}, \quad
C_{cut} = 0.01 \nonumber\\
C_{C} = 0.02425944 \cdot Q_{Source} + 16.32170162 \cdot v \nonumber\\
- 6.69688692 \cdot l + 5.16127479 \cdot \omega
\end{gather}    

%% file: chapters/thesis-chapter-6.tex
\lhead{Chapter 6: Ultra-thin-walled tube welding}
\section{Introduction}
This chapter answers the main research question of this thesis: is there a general case procedure for ultra-thin-walled tube welding? To achieve this, initially an experimental investigation undertaken to develop a welding procedure for the cooling system in the ATLAS ITk is described. This is followed by an analysis of the results from this investigation to understand the effects of the welding parameters within the welding procedure. This analysis looks at both the physical results as well as the data as a whole and motivates the use of \gIF to simulate the welding procedure to obtain a match between simulation and experiment. With the match established, a general case for ultra-thin-walled tube welding is found through running a large batch of simulations. Finally, this general case is presented to answer to main research question of this thesis.

Parts of the chapter are adapted from the paper `\emph{In situ micro gas tungsten constricted arc welding of ultra-thin walled \SI{2.275}{\milli\meter} outer diameter grade 2 commercially pure titanium tubing}' principally written by the author \cite{JINST}. The author was involved in the experimental welding trials although these were principally performed by Sam Edwards and Paul Kemp-Russell. The metallographs, UTS tests and hardness tests were performed by Micka\"{e}l Crouvizier at CERN. The sleeve fitting used was designed by Fred Gannaway with input from Sam Edwards and Paul Kemp-Russell. Ben Kitchener manufactured the custom shunt box used to measure welding current. The demonstration cooling circuit featured in this chapter was proposed and overseen by George Viehhauser. The CAD model of this demonstration cooling was created by Liam Cooper. All other figures were created by the author, all of the analysis was performed by the author and all of the simulation work was performed by the author.

\FloatBarrier
\section{ATLAS ITk}
The core application of ultra-thin-walled tube welding addressed in this chapter is for the cooling system of the ATLAS ITk strip detector stave. Breaking this down, the ATLAS detector \cite{ATLAS-TDR} is one of the four main particle detector systems positioned around the Large Hadron Collider at CERN in Geneva, Switzerland. The ATLAS detector is made of multiple subsystems one of which is the Inner Detector. Given the Inner Detector was installed in 2007, it has come to the end of its service life due to accumulated radiation damage amongst other deterioration. Thus, along with the entirety of the ATLAS detector, the Inner Detector is being replace with a more modern system \cite{ATLAS-upgrade}. The replacement for the Inner Detector in the upgraded ATLAS detector is the Inner Tracker or - as it is referred to in this thesis - the ITk. The ITk will have an all-silicon semiconductor tracking system consisting of an inner 5-layer pixel detector surrounded by a 4-layer strip detector. The ITk Strip Detector \cite{ITk-Strip-TDR} has small read-out systems built into staves for the cylindrical portion of the detector and into petals for the disk portion. 

\begin{figure}[!htbp]
\centering
\includegraphics[width=12cm]{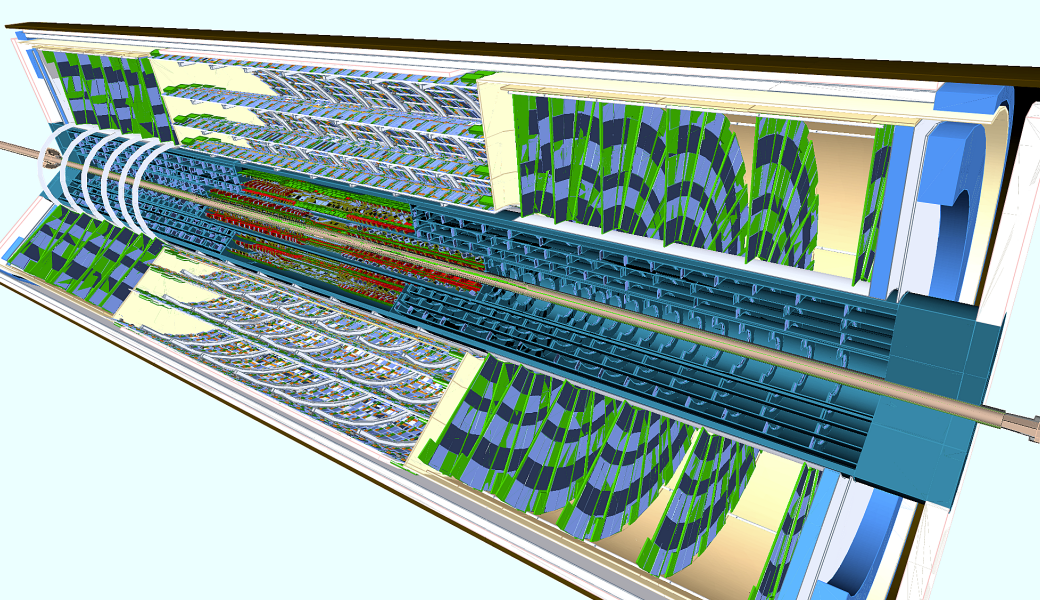}
\caption{Illustration of the ITk from the ITk technical design report \cite{ITk-Strip-TDR}. This image shows a portion of the ITk cut away to reveal the central beam tube that the ITk surrounds. The electronics structures parallel to the beam pipe are the stip detector staves and those perpendicular are petals.}
\label{5-fig-ITk-stip}
\end{figure}

\begin{figure}[!htbp]
\centering
\includegraphics[width=12cm]{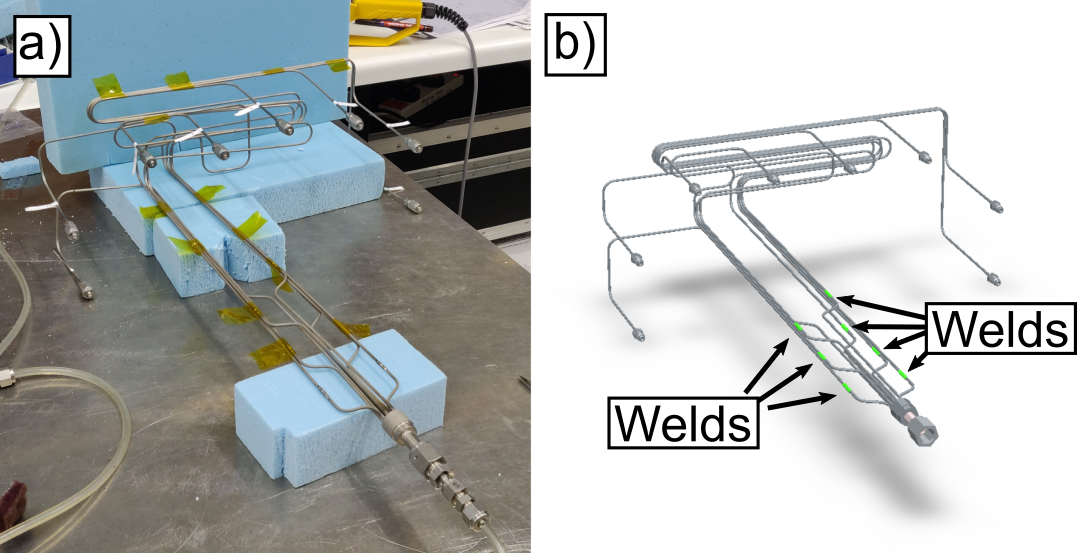}
\caption{Produced part (a) and CAD model (b) of high-value cooling circuit for a two phase CO$_2$ cooling plant presented in \cite{JINST}. The joints are highlighted green in (b). Note the complex curving geometry of the tubes. The cooling plant was used as a demonstration for the forthcoming ATLAS upgrade.}
\label{5-fig-JINST-babydemo}
\end{figure}

Due to leakage current, where unwanted current flows through the semiconductors in the read-out systems, thermal runaway can occur. Here, small temperature increases from the leakage current cause more current to flow further increasing the temperature in a positive feedback loop. To alleviate this, the ITk features a cooling system. The cooling system adds material to the ITk and non-detecting material reduces the efficacy of the ITk's particle detection ability. Briefly, this is due to the non-detecting material affecting the particles produced in collisions by the LHC. This interference can be quantified in terms of radiation length which is the mean length in centimetres of a material to reduce the energy of an electron by $\frac{1}{e}$\footnote{To avoid confusion, this is the mathematical constant $e = 2.7182818$ and not the charge of an electron.}. In the Inner Detector, the cooling system was stainless steel. However, considerably less titanium than stainless steel is required to achieve the same mass flow handling ability. The ultra-thin-walled commercially pure titanium tubing within the ITk provides a 42\% weight saving comparing to stainless steel. Titanium does have a longer radiation length than stainless steel yet due to the material reduction it is the better choice. Thus, the cooling circuits in ITk feature titanium tubing. The cooling circuits within the carbon fibre structure of the Strip Detector stave are made of \SI{2.275}{\milli\meter} outer diameter (OD), \SI{160}{\micro\meter} wall thickness, grade 2 commercially pure titanium (CP-2 Ti) tubing. This is in contrast to the more common application of thin-walled tube welding in condenser units where wall thickness of the tubes are rarely below \SI{500}{\micro\meter}. The joints within the cooling circuits are required to be robust to prevent the potential catastrophic failure of the ATLAS detector. Secondly, to enable consistent mass flow within the cooling circuits the inner diameter of the cooling tubes need to remain constant. The best way to fulfil these requirements is through a carefully welded joint. Hence, this is an application of ultra-thin-walled tube welding. Figure \ref{5-fig-JINST-babydemo} shows a demonstration two phase CO$_2$ cooling plant for the ATLAS ITk presented in \cite{JINST}.

Despite the attractive properties of titanium (high strength to weight ratio and excellent corrosion resistance), welding challenges have encumbered its industrial application in ultra-thin-walled tubing. Specifically, as welding procedures do not scale down proportionally and titanium is relatively more difficult to join compared to other metals, ultra-thin-walled tube welding is particularly challenging. ASTM standard  B862 \cite{ASTM-pipe} for welding Titanium Pipe stops above \SI{1000}{\micro\meter} in wall thickness and 10 mm in outer diameter and the granularity for outer diameter tolerance in titanium condenser and heat exchanger tubes in ASTM standard B338 \cite{ASTM-tube} stops at \SI{25.4}{\milli\meter}. Thus, welding the ultra-thin-walled CP-2 Ti tubing for the ATLAS ITk Strip Detector Stave cooling circuits requires a custom welding procedure. The most popular and reliable thin sheet and thin-walled tube welding methods for titanium include Electron Beam Welding (EBW), Laser Beam Welding (LBW) and variations of Gas Tungsten Arc Welding (GTAW). 
%As mentioned previously, ASTM standard B862 \cite{ASTM-pipe-spec} for welding Titanium Pipe stops above \SI{1000}{\micro\meter} in wall thickness and \SI{10}{\milli\meter} in outer diameter and the granularity for outer diameter tolerance in titanium condenser and heat exchanger tubes in ASTM standard B338 \cite{ASTM-tube-spec} stops at \SI{25.4}{\milli\meter}.

A key challenge in orbital welding of ultra-thin-walled tubing is to ensure that the tube does not collapse under the pressure from the welding heat source. A solution to this is to create an inner gas flow to counteract the pressure from the arc. This inner gas flow differs from conventional shielding to prevent oxidation as its role is to provide a physical internal buttressing force for the molten metal during welding. This gas flow needs to be sufficiently high to prevent the tube from collapsing inwards but also not high enough so as to blow the weld outwards. Further, for the ATLAS ITk maintaining a consistent inner diameter of the tubes to ensure consistent mass flow within the cooling circuit is critical. Thus the acceptable range of gas flows is reduced as the inner gas flow mustn't cause the weld to even lean inwards or outwards.

\begin{figure}[!htb]
\centering
\includegraphics[width=10cm]{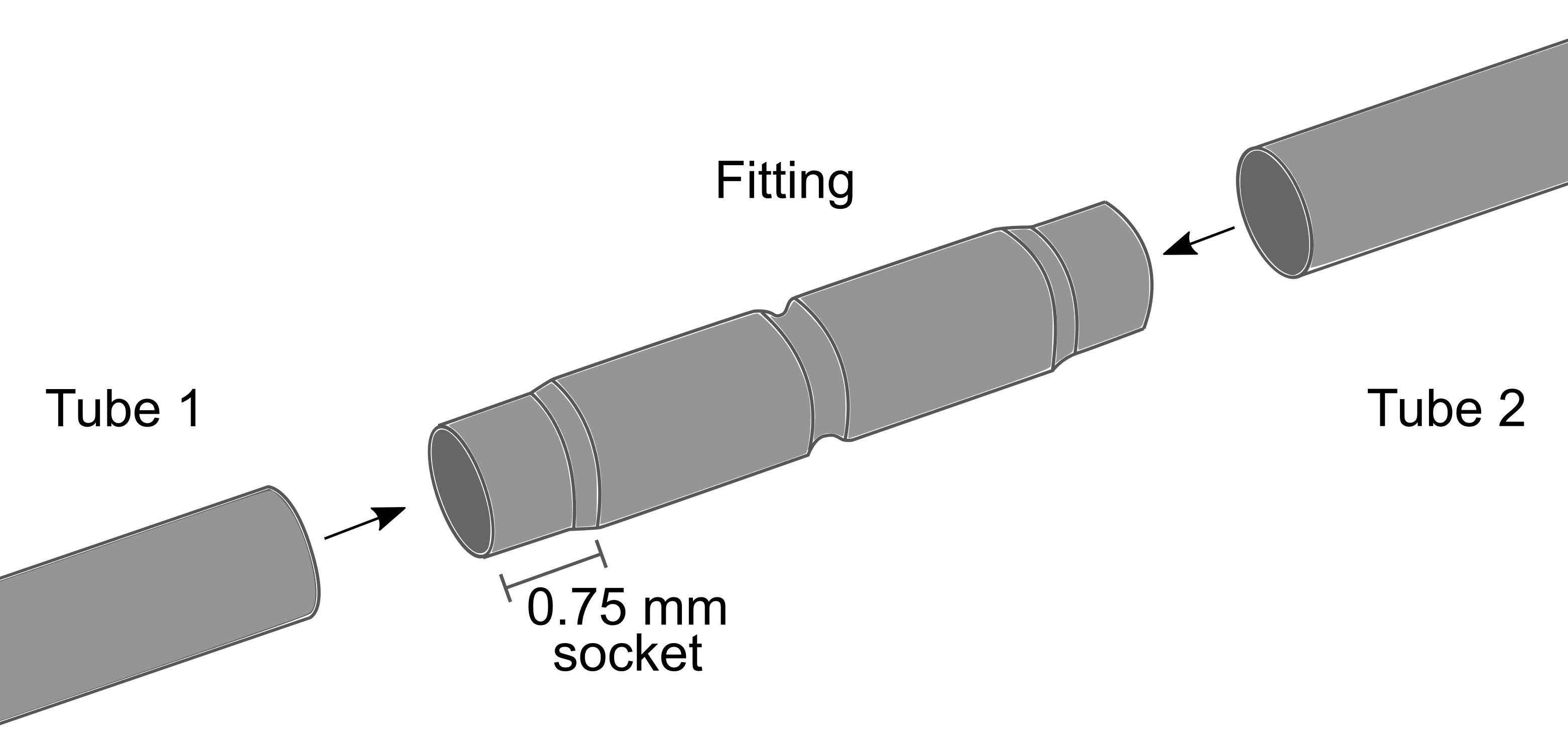}
\caption{Illustration of the sleeve fitting and socket insertion process used to join two tubes. The central groove in the fitting is designed to accommodate O-rings to seal connections; it also allows for a standard purchase point.}
\label{5-fig-JINST-fitting}
\end{figure}

\begin{figure}[!htb]
\centering
\includegraphics[width=10cm]{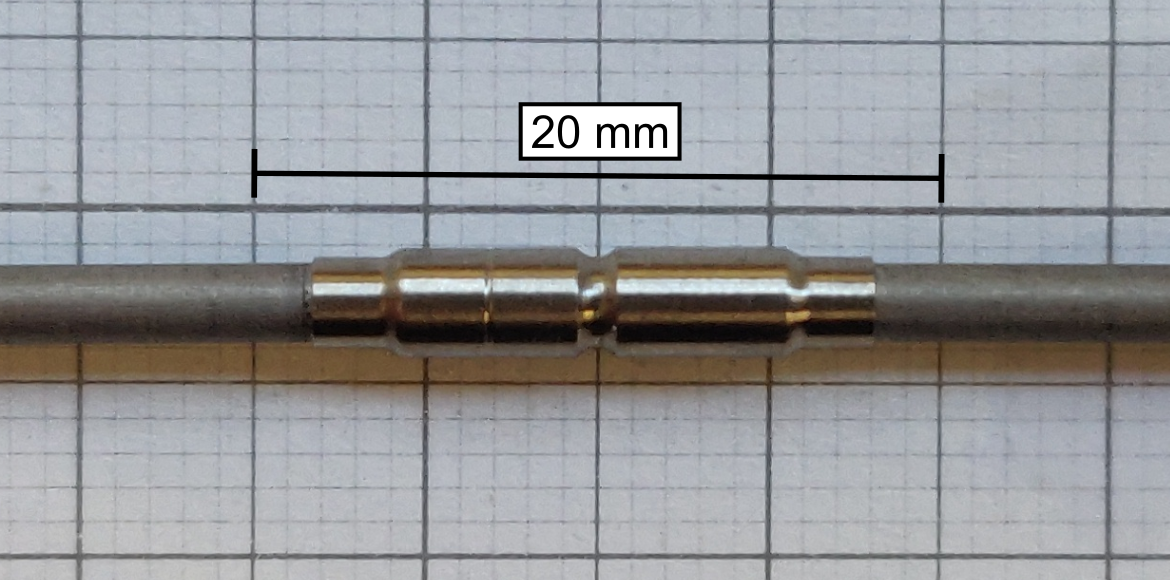}
\caption{Photo of two tubes inserted into a sleeve to create a mechanical joint.}
\label{5-fig-JINST-mechanical}
\end{figure}

Given the thinness of the tube walls, when cut the ends of the tubes are often slightly imperfect meaning rotation and reorientation is required to achieve face-to-face contact between them. Whilst possible in a lab environment, this is completely unfeasible in a fixed orientation (and sometimes in situ) environment such as the ATLAS ITk. For instance, the cooling circuit shown in Figure \ref{5-fig-JINST-babydemo} requires specific angles for all of the tubes to fit together. To address this challenge, a custom sleeve fitting - shown in Figure \ref{5-fig-JINST-fitting} - that features a \SI{0.75}{\milli\meter} tube insertion socket facilitates exact positioning of a tube. This socket thickens the effective tube wall at the point of weld to \SI{0.3}{\milli\meter}. The fitting consists of the socket ends followed by a thicker section before a return to a thinner central region. This specific shape is to enable accommodation of O-rings to seal connections for quality assurance testing equipment. There is also an additional benefit of the thinner central region - it allows for a standard purchase point for setting the cooling manifolds in a jig during production. Using this custom enveloping fitting adds unavoidable mass to the cooling circuit. However, this is a necessary trade-off for production and the increased wall thickness adds additional reliability. Once inserted - shown in Figure \ref{5-fig-JINST-mechanical} - the inner diameter of the fitting is identical to the base tubes enabling unconstricted coolant mass flow.

Finally, the experimental welds covered in this chapter were in a large part created in a research and development setting where the specific dimensions and geometry of the tubing was subject to change. As an example, the sleeves were not apart of the initial plan and were introduced after a successful tube-to-tube procedure had already been established rendering the tube-to-tube procedure obsolete. This made approaching the optimization of the welding procedure through a design of experiments approach challenging as design changes within the rest of the cooling system of the ATLAS ITk could make the optimization of specific joinery scenarios obsolete. Instead, the approach was to start with a reasonable initial procedure and then use knowledge of welding principles to adapt the procedure. For instance, a starting point for current is \SI{1}{\ampere} for every thousandth of an inch so a \SI{160}{\micro\meter} thick work piece equates to around \SI{6.3}{\ampere}. Using this value with other reasonable parameter settings may cause the tube to collapse as the welding current is too high. Therefore, \SI{5.3}{\ampere} could be tried next which may not be sufficiently high to melt the tube fully resulting in insufficient welding. The parameters involved (and there approximate effect) in this optimization process are listed in Table \ref{5-tab-exp-param}. Combined with the accommodation of changes to the design if the cooling circuits this optimization process required approximately 300 welds to complete.      

\begin{center}
\begin{table}
\caption{\label{5-tab-exp-param}Welding parameter definitions (and their approximate effects) used during the experimental investigation into ultra-thin-walled tube welding. An illustration of the physical meaning of some of these parameters is shown in Figure \ref{5-fig-JINST-set-up}.}
\begin{tabular}{|p{4cm}|p{11cm}|}
\hline
\textbf{Welding Parameter} & \textbf{Definition and approximate effect}\\
\hline
Arc gap & Governs the distance between the electrode tip and work metal along the z-axis which, as GTAW is a constant current process, dictates the electrical voltage during welding. In general, a larger voltage results in more heat input. \\
\hline
Main current & Current flowing through the GTAW arc, larger currents input more heat into the work metal.  \\
\hline
Interbackground current (Background current) & Specific parameter for the welding machine used (VBCie IP50) recommended by the manufacturer to be $\approx 80\%$ of the main current. This is effectively equivalent to a background current in pulsed current GTAW which is the lower level of current that a pulsed current pulses between.\\
\hline
Rotational head speed & Determines the travel speed of the arc around the work metal in a clockwise rotation around the y-axis. Together with current and voltage, this parameter determines the heat input. A high rotational head speed can cause porosity due to solidification occurring before gases can escape. \\
\hline
Initial current & Controls the current amperage used when the arc is generated. A large initial current can potentially damage the thin walls of the tubing. \\
\hline
Machine gas flow rate & Controls the amount of shielding gas that engulfs the outside of the work metal. Due to the shape of the orbital weld head, this gas flow envelops the tubing in all directions and it does not result in any pressure on the weld. The main role of this flow is to remove impurities from the weld, excessive gas flow can result in arc strike failures. \\
\hline
Internal gas flow rate & Controls the amount of shielding gas that flows inside of the CP-2 Ti tube thereby determining the internal gas pressure at the joint. Precise control is required to ensure consistent internal diameter in the cooling tubes with high flow risking outer blowout and low flow risking inner collapse.  \\
\hline
Electrode Position & Controls the y-axis position of the arc along the work metal, thus determining the location of the heat input. \\
\hline
Socket depth & Controls how far the tube sits within the fitting. Together with the electrode position, the socket depth is essential in positioning the weld bead in the optimum position for the strongest joint. Whilst it is crucial to limit the HAZ, the weld bead needs to be sufficiently large to achieve full fusion between the tube and the sleeve. \\
\hline
\end{tabular}    
\end{table}
\end{center}

\section{Experimental investigation}
\subsection{Experimental parameters}
\subsubsection{Overview}
As detailed in Table \ref{5-tab-exp-param}, there are many parameters involved in the welding of ultra-thin-walled tubing. However, with a view for simulations in mind, two parameters are of particular importance: the welding current and the inner gas flow. The welding current plays a critical role in the heat input into the work piece. When calculating heat input, voltage and travel speed are the same power order as current (e.g. double the current and half the voltage gives the same heat input) thus with the VoF-enthalpy-porosity methodology of \gIF they are somewhat interchangeable. Note, voltage, current and travel speed - both from a simulation and experiment perspective - are not literally interchangeable. However there are parameters in \gIF such as arc power where there is only one value so $\SI{1}{\ampere} \times \SI{100}{\volt} \equiv \SI{20}{\ampere} \times \SI{5}{\volt}$. Further, there are multiple parameters that define the current waveform shown in Figure \ref{5-fig-cur-shape} which in \gIF are reduced to a single value. With this in mind, interchangeable in this context refers to the flexibility in \gIF for achieving similar results with different parameter values. Therefore, having high confidence in the exact current used in experiment enables the simulation to be `anchored'. The other parameter of particular importance is the inner gas flow. Given the outer gas flow is dispersed around the work piece it is the inner gas flow which plays the critical role in determining the shape of the weld. This is assuming a reasonable level of outer (machine) gas flow rate high enough to prevent impurities but not so high as to cause arc strike failures or arc wandering. Further, it too has proxy parameters in \gIF that it can be `anchored' to and is therefore also important to measure and model accurately.

\subsubsection{Welding current}
\begin{figure}[!htb]
\centering
\includegraphics[width=12.5cm]{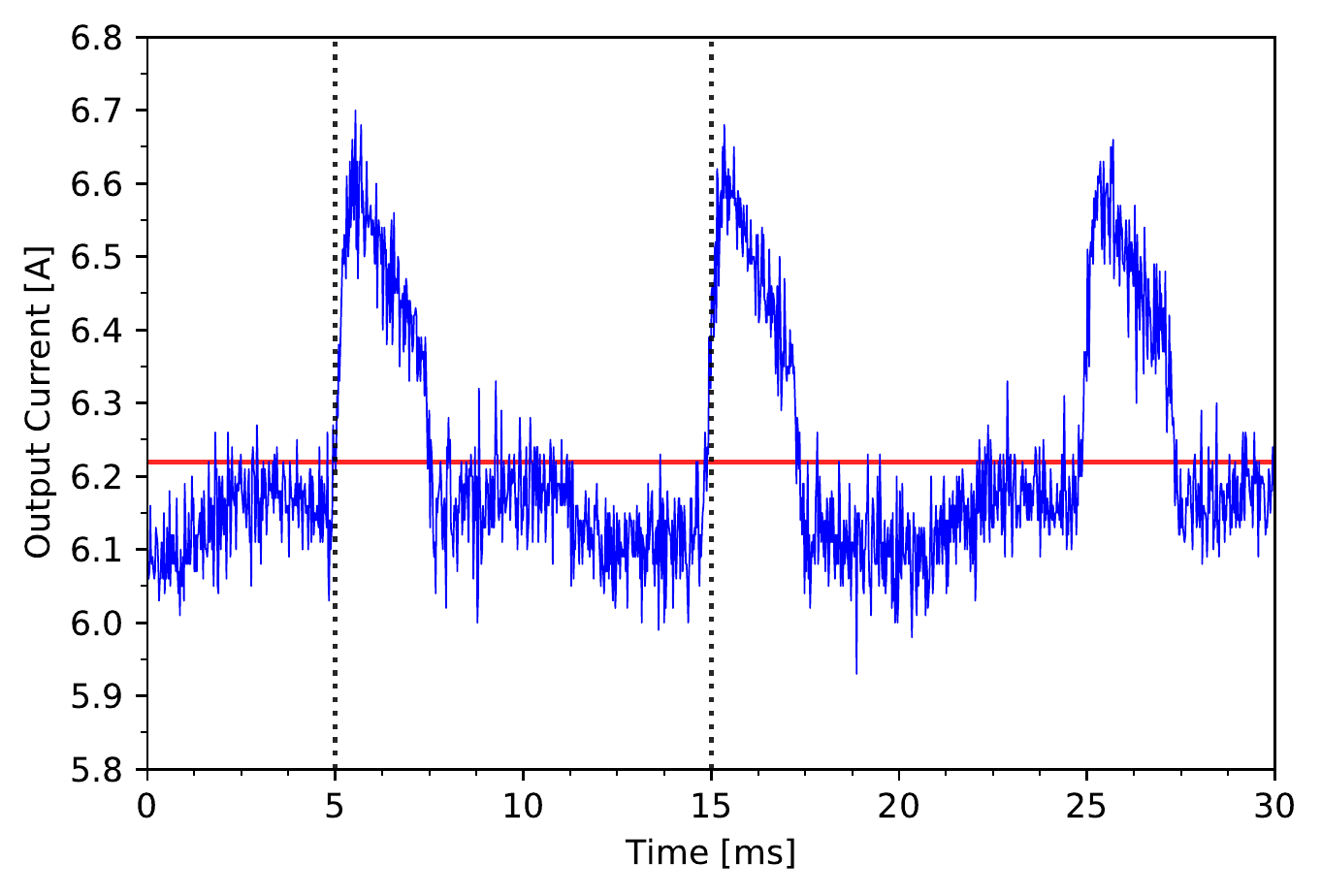}
\caption{Outputted current when the welding machine main current is set to \SI{6}{\ampere} and interbackground to \SI{6.8}{\ampere}. Red line shows average current level of \SI{6.22}{\ampere}. Dotted black lines show a pulse width of \SI{10}{\milli\second} (\SI{100}{\hertz}). Note the `noise' is actually the \SI{20}{\kilo\hertz} InterPulse current superimposed over the \SI{100}{\hertz} peak and background current.}
\label{5-fig-cur-shape}
\end{figure}

Pulsed current GTAW is a popular variant of GTAW where the current of the electric arc is alternated between a higher peak current and a lower background current. Pulsed GTAW brings several advantages such as enhanced arc stability, increased depth/width ratio and a narrower heat affected zone. Enhanced arc stability helps to reduce the chance of interruption of the welding process due to arc deflection. An increased weld depth/width ratio produces more integral joints, and a narrower heat affected zone limits the detrimental effects of heat on the work metal \cite{pulsedCurrentYous, pulsedCurrentTraida}. The novelty of the model of welding machine used for the experimental investigation - an IP50 - is the inclusion of a \SI{20}{\kilo\hertz} \emph{InterPulse} current that is superimposed on top of a typical pulsed current waveform for a GTAW arc. The \emph{InterPulse} waveform is shown in Figure \ref{5-fig-cur-shape}. 

The role of the \emph{InterPulse} current is to create enhanced arc constriction. This is in contrast to regular pulse current GTAW which is often understood as the peak current providing more heat and increasing the size of the weld pool whilst the background current maintains arc stability and provides a `cooling period' for the metal to solidify producing a series of spot welds \cite{pulsePaper1, pulsePaper2, pulsePaper3, pulsePaper4}. This `heating and cooling' description does not consider the response time of the arc plasma to an increase in the current control signal or the inertia of the arc plasma as the current magnitude changes. In fact, at higher pulsing rates, the `heating' and `cooling' begin to mix (a mixed state at \SI{3}{\kilo\hertz} pulsing rate was shown as early as 1988 by Saedi and Unkel \cite{saedi1988arc}). 

At high current pulsing rates GTAW arcs begin to constrict. Due to the Lorentz force, the current flow within the arc results in a radially inward ‘pinch’ effect with a force proportional to the square of the current (as both the current density and magnetic field in the arc are each dependent on the current). At the peak pulsing current the Lorentz force is quadratically increased but the response time of the arc plasma to the increase in current lags behind the increase in the Lorentz force. When the current subsequently decreases to the background level, the response of the arc plasma to the relaxation of the ‘pinch’ effect is again slower than the forthcoming increase back to the peak pulse level. Thus, in a `mixed' state the arc constricts which focuses the conduction path between the electrode and work metal resulting in increased current density in the centre of the arc compared to a conventional arc for the same current. Thus, less heat is ‘wasted’ in the arc plume and surrounding metal, and more is focused into the weld pool. Constriction at high pulsing rates has been well reported \cite{hfPulse1, hfPulse2, hfPulse3}. 

The current fluctuations in the \emph{InterPulse} waveform in Figure \ref{5-fig-cur-shape} within the main and the background current are $< \pm 5\%$ of the current's value which is not a significant amount compared to a $\pm 50\%$ or $100\%$ change for a typical pulsing current (e.g. a main current of \SI{80}{\ampere} and a background current of \SI{40}{\ampere} which thus limits the effects of the constriction. This was due to the setup of the equipment chosen after consultation with the manufacturer on how best to weld ultra-thin-walled titanium tubing. However, as a broader novelty there is experimental evidence explored in detail in \cite{interpulseLeary} that supports the view that \emph{InterPulse} limits the heat input when welding commercially pure titanium in a way consistent with a focused heat input due to arc constriction. There is also experimental support for the benefits of \emph{InterPulse} in other welding procedures \cite{interpulse-Michalis, interpulse-alt}.

When dealing with experimental equipment a predictable response to parameter changes allows for operator confidence when optimizing the welding procedure. The Insulated-Gate Bipolar Transistors (IGBTs) in many welding power supplies do not provide a linear output at particularly low current levels. Whilst the IP50 (the terms `IP50' and 'VBCie IP50' are used interchangeably in this thesis.) is optimized specifically for low currents quality assurance tests were performed to ensure the output current was equal to the current the machine was set to. Here, a custom shunt box was used to measure the actually outputted current compared to the current the machine was set to output. Two separate tests were performed over a year apart. The initial test was performed prior to the incorporation of the sleeve when slightly lower ($0 - \SI{5}{\ampere}$) amperages were used to join the tubes. At that point, manufacturing of the ATLAS ITk cooling system was planned to take place at two production sites with two separate IP50s: one at Sheffield and one at the Rutherford Appleton Laboratory (RAL). The second test was performed after the sleeves had been incorporated and Sheffield selected as the sole production site. At this point the manufacturer had tuned the IP50 to better deal with the higher ($4 - \SI{8}{\ampere}$) amperages required. 
\begin{figure}[!htb]
\centering
\includegraphics[width=12.5cm]{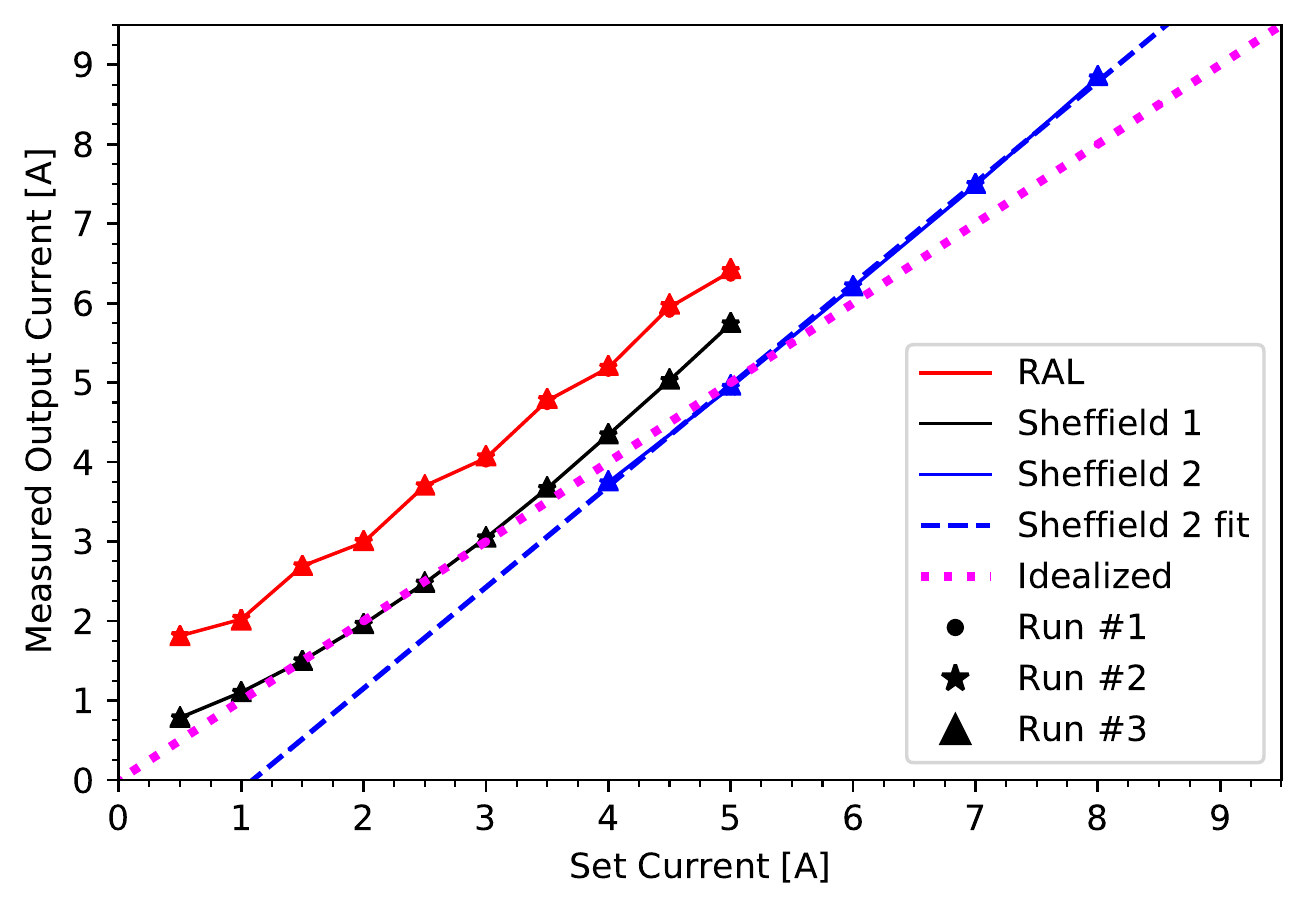}
\caption{Current values the welding machine is set to against measured average output current in amps. Dotted magenta line shows an idealized system where setting the current to \SI{5}{\ampere} would output \SI{5}{\ampere}. Red line shows the results for a IP50 welding machine based at RAL, black line shows the IP50 based at Sheffield before maintenance and the solid blue line shows the results for this machine post maintenance. The dashed blue line shows a linear regression fit on the `Sheffield 2' results producing $y = 1.272 \cdot x - 1.392$. The maintenance specifically optimized for the range of currents used to weld ultra-thin-walled tubing. Each data set was repeated three times shown with three different overlaying markers to illustrate how consistent the output is.}
\label{5-fig-cur-comp}
\end{figure}

All three results are plotted on Figure \ref{5-fig-cur-comp} to illustrate the variety in actual outputted current compared to the current the machine is set to. For all cases, the measured output current does not exactly match the set current for their full range e.g. they are biased. However, as long as this bias is known it can be accounted for. This issue is with the confidence in the linearity of the current settings. When manually tuning the parameters operators need confidence a change from \SI{5}{\ampere} to \SI{6}{\ampere} will cause the same increase as from 6 to \SI{7}{\ampere}; looking at the `Sheffield 1' and `RAL' results this is not trivial. To investigate this, the results for `Sheffield 2' are linearly fitted as $y = 1.272 \cdot x - 1.392$. The fit has an $r^2 = 0.9998$ thus there is a high degree of confidence in the linearity of the current settings. The fit is used as a transformation to find the actual output current. As an interesting side note, the `wobble' on the response from the RAL model is probably due to the configuration of the capacitors in the machine. This is difficult to conclude absolutely however as it is legacy machine as the manufacturer has gone out of business.

\subsubsection{Gas flow}
The gas flow delivered to the tubing plays a critical role in the integrity of the produced weld. As will be covered in the section \ref{5-exp-procedure} the gas flow is controlled through adjusting external flow meters and delivered to the tubing through an external manifold whereupon its pressure was measured. Therefore, the same concerns about linearity do not apply as the only relevant value was the output pressure at the weld and not how much gas either was inputted into the manifold through adjusting the dials on the flow meters or leaked out of the manifold en route to the weld location. In fact, there was a challenge in production consistency; the gas flow manifold demanded regular inspection to ensure consistent flow rate. The key issue with the gas flow is assessing operator confidence in the values measured for it.

The precise milling of the fittings used means that when the tubes are inserted (shown in Figure \ref{5-fig-JINST-mechanical}) there is an effective `mechanical joint' created. Before any welding has taken place these mechanical joints handle the inner gas flow. Therefore, understanding the differences between the mechanical joints (pre-weld) and welded joints (post-weld) enables confidence\footnote{Confidence of consistency in measurement; there is still the potential for a large systematic offset.} in assessment of the internal pressure. For instance, if there was a large difference in the internal pressure before and after welding then it would be extremely difficult to understand the exact internal pressure conditions around the weld pool during welding. 
\begin{figure}[!htb]
\centering
\includegraphics[width=12.5cm]{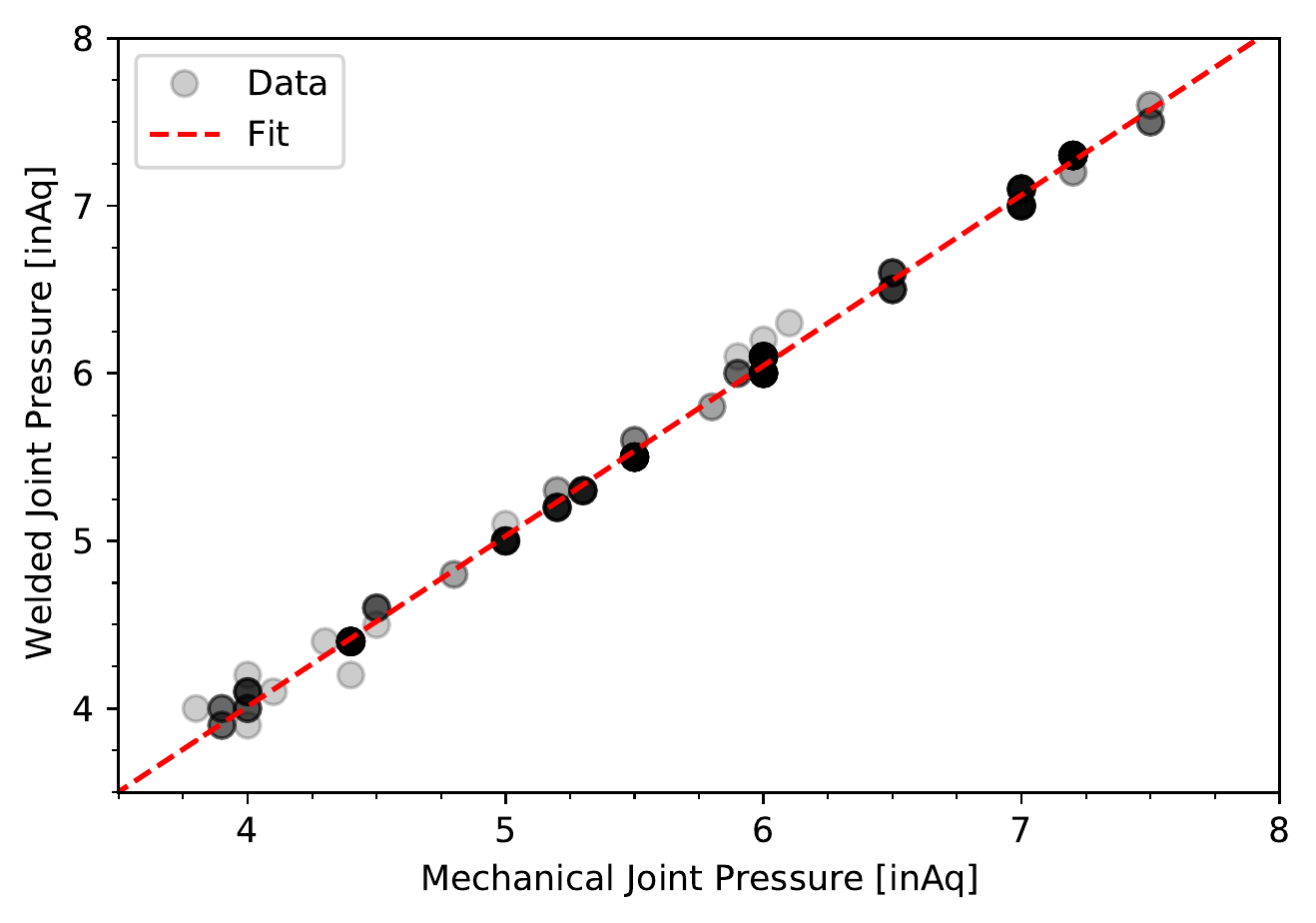}
\caption{Inner gas pressure of mechanical joints against welded joints for 292 welds performed during research and development that did not burst due to inapt parameters. Data points are shown in black and have a 20\% opacity to illustrate duplicate values. Dashed red line shows a linear least squares fit of $y = 1.017 \cdot x - 0.058$ where the coefficient of determination $r^2 = 0.999$. The least squares fit was performed using the SciPy \cite{2020SciPy-NMeth} function \emph{scipy.stats.linregress}. The coefficient of determination is the proportion of the variance in the dependent variable (Welded Joint Pressure) that is explained by this linear fit \cite{colman2009dictionary} ranging between 0 (no explanation) to 1 (total explanation). The value of 0.999 indicates a very good fit. Note as $1.017 \approx 1$ and $-0.058 \approx 0$ the mechanical joints have only very slightly less pressure handling ability compared to the welded joints.}
\label{5-fig-gas-pres}
\end{figure}

Fortunately, as shown in Figure \ref{5-fig-gas-pres}, the tight fit between the tube and fitting of the mechanical joints is capable of handling the inner pressure well. The relationship between the mechanical joints and welded joints can be linearly fitted as $y = 1.017 \cdot x - 0.058$ where $r^2 = 0.999$. As $1.017 \approx 1$ and $-0.058 \approx 0$ it can be concluded that the mechanical joints are essentially as good as the welded joints in terms of pressure handling ability and therefore an operator can have confidence in the values measured for inner pressure. Finally, because the mechanical joints and welded joints essentially have the same pressure handling ability they are not independent. This is useful when analysing the experimental data as one set of measurements can be dropped from the data set simplifying the analysis.  

\newpage
\subsection{Experimental procedure}
\label{5-exp-procedure}
\begin{figure}[!htbp]
\centering
\includegraphics[width=15cm]{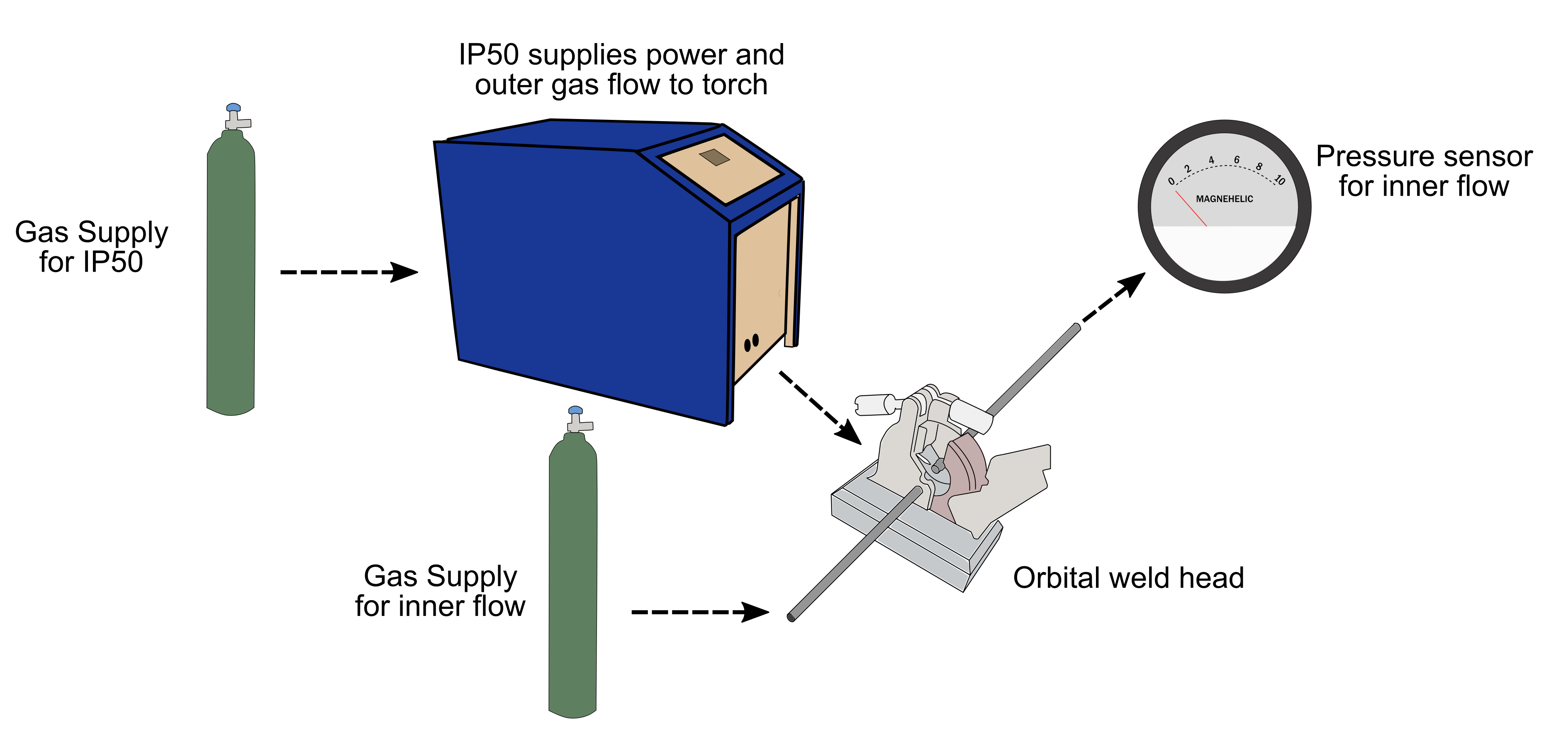}
\caption{Illustration of the equipment setup used to create the welds. Note two separate gas supplies are used to create an outer and inner gas flow for the weld.}
\label{5-fig-JINST-set-up}
\end{figure}

For the experimental investigation, the setup involved a VBCie IP50 used to power a Polysoude UHP 250 torch with a closed chamber orbital weld head attached. The orbital weld head is fitted with a \SI{2.275}{\milli\meter} tube-clamping insert that encloses the joint during welding. The Polysoude torch and weld head measures $225 \times 25 \times$ \SI{50}{\milli\meter} at its greatest extent allowing it to fit easily within small spaces for in situ production. Two 99.9999\% pure Argon gas bottles are used: one connected to the IP50 to provide external shielding and the other fed into the CP-2 Ti tube through an external manifold to provide shielding and internal pressurization. External flow meters control the gas flows and a pressure gauge connected to the outflow end of the CP-2 Ti tubing measures the outflow pressure from the inner gas. This setup is illustrated in Figure \ref{5-fig-JINST-set-up}. 

\begin{figure}[!htbp]
\centering
\includegraphics[width=12.5cm]{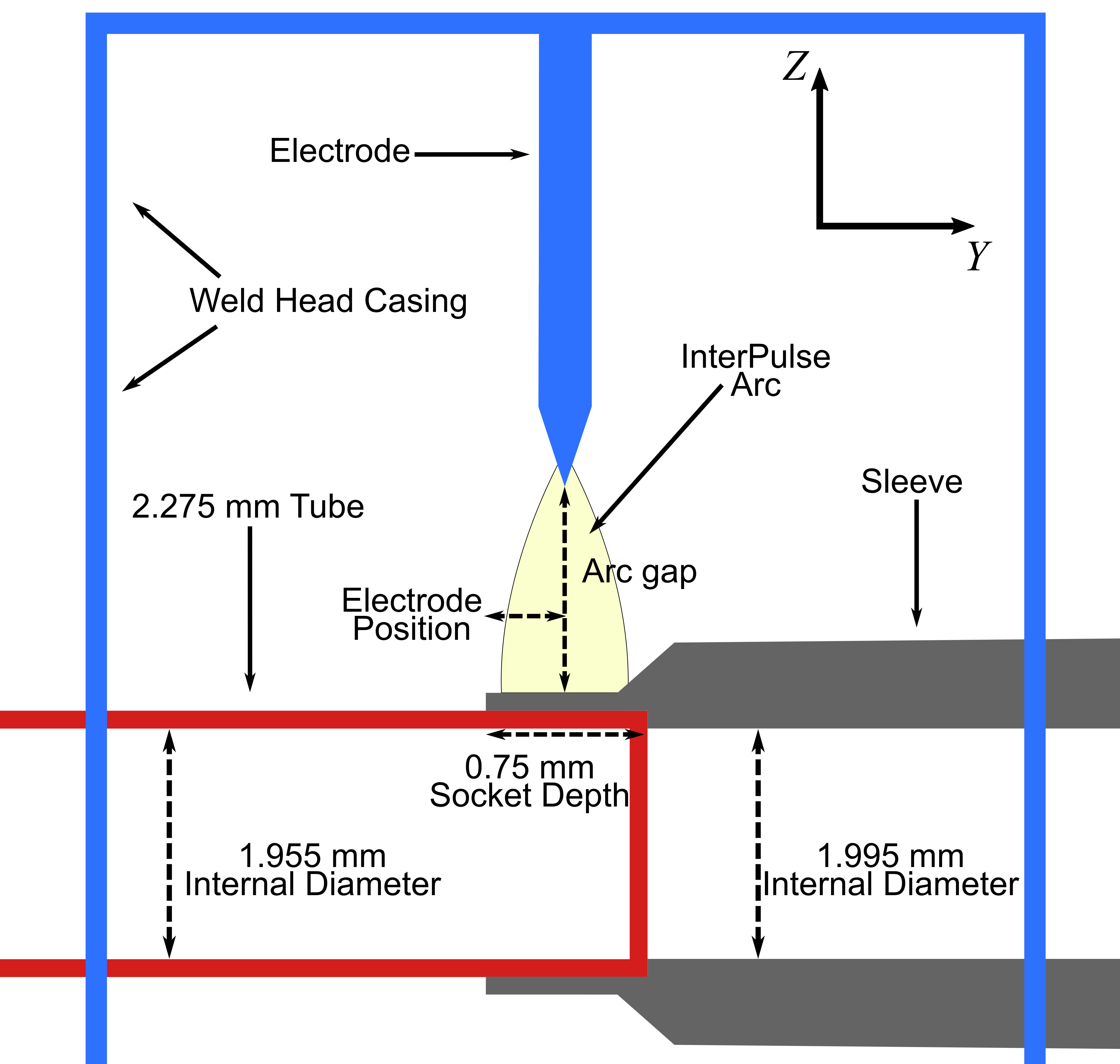}
\caption{Illustration of a cross-sectional view inside the orbital weld head showing the various elements and parameters. The blue lines are parts of the Polysoude UHP 250 weldhead and the electrode, the red line represents the CP-2 Ti tubing, and the grey areas show the sleeve fitting.}
\label{5-fig-JINST-schematic}
\end{figure}

Ensuring high reliability in produced welds is vital in delivering a long service life for the cooling manifolds. Thus, before welding both the sleeve and tube are cleaned with an abrasive pad and then washed thoroughly in Isopropyl alcohol (IPA). Further, the inner and machine gas supply manifolds use standard Swagelok fittings to reduce the chance of leaks — increasing the operator confidence in gas supply reliability. Finally, the electrodes are regularly replaced (after $15 - 20$ welds) to ensure that the electrode tip profile is consistent.  Figure \ref{5-fig-JINST-schematic} shows a cross sectional view of the welding torch during production illustrating the process parameters: the blue lines are parts of the Polysoude torch head and electrode, the red line represents the 2.275 mm CP-2 Ti tubing and the grey shape shows the custom sleeve with the insertion socket.

\FloatBarrier
\section{Experimental results}
\label{5-sec-exp-res}
\subsection{Physical analysis}
\subsubsection{Mechanical examination}
\begin{figure}[!htb]
\centering
\includegraphics[width=12.5cm]{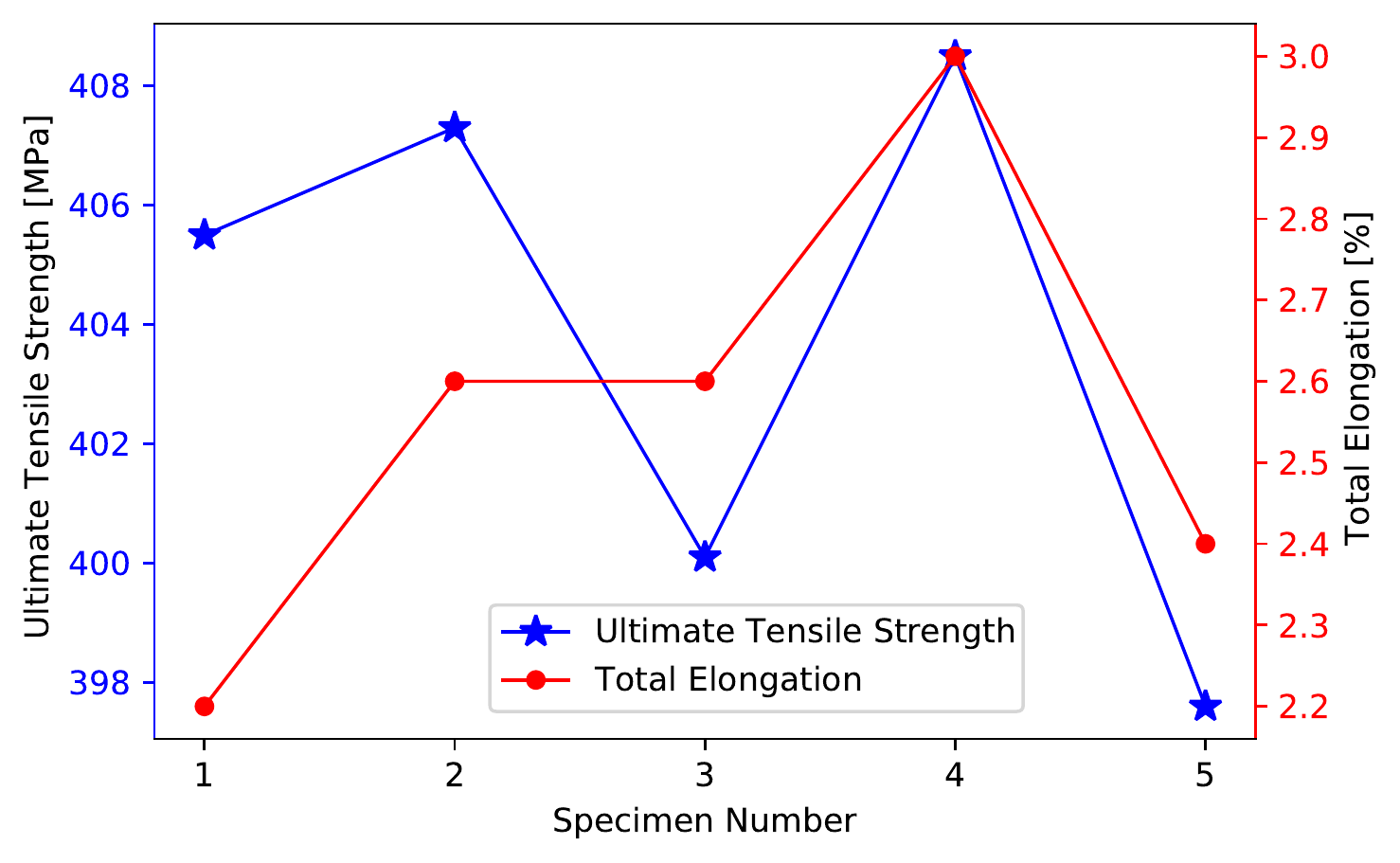}
\caption{Ultimate Tensile Strength and Total Elongation for five CP-2 Ti tube welds.}
\label{5-fig-JINST-micro-UTSel}
\end{figure}

The Ultimate Tensile Strength (UTS) and Total Elongation (TE) of five welded samples were tested with a ZPM machine equipped with a 1 kN load cell. Clamped vertically with pressure grips 35 mm apart, the samples were elongated at 0.5 mm per minute (Method A2 of ISO 6892-1 \cite{ISOA2}). As shown in Figure \ref{5-fig-JINST-micro-UTSel}, the five weld samples have UTS in the 398.0 – 408.9 range with TE varying between 2.2 and 3.0\%. The average UTS of the samples was 403.8 ± 4.2 MPa with an average elongation of 2.5 ± 0.3\%. This average elongation of 2.5 ± 0.3\% is quite small compared to the base metal level of 24 \% \cite{elongation-1}. This in fact suggests the welds have a restricted ability to sustain plastic deformation meaning fractures are more likely to be severe. 

Surprisingly, this average exceeds the UTS of the base CP-2 Ti material, \SI{340}{\mega\pascal} \cite{intro2aero}. The UTS of the specific tubing used was not tested for due to limited resources however prior research indicates that in fact the UTS of the welded samples should be less than that of the base material \cite{UTS-1, UTS-2}. Other research has reported a UTS of commercially pure titanium of \SI{539}{\mega\pascal} \cite{elongation-1}, given the measured UTS of the welded sample it is likely the base CP-2 Ti tubing is closer to the value in \cite{elongation-1} than \cite{intro2aero}. All samples broke in the tube HAZ indicating reasonable weld quality. Ideally, a break in the base metal (BM) would be achieved yet the break in the HAZ is still at a larger UTS than the minimum specified UTS of CP-2 Ti validating the procedure.       

\begin{figure}[!htb]
\centering
\includegraphics[width=12.5cm]{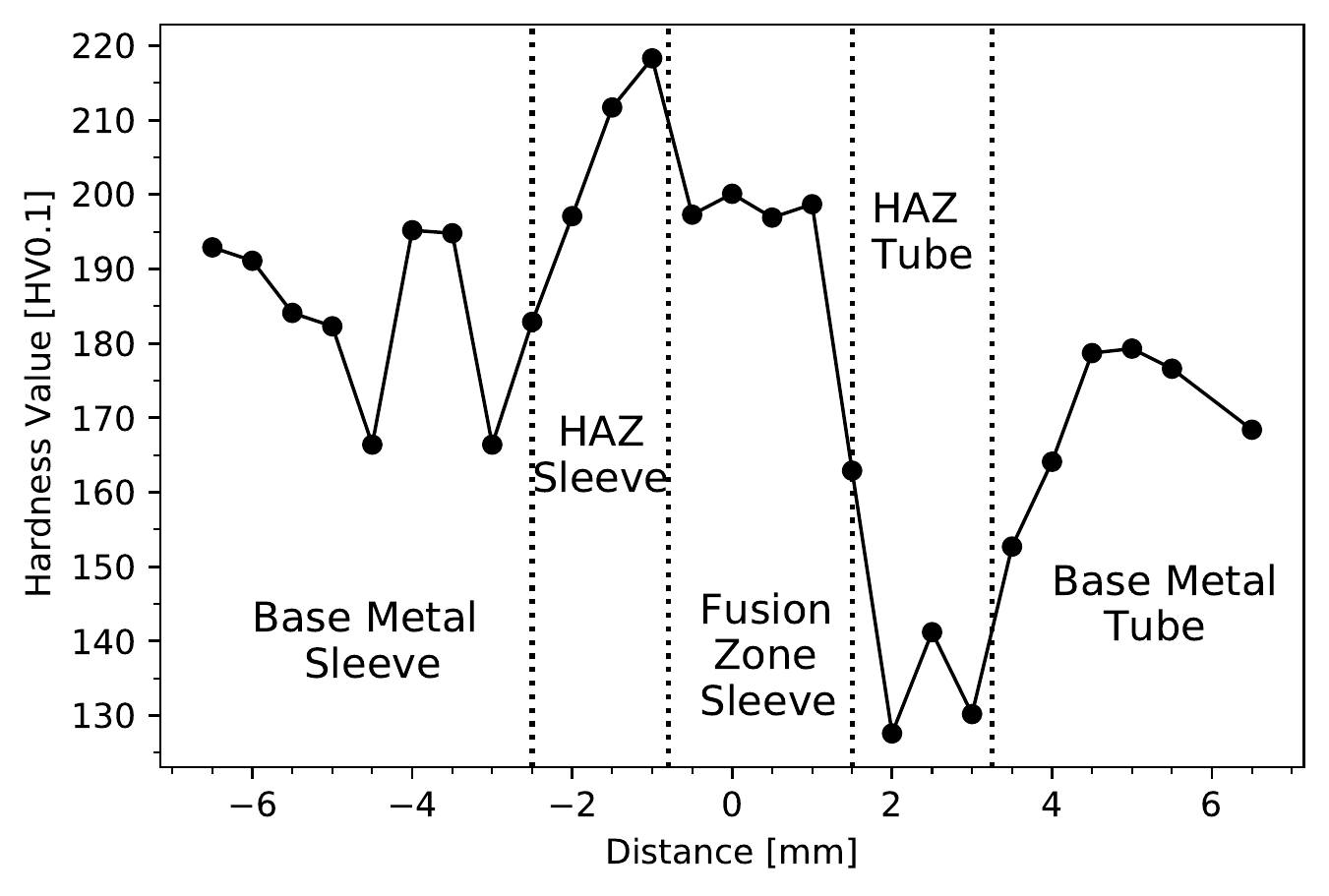}
\caption{Micro Hardness measurements along the full length of the sleeve-weld-tube for the CP-2 Ti tubing.}
\label{5-fig-JINST-micro-HV}
\end{figure}

A microhardness survey was carried out under a load of 100 g with a Vickers indenter to assess the hardness throughout the welded CP-2 Ti tubes. Figure \ref{5-fig-JINST-micro-HV} shows the hardness profile across the full length of the sleeve-weld-tube regions of a 2.275 mm CP-2 Ti tube sample. Here, the Vickers number (HV) varies between 127.6 (in the tube HAZ) and 218.3 (in the sleeve HAZ). Intrinsic variations in the microstructure account for the fluctuations within each region however, each region exhibits its own hardness characteristics. The martensitic acicular structure in the fusion zone region accounts for its higher HV matching previous reports \cite{cpTiGTAW1, cpTiGTAW2} of a higher HV in the weld fusion zone compared to base metal for GTAW welds of commercially pure titanium. Yet the HAZ in the tube has a noticeable lower HV than the HAZ in the sleeve. As the sleeve and tube are from different manufacturers some variation in the base metal hardness is to be expected. The variation in the direction of HV change is due to the manufacturing process; the tubes are cold drawn and the sleeves machined. The cold drawn tubes are effectively in a compressed state that is relaxed by the welding arc heat, whereas, the sleeve does not appear to have any residual stresses with its HAZ with HV variation due variations in $\alpha$ and $\beta$ phases. Another possible cause of the variation in HV is differences in heat flow, the sleeve area has physically more mass meaning it is likely to stay hotter for longer compared to the base ultra-thin-walled tube.

\subsubsection{Microstructure examination}
\label{5-sec-ms-exam}
\begin{figure}[!h]
\centering
\includegraphics[width=13.5cm]{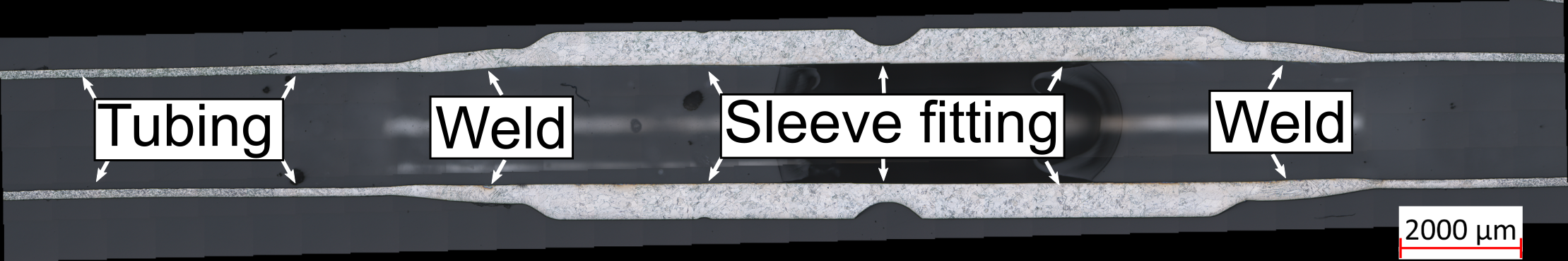}
\caption{A mosaic micrograph of full tube-sleeve-tube joints taken under polarized light. The tube, weld and sleeve is indicated. A \SI{2000}{\micro\meter} scale is indicated on the right hand side.}
\label{5-fig-JINST-micro-mosaic}
\end{figure}

For general microstructural characterization, welded tube-sleeve-tube samples were mounted in cold set resin. Then, the samples were ground down to a longitudinal axis plane with progressively finer grit paper and diamond suspensions before a final polish with colloidal silica suspension and etching with Etchant \#192 of ASTM E407 \cite{ASTM-etch}. To illuminate the microstructure, imaging was performed under polarized light. As the tube is inserted into a socket within the sleeve and, as the weld is orbital, it is unfeasible to identify the fusion line. A mosiac micrograph of a complete tube weld is shown in Figure \ref{5-fig-JINST-micro-mosaic}. 

\begin{figure}[!h]
\centering
\includegraphics[width=13.5cm]{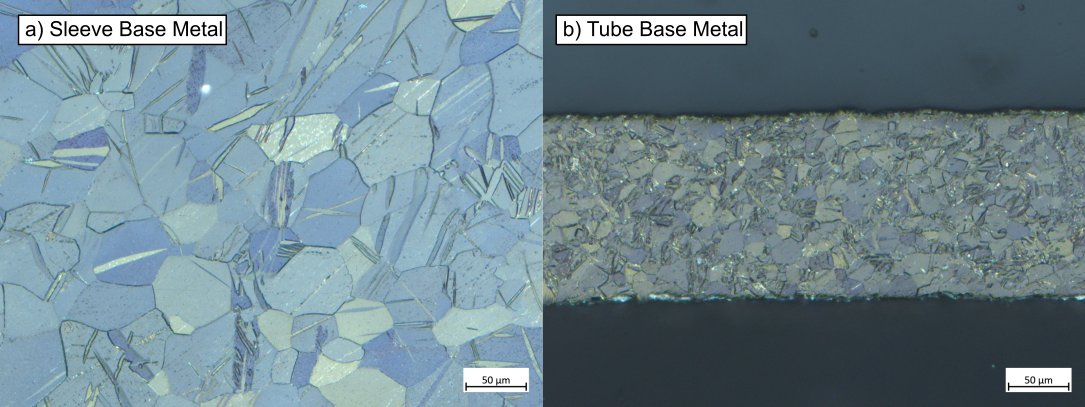}
\caption{Optical micrographs of a) base metal microstructure of the sleeve fitting, b) base metal microstructure of the tube. The micrographs are taken under polarized light revealing the grain structure. Images are essentially zoomed in regions for the sleeve and tube indicated in Figure \ref{5-fig-JINST-micro-mosaic}}
\label{5-fig-JINST-micro-bm}
\end{figure}

The sleeve and tube have different base microstructures; as highlighted in Figure \ref{5-fig-JINST-micro-bm} the base sleeve has an ASTM Grain size number \cite{ASTM-grain} of 4.5 whereas the tube has a ASTM Grain size number of 8.6 thus the tube has considerably smaller grains than the sleeve. The base metal structure in the sleeve clearly shows a typical equiaxed $\alpha$ titanium microstructure with multiple in-grain dislocations of elongated $\alpha$ microstructure. The structure of the tube base metal is similar except for a noticeably smaller average grain size.

\begin{figure}[!h]
\centering
\includegraphics[width=13.5cm]{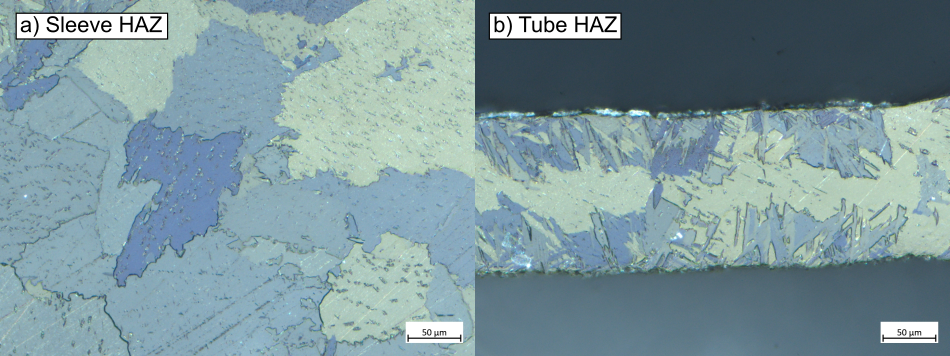}
\caption{Optical micrographs of a) heat affected zone of the sleeve fitting post welding, b) heat affect zone of the tube post welding. The bright spots on the edges of the sleeve are colloidal silica particles embedded during polishing of the sample.}
\label{5-fig-JINST-micro-HAZ}
\end{figure}

\begin{figure}[!h]
\centering
\includegraphics[width=7cm]{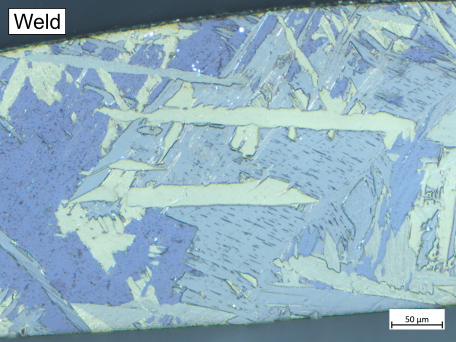}
\caption{Optical micrograph of the fusion zone region of a ultra-thin-walled tube weld with acicular grain structure.}
\label{5-fig-JINST-micro-weld}
\end{figure}

The solid-state transformations in HAZ play a critical importance in weld integrity. As the sleeve and tube have different base metal microstructure, they have different HAZ microstructures. The HAZ in both the sleeve and the tube is shown in Figure \ref{5-fig-JINST-micro-HAZ}. The HAZ in the sleeve shows increased grain growth reaching an ASTM Grain size number of 3.1 with a noticeable reduction of in-grain dislocations compared to the BM. The HAZ in the tube also shows significant grain growth reaching an ASTM Grain size number of 5.5 whilst also retaining some in-grain dislocations. Predictably, both HAZ regions exhibit decreasingly coarse grains at an increased distance from the fusion zone. A micrograph of the fusion zone is shown in Figure \ref{5-fig-JINST-micro-weld} displaying complete fusion of the tube and sleeve. Typically, the grain size in the fusion zone of a weld is the largest due to the longest dwell time at elevated temperatures. However, in the present work with the ultra-thin walls of the tube, the complex heat flow in orbital welding, the difference in grain sizes of the base metal and, the GTCAW process used, the grain size in the weld does not exceed that of the larger sleeve grain size instead manifesting as acicular $\alpha$ phase structure. This acicular structure is typical of a higher cooling rate despite the longer dwell time at the elevated temperatures used in this process.

\subsubsection{Parameter space examination}
\label{5-sec-parameter-space-exam}
Over the course of the parameter tuning it was established experimentally that moderating the inner pressurization was key to ensuring a consistent internal diameter of the produced welds. With a welding procedure that consistently passed visual inspection established, a modified NASA arc welding optimization methodology \cite{naseWeld} was applied to the inner pressure management to investigate the range of acceptable inner pressures. Once a base procedure is established, the NASA optimization technique essentially involves performing a minimum of three welds at the nominal procedure then a minimum of three welds at a `limit low' heat input setting and a corresponding three welds at a `limit high' heat input setting. In the present application, a sufficient heat input can produce a variety of welds depending on the other parameters. Through altering inner gas pressure as well as heat input location, both acceptable and rejected welds can be produced. Therefore, the limit low and high procedure can equally be applied to the inner pressurization. However, this does not negate the importance of heat input limits. In fact ideally a design of experiments approach could have been taken to identify the limits of, and interaction between, heat input and inner pressurization. This approach could even have been extended to all parameters. However, due to limit lab resources only a set of six welds focusing on the limits of inner pressurization were produced.  

\begin{figure}[!h]
\centering
\includegraphics[width=14cm]{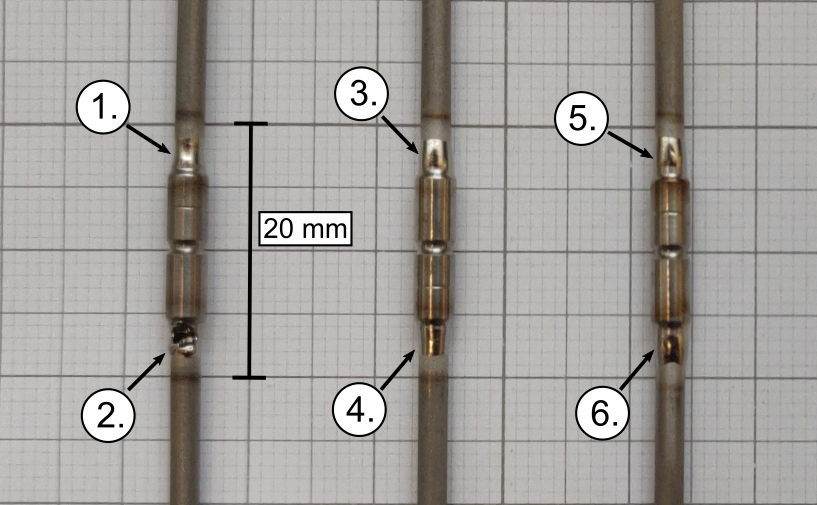}
\caption{Photograph of a selection of welds with ‘low limit’ inner diameter pressurization (1,2), correct
inner diameter pressurization (3,4) and ‘high limit’ inner diameter pressurization (5,6).}
\label{5-fig-JINST-pass-fail}
\end{figure}

Through gradually lowering or raising the gas pressure whilst keeping the rest of the procedure the same, the low limit was found to be 1194 Pa whereas the high limit was found to be 1742 Pa. Figure \ref{5-fig-JINST-pass-fail} shows a selection of the produced welds. Welds 1 and 2 show the lower ‘low limit’ of inner pressure; whilst Weld 1 nearly buckled but successfully fused, Weld 2 completely collapsed due to the insufficient gas pressure. This shows 1194 Pa to be the effective low limit. Welds 3 and 4 are with an ideal procedure and an inner diameter pressure of 1448 Pa displaying excellent exteriors. In Welds 5 and 6 the inner gas pressure is increased beyond the acceptable range showing bulging as the gas pressure pushes the molten metal out. 

These results raise interesting questions about the parameter balance of the inner pressure. The force pushing the weld inwards against the inner pressure is the stagnation pressure from the impinging plasma arc. The clear difference in microstructure of the fusion zone compared to the HAZ and base metal (covered in Section \ref{5-sec-ms-exam}) indicate that the weld penetrated through the entire wall thickness. This suggests that during the welding process a liquid metal disk is suspended between the two competing forces. As the pressure from GTAW arcs peaks in the centre of the arc it therefore follows that this liquid metal disk is subject to a Laplace pressure as it curves to accommodate the competing pressures. However, using a classic estimation of arc pressure \cite{linEagarGtawPressure} given by 

\begin{equation}
p_{arc} = \frac{\mu_{0} I^{2}}{4 \pi^{2} R^{2}}    
\end{equation}

for arc pressure $p_{arc}$ calculated from current $I$ and arc radius $R$ where $\mu_0$ is the vacuum permeability. Using a known successful average current of \SI{6.22}{\ampere} and taking a generous $\text{arc radius} = 1.1 \times \text{weld radius}$ the arc pressure can be estimated as  

\begin{equation}
p_{arc} = \frac{4\pi \times 10^{-7} \times (6.22)^{2}}{4 \pi^{2} \times (0.001)^{2}} = \SI{1.345}{\pascal}    
\end{equation} 

which is several orders of magnitude below the measured inner pressure for even the `low limit' case. Such a drastic mismatch in the pressures suggests the competing Laplace pressure hypothesis is invalid. One solution to this is that perhaps in successful welds the liquid weld pool solidifies before it can burst and the pressure need only be `balanced enough' rather than completely even. Knowing the arc pressures produced by the IP50 welding machine at low current amperage would elucidate this. It was planned to measure these pressures experimentally using a custom made water cooled copper block with an internal pressure sensor similar to that presented in \cite{arcPressureExp}. However, due to unforeseen circumstances, only an initial trial block without an internal pressure sensor was tested. The fact that there is a `low limit' shows that the inner pressure is required to push against \emph{something}. And the `high limit' shows that too much inner pressure can lead to the liquid weld pool rupturing. Given that both the liquid metal weld pool is very small and the successful experimental welds are in general consistent around their circumference, gravity can be excluded as a contributing factor. The remaining candidate is thus surface tension; perhaps on the curved surface of the tube the surface tension causes the inward collapse without a buttressing internal gas force. However, with a closed weld head, the surface of the liquid weld pool cannot be seen. This therefore motivates a use of \gIF whereby the effect of surface tension could be quantified but this is not the focus of this thesis. For the purposes of establishing a general case, the key piece of information from this parameter space investigation is that \emph{something} is pushing the weld inwards and to simulate this in \gIF will thus require a pressure gradient.  

\subsection{Data analysis}
\label{5-sec-data-analysis}
% N.B one of the rot speed values is 6.8 (e.g. current) and not 25 like all the runs 
\subsubsection{Raw data}
\label{5-sec-raw-data}
The large amount of experimental runs presents the opportunity of possible insights into ultra-thin-walled tube welding through data analysis. Overall, 310 welds were performed with a complete set of values being recorded for 292 of them. Of these 292, 139 were welded successfully and were visually flawless from the outside, 103 were welded successfully but had some sort of visual floor to them and in all 50 welds failed. For the data analysis, the failed welds are scored 0, the flawed welds 0.5 and the successful welds as 1; these values are collectively referred to as the welds `score'. A variety of welding parameters were used but due to the approach taken the parameter space investigation was unsystematic. Specifically, the welds were not produced for an investigation into the physics of ultra-thin-walled tubing but were instead an optimization of the production of a specific weld. Development was performed through an iterative process starting from a procedure suggested by an experienced welding engineer. There were repeated design changes, specification changes, and equipment failures leading to the somewhat unsystematic variety of parameters used. From the perspective of ultra-thin-walled tube welding ideally a design of experiments or other systematic approach would have been taken however this data is what is available. 

The raw data does not include a 1:1 relation with the actual welding parameters detailed in Table \ref{5-tab-gtaw-param}. Rather, ancillary or corresponding parameters were recorded instead. This is not ideal however it is how the data was recorded. These corresponding `data' parameters are summarised in Table \ref{5-tab-data-param}. Two slightly ambiguous ones here are the electrode position and the electrode tool setting. The electrode position is the horizontal distance from the edge of the socket to the electrode tip which is shown in Figure \ref{5-fig-JINST-schematic}. And the electrode tool setting is the length of the final section of a bespoke tool that was created to position the sleeve within the weld head. The socket depth (the distance the tube goes into the sleeve) is the same length as the electrode position and was altered a small number of times throughout the experiment. Irrespective of these parameters for the electrode, the arc gap was always set to a constant \SI{1.25}{\milli\meter}.

\begin{table}[ht]
\setlength{\tabcolsep}{10pt}
    \centering
        \caption{\label{5-tab-data-param} Definitions of parameters recorded in the raw data and their (if present) correspondence with the experimental welding parameters listed in Table \ref{5-tab-exp-param}. (*) The gas supply manifold used was a few meters long so this is an upper estimate given the leaks that will occur en route.}
        \begin{tabular}{p{5cm} p{9cm}}
        \toprule
        Data parameter & Definition / correspondence \\
        \hline
        Electrode tool setting & Length of tool head used to position sleeve. \\
        Machine gas & Setting on the flow meter(*) for the machine gas flow rate in Table \ref{5-tab-exp-param}. \\
        Gas flow rate & Setting on the flow meter(*) for the internal gas flow in Table \ref{5-tab-exp-param}. \\
        Mechanical gas pressure & Pressure measured with sensor shown in Figure \ref{5-fig-JINST-set-up} pre-welding. \\
        Weld gas pressure & Pressure measured with sensor shown in Figure \ref{5-fig-JINST-set-up} post-welding.  \\
        Main current & Identical as Table \ref{5-tab-exp-param}.  \\
        Background current & Identical as Table \ref{5-tab-exp-param}. \\
        Rotational head speed & Identical as Table \ref{5-tab-exp-param}. \\
        Initial current & Identical as Table \ref{5-tab-exp-param}. \\
        Electrode Position & Horizontal distance of electrode tip to edge of sleeve, shown in Figure \ref{5-fig-JINST-schematic}. \\
        Warmup welds & The number of welds performed at a low amperage to get the electrode warm.  \\
        Total welds with electrode & As stated - the total number of welds with the current electrode.  \\
        \bottomrule
    \end{tabular}
\end{table}

\begin{figure}[!htbp]
\centering
\includegraphics[width=15cm]{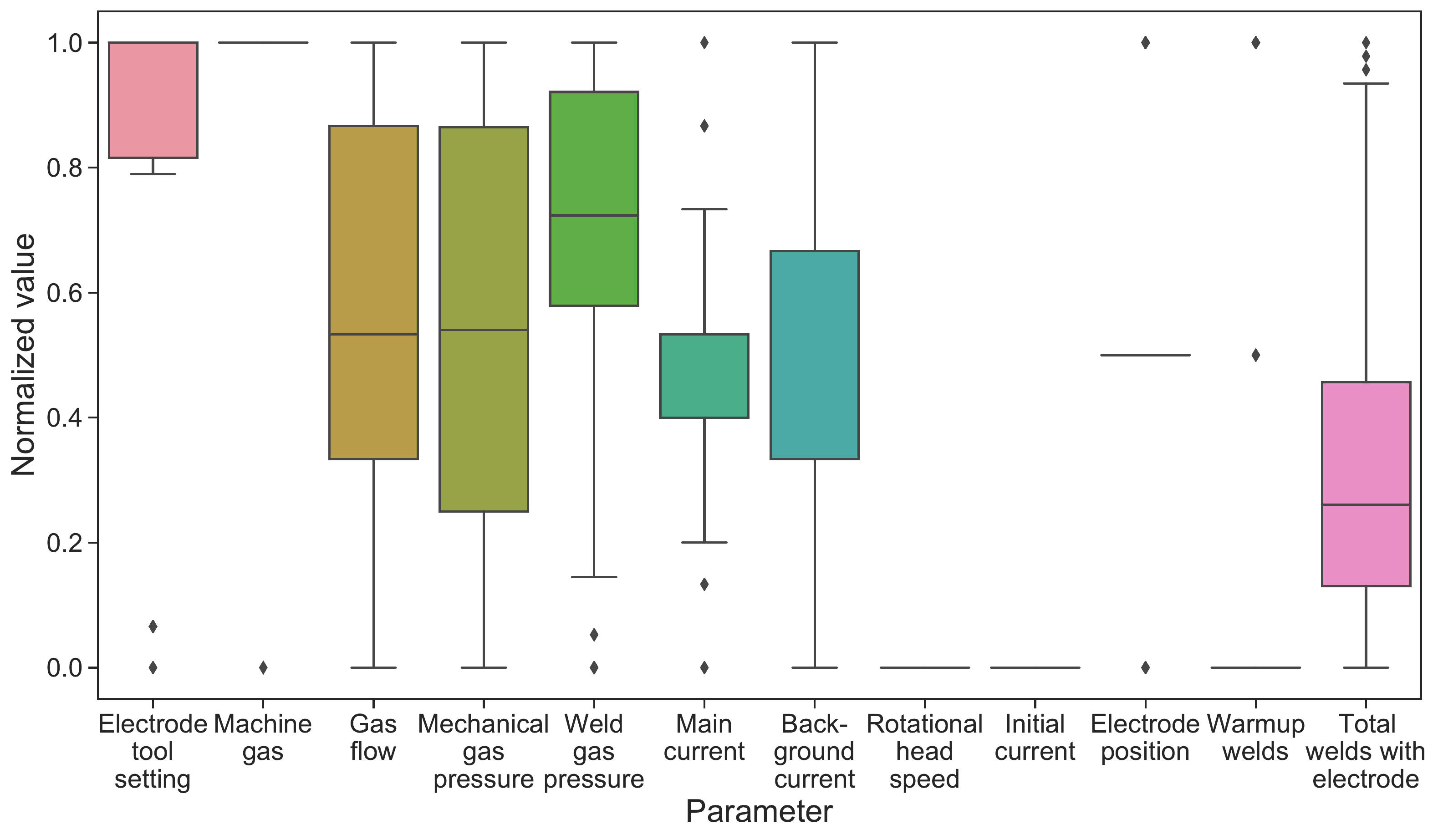}
\caption{Box plot of the data parameters recorded for the 292 recorded welds.}
\label{5-fig-parameters}
\end{figure}

To gain an overall view of the data, a box plot of the values for all of the parameters is presented in Figure \ref{5-fig-parameters}. This shows that for many variables there is little or zero change in the entire data set. For instance, only one initial current value and rotational speed are used, these values can thus be dropped from the data set as they are not different for successful and failed welds. Next there are parameters that have effectively zero interquartile range so it is doubtful whether their is enough variation for their effect to be quantified. These are the machine gas, the electrode position and the warm up welds and they have correlation coefficients with the weld score of 0.03, 0.11 and -0.09 respectively. Given this, these parameters can also be dropped from the data set. The remaining parameters can thus be used for the analysis. 

% mac-gas	0.033995935
% elec-pos	0.113652659
% warm-up	-0.094973788

\subsubsection{Principal component analysis}
Given there is a set of candidate predictor variables (the parameters) and categorical scores (failed, flawed and successful) this appears as a classification problem. However, the relatively low number of samples (292), the unequal distribution between the categories, and the low variability in some parameters means that there will potentially be a lot of variance in the fit. A solution to this would be to get more experimental data however as this is not possible\footnote{Due to COVID-19 and lack of lab access.} the next best route is data synthesis via running \gIF with a variety of parameters. However, there is still value in exploratory analysis of the experimental data set. With so many variables, a good choice of method for this is principal component analysis.

Principal component analysis (PCA) is a standard dimensional reduction technique where variables are linearly combined into new principal components. For a set of variables \textbf{X}, the principal components are the eigenvectors of the covariance matrix of \textbf{X}. In this case these are the variables in Section \ref{5-sec-raw-data} that were not dropped. The PCA was performed in Python using inbuilt PCA function in the scikit-learn package \cite{scikitPackage}. Summing the variance of the first two principal components (e.g. the eigenvalues from the covariance matrix and eigenvectors) accounts for only 72.2\% of the total variance. This is not particularly high but it is high enough to illustrate the data in a biplot for exploratory analysis. 

The biplot plot shows that the background current is strongly correlated with the main current and both have a strong influence on principal component 1. Further it suggests that there is a lot of redundancy between the two currents that could potentially justify a combined current variable. Similarly, gas flow rate, mechanical gas pressure and weld gas pressure are all strongly correlated together and strongly influence principal component 2 - again suggesting redundancy between the variables. Interestingly this is also true for the total welds with the electrode. The orthogonality between these two groups (currents and pressures) indicates that they are not correlated. This information is useful as it shows that both apt current and apt pressure are required for a successful weld.  

The biplot seems to suggest that the electrode tool setting is negatively correlated with the currents and principal component 1. Curiously this is an artefact is actually due to design changes in the welding set up coinciding with improved operator competence. One electrode tool setting was used for the first $\approx 200$ welds of these 37\% were successful, 40\% were flawed and the rest failed. For the subsequent $\approx 100$ welds after the electrode tool setting was progressively changed, only two were complete failures whereas 75\% were successful and 23\% flawed. \footnote{Here $75\% = \frac{62}{83}$ and $23\% = \frac{19}{83}$}. Therefore, it appears as though electrode tool setting has an effect when in fact it was due to a higher success rate being established when the change was implemented.

\begin{figure}[!htbp]
\centering
\includegraphics[width=15cm]{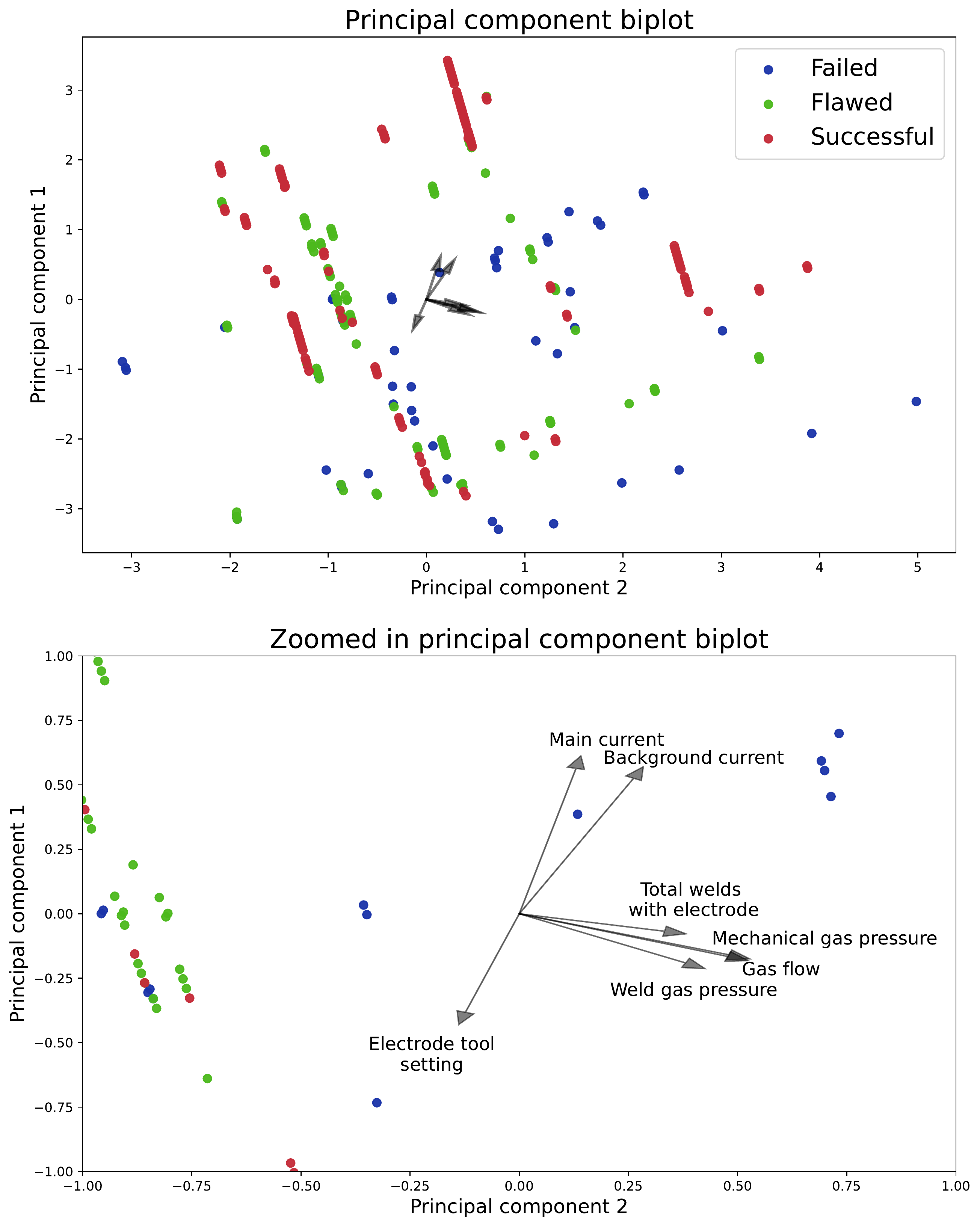}
\caption{Principal component analysis biplot. The bottom image is a zoomed in version of the top to allow the loading plot to be seen more easily. Clockwise, the variable labels read `Main current', `Background current', `Total welds with electrode', `Mechanical gas pressure', `Gas flow', `Weld gas pressure', and `Electrode tool setting'.}
\label{5-fig-pca}
\end{figure}

\subsection{Analysis conclusion}
The physical analysis shows the efficacy of the procedure outlined in Section \ref{5-exp-procedure} through a mechanical and microstructural examination. Further, the analysis reveals that not only is the inner buttressing pressure required for ultra-thin-walled tube welding but critically this pressure needs to be moderated to a `goldilocks' level. These low and high limits can in theory be employed by \gIF to benchmark the required pressures. 

Following this, the data analysis first demonstrates that a fair amount of the variables can be dropped from the data as they have either zero or effectively zero variation. From the produced PCA, the remaining variables can largely be grouped into currents and pressures. The exception here is the electrode tool setting although this can be negated for the reasons given. Overall, both the physical and data analysis show the importance of current to form the weld pool and pressure to orchestrate its position. This information can be applied to the simulation work. 

\section{Simulation}
\subsection{Case outline}
\label{5-sec-case-outline}
\subsubsection{Domain}
To find a general case for ultra-thin-walled tubing through simulation an appropriate domain is required. However, whilst it is possible to simulate a cylindrical domain with a rotating heat source, this comes with considerable disadvantages:

\begin{itemize}
    \item The implementation of the heat source covered in Chapter 4 lends itself more to linear translational movement whereby the heat source position is updated as simply $\text{vector} \times \text{time}$.
    \item Cylindrical domains have only three boundaries: the top and bottom of the cylinder and the cylinder wall. This means it is difficult to create an internal buttressing pressure without creating an accurate internal pipe flow simulation on top of an accurate GTAW simulation.  
\end{itemize}

\begin{figure}[!htbp]
\centering
\includegraphics[width=15cm]{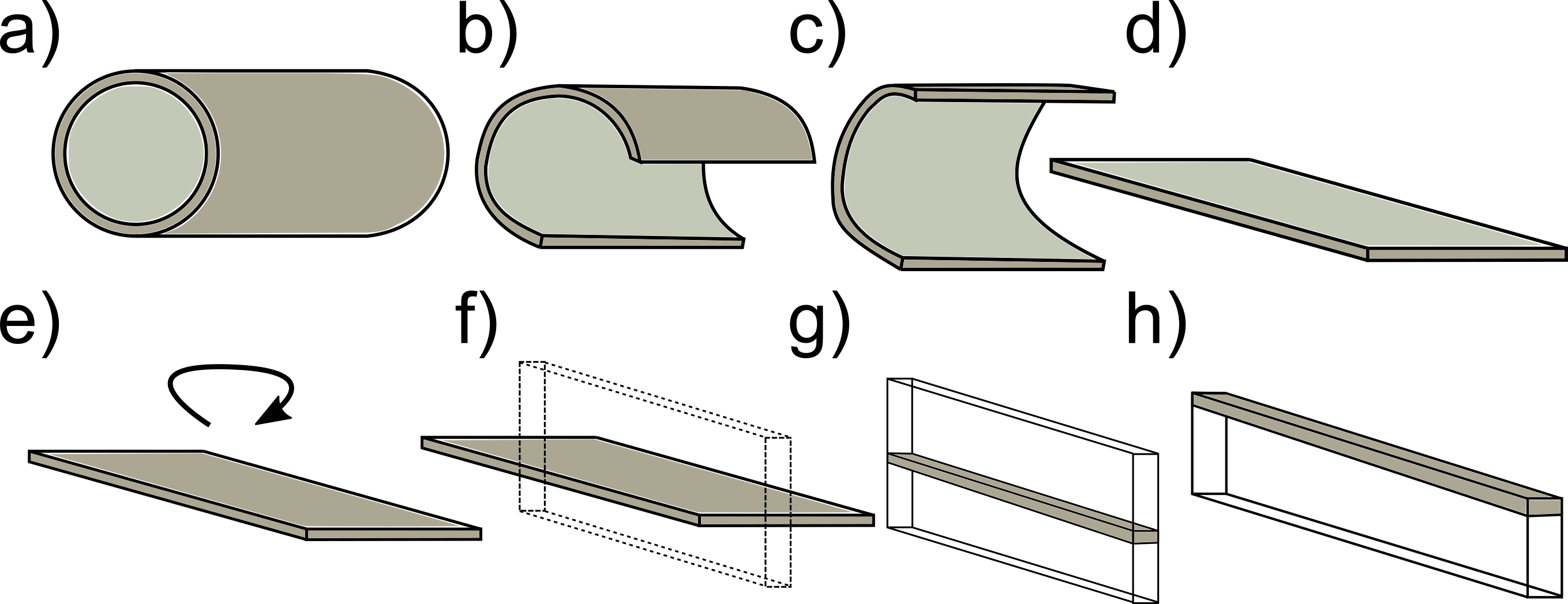}
\caption{Illustration of a geometric transformation of ultra-thin-walled tube, a), into the representative domain used for the simulations, h). This unravelling enables a flat geometry to be used to simulate orbital welding although information about the curvature is lost. Note how the wall of the tube is `cut' creating a discontinuity, this is accounted for in \gIF through the use of multiple heat sources shown in Figure \ref{5-fig-tube-unravel}.}
\label{5-fig-tube-unravel}
\end{figure}

Given this, it is preferred to simulate a `flattened' tube. Further, if the domain can be reduced to two dimensions without impacting the physics better performance can be achieved. With these preferences, the domain chosen is a transformation of a ultra-thin-walled tube into a square shape shown in Figure \ref{5-fig-tube-unravel}. Here, a representative `flattened' domain is created to mimic tubular geometry. This technique approximates the curved surface of the tube as completely flat thus losing physical information however it provides substantial performance benefits. In principle the inside and outside of the tube can be either the top or bottom but it was found that using the top of the domain as the outside of the tube (and the bottom as the inside) resulted in the best performance.

As \gIF does not simulate radiation, within the domain the only heat transfer mechanisms are convection and conduction. This can be compensated for by including the custom radiative boundary conditions detailed in Chapter 3. However, these boundary conditions necessitate the appropriate phase (the metal) to be at the boundary thus limiting the surface evolution of the liquid fraction which needs a gas phase to move into / away from. Simulating heat transfer and liquid fraction surface evolution are both important thus both should be employed. This is achieved by modifying domain g) in Figure \ref{5-fig-tube-unravel} to have one surface of the tube wall a boundary and the other exposed to the gas phase. Whilst the heat source is largest at the top (outside) surface of the tube, the large convective heat transfer present at the bottom (inside) surface during tube welding due to the inner gas flow should theoretically be important also. After trying both, it was found that setting the top of the tube to a boundary to enable accurate radiative heat and convective loss modelling was better. This was chiefly because it achieve superior computation performance and thus accurately simulating convective heat loss inside the tube was largely neglected in favour of accurately simulating the outside of the tube.  

Flattening the tube required the wall to be `cut' thus one end of the cut is theoretically connected to the other. Therefore, as the heat source leaves one side it should be entering the other. The inbuilt \emph{symmetry} boundary conditions in \of unfortunately do not simulate this transitions accurately. To address this, the domain length was tripled and additional trailing and leading heat source terms were added (e.g. additional class instances) to the domain. These additional heat sources start in front of and behind a `core region' creating a more accurate transition. This core region is the only section of the full domain that counts in the assessment of the simulations. 

\begin{figure}[!htbp]
\centering
\includegraphics[width=15cm]{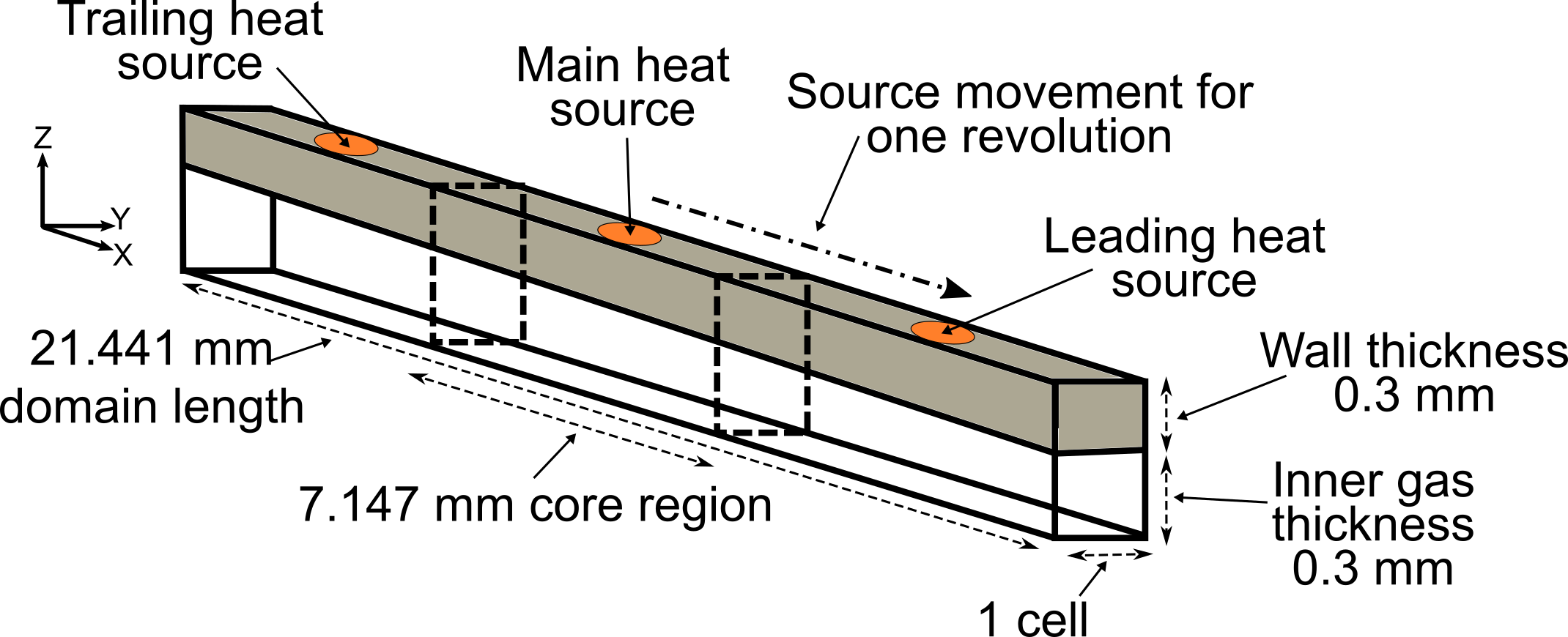}
\caption{Labelled illustration of the domain used for the tube simulations. Here  h) shown in Figure \ref{5-fig-tube-unravel} is extended forward and backward to enable the orbital movement of the tube welding process to be simulated. The domain is $200 \times 1 \times 30$ cells in size.} 
\label{5-fig-tube-domain}
\end{figure}

Looking again at Figure \ref{5-fig-JINST-micro-mosaic} reveals the thickness of the weld varies as the tube transitions into the sleeve. To capture this within the flattened domain, an average value of \SI{300}{\micro\meter} is used as a constant wall thickness. Combining this wall thickness with the heat source and boundary features created the domain illustrated in Figure \ref{5-fig-tube-domain}. This is the final domain used for the tube welding simulations.

\subsubsection{Parameter mapping}
\label{5-sec-parameter-mapping}
The experimental results covered in Section \ref{5-sec-exp-res} can be used to benchmark the solver. In order to do this, the experimental parameters need to be linked to variables in \emph{gtawFoam}. Table \ref{5-tab-gtaw-param} lists the recorded parameters covered in Section \ref{5-sec-data-analysis} and whether there is a corresponding variable in \gIF and what that variables is. Here, the candidate \gIF variables are the pressure field, the $Q_{Source}$ value in the heat source term, and the velocity of the heat source term. However, there are a few many-to-one relationships. A key target element for the simulation is that when a successful procedure is created in \gIF it can be converted back into an experimental procedure. Therefore, some variables will need to be combined and others dropped if there is no possible correspondence. Secondly, from an assessment perspective, there needs to be apt experimental data to compare with. As covered in Section \ref{5-sec-raw-data}, some data has so little variety that it cannot be easily assessed against. The most prominent example of this is the rotational speed which was set to 25 RPM for all experiments. However, this can be leveraged as a known parameter for an experimental welding procedure reducing the amount of predictions required by \emph{gtawFoam}. Converting the experiment and simulation results into a common set of variables that can be transferred between is termed `mapping'. This mapping system can thus be used to covert \gIF results into an experimental welding procedure.

\begin{table}[ht]
\setlength{\tabcolsep}{10pt}
    \centering
    \caption{\label{5-tab-gtaw-param} Correspondence (C) of the raw data parameters to a variable in \emph{gtawFoam} and whether there is apt experimental data to compare with. Note this is for the solver as described in Chapter 3 and not whether additional features could be added.}
    \begin{tabular}{l l l l}
    \toprule
    Parameter & C & Data & \gIF Variable\\
    \hline
    Electrode tool setting\rule{0pt}{2.6ex} & No & Yes & - \\
    Machine gas & Yes & No & Pressure field \\
    Gas flow & Yes & Yes & Pressure field \\
    Mechanical gas pressure & Yes & Yes & Pressure field \\
    Weld gas pressure & Yes & Yes & Pressure field \\
    Main current & Yes & Yes & $Q_{Source}$ in the heat source term \\
    Background current & Yes & Yes & $Q_{Source}$ in the heat source term \\
    Rotational speed & Yes & No & Velocity of the heat source term \\
    Initial current & No & No & - \\
    Electrode position & No & No & - \\
    Warm up welds  & No & No & - \\
    Total welds with electrode & No & Yes & - \\
    \bottomrule
    \end{tabular}
\end{table}

For the mapping, Table \ref{5-tab-gtaw-param} shows that unfortunately there are some experimental parameters that simply cannot be translated into \emph{gtawFoam}. There is nothing close to a \gIF parameter for the number of warm-up welds, the total welds with an electrode or the electrode tool setting. The initial current and electrode position could potentially be implemented into \gIF but there is not enough experimental data for these parameters to enable an assessment. Thus, these parameters cannot be mapped directly between experiment and \gIF and instead they must be set to fixed amounts.    

Both the main and background current values can be directly mapped to the current portion of $Q_{Source}$. This is because - as mentioned above - the same arc gap, and thus the same voltage, was used for all experiments. It can be safely assumed the efficiency is the same for all experimental runs thus $Q_{Source} \propto \text{Current}$ (from $Q_{Source} = \eta V I$ for an efficiency $\eta$, voltage $V$ and current $I$). From Figure \ref{5-fig-cur-shape} the welding machine spends approx $3\times$ as much time at the peak current compared to the background current. With this, a weighted average can be calculated as Mapped $Q_{Source}(I) = 0.75 \times \text{Main current} + 0.25 \times \text{Background current}$. The final value of $Q_{Source}$ will then be calculated by multiplying `Mapped $Q_{Source}$(I)' by the voltage (the efficiency is rolled up into the computational constant $C_C$). The extensive benchmarking in Chapter 4 means that there is a reasonably high amount of confidence in the efficacy of \gIF in modelling $Q_{Source}$. Given the straightforwardness of this mapping it follows that there should also be confidence in Mapped $Q_{Source}(I)$.       

The table shows four pressure fields that can all be mapped to the \gIF pressure field. This could potentially be done through amalgamating all four but there are reasons not to do this. Firstly, as shown in Figure \ref{5-fig-parameters} the machine gas used is the same for all experiments (bar two). Adding this into an amalgamated pressure variable would cause extra complication without a clear benefit. Secondly, the weld gas pressure is more of a dependent than independent variable. This is because if the weld is successful it is the same as the mechanical gas pressure (shown in Figure \ref{5-fig-gas-pres}) and if the weld is flawed or unsuccessful by definition it doesn't have the correct pressure handling ability. This can thus be dropped leaving gas flow and mechanical gas pressure. These two variables essentially describe the same physical phenomenon in different ways. Physically, the gas flow is the value the flow meter that is connected to the gas supply manifold near the gas bottle was set to. Conversely, mechanical gas pressure is the value measured by the pressure sensor at the gas flow outlet. Essentially this means the gas flow is the independent variable that set the dependent variable of the mechanical gas pressure. However, as there was a few meters of steel manifold and rubber tubing between the pressure sensor and flow meter there will be some gas leakage between the two. Given this leakage is completely dependent on the set up, the more transferable (between experimental setups) value is the mechanical gas pressure. This means that it is reasonable for the mechanical gas pressure alone to be used to set the pressure field used in \emph{gtawFoam}.  

Simply applying the mechanical gas pressure value throughout the simulation domain will not affect the simulation as the pressure gradient will be zero. To create a gradient, some portion of the domain - realistically a boundary - should be set to a different pressure. With this, the challenge becomes what this different pressure should be. As detailed in Chapter 3, \gIF uses a transformed pressure $p_{rgh} = p - \rho(g \cdot \hat{x})$. However, as gravity will be set to zero (detailed further in Section \ref{5-sec-no-gravity}), this term will reduce to $p_{rgh} = p$. Therefore, this different pressure can be set to a value that would cause inward collapse if the internal pressure is too low and will result in outward bursting if the internal pressure high. Unlike $Q_{Source}$, there is not an extensive benchmark available as to what the different pressure should be. For now, it can be simply characterized as a function of the internal pressure e.g. $f($Mechanical gas pressure$)$.   

Finally, looking again at Figure \ref{5-fig-pca} demonstrates the orthogonality between the variables associated with $Q_{Source}$ and the variables associated with the pressure field in Table \ref{5-tab-gtaw-param}. Therefore, this mapping should describe a large proportion of the ultra-thin-walled tube welding process. Whilst this is not a complete description of the process, it should be enough for \gIF to be both validated and be used for prediction. To perform both of these requires an assessment mechanism to be established. 

\begin{table}[!htbp]
\setlength{\tabcolsep}{10pt}
    \centering
    \caption{\label{5-tab-sim-to-exp-mapping} Parameter mapping between experimental data and \emph{gtawFoam}. Here, N/A in \gIF means that the mapping only refers to the experimental requirements. Further, fixed means that the value is constant for all experiments and simulations. Finally, as detailed in Section \ref{5-sec-parameter-mapping}, the machine gas, gas flow and weld gas pressure are dropped for \emph{gtawFoam}.}
    \begin{tabular}{l l l}
    \toprule
    Data & \gIF & Mapping\\
    \hline
    Electrode tool setting\rule{0pt}{2.6ex} & N/A & Fixed $= \SI{5.65}{\milli\meter}$ \\[0.15cm]
    Machine gas & Pressure field & \begin{tabular}{@{}l@{}}Fixed $= \SI{2.5}{\litre\per\minute}$\\Dropped for \gIF\end{tabular}\\[0.45cm]
    Gas flow & Pressure Field & \begin{tabular}{@{}l@{}}Adjust during experiment\\Dropped for \gIF\end{tabular}\\[0.45cm]
    Mechanical gas pressure & Pressure Field & \begin{tabular}{@{}l@{}}Mapped Pressure = \\ $f$(Mechanical gas pressure)\\\end{tabular}\\[0.45cm]
    Weld gas pressure & Pressure Field & \begin{tabular}{@{}l@{}}Check same as mechanical \\ pressure for experiment\\Dropped for \gIF \\\end{tabular}\\[0.75cm]
    Main current & $Q_{Source}$ & \begin{tabular}{@{}l@{}}Mapped $Q_{Source} =$\\$0.75 \times \text{Main current}$ \\ $+ 0.25 \times \text{Background current}$\end{tabular}\\[0.75cm]
    Background current & $Q_{Source}$ & \begin{tabular}{@{}l@{}}Mapped $Q_{Source} =$\\$0.75 \times \text{Main current}$ \\ $+ 0.25 \times \text{Background current}$\end{tabular}\\[0.75cm]
    Rotational speed & Source velocity & Fixed $= \SI{2.98}{\milli\meter\per\second}$\\[0.15cm]
    Initial current & N/A & Fixed $= \SI{2.4}{\ampere}$\\[0.15cm]
    Electrode position & N/A & Fixed $= \SI{0.75}{\milli\meter}$\\[0.15cm]
    Warm up welds  & N/A & \begin{tabular}{@{}l@{}}Two for a new electrode \\ zero otherwise\end{tabular}\\[0.45cm]
    Total welds with electrode & N/A & Maximum 15 uses each \\
    \bottomrule
    \end{tabular}
\end{table}

\subsubsection{Assessment mechanism}
With the domain and parameter mapping in place, an assessment mechanism is required to established whether a procedure run in \gIF produces a successful weld. Whilst there are three experimental scores (failed, flawed and successful) it is prudent to reduce these to two (failed and successful) for simulation assessment. This is because as \gIF only calculates phases and not metallurgy thus a flaw that would be picked up experimentally such as anatese formation would not be simulated. Secondly, flawed welds are \emph{effectively} failed welds given that they could fail in the future due to their flaws and they are not successful welds. 

\begin{figure}[!htb]
\centering
\includegraphics[width=14.5cm]{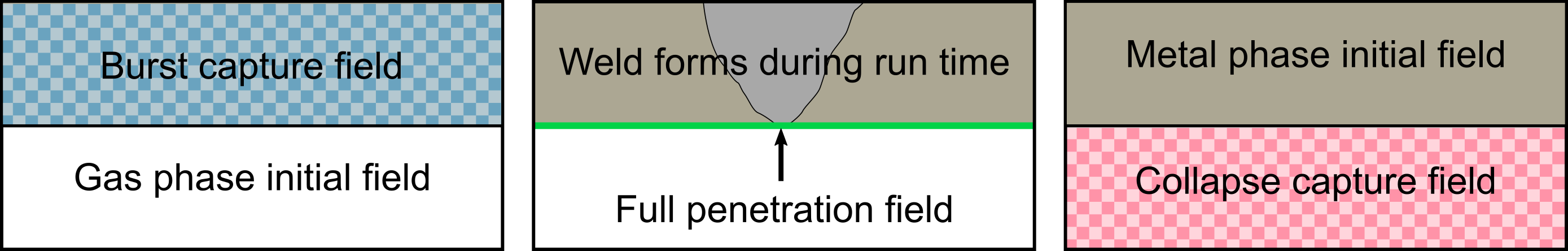}
\caption{Illustration of the assessment fields for tube welding.}
\label{5-fig-tube-assessment}
\end{figure}

The welds created by \gIF need to both not collapse and not burst. Further, the weld is required to fully penetrate the initial metal phase. To implement this, a series of assessment fields $\gamma$ are defined at the initial time ($t = t_0$). These fields are based off the initial fields for the gas and metal phase with an interface region subtracted. These assessments are exclusive to the core region of the domain as it corresponds only to the physical portion. These fields are defined in equation \ref{5-eq-assessment-fields} and highlighted in Figure \ref{5-fig-tube-assessment}. 

\begin{equation}
\begin{gathered}
\label{5-eq-assessment-fields}
\alpha_1(t = t_0) - (\alpha_1(t = t_0) * \alpha_3(t = t_0)) = \gamma_{Collapse} \\
\alpha_3(t = t_0) - (\alpha_1(t = t_0) * \alpha_3(t = t_0)) = \gamma_{Burst} \\
(\alpha_1(t = t_0) \land\alpha_3(t = t_0) \geq 0.1) = \gamma_{Penetration}.
\end{gathered}
\end{equation}

With the assessment fields established, at every time step a series of logical tests - shown in equation \ref{5-eq-logics} - are applied. Critically, these tests are one-way `flip switches' so that once they change they cannot change back. This means that, for example, if the tube bursts at any point and $\zeta_{Burst}$ is set to True it will remain that way for the rest of the time steps.

\begin{equation}
\begin{gathered}
\label{5-eq-logics}
\zeta_{Collapse} = \begin{cases}
    \text{True}, & \text{if}\ \gamma_{Collapse} = 1 \land \alpha_{2} = 1 \\
    \text{False}, & \text{otherwise}        
    \end{cases} \\
\zeta_{Burst} = \begin{cases}
    \text{True}, & \text{if}\ \gamma_{Burst} = 1 \land \alpha_{Penetration} = 1 \\
    \text{False}, & \text{otherwise}        
    \end{cases} \\
\zeta_{Penetration} = \begin{cases}
    \text{True}, & \text{if}\ \gamma_{Penetration} = 1 \land \alpha_{weld} \geq 0.5 \\
    \text{False}, & \text{otherwise}        
    \end{cases}. 
\end{gathered}
\end{equation}  

Note in equation \ref{5-eq-logics}, $\alpha_{weld}$ is the final produced weld. This is found from the sum of $\alpha_2$ over all time steps constrained between 0 and 1. Finally, these logical flags can be used to give the weld a score according to equation \ref{5-eq-score}. Here, each case is evaluated successively and once a score is given \gIF terminates. This four option scoring gives additional information than simply a pass / fail. Note there is no default option, this is a feature not a bug as it identifies if there has been a simulation error. This scoring system can then be compared to the experimental results to assess the match between \gIF and experiment.

\begin{equation}
\label{5-eq-score}
\text{Score} = \begin{cases}
    \text{Fail (Burst)}, & \text{if}\ \zeta_{Burst} \\
    \text{Fail (Collapsed)}, & \text{if}\ \zeta_{Collapse} \\
    \text{Fail (Lack of penetration)}, & \text{if}\ !\zeta_{Penetration} \\
    \text{Successful}, & \text{if}\ \zeta_{Penetration} \land (!\zeta_{Burst} \ \lor \ !\zeta_{Collapse})
    \end{cases}.
\end{equation}    

\subsubsection{Gravity}
\label{5-sec-no-gravity}
\begin{figure}[!htb]
\centering
\includegraphics[width=10cm]{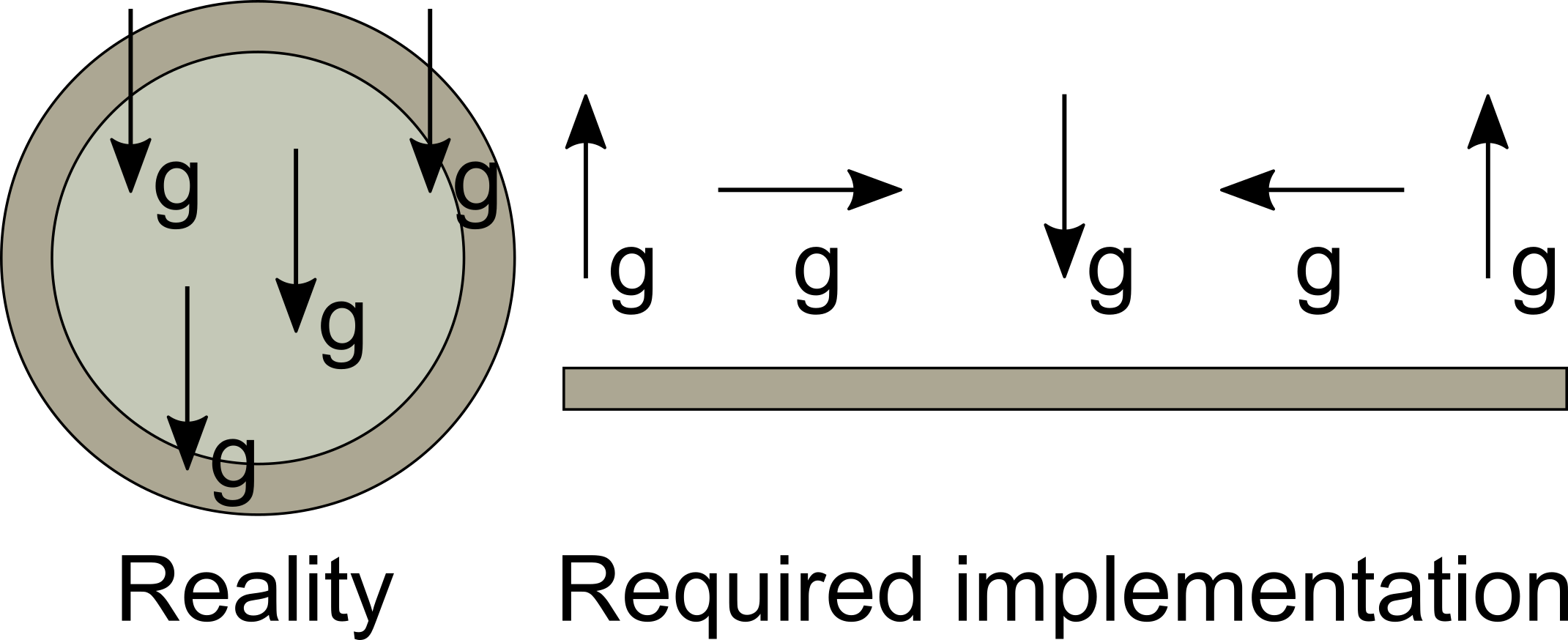}
\caption{Illustration of how gravity affects the tube welding process (reality) compared to the required implementation to apply to the flattened domain used for the simulations.}
\label{5-fig-tube-gravity}
\end{figure}

Acceleration due to gravity is set to zero for the tube weld simulations. This is for a few reasons reasons. Firstly gravity creates a significant performance penalty with runs taking over $10 \times$ as long to complete. Secondly, the default implementation of gravity in \of is as a body force acting on the whole domain. As shown in Figure \ref{5-fig-tube-gravity}, to accurately simulate the gravity vector would require a custom implementation whereby a field of gravity vectors is imposed on the domain; this would be quite challenging to create. Finally, turning gravity on with it acting downwards on the whole domain does not actually affect the results at all as the pressure gradients required to overcome the relatively high surface tension of liquid titanium mean that gravity contributes relatively little to the momentum equation. Turning gravity on simply shifts up the absolute values of the pressures.

\subsection{Trial results}
\label{5-sec-trial-results}
As a trial into the efficacy of the tube simulation, three parameter sets - passing, bursting and collapsing - are tested. Mapping wise, the pressure gradient is centered around an arbitrary value which is chosen as \SI{1}{\pascal}. Again, this is in a environment without gravity and where only the gradient matters thus an arbitrarily small value can be used. To emphasis this the pressures will be given in arbitrary units a.u.. Given the inner pressure is the parameter adjusted experimentally, the top boundary is set to a fixed value of 1 a.u. with the bottom boundary used to create the gradient. A collapsing case is where there is a positive gradient and a bursting case where there is a negative gradient. Here positive means that the pressure increases along the z-axis shown in Figure \ref{5-fig-tube-domain} whereas negative means that the pressure decreases along the z-axis. The transformation of the tube shown in Figure \ref{5-fig-tube-unravel} means that the vectors within the velocity field will not technically be accurate. This affects surface tension strongly. Given the pressure is already arbitrary, the surface tension force can thus also to be reduced to an apt level for the 1 a.u. arbitrary pressure: \SI{0.001}{\newton\per\meter}. This helps to limit the characteristic spurious currents in VoF \cite{spuriousVoF}. 

To limit the variables between runs to simply the gradient pressure and $Q_{Source}$ it is prudent to use one computational constant for all runs. The final procedure used a measured average current of \SI{6.22}{\ampere}. For the \SI{1.25}{\milli\meter} arc gap, an arc voltage of $\approx$ \SI{5}{\volt} was estimated. For convenience, the exact value is set to give $Q_{Source} = 32$. Experimentally, the weld width for the vast majority of the successful samples was \SI{1.5}{\milli\meter}; see welds 3 and 4 in Figure \ref{5-fig-JINST-pass-fail} for an example. Thus the parameter $\omega = \SI{1.5}{\milli\meter}$. Taking a reasonable depth to width ratio of $\frac{1}{3}$ gives a weld depth of $l = \SI{0.5}{\milli\meter}$. The travel velocity is 25 RPM for all welds. For a \SI{2.275}{\milli\metre} outer diameter tube, this gives a linear travel velocity of $\frac{25 \times 2.275 \times \pi}{60} = \SI{2.977968}{\milli\meter\per\second}$. This is probably too precise given the inevitable inaccuracy of physics mechanisms so the travel velocity is determined to be \SI{2.98}{\milli\meter\per\second}. This means that one revolution takes \SI{2.4}{\second}. Using the equation for the computational $C_C$ found from the welding benchmarks in Chapter 4 (reprinted as equation \ref{5-eq-CC} for convenience) $C_C$ is found to be 54.13. The remaining thermophysical properties are given in Table \ref{5-tab-tube-TPP}. 

\begin{equation}
\label{5-eq-CC}
\begin{split}
C_{C} &= 0.02425944 \cdot Q_{Source} + 16.32170162 \cdot v \nonumber\\
&- 6.69688692 \cdot l + 5.16127479 \cdot \omega \\
\end{split}
\end{equation}

\begin{center}
\begin{table}
\caption{\label{5-tab-tube-TPP}Thermophysical properties used for the tube welding case. Values taken from \cite{thermoPhysProp} unless noted else wise.}
\begin{tabular}{l l l l}
\toprule
Phase & Property & Value & Units\\
\hline
$\alpha_1$\rule{0pt}{2.6ex} & Density, $\rho_1$ & 1 & \si{\kilogram\per\metre\cubed} \\
 & Specific Heat Capacity, $c_{p,1}$ & 1000 & \si{\metre\squared\per\second\squared\per\kelvin} \\
 & Thermal Conductivity, $k_1$ & 0.02 & \si{\kilogram\metre\per\second\cubed\per\kelvin} \\
 & Kinematic Viscosity, $\nu_1$ & \num{1.48e-5} & \si{\metre\squared\per\second} \\
 & Emissivity, $\epsilon$ & 0 & None required \\
 & $h_{Convection}$ & 50 & None required \\
\hline
$\alpha_2$\rule{0pt}{2.6ex} & Density, $\rho_2$ & 4500 & \si{\kilogram\per\metre\cubed} \\
 & Volumetric Thermal Expansion Coefficient, $\beta$ & 0 & \si{\per\kelvin} \\
 & Specific Heat Capacity, $c_{p,2}$ & 964.9 & \si{\metre\squared\per\second\squared\per\kelvin} \\
 & Thermal Conductivity, $k_2$ & 31 & \si{\kilogram\metre\per\second\cubed\per\kelvin} \\
 & Melting Point, $T_{melt}$ & 1940 & \si{\kelvin} \\
 & Reference Temperature, $T_{ref}$ & 1940 & \si{\kelvin} \\
 & Latent Heat of Fusion, $L_f$ & \num{2.95e5} & \si{\metre\squared\per\second\squared} \\
 & Kinematic Viscosity, $\nu_2$ & \num{1.15e-6} & \si{\metre\squared\per\second} \\
 & Emissivity, $\epsilon$ & 0.525 & None required \\
 & $h_{Convection}$ & 10 & None required \\
\hline
$\alpha_3$\rule{0pt}{2.6ex} & Density, $\rho_3$ & 4500 & \si{\kilogram\per\metre\cubed} \\
 & Specific Heat Capacity, $c_{p,3}$ & 600.9 & \si{\metre\squared\per\second\squared\per\kelvin} \\

 & Thermal Conductivity, $k_3$ & 31 & \si{\kilogram\metre\per\second\cubed\per\kelvin} \\
 & Kinematic Viscosity, $\nu_3$ & \num{1.15e-6} & \si{\metre\squared\per\second} \\
 & Emissivity, $\epsilon$ & 0.5 & None required \\
 & $h_{Convection}$ & 10 & None required \\
%\hline
%All & Specific Heat Capacity & 381.5 & \si{\kilogram\per\metre\cubed} \\
\bottomrule
\end{tabular}    
\end{table}
\end{center}

\begin{figure}[!htb]
\centering
\includegraphics[width=15cm]{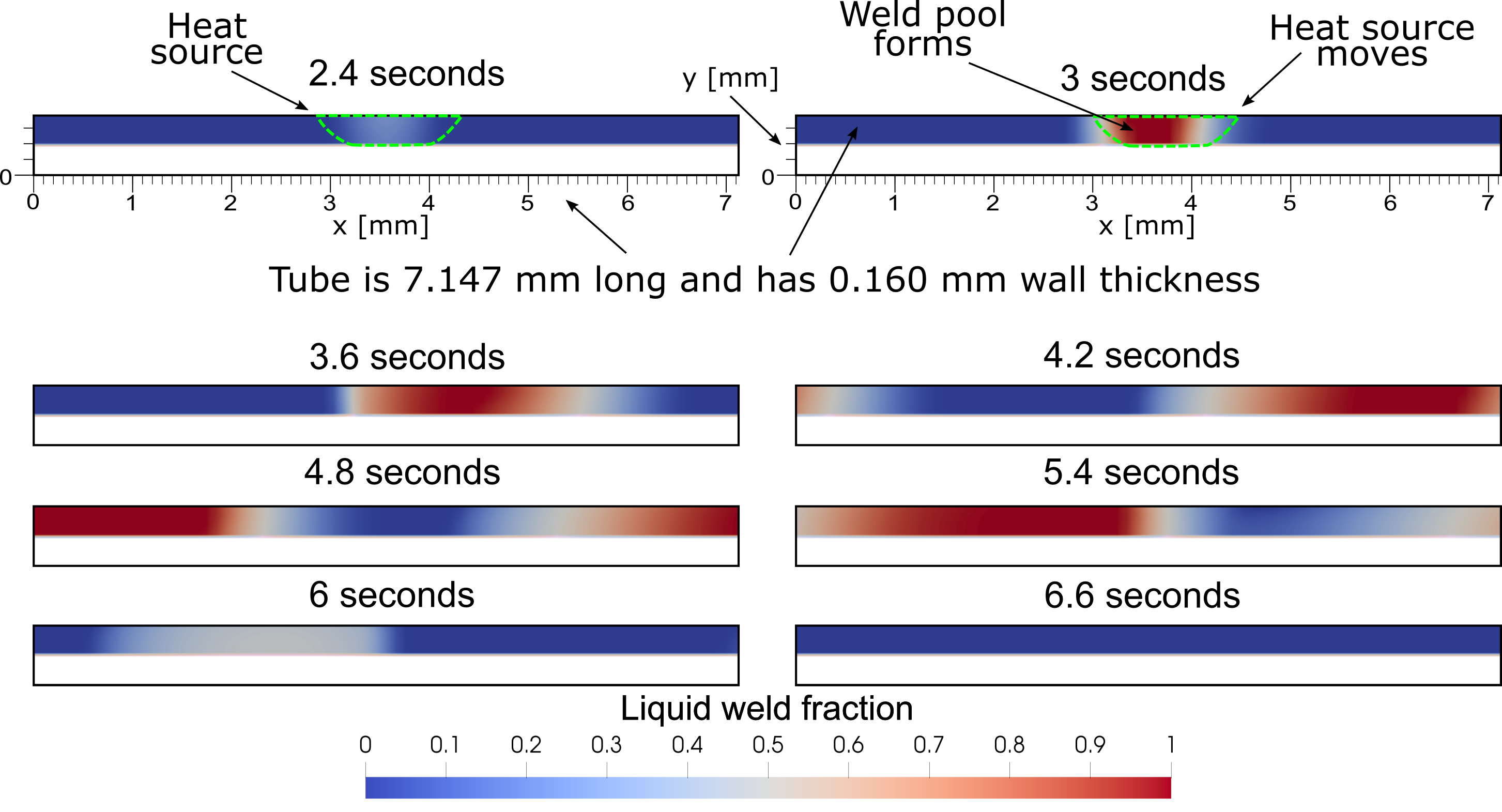}
\caption{Simulation output from \gIF of a successful weld in the core region. The top two images show the location of the heat source as the weld pool forms along with a set of axis showing the \SI{7.147}{\milli\meter} long core region with \SI{160}{\micro\meter} thick tube walls.} 
\label{5-fig-sim-trial-base}
\end{figure}

With the case outlined, the aforementioned passing, bursting and collapsing parameters are tested. As the top pressure is 1 a.u., setting the bottom pressure to 1 a.u. gives the zero pressure gradient base case shown in Figure \ref{5-fig-sim-trial-base}. Here, the evolution of the liquid fraction is shown as the heat source starts in the centre of the domain and remains there for \SI{2.8}{\second} before travelling for \SI{2.4}{\second} along the positive x direction. The trailing heat source shown in Figure \ref{5-fig-tube-domain} enters the domain as the main heat sources leaves whilst the leading heat source warms the metal the main heat source enters into. Note the liquid weld fraction fully penetrates the metal ensuring a successful weld.  

The next step is to introduce pressure gradients. It was found that as the liquid weld pool is forming and only a small amount of cells are exposed to the gas phase the weld can easily collapse or burst before the surface tension has a chance to fully affect the momentum equations. Therefore, the pressure gradients are only activated at \SI{3}{\second} into the simulation. For the bursting case the gradient is created through setting the bottom boundary to 2 a.u. whilst in the collapsing case it is set to 0 a.u. 

\begin{figure}[!htb]
\centering
\includegraphics[width=15cm]{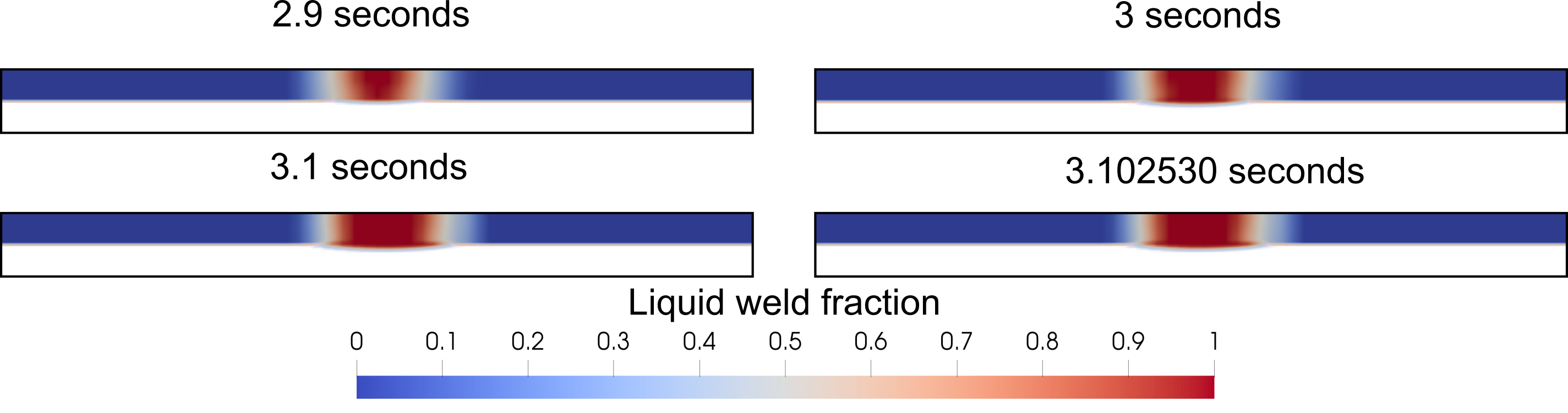}
\caption{Results from the simulation of a failed weld due to collapse in the core region} 
\label{5-fig-sim-trial-collapse}
\end{figure}

\begin{figure}[!htb]
\centering
\includegraphics[width=15cm]{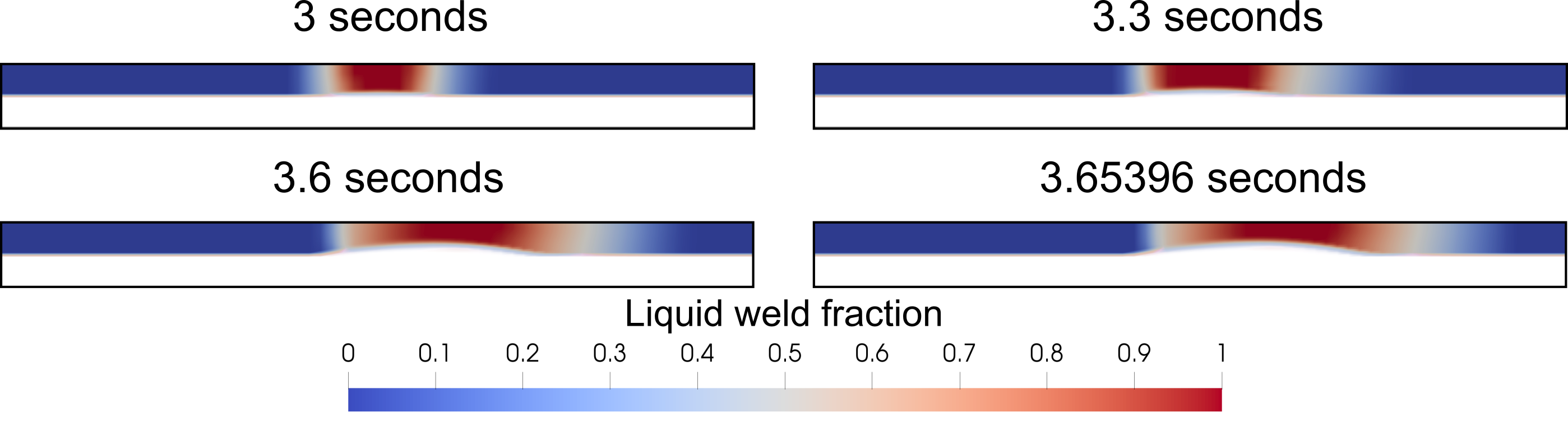}
\caption{Results from the simulation of a failed weld due to bursting in the core region.} 
\label{5-fig-sim-trial-burst}
\end{figure}

\FloatBarrier
\section{General case}
\subsection{Batch simulation results}
\label{5-sec-batch-sim}
\begin{figure}[!htb]
\centering
\includegraphics[width=14cm]{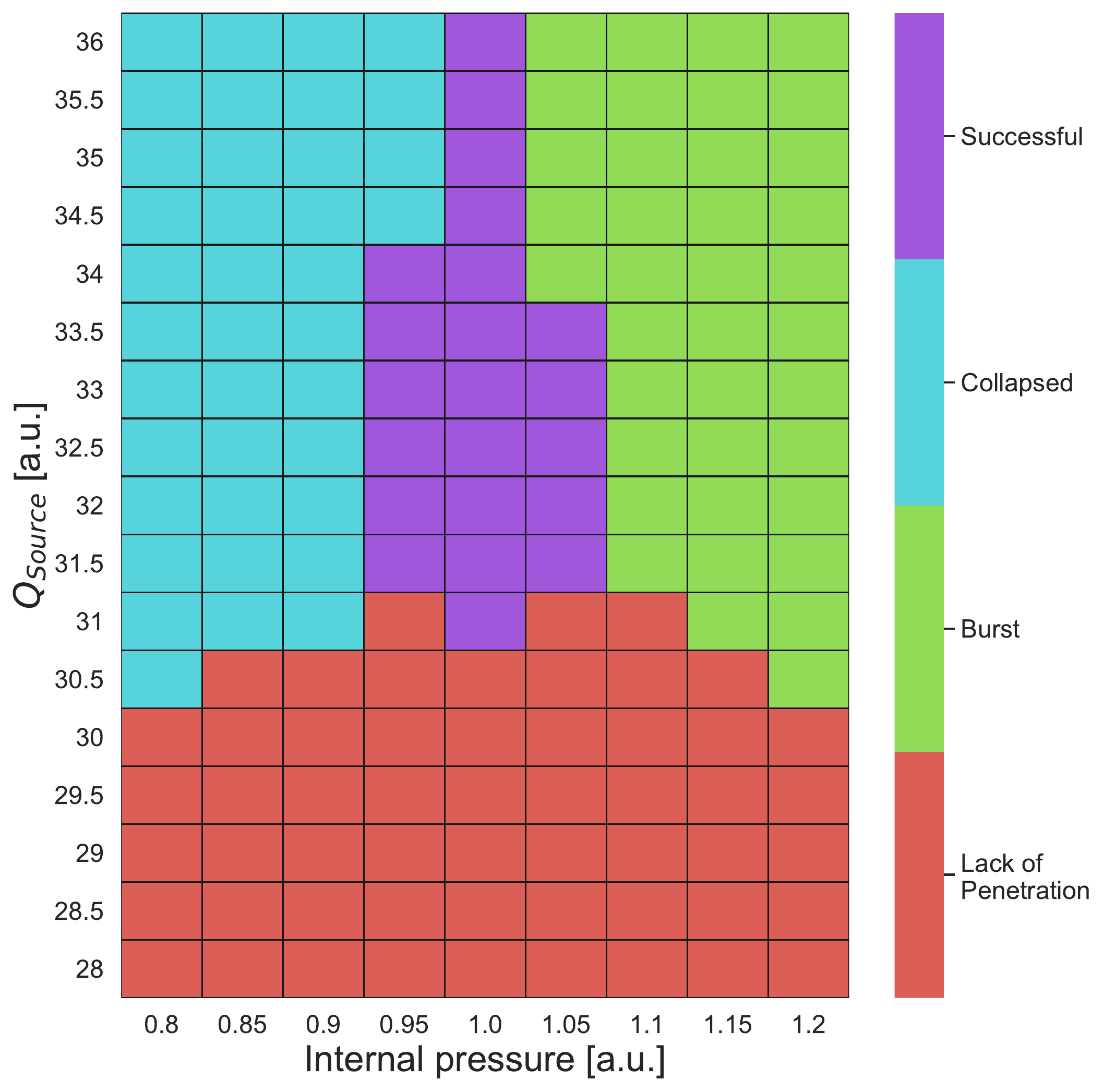}
\caption{Scores for a variety of mapped parameter pairings for the tube welding case.} 
\label{5-fig-sim-300}
\end{figure}

With the trial welds established, \gIF can run with a broad range of pairings of the two mapped parameters. Starting from the base case trial results established in Section \ref{5-sec-trial-results}, the internal pressure for the bottom boundary is altered along with the value for $Q_{Source}$. The scored results from a batch of such simulations is shown in Figure \ref{5-fig-sim-300}. Here, there is a clear `goldilocks zone' where the internal pressure is not too low as to collapse the weld but not too high as to cause it to burst. This pressure balance is somewhat expected given the trial results. However, critically there is also apparent too high and too low values for $Q_{Source}$. When $Q_{Source}$ is insufficiently high the weld does not fully penetrate whilst when it is excessively high the liquid weld pool becomes too temperamental and easily bursts or collapses. Also, note that when the bottom boundary is set to the same as the top (at 1 [a.u.]) if $Q_{Source}$ is large enough to penetrate fully then the simulation always passes. This is as there is no pressure gradient and thus no force to perturb the liquid weld pool ensuring stability. This is because gravity is turned off and the surface tension cancels itself out with the lack of an anisotropic force. Given this, the top of the acceptable $Q_{Source}$ range is taken from adjacent cells.  

\subsection{Experimental parameter space}
\label{5-sec-exp-param-space}
\begin{figure}[!htb]
\centering
\includegraphics[width=14cm]{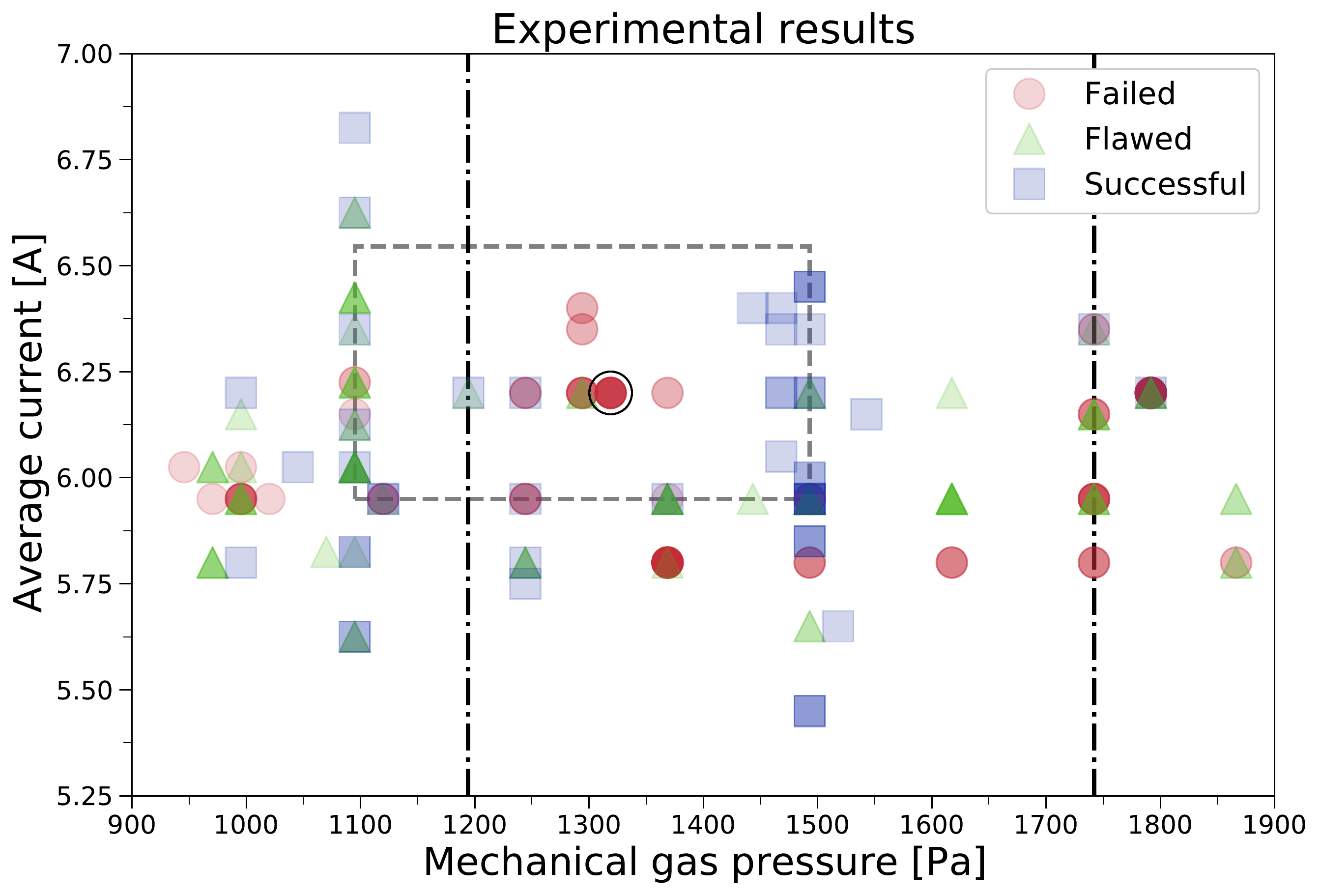}
\caption{Experimental results expressed in terms of simply mechanical gas pressure and current. Each marker for failed, flawed and successful has a 20\% opacity to show repeated parameters. The black circle indicates the final welding procedure used for production. The dash-dotted black lines shown the calculated limit low and limit high detailed in Section \ref{5-sec-parameter-space-exam}. The dashed grey box is the focus of the simulation work.} 
\label{5-fig-exp-mapping}
\end{figure}

In order to establish a general case for ultra-thin-walled tubing, the \gIF results needs to match the experimental results detailed in Section \ref{5-sec-exp-res}. Then, the mapped parameters can be used to predict the general case. The results presented in Section \ref{5-sec-batch-sim} show an upper and lower limit for the pressure gradient for tube welding in a similar manner to the experimental results for pressure values in Section \ref{5-sec-parameter-space-exam}. The corollary of this is that the pressure gradient should be directly related to these values. However, Figure \ref{5-fig-exp-mapping} shows all of the experimental data in terms of average current and the mechanical gas pressure. Here, the values for the upper and low pressure limits do not seem to separate the failed, flawed and successful welds accurately. Therefore, deriving a pressure gradient from upper and lower limits presented in Section \ref{5-sec-parameter-space-exam} would result in the same problem. Along with this, the following points should be considered:      

\begin{itemize}
    \item The assessments in Figure \ref{5-fig-exp-mapping} are exterior assessments. Some welds that were initially judged successful from their exterior were actually flawed due to incomplete internal fusion. An image of a weld cross section with incomplete internal fusion is shown in Appendix D. This may be why some successful markers in Figure \ref{5-fig-exp-mapping} are far from the final welding procedure. 
    \item Failing and flawed welds can be due to operator error rather than the procedure. This is why some failed and flawed markers overlap successful ones.
    \item There is unavoidable experimental variation. Even if the same procedures are followed occasionally there will be some variation. 
    \item The aim is for \emph{a} general case and not necessarily \emph{the} general case. There may be stable `islands' within the mapped parameter space but ultimately a successful weld is one that meets the required standards rather than one that came from a particular procedure. 
\end{itemize}

With these points in mind, an alternative method for the pressure gradient is to create the limits around the most common failure pressures relative to the final optimized procedure. Shown by the vertical edges of the grey box in Figure \ref{5-fig-exp-mapping}, there are strong vertical lines all of majority flawed and failed welds at \SI{1095}{\pascal} and \SI{1493}{\pascal}. In addition to these pressure edges, from knowledge of GTAW there must be a lower limit for $Q_{Source}$ that results in incomplete fusion and a upper limit that results in excessive melting that compromises the tube's shape. Therefore, upper and lower horizontal lines can be added to the vertical lines to form an experimental `goldilocks zone' of weld success. From the available experimental data, this alternative method is preferred.

\subsection{Combined model}
\label{5-sec-combined-model}
Both the simulation and experimental results given in sections \ref{5-sec-batch-sim} and \ref{5-sec-exp-param-space} respectively indicate a `goldilocks zone'. Therefore, the mapping between experiment and simulation can be found through overlaying them. The pressure limits with the alternative method outlined in \ref{5-sec-exp-param-space} are \SI{1095}{\pascal} and \SI{1493}{\pascal} yet to avoid the perception of precision these two values are reevaluated as \SI{1100}{\pascal} and \SI{1500}{\pascal}. The corresponding limits suggested by Figure \ref{5-fig-sim-300} are 0.95 a.u. and 1.05 a.u. Linearly interpolating between these limits gives a relationship of $\text{Experiment} = 4000 \times \text{Simulation} - 2700$ for the pressure.

Whilst the upper and lower pressure limits has strong evidence in sections \ref{5-sec-parameter-space-exam} and \ref{5-sec-exp-param-space} this is not true for the current. It is reasonable to assume that the current needs to be sufficient to penetrate through the tube wall but not too high as to destroy the geometrical integrity of the tubing but there is not the clear experimental evidence as with pressure. Looking closely at Figure \ref{5-fig-exp-mapping} suggests that at $\approx \SI{5.95}{\ampere}$ the welds begin to fail however there is no obvious corresponding top edge. The simulation results suggest that welds begin to fail at $\approx 1.1 \times$ the lower limit of $Q_{Source}$. Applying this to the experimental results suggests an upper limits of $\approx \SI{6.545}{\ampere}$; this value is used as the top edge of the dashed grey box in Figure \ref{5-fig-exp-mapping}. Using the same linear interpolation method as used for pressure with upper and lower limits of 31 a.u. and 34 a.u. gives a relationship of $\text{Experiment} = 0.1983 \times \text{Simulation} - 0.1983$. 

\subsection{Predictions}
The relationships between experiment and simulation established in Section \ref{5-sec-combined-model} can be used to make predictions for welding on other ultra-thin-walled tubing. Given the experimental evidence available is exclusively for \SI{2.275}{\milli\meter} outer diameter tubing predictions using this evidence should be more accurate with similar tubing. Therefore, hypothetical cases are considered where the wall thickness is increased or decreased by up to \SI{50}{\micro\meter}. With these specifications, the experimental parameters eliminated from the raw data in Section \ref{5-sec-raw-data} should still work leaving the current and pressure levels to be tuned. 

\begin{figure}[!htbp]
\centering
\includegraphics[width=15cm]{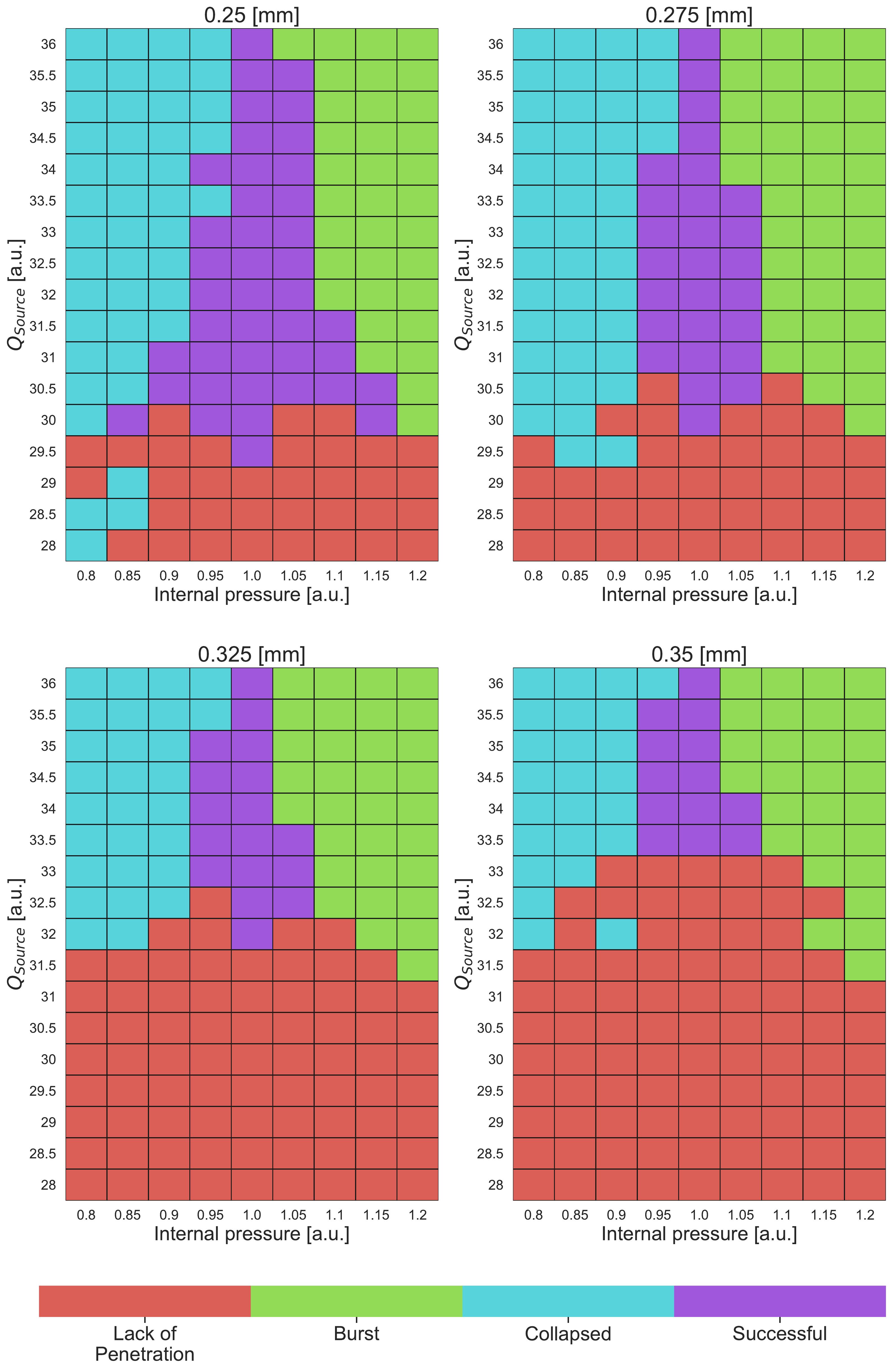}
\caption{Simulation scores for tube weld simulation using a hypothetical thicker and thinner walled tubing.}
\label{5-fig-sim-predict}
\end{figure}

In order to identify the required pressure, the same batch procedure used to produce Figure \ref{5-fig-sim-300} is run for the hypothetical thicker or thinner tubing. The same domain and parameters are used as described in Section \ref{5-sec-case-outline} except the wall thickness of the tube is altered to one of \SI{250}{\micro\meter}, \SI{275}{\micro\meter}, \SI{325}{\micro\meter}, or \SI{350}{\micro\meter}. The results for these procedures are all shown in Figure \ref{5-fig-sim-predict}.

Here, the `goldilocks zone' is shifted upwards are downwards depending on whether the tube wall is thickened or thinned respectively. For the y-axis, this is readily interpreted as a larger current being required to fully penetrate the thicker walls and less current required to penetrate thinner walls. Another way to confirm this is to look how the percentage of procedures that fail due to lack of penetration (in red) increases with the wall thickness. On the $x$-axis the maximum width of the purple `goldilocks zone' is similar for the majority of wall thicknesses. The exception here is for \SI{250}{\micro\meter} at low current values. Here the width of the purple `goldilocks zone' noticeably widens at lower $Q_{Source}$ values. Rather than provide a physical interpretation for this it is better to consider that these values are at the extremes of predictions with no experimental tests to verify them. Thus this feature may well be an artefact of the assessment mechanism. Given this, the $x$-axis interpretation is that the width of the `goldilocks zone' for possible pressures roughly remains constant irrespective of tube wall thickness.  

With these new zones established, the final step is to convert them into an experimental welding procedure. Intuitively, the middle of the `goldilocks zone' should provide the best weld however the final welding procedure used for production (shown inside the black circle in Figure \ref{5-fig-exp-mapping}) is not in the middle of the identified `goldilocks zone' but slightly to the right. Additionally, due to the lack of pressure gradient the simulation will always favour an internal pressure of 1 a.u. Given this information, it is prudent to suggest a slightly larger value of internal pressure for the thicker wall and a slightly smaller one for the thinner wall. However this does require assuming the pressures from the different $Q_{Source}$ values are roughly identical. As mentioned previously, there was an initial plan to measure these pressure change with increasing current for the welding machine used in the experimental work but this was impacted by a lack of access to the required lab space. It is worth noting both that the formula for arc pressure and the experiment in Section \ref{5-sec-parameter-space-exam} both suggest that higher currents mean higher internal pressures are required. 

The value for current in Figure \ref{5-fig-exp-mapping} is at the lower end of the `goldilocks zone'. As this is the only experimental evidence the current choice should thus also be from the lower end of the simulation `goldilocks zone'. This is arbitrarily defined by the author as 0.5 a.u. higher from where failure due to lack of penetration ends. These values will thus be 30.5 a.u. for \SI{250}{\micro\metre}, 31 a.u. for \SI{275}{\micro\metre}, 33 a.u. for \SI{325}{\micro\metre} and 33.5 a.u. for \SI{350}{\micro\metre}. Combining these predictions with the experimental procedure for the \SI{300}{\micro\meter} wall thickness tubing, a suggested experimental parameters for the thicker and thinner tube is shown in Table \ref{5-tab-exp-predictions}.

\begin{table}[!htb]
\setlength{\tabcolsep}{10pt}
    \centering
    \caption{\label{5-tab-exp-predictions} Experimental parameters to use for welding \SI{2.275}{\milli\meter} titanium tubing with wall thicknesses between \SI{250}{\micro\metre} and \SI{350}{\micro\metre}.}
    \begin{tabular}{c c}
    \toprule
    Parameter & Value\\
    \hline
    \begin{tabular}{@{}c@{}}Electrode tool\\setting\end{tabular}\rule{0pt}{2.6ex} & $\SI{5.65}{\milli\meter}$ \\
    Machine gas & $\SI{2.5}{\litre\per\minute}$ \\%[0.45mm]
    Gas flow & \begin{tabular}{@{}c@{}}Adjust during experiment to achieve mechanical\\ gas pressure\end{tabular}\\%[0.45cm]
    \begin{tabular}{@{}c@{}}Mechanical gas\\pressure\end{tabular}\rule{0pt}{2.6ex} & Parameter Y from Table \ref{5-tab-exp-predictions-sim}\\%[0.45cm]
    Weld gas pressure & \begin{tabular}{@{}c@{}}Check same as mechanical \\ pressure post weld \\\end{tabular}\\[0.45cm]
    Main current & \begin{tabular}{@{}c@{}}Main and background\\average to value X from Table \ref{5-tab-exp-predictions-sim}\end{tabular}\\[0.45cm]
    Background current & \begin{tabular}{@{}l@{}}Main and background\\average to X from Table \ref{5-tab-exp-predictions-sim}\end{tabular}\\%[0.75cm]
    Rotational speed & 25 RPM\\%[0.25mm]
    Initial current & $\SI{2.4}{\ampere}$\\%[0.25mm]
    Electrode position& $\SI{0.75}{\milli\meter}$\\%[0.25mm]
    Warm up welds  & \begin{tabular}{@{}l@{}}Two for a new electrode \\ zero otherwise\end{tabular}\\[0.25mm]
    \begin{tabular}{@{}l@{}}Total welds\\with electrode\end{tabular}\rule{0pt}{2.6ex} & Maximum 15 uses each \\
    \bottomrule
    \end{tabular}
\end{table}

\begin{table}[!htb]
\setlength{\tabcolsep}{10pt}
    \centering
    \caption{\label{5-tab-exp-predictions-sim} Predicted parameters to use for welding \SI{2.275}{\milli\meter} titanium tubing with wall thicknesses between \SI{250}{\micro\metre} and \SI{350}{\micro\metre}. To avoid the perception of precision, the predicted current values are given to one decimal place and the pressures rounded to the nearest 50. The final experimental procedure is given for \SI{300}{\micro\metre}.}
    \begin{tabular}{l l l l l l}
    \toprule
    Parameter & \SI{250}{\micro\meter} & \SI{275}{\micro\meter} & \SI{300}{\micro\meter} & \SI{325}{\micro\meter} & \SI{350}{\micro\meter}\\
    \hline
    Current Average (X) & \SI{5.8}{\ampere} & \SI{5.9}{\ampere} & \SI{6.22}{\ampere} & \SI{6.4}{\ampere} & \SI{6.5}{\ampere} \\
    Pressure Value (Y) & \SI{1200}{\pascal} & \SI{1250}{\pascal} & \SI{1318.85}{\pascal} & \SI{1350}{\pascal} & \SI{1400}{\pascal} \\
    \bottomrule
    \end{tabular}
\end{table}

\section{Chapter Summary}
This chapter presented a combined experimental and numerical investigation into ultra-thin-walled tubed welding. An experimental procedure for ultra-thin-walled tube welding was outlined and validated through a physical examination on specific welds. Then, the experimental results were analysed in their entirety revealing the importance and orthogonality of the welding current and inner pressure. The solver outlined in Chapter 3 and benchmarked in Chapter 4 was then applied to ultra-thin-walled tube welding and matched against experiment. Both the experiment and simulation revealed a `golidlocks zone' of successful welding. This was identified experimentally for the \SI{2.275}{\milli\meter} titanium tubing to be where the inner pressure is between 1100 and \SI{1500}{\pascal} and the current is between 5.95 and \SI{6.545}{\ampere}. These values were then used to create a map between the experimental and simulation results. The simulation was then extended to predict the behaviour of ultra-thin-walled tube welding with thicker and thinner tube walls and mapped back to create a proposed experimental procedure. These predicted results both featured an apparent `golidlocks zone' thus identifying this `golidlocks zone' is a generalization of ultra-thin-walled tube welding.

%% file: chapters/thesis-chapter-7.tex
\lhead{Chapter 7: Thin structure welding}
\section{Introduction}
This chapter details another application of \emph{gtawFoam}. \gIF is specifically designed to solve the ultra-thin-walled tube welding case outlined in Chapter 5. However, there are adjacent problems that can also utilize the inbuilt strengths of \emph{gtawFoam}. A good candidate adjacent problem is turbine blade repair. This repair process involves multiple passes of homogeneous welding on damaged turbine blades to weld on layers of filler metal and rebuild them. This is a form of direct energy deposition additive manufacturing. However, this term only applies when there is a second pass of homogeneous welding (hence\emph{additive}) so for clarity the whole process is referred to as homogeneous welding. Through applying \gIF to turbine blade repair, the wider applicability of \gIF can be assessed. 

\section{Case outline}
\label{6-sec-outline}
\subsection{Background}
The geometrical integrity of aeroengine turbine blades degrades during their service life due to wear and foreign object damage. Given the manufacturing process of turbine blades is expensive ($\approx$ £5000 per blade) it is economical to repair these blades. The welding of multiple passes of homogeneous welding to repair these blades is typically done manually achieving only a 60\% successful repair rate. Note, this figure comes from in-person conversations with welding engineers specializing in turbine blade repair. Some blades are damaged beyond repair but around 90\% are potentially repairable. It is $\approx$ 60\% of this $\approx$ 90\% that are typically repaired for an average batch of damaged blades. This is in part due to the curving aerofoil shape of the blades with thinner and thicker sections requiring different heat input over the meandering length of the blade. The low rate of successful repair of these high-value blades coupled with the shortage of skilled welding engineers \cite{welder-shortage} makes turbine blade repair a good candidate for automation.  

The creation of an automated repair process for turbine blades requires knowledge of an apt welding procedure. This is challenging as the blades could be damaged in a variety of ways and within an engine such as the Rolls Royce Trent 1000 each blade will have a unique aerofoil shape. Therefore, automation will require not only computer vision and robotic manipulation of the blades but a potentially different welding procedure for each blade. Neglecting the computer vision and robotic manipulation challenges, to address the blade welding problem it is prudent to first address a simpler problem such as welding repair of a simple flat plate on its side. A flat plate on its side can be used as a blade proxy enabling the multiple layers of homogeneous welding challenge to be addressed first. This was the approach took during the development of an autonomous robotic turbine blade repair system constructed at The University of Sheffield. 

\begin{figure}[!htb]
\centering
\includegraphics[width=14cm]{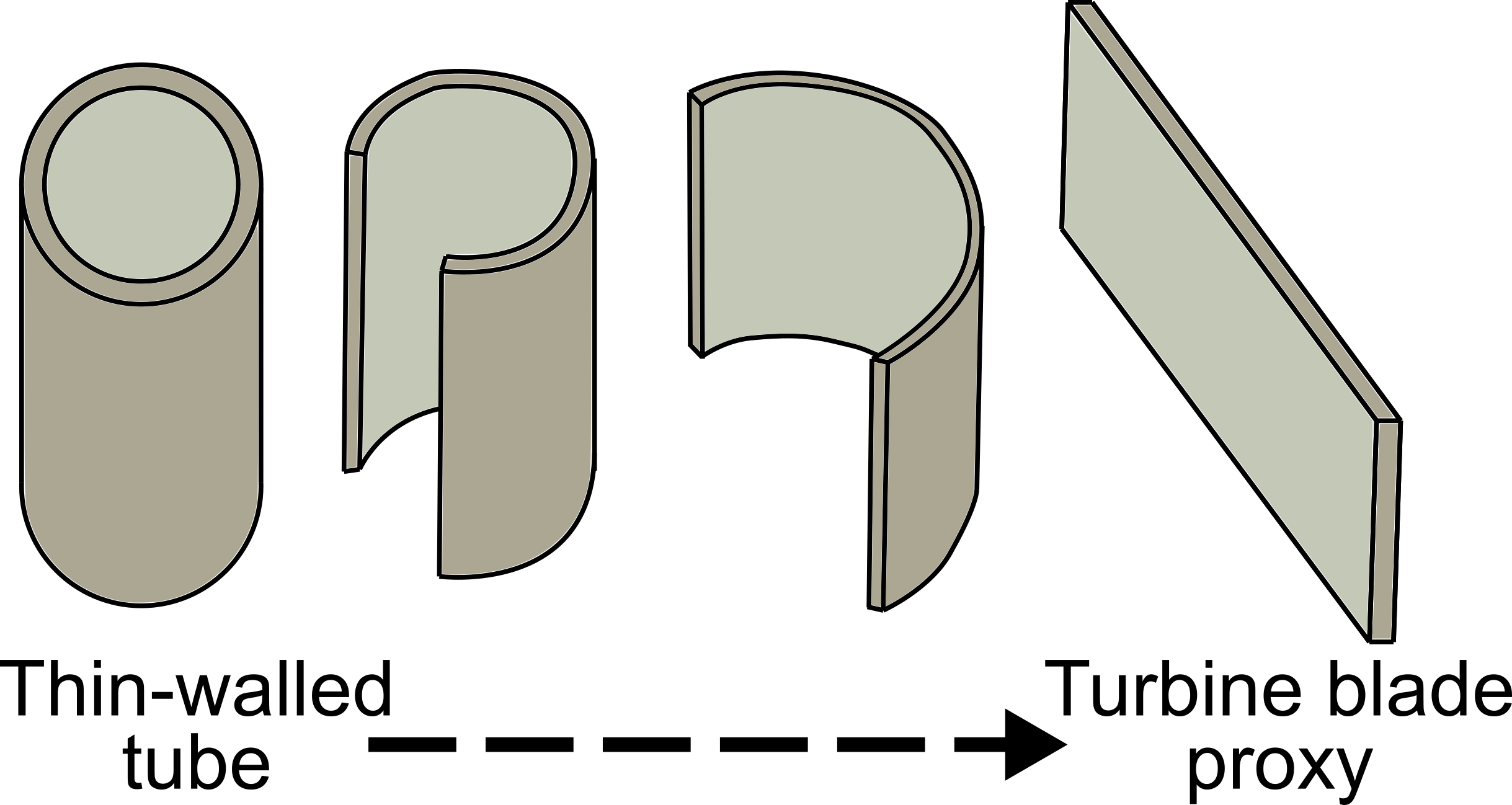}
\caption{Illustration of the similarity of thin-walled tubing to a turbine blade proxy. Image shows the geometrical transformation of a ultra-thin-walled tube into a turbine blade proxy through `\emph{slicing}' through the tube wall and unravelling the tube. Image is purely for illustration purposes to show the similarity between the two shapes.}
\label{6-fig-morph}
\end{figure}

The development of the robotic welding repair system was the product of a collaboration with the welding machine manufacturer that designed the IP50 featured in Chapter 5 to see if welding insights from ultra-thin-walled tube welding could be transferred to turbine blade repair. Due to unforeseen challenges, there is limited experimental data available from this collaboration however, thermal camera footage of trial welding runs is available from a prototype of this system\footnote{Courtesy of Jon Willmott.}. Here, \SI{60}{\milli\meter} by \SI{40}{\milli\meter} by \SI{1.5}{\milli\meter} 316L stainless steel trial blades are ``repaired''. This thickness is beyond the \SI{500}{\micro\meter} boundary for ultra-thin wall but it is what is available. Similarly, 316L is very different from the typical nickel-based superalloys used for aero-engine turbine blades but it was the only material available for these tests. This said, it is considerably cheaper to use 316L than nickel-based superalloys so the use of 316L did have a cost advantage. As shown in Figure \ref{6-fig-morph}, the ultra-thin-walled tubing case can be easily transformed into a flat plate case. This shows the problems are not too dissimilar and thus \gIF can be applied to turbine blade repair - the efficacy of which can be assessed with the aforementioned thermal camera videos.

\subsection{Specification}
To apply \gIF to turbine blade repair an appropriate domain must be created. Creating a large rectangular domain full of a gas phase and then defining a solid phase in the shape of a blade inside runs into a few issues. Firstly, due to how the VoF method resolves interfaces, the cells at the corners of the blade that have $< 50\%$ solid fraction will gradually `smooth' out losing phase fraction and creating rounded corners. This `smoothing' is due to numerical diffusion. This compromises the geometric integrity of the blade. Secondly, large domains are computationally intensive. Finally, as discussed in Chapter 5, radiation is not implemented for \gIF so heat transfer in the gas phase is hamstrung. This final issue is exacerbated by how \gIF struggles to accurately model convection currents. This is due to the spurious currents \cite{spuriousVoF} present in the VoF implementation used.

\begin{figure}[!htb]
\centering
\includegraphics[width=9cm]{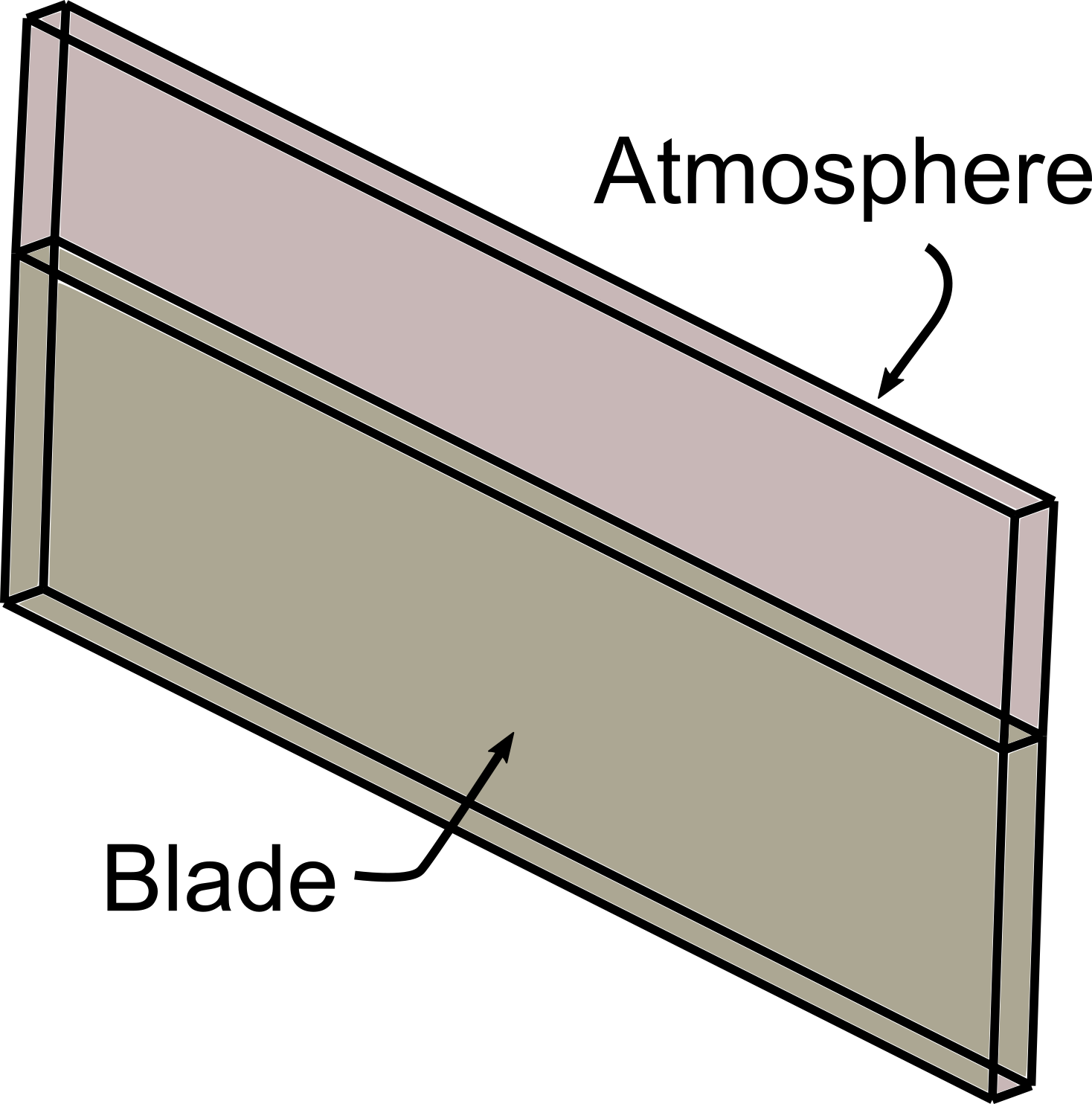}
\caption{Illustration of the domain used in the turbine blade repair case. Note the separate top and bottom regions for the blade and atmosphere. The domain is meshed with 120 by 5 by 90 cells with a gradient of 4 and $\frac{1}{4}$ for the top and bottom regions respectively. The gradient weights the mesh cells towards the interface between the atmosphere and blade.}
\label{6-fig-blade-dom}
\end{figure}

The solution to this, is to limit the size of the domain, keep the corners of the solid fraction touching a boundary and leverage the convective / radiative boundary condition presented in Chapter 3. The result of this combination is shown in Figure \ref{6-fig-blade-dom}. Here, the domain is split into two blocks with the lower block modelling the blade with radiative boundary conditions and the upper block modelling the atmosphere. A heat source term can then be applied onto to the metal (blade) portion of the domain enabling simulation of the root pass during blade repair. For convenience, the convective / radiative boundary condition is reprinted in equation \ref{6-boundary-eq-robin}. Here, different values are used for $h_{Rad}$ (depending on the emissivity) in the atmosphere and blade portion of Figure \ref{6-fig-blade-dom}. 

\begin{equation}
\begin{gathered}
\label{6-boundary-eq-robin}
T_{Boundary} = f \cdot T_{\infty} + (1 - f) \cdot T_{Centre}, \;\;\;\; f  = \frac{1}{\left( 1 + \frac{k}{\delta (h_{Conv} + h_{Rad})} \right)} \\
\text{where} \ h_{Rad} = \epsilon \sigma (T_{\infty}^2 + T_{Boundary}^2)(T_{\infty} + T_{Boundary}). 
\end{gathered}
\end{equation}

To account for the lack of radiation and limited convection within the gas phase, an arbitrary $10 \times$ multiple is applied to the conductivity of the gas phase. To compensate for this, the formula used in the previous chapters to calculate $C_C$ is dropped and it is instead fixed to 200. This value was found through a process of trial and error where the value that best matched the experimental work was chosen. Further, surface tension and gravity are turned off to improve performance. This change likely degrades the physical accuracy of the simulation as surface tension and gravity are key factors in material deposition. However this is an example case to show how \gIF can be applied to other situations so this change is judged apt. 

Given the features are available, there is the opportunity to use temperature dependent thermophysical properties for thermal conductivity and specific heat capacity. The ASM formulation \cite{thermoPhysProp} is used for both and, for convenience, the equations for which are reprinted in equations \ref{6-eq-t-dep-k} and \ref{6-eq-t-dep-cp}. The properties used for the different phases in the simulation are shown in Table \ref{6-tab-blade-TPP} and the boundary conditions used are shown in Table \ref{6-tab-blade-bc}.

\begin{table}[!htb]
\caption{\label{6-tab-blade-bc}Boundary conditions for blade repair case. All alpha phases have the same \emph{zeroGradient} boundary condition.} 
\centering
\begin{tabular}{l l l l l}
\toprule
Boundary & Velocity & Pressure & Temperature & $\sum\limits_{i=1} \alpha_i$ \\
\hline
Top & \emph{PIOV} & \emph{TP} & \SI{300}{\kelvin} & $\partial_n = 0$ \\
Upper Left & \emph{NS} & \emph{FFP} & \emph{Radiative Air} & $\partial_n = 0$ \\
Upper Right & \emph{NS} & \emph{FFP} & \emph{Radiative Air} & $\partial_n = 0$ \\
Upper Front & \emph{NS} & \emph{FFP} & \emph{Radiative Air} & $\partial_n = 0$ \\
Upper Back & \emph{NS} & \emph{FFP} & \emph{Radiative Air} & $\partial_n = 0$ \\
Lower Left & \emph{NS} & \emph{FFP} & \emph{Radiative Metal} & $\partial_n = 0$ \\
Lower Right & \emph{NS} & \emph{FFP} & \emph{Radiative Metal} & $\partial_n = 0$ \\
Lower Front & \emph{NS} & \emph{FFP} & \emph{Radiative Metal} & $\partial_n = 0$ \\
Lower Back & \emph{NS} & \emph{FFP} & \emph{Radiative Metal} & $\partial_n = 0$ \\
Btm & \emph{NS} & \emph{FFP} & \emph{Radiative Metal} & $\partial_n = 0$ \\
\bottomrule
\end{tabular}
\end{table}

\begin{equation}
\label{6-eq-t-dep-k}
    \kappa(T) = a + b\left(\frac{T}{273.15}\right) + c\left(\frac{T^2}{273.15}\right)
\end{equation}

\begin{equation}
\label{6-eq-t-dep-cp}
    c_p(T) = a + bT + cT^{-2}
\end{equation}

\begin{center}
\begin{table}
\caption{\label{6-tab-blade-TPP}Thermophysical properties used for the blade repair case. Values taken from \cite{thermoPhysProp} unless noted else wise. Note the `Base' thermal conductivity is the value used for the radiative boundary conditions.}
\begin{tabular}{l l l l}
\toprule
Phase & Property & Value & Units\\
\hline
$\alpha_1$\rule{0pt}{2.6ex} & Density, $\rho_1$ & 1 & \si{\kilogram\per\metre\cubed} \\
 & Specific Heat Capacity, $c_{p,1}$ & 1000 & \si{\metre\squared\per\second\squared\per\kelvin} \\
 & Thermal Conductivity, $k_1$ & 0.02 & \si{\kilogram\metre\per\second\cubed\per\kelvin} \\
 & Kinematic Viscosity, $\nu_1$ & \num{1.48e-5} & \si{\metre\squared\per\second} \\
 & Emissivity, $\epsilon$ & 1.0 & None required \\
 & $h_{Convection}$ & 25 & None required \\
\hline
$\alpha_2$\rule{0pt}{2.6ex} & Density, $\rho_2$ & 6881 & \si{\kilogram\per\metre\cubed} \\
 & Volumetric Thermal Expansion Coefficient, $\beta$ & 0.77 & \si{\per\kelvin} \\
 & Specific Heat Capacity \cite{Mills-316L}, $c_{p,2}$ & 830 & \si{\metre\squared\per\second\squared\per\kelvin} \\
 & Thermal Conductivity Base, $k_2$ & 24 & \si{\kilogram\metre\per\second\cubed\per\kelvin} \\
 & Thermal Conductivity Coefficient $a$, $k_{2, A}$ & 6.6 & \si{\kilogram\metre\per\second\cubed\per\kelvin} \\ 
 & Thermal Conductivity Coefficient $b$, $k_{2, B}$ & \num{12.14e-3} & \si{\kilogram\metre\per\second\cubed\per\kelvin\squared} \\ 
 & Melting Point, $T_{melt}$ & 1690 & \si{\kelvin} \\
 & Reference Temperature, $T_{ref}$ & 1690 & \si{\kelvin} \\
 & Latent Heat of Fusion, $L_f$ & \num{2.6e5} & \si{\metre\squared\per\second\squared} \\
 & Kinematic Viscosity, $\nu_2$ & \num{2.97e-7} & \si{\metre\squared\per\second} \\
 & Emissivity \cite{316ssEmiss}, $\epsilon$ & 0.4 & None required \\
 & $h_{Convection}$ & 25 & None required \\
\hline
$\alpha_3$\rule{0pt}{2.6ex} & Density, $\rho_3$ & 7950 & \si{\kilogram\per\metre\cubed} \\
 & Specific Heat Capacity $a$, $c_{p,3a}$ & 412 & \si{\metre\squared\per\second\squared\per\kelvin} \\
 & Specific Heat Capacity $b$, $c_{p,3b}$ & 0.2 & \si{\metre\squared\per\second\squared\per\kelvin\squared} \\
 & Specific Heat Capacity $c$, $c_{p,3c}$ & \num{-2e-5} & \si{\metre\squared\per\second\squared\per\kelvin\cubed} \\
 & Thermal Conductivity Base, $k_3$ & 40 & \si{\kilogram\metre\per\second\cubed\per\kelvin} \\
 & Thermal Conductivity Coefficient $a$, $k_{3a}$ & 6.31 & \si{\kilogram\metre\per\second\cubed\per\kelvin} \\ 
 & Thermal Conductivity Coefficient $b$, $k_{3b}$ & \num{27.2e-3} & \si{\kilogram\metre\per\second\cubed\per\kelvin\squared} \\
 & Thermal Conductivity Coefficient $c$, $k_{3c}$ & \num{-7e-6} & \si{\kilogram\metre\per\second\cubed\per\kelvin\cubed} \\
 & Kinematic Viscosity, $\nu_3$ & \num{3.06e-7} & \si{\metre\squared\per\second} \\
 & Emissivity \cite{316ssEmiss}, $\epsilon$ & 0.4 & None required \\
 & $h_{Convection}$ & 25 & None required \\
%\hline
%All & Specific Heat Capacity & 381.5 & \si{\kilogram\per\metre\cubed} \\
\bottomrule
\end{tabular}    
\end{table}
\end{center}

\subsection{Filler metal term}
In order to simulate the repair process fully \gIF needs to model the addition of filler metal. To achieve this, additional source terms need to be added to the phase fraction equation in \emph{gtawFoam}. Given the heat source term  $Q_{Source}$ acts as a Goldak-like \cite{goldak1984} volumetric heat source within the base metal to implement a filler metal term as a solid wire to be melted would be quite involved. Instead the filler metal should be added as liquid metal in a manner reminiscent of MIG/GMAW. As $Q_{Source}$ already features a geometric field a sensible first step is to reuse this. Although, the paraboloid shape will be replaced with a sphere to better mimic a drop of liquid filler metal. This is expressed as

\begin{subequations}
\label{6-eq-alpha-shape}
\begin{eqnarray}
&& \alpha_{exp}(x,y,z)  = \exp\left(\ln{C_{\alpha}}(\frac{(x^2 + y^2 + z^2)}{\Omega^2})\right) \\
&& g_{\alpha}(x, y, z) = \begin{cases}
                    1, & \text{if}\ \alpha_{exp}(x,y,z) \geq C_{\alpha} \\
                    0, & \text{otherwise}        
                    \end{cases}.
\end{eqnarray}
\end{subequations}     

where - compared to the paraboloid heat source detailed in Chapter 3 - $\alpha_{exp}$ is the replacement exponential function, $C_{\alpha}$ replaces $C_{Cut}$, and the separate $\omega$ and $l$ terms are replaced with $\Omega$. Given the phase fraction for each phase is already represented as $0 \leq \alpha \leq 1$ the geometric field in equation \ref{6-eq-alpha-shape} can be used directly as a source term for the liquid metal. However, compared to the source terms for phase change ($S_{\alpha_{2}}$ and $S_{\alpha_{3}}$) the source terms for the filler metal $S_{Filler}$ will act on the liquid phase and the gas phase. This is because adding liquid metal to the domain necessitates the removal of the gas phase so each cell can maintain a total phase fraction equal to one: $\sum_{i=1}^{3} \alpha_i = 1$. Therefore $S_{Filler}$ is added to the liquid metal phase and subtracted from the gas phase. This gives a new phase fraction equation of

\begin{equation}
\label{6-alpha-source-equation}
\begin{gathered}
    \partial_{t}\alpha_{1} + \nabla \cdot \boldsymbol{\alpha_{1} U} + \nabla \cdot (\alpha_{1} \alpha_{2} \boldsymbol{U_{c_{12}}}) + \nabla \cdot (\alpha_{1} \alpha_{3} \boldsymbol{U_{c_{13}}})  = - S_{Filler} \\ 
    \partial_{t}\alpha_{2} + \nabla \cdot \boldsymbol{\alpha_{2} U}  + \nabla \cdot (\alpha_{2} \alpha_{3} \boldsymbol{U_{c_{23}}}) = S_{\alpha_{2}} + S_{Filler}\\
    \partial_{t}\alpha_{3} + \nabla \cdot \boldsymbol{\alpha_{3} U}  = S_{\alpha_{3}}
\end{gathered}
\end{equation}

where $S_{Filler}$ is defined so as to add liquid metal just above the weld pool mimicking the blade repair process. Similar to the heat source term the filler metal source term can also be defined wherever the user wishes. Although it has only been used in this thesis for the blade repair process. The $S_{Filler}$ term is implemented in essentially the same way as the $Q_{Source}$ term except that it affects the phase fraction equation. 

When $S_{Filler}$ introduces liquid metal into the phase fraction equations it does not affect the temperature equation. Given $Q_{Source}$ term is a volumetric heat source within the base metal and the aforementioned lack of radiation and limited convection the gas phase ends up being substantially below the melting temperature of the liquid metal. This causes the liquid phase fraction to rapidly solidify within the gas phase when it is introduced. Intuitively, a solution would be to introduce a second heat source to warm the liquid metal but this negatively impacts performance substantially. This is due to incomplete melting / solidification leaving many cells with $\approx 5 - 10\%$ solid or liquid fraction making the discretized equations difficult to solve. Instead, the temperature of the liquid fraction introduced by $S_{Filler}$ can be set to $1.75 \times$ the melting temperature. Similarly to the value for $C_C$ in this case, this $1.75$ multiple was chosen as it produced the best results. This temperature is applied to cells with more than 20\% $S_{Filler}$. The final element to add is that - similar to the liquid bismuth benchmark covered in Chapter 4 - a change in liquid metal fraction in a cell automatically adds latent heat to it. Therefore the $S_{Filler}$ will add unphysical latent heat to the system at the time-step that it is first defined in. This can resolved by simply multiplying the latent heat term by $1 - S_{Filler}$.

\section{Results}
\subsection{Overview}
\label{6-sec-results}
There are experimental results in the form of thermal camera videos \cite{nickBooneThesis} taken of the blade repair process on proxy blades. These videos can be used to evaluate the efficacy of \gIF to simulate blade repair. In fact, these videos can serve as a benchmarking case to find the apt variable configurations that achieve a close match between experiment and simulation. Unfortunately, only limited details about the procedure used in the thermal camera videos are available. Further, no metallographs showing the layers of additive material on the repaired blades are available. This therefore limits the comparison to the temperature field and - as detailed in Chapter 4 - \gIF prioritises the modelling of the phase fraction. Nevertheless, they are still adequate in providing some evaluation. 

The videos consist of the root pass as well as the first pass (first layer of homogeneous welding) of welding on \SI{60}{\milli\meter} by \SI{40}{\milli\meter} by \SI{1.5}{\milli\meter} 316L stainless steel plates. To improve performance, only \SI{20}{\milli\meter} of the blade is simulated, as shown in Figure \ref{6-fig-blade-dom}. No video of a second - additive manufacture - pass is available however this can still be simulated as a showcase of the features available in \emph{gtawFoam}. In the videos, the camera remains static and zoomed into a region left of the electrode whilst the blade is moved horizontally. The thesis for which the images were gathered \cite{nickBooneThesis} states the camera was calibrated for a radiance temperature of 700 to \SI{1500}{\degreeCelsius}. However, only the raw unprocessed footage is available from which it can be deduced (knowing the melting point of 316L) the absolute values in the temperature colour bar are off by around 600. The calibrated images in the thesis suggest the gradient is still correct in the raw footage so the values in the colour bar can simply be shifted up by 600. Without the final produced calibrated images it is unlikely that the 316L thermal data can be used for comparison to normal welding. Thus these results should be interpreted as a approximate comparison to see if whether \gIF could in principle be used for AM.

From assessment of the videos and conversations with the experimenters the specifications for the heat source used for the root pass can be deduced. Unfortunately, these have to then be estimated for the first pass (and of course for the hypothetical second pass also). These specification are shown in Table \ref{6-tab-blade-source-heat}. In all the videos, there is a \SI{20}{\second} pause time before the blade begins to move enabling the weld pool to form; this is implemented in all of the simulations.

\begin{table}[!htb]
\setlength{\tabcolsep}{10pt}
    \centering
    \caption{\label{6-tab-blade-source-heat} Heat source specifications for blade repair case. Note the only $Q_{Source}$ available is for the root pass which is calculated from the \SI{17.5}{\ampere} and \SI{7.5}{\volt} values used for the main (neither ramp up or down) portion of welding. Thus the $Q_{Source}$ values for the first and second pass are inferred. The same value for $C_C$ is used for all cases to limit possible `over-fitting' given it was chosen to give the best results for the root pass rather than being found from the formula presented in Chapter 5.}
    \begin{tabular}{l l l l l l l}
    \toprule
    Pass & $Q_{Source}$ [\si{\watt}] & $l$ [\si{\milli\meter}] & $\omega$ [\si{\milli\meter}] & $v$ [\si{\milli\meter\per\second}] & $C_{cut}$ & $C_C$ \\
    \hline
    Root & 131.25\rule{0pt}{2.6ex} & 2 & 4 & 0.5 & 0.01 & 200 \\
    First & 150 & 2 & 4 & 0.5 & 0.01 & 200 \\
    Second & 165 & 2 & 4 & 0.5 & 0.01 & 200 \\
    \bottomrule
    \end{tabular}
\end{table}

\subsection{Root pass}
\begin{figure}[htb]
\centering
\includegraphics[width=13cm]{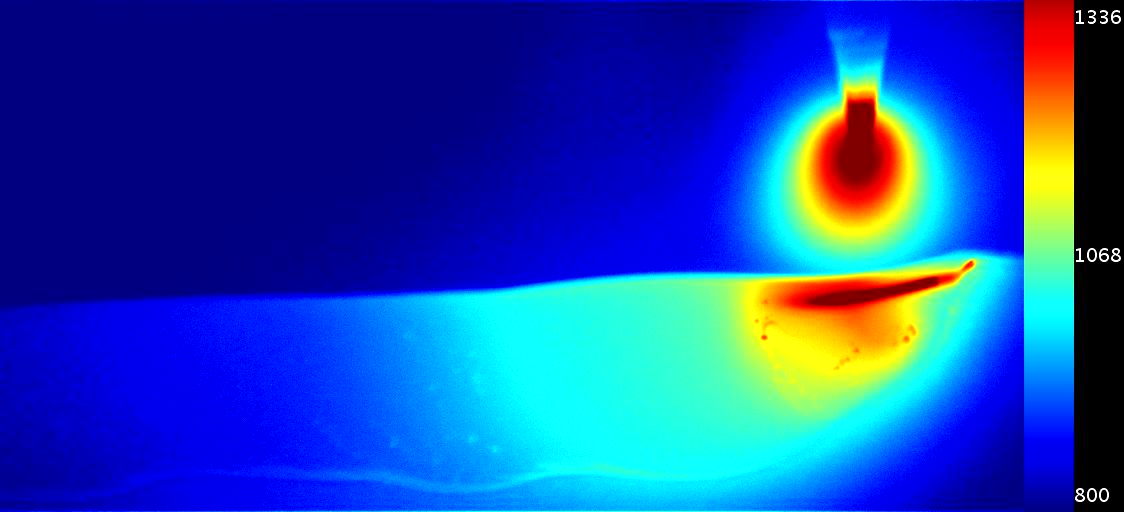}
\caption{Raw thermal camera output from the root pass. As covered in Section \ref{6-sec-results}, the values in the colour bar at $\approx \SI{600}{\kelvin}$ below their real values.}
\label{6-fig-thermal-root}
\end{figure}

\begin{figure}[htb]
\centering
\includegraphics[width=13cm]{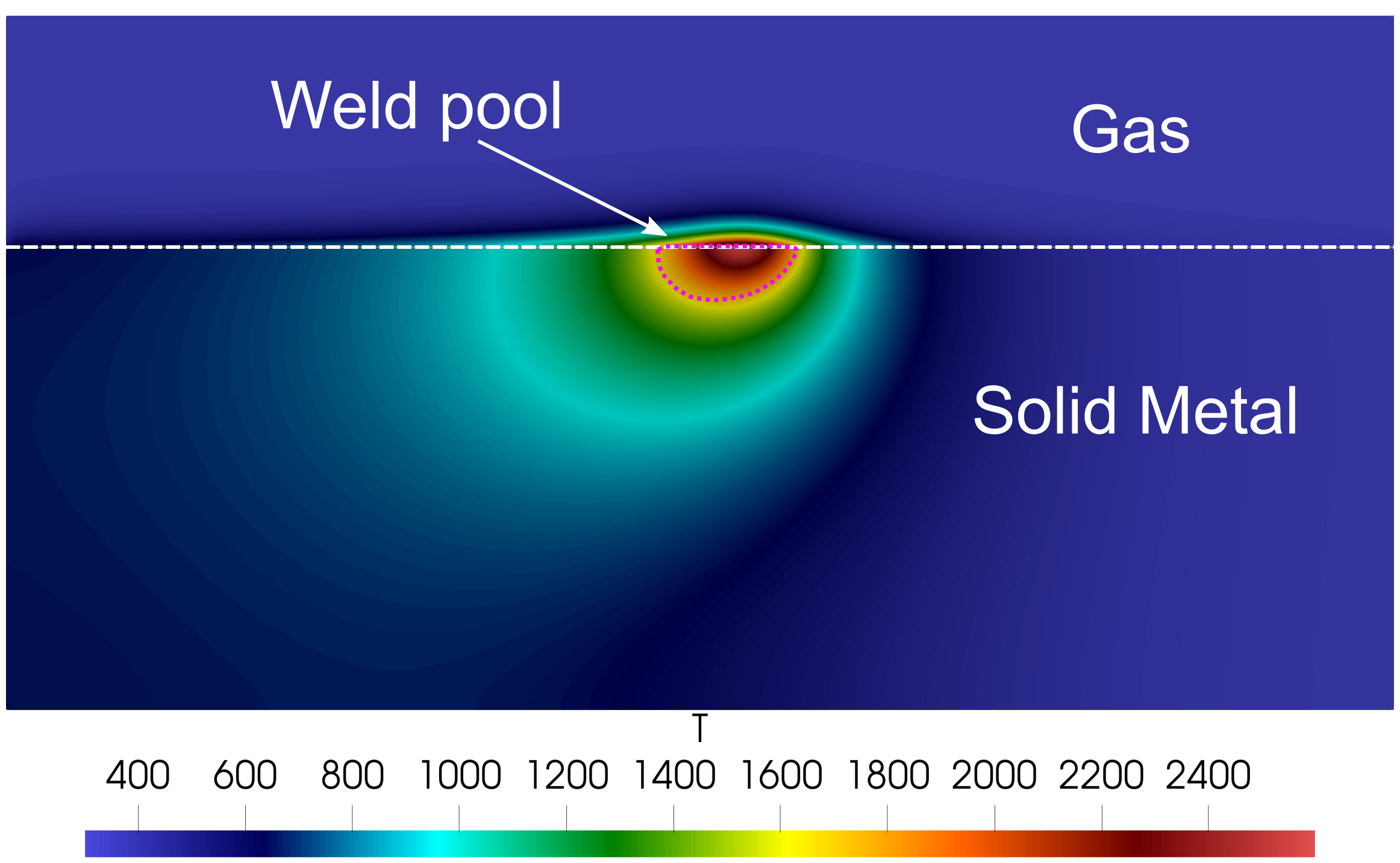}
\caption{Full domain view of temperature field for the root pass simulation. The dotted white line shows the separation between the gas phase and base metal. The weld is highlighted by a dotted magenta line.}
\label{6-fig-sim-root-temp-dom}
\end{figure}

\begin{figure}[htb]
\centering
\includegraphics[width=13cm]{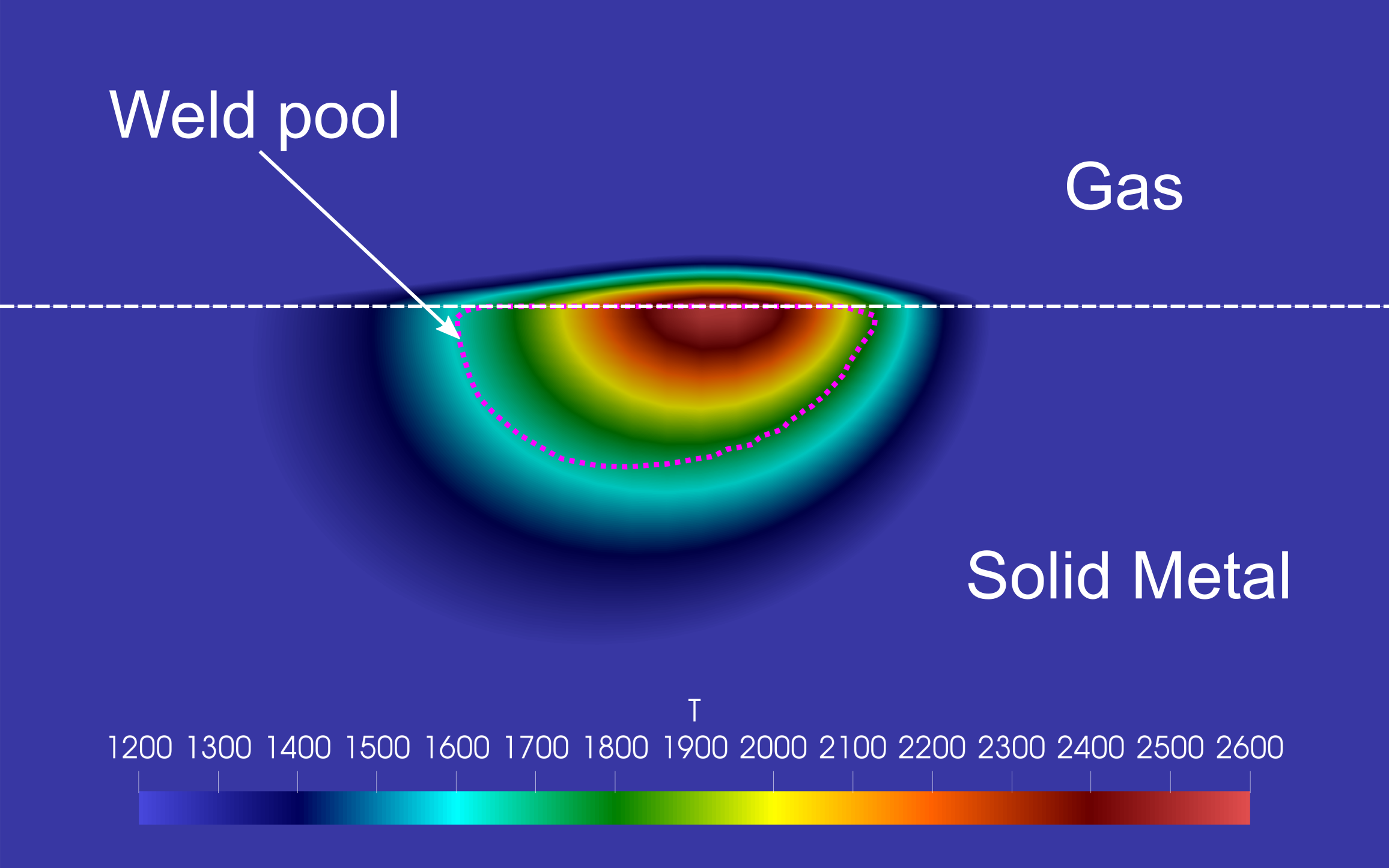}
\caption{Zoomed in view of weld. Note the change in color bar to better illustrate the temperature gradient within the weld. As with Figure \ref{6-fig-sim-root-temp-dom}, the dotted white line shows the separation between the gas phase and base metal and the weld is highlighted by a dotted magenta line.}
\label{6-fig-sim-root-temp-zoom}
\end{figure}

\begin{figure}[htb]
\centering
\includegraphics[width=13cm]{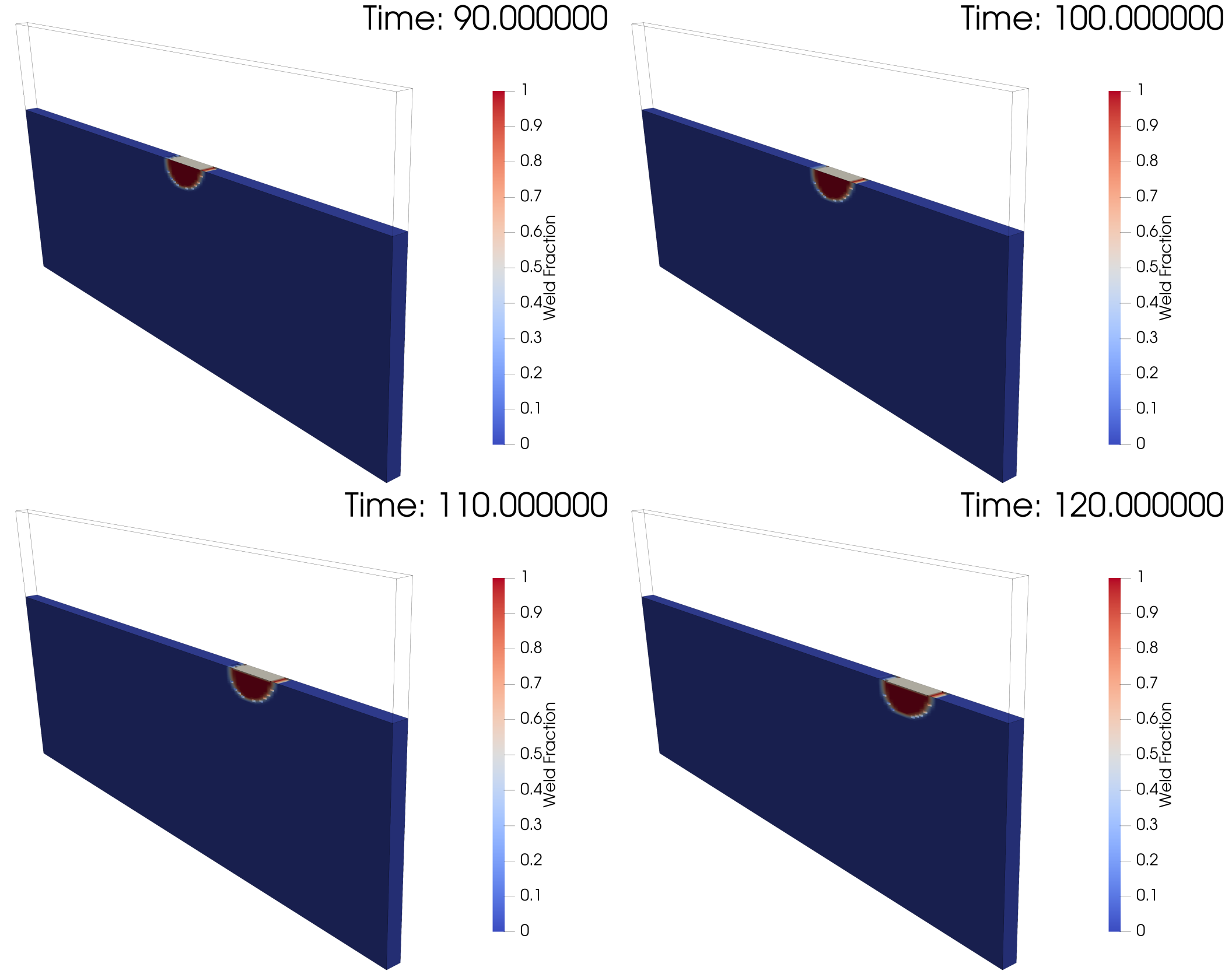}
\caption{Evolution of the weld fraction for a selection of times.}
\label{6-fig-sim-root-weld}
\end{figure}

The thermal camera footage for the root pass is shown in Figure \ref{6-fig-thermal-root}. To compare to this, a full domain image of temperature is shown in Figure \ref{6-fig-sim-root-temp-dom} and a zoomed in view of the weld pool is shown in Figure \ref{6-fig-sim-root-temp-zoom}. These figures show a reasonable match between simulation and experiment. The evolution of the liquid weld fraction with time is shown in Figure \ref{6-fig-sim-root-weld}. Here, the figure shows the weld pool moving along the blade proxy melting new metal in the front of the pool as metal at the back of the pool solidifies. Therefore, it is concluded that \gIF successfully simulates the root pass.  

\subsection{First pass}
\begin{figure}[htb]
\centering
\includegraphics[width=13cm]{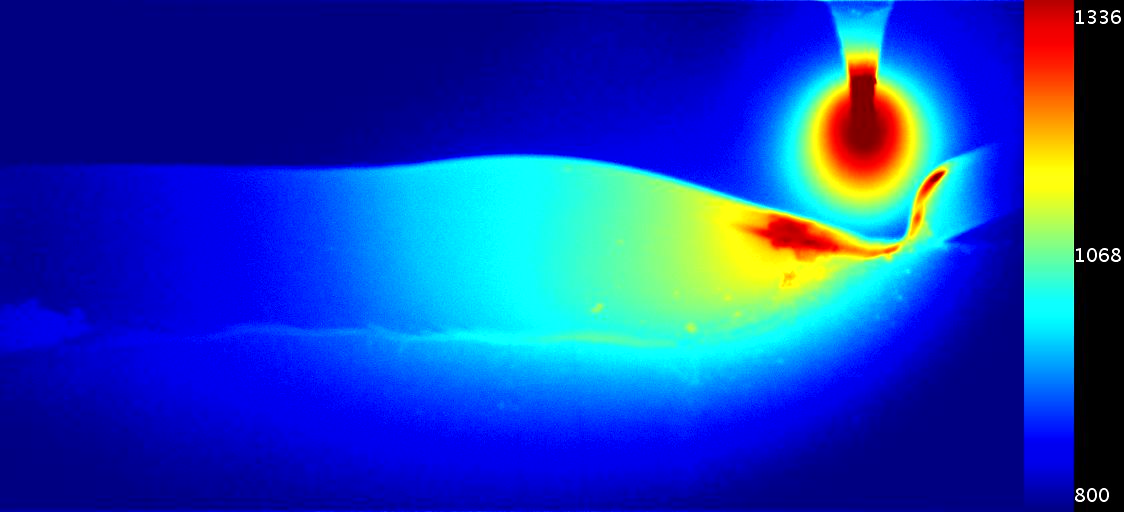}
\caption{Raw thermal camera output from the first pass. Unlike the simulation, the filler metal comes in solid and is melted by the arc. As covered in Section \ref{6-sec-results}, the values in the colour bar at $\approx \SI{600}{\kelvin}$ below their real values.}
\label{6-fig-thermal-first}
\end{figure}

Shown in the thermal camera footage in Figure \ref{6-fig-thermal-first}, the first pass introduces filler metal into the liquid weld pool. The video shows the filler metal is fed into the front of the weld pool whereupon it melts into the weld pool. Although, to simulate this, the $S_{Filler}$ term is set to directly above the heat source term. This is because it was found to affect the weld pool dynamics if it was placed on the leading edge of the weld pool. The exact specifications of the source term are shown in Table \ref{6-tab-alpha-source}.

\begin{table}[!htb]
\setlength{\tabcolsep}{10pt}
    \centering
    \caption{\label{6-tab-alpha-source} Alpha source specifications for first pass in the blade repair case. The position is relative from the top left corner of the blade where the heat source starts and with a $y$-axis going into the thermal camera images.}
    \begin{tabular}{l l l l l }
    \toprule
    Pass & $\Omega$ [\si{\milli\meter}] & Position [\si{\milli\meter}] & Start time [\si{\second}] & $v$ [\si{\milli\meter\per\second}] \\
    \hline
    First & 1\rule{0pt}{2.6ex} & (0 0 2.5) & 20 & 0.5 \\
    \bottomrule
    \end{tabular}
\end{table}

\begin{figure}[htb]
\centering
\includegraphics[width=13cm]{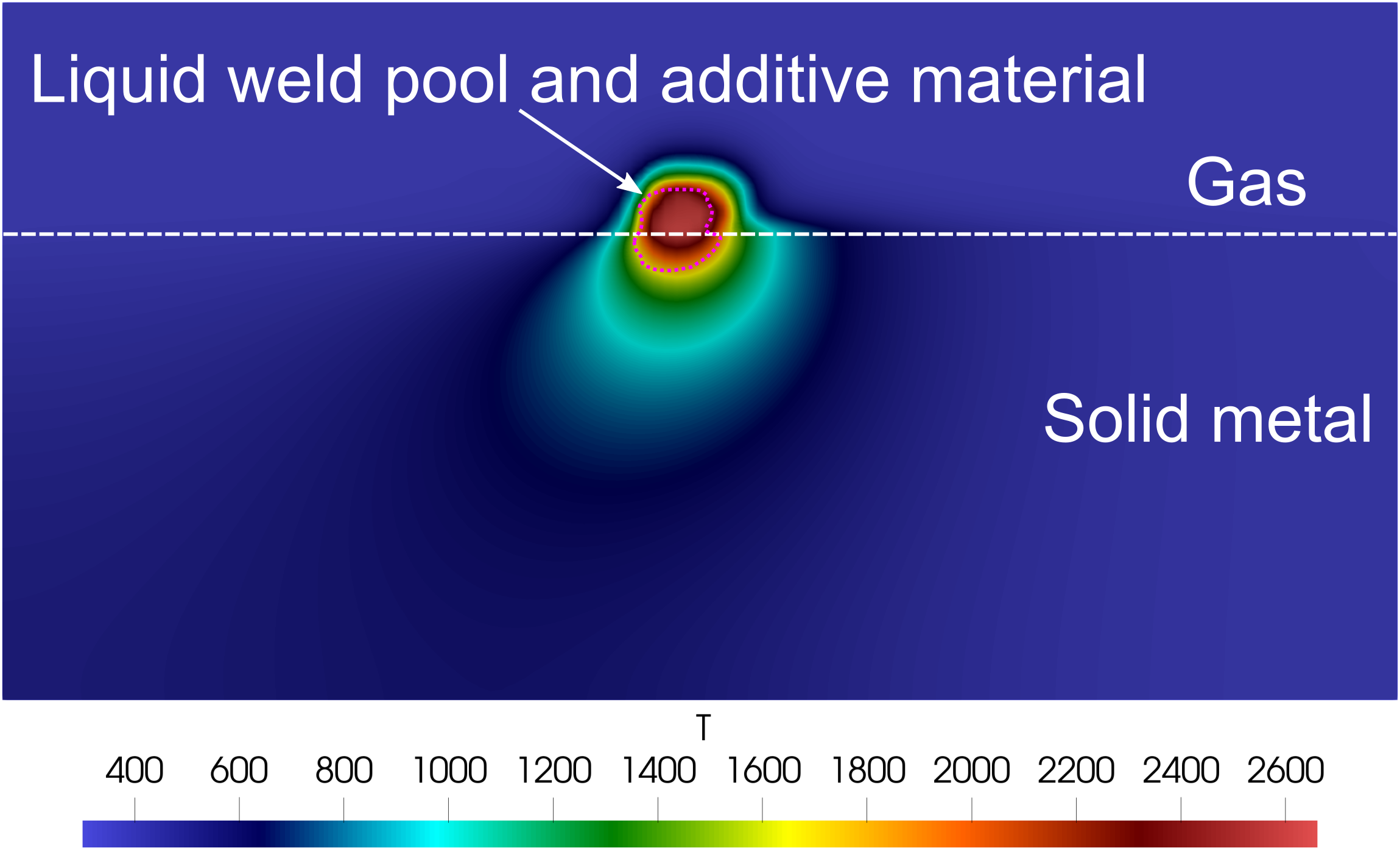}
\caption{Full domain view of temperature field for the first pass simulation. The dotted white line shows the separation between the gas phase and base metal. The combined liquid weld pool and additive material is highlighted by a dotted magenta line.}
\label{6-fig-sim-first-temp-dom}
\end{figure}

\begin{figure}[htb]
\centering
\includegraphics[width=13cm]{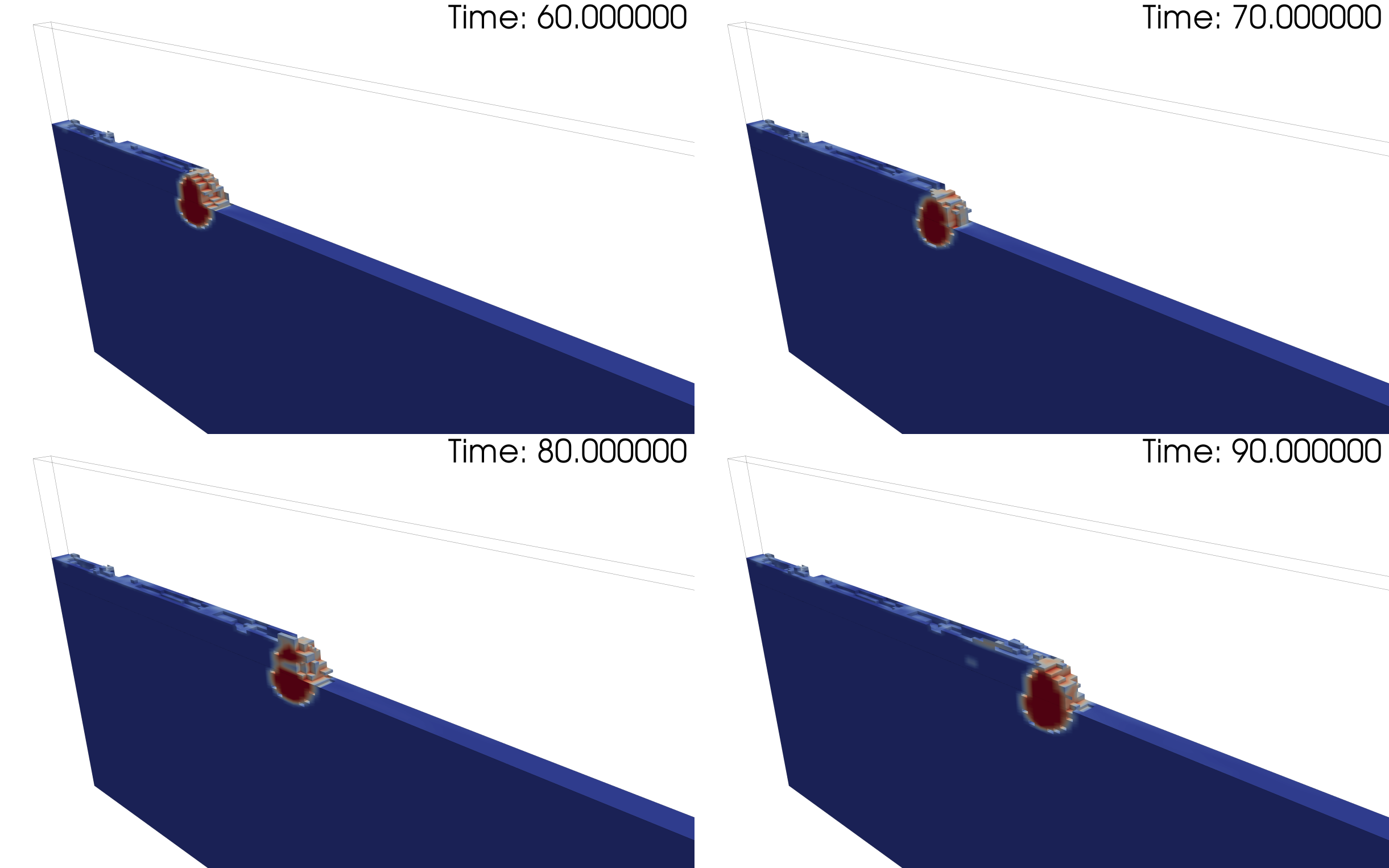}
\caption{Liquid weld fraction for the deposition of the additive material for the first pass for a selection of times. Note the boundaries in the gas phase cause rapid solidification in the boundary cells so these cells have been removed for this visualization so that the liquid fraction can be seen.}
\label{6-fig-sim-first-weld}
\end{figure}

The results from the simulation for the temperature field are shown in Figure \ref{6-fig-sim-first-temp-dom}. Here, the liquid filler metal enters the domain at $1.75 \times$ the melting temperature of 316L causing a larger region within the domain to be at the peak temperatures compared to in the root pass. Due to this, the match with the thermal camera footage in Figure \ref{6-fig-thermal-first} is not as good as with the root pass. However, \gIF does successfully simulate the deposition of filler metal for the first pass as shown in Figure \ref{6-fig-sim-first-weld}. This serves to demonstrate that the blade repair process can be simulated with \emph{gtawFoam}. 

\subsection{Additive manufacture}
\begin{figure}[htb]
\centering
\includegraphics[width=15cm]{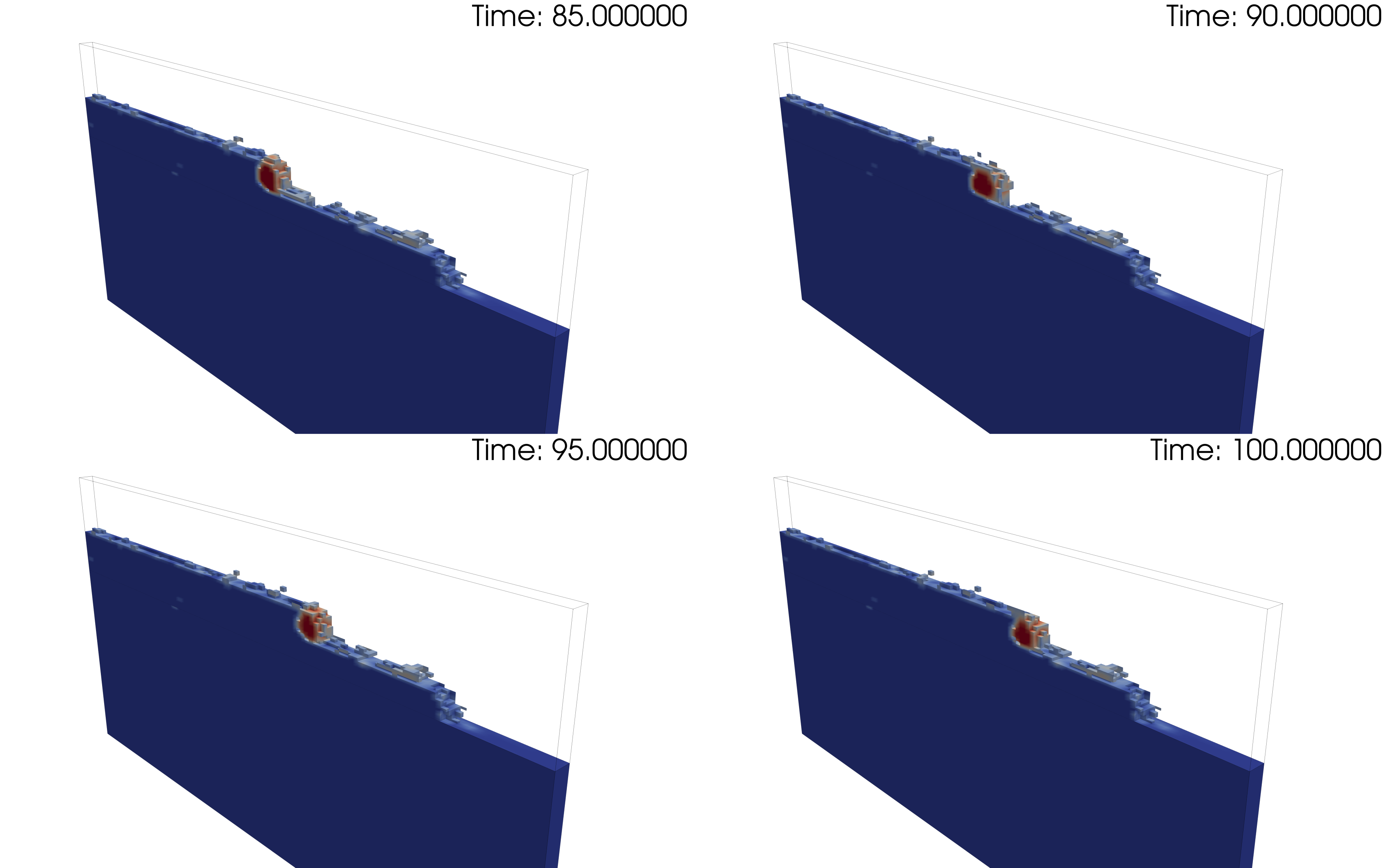}
\caption{Liquid weld fraction for the deposition of the additive material for the second pass for a selection of times. With this second pass shown, the process can be repeated $n$ times depending on the required additive material build up. Note the boundaries in the gas phase cause rapid solidification in the boundary cells so these cells have been removed for this visualization so that the liquid fraction can be seen.}
\label{6-fig-sim-second-weld}
\end{figure}

Additive manufacture can be simulated largely through repeating the first pass with updated coordinates for the alpha and heat sources. However, after the first pass the blade is now larger and so occupies a portion of the upper block in Figure \ref{6-fig-blade-dom}. This means that the region of the blade in the upper block has atmospheric boundary conditions rather than blade boundary conditions. The result of this is that the positioning of the heat source in this block will result in considerably more radiative heat loss through the boundaries. The solution to this is either to change the boundaries or increase the power of the heat source. Redefining the boundary or domain somewhat defeats the point as it would effectively create an additional first pass thus the solution is to increase the power of the heat source. Further, as covered in Chapter 3, the heat source and the alpha source update their position through multiplying the current simulation time by the specified velocity. Therefore repositioning each source from the first pass will not work - a second source for each is required. This can be achieved through copying the time directory at the end of the first pass and resetting the time to zero. In fact, given the only file that matters is the phase fraction just these files can be copied and the initial conditions replaced. To avoid resetting the time, two new `AM' class instances for the heat source and alpha source terms can be added. Given this will require different temperature and phase fraction equations (ones with two source terms each) in \gIF this is implemented as separate equations. Thus `TEqn.H' becomes `AMTEqn.H' with the same for the phase fractions. A user can alternate between the two using a simple compilation flag with conditional statements then choosing the files to compile with. The results from the AM case using the time resetting method are shown in Figure \ref{6-fig-sim-second-weld} and the alpha source details are given in Table \ref{6-tab-alpha-source-am}.

% Table \ref{6-tab-blade-source-heat} and the alpha sources in Table \ref{6-tab-alpha-source-am}.
\begin{table}[!htb]
\setlength{\tabcolsep}{10pt}
    \centering
    \caption{\label{6-tab-alpha-source-am} Alpha source specifications for additive manufacture in the blade repair case. The position is relative from the top left corner of the blade where the heat source starts and with a $y$-axis going into the thermal camera images.}
    \begin{tabular}{l l l l l }
    \toprule
    Pass & $\Omega$ [\si{\milli\meter}] & Position [\si{\milli\meter}] & Start time [\si{\second}] & $v$ [\si{\milli\meter\per\second}] \\
    \hline
    First & 1\rule{0pt}{2.6ex} & (0 0 2.5) & 20 & 0.5 \\
    Second & 1\rule{0pt}{2.6ex} & (0 0 5) & 20 & 0.5 \\ 
    \bottomrule
    \end{tabular}
\end{table}

\FloatBarrier
\section{Future outlook}
\begin{figure}[!htb]
\centering
\includegraphics[width=13cm]{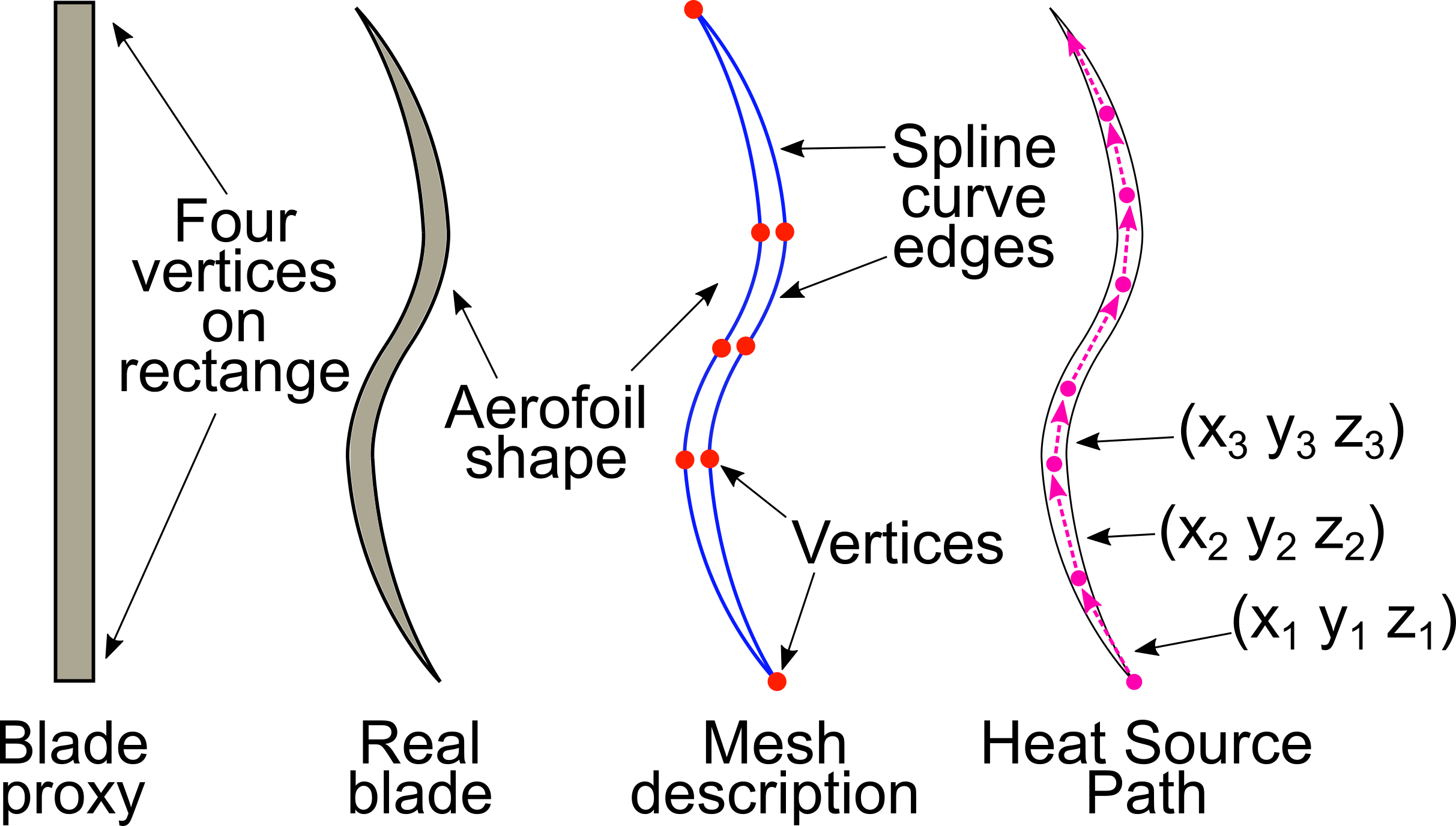}
\caption{Breakdown of the challenges required to implement \gIF on a real blade. Note the difference in shape between the blade proxy and real blade. A case with a domain accurate to a real turbine blade shape would require more vertices, edges and a convoluted heat source path specification.}
\label{6-fig-aero}
\end{figure}

The obvious next step in developing the turbine blade repair case is to run simulations on an aerofoil blade. Successfully predicting the required welding procedures or success chance of blade repair would be a useful addition to any turbine blade repair process. In principal this is not too different from the blade proxy case however the implementation could potentially be quite involved. For a start, as shown in Figure \ref{6-fig-aero}, specifying the aerofoil shape of a blade in \of will require multiple extra vertices and will require spline edges to specify curvature. Further, the heat source term in \gIF updates its position through applying a vector to its current position multiplied by a time interval. For a straight line only one vector will be required for all time but, as also shown in Figure \ref{6-fig-aero}, for an aerofoil multiple successive vectors will be required. The implementation of this (in pseudocode) would be if time $t \leq t_1$ use vector $v_1 \times t$ else if $t_1 <$ time $\leq t_2$ use vector $v_1$ for $t_1$ seconds then add $v_2 \times (t - t_1)$ and so on. Given each blade could have a unique shape it would be arduous to implement a unique case for each blade. Two possible routes to address the meshing challenge are a `3D scan to mesh' process or a `categorize blade' process. The heat source path issues can be fixed with an improved implementation.  

All the meshes presented in this thesis so far have been created using the \of utility \emph{blockMesh} whereby vertices and edges are specified in a \emph{blockMeshDict} file. Whilst Figure \ref{6-fig-aero} shows how this could be done for a real blade \of also features various utilities to import ready made meshes. Most CFD programs (\of included) have the ability to import and export STL files. Therefore, an STL file of a volumetric mesh generated from a 3D scan of a blade can be converted into a mesh \of can run \gIF on. This `3D scan to mesh' process is somewhat dependent on the hardward used. However there are software tools as Autodesk Meshmixer (https://www.meshmixer.com) that aid the process of turning a 3D scan into a good mesh.          

The `categorize blade' process would involve creating `buckets' of blade shapes with \emph{blockMeshDict} using vertices and curved edges as shown in the `Mesh description' image on Figure \ref{6-fig-aero}. Through changing the blade length, blade width along the length and the curvature of the spline edges joining the vertices a variety of aerofoils can be created. Various blades could then be categorized and bucketed according to their geometry. A mesh from one of these buckets could then be employed by \gIF as a `match' for any specific blade. Whilst the mesh may not be an exact match, it would be guaranteed to be well defined as it will be created through \emph{blockMeshDict} and not be subject to any errors that can occur during a 3D scan. The amount of buckets used has no reasonable limit as with a manufacturing price of $\approx$ £5000 per blade it is not an uneconomical idea to have 1000s of buckets if the simulations enable perfect or near-perfect repair rates.  

Finally, for the heat source path the best solution would be to improve the implementation of the heat source movement to make it more user friendly. Currently, the position is simply updated through adding on a time scaled multiple of a `movement' vector to the initial position. For example with an initial position of (0 0 0) and a movement vector of (1 0 0), after one second the new position is (0 0 0) + (1 0 0) $\times$ 1 = (1 0 0) and after two seconds it is (0 0 0) + (1 0 0) $\times$ 2 = (2 0 0) etc. Implementing movement with a different vector after $t = 10$ seconds such as (0 0 0) + (1 0 0) $\times$ \textbf{F} + (0 1 0) $\times$ \textbf{G} $\cdot (t - 10)$ where $\textbf{F} = t$ and $\textbf{G} = 0$ when $t \leq 10$ and $\textbf{F} = 10$ and $\textbf{G} = 1$ otherwise is feasible but inelegant. A possible better implementation would be to update the position of the heat source through a discrete list of points linearly moving the source between them with time in a similar manner to \cite{holtzmannLaserBC}. Although this still requires the user to enter the list of points. An elegant solution to this would be to create a path finding algorithm that creates the list of points given a specific blade shape - this could even be tied into the path planning system of a turbine blade repair robot.       

With these issues solved, the final problem would be tying a result from \gIF to a concrete physical welding procedure. In a similar manner to the welding of ultra-thin-walled tubing covered in Chapter 5, the phase space of blade welding could be explored through finding \gIF welding procedures appropriate for a large amount of blades. However, there are multiple issues with this approach. For instance, as previously covered, the heat source in \gIF has only a single value and not a current and voltage so a certain amount of judgement would be required for finding appropriate values for a physical welding procedure that is mimicking results generated in \emph{gtawFoam}. Similarly, physical variables such as the arc gap and gas flow rate do not have direct analogous values in \emph{gtawFoam}. Furthermore, the extremely thin corners of the aerofoil turbine blades can easily blow away during the striking of a GTAW arc. In fact, the best/only way to address this destruction of the blade corners is to feed excessive filler metal into the weld pool to create `\emph{mickey mouse ears}' on the corners of the blade that can then be milled off. The best solution to these issues is actually to reframe the role \gIF can play in creating a welding procedure. Instead of relying on \gIF to solve all problems it instead can provide a solid starting point from which physical tinkering can proceed.   

\section{Chapter summary}
This chapter shows the wider application of \emph{gtawFoam} through the simulation of the GTAW process in turbine blade repair. The validity of the simulation results is assured through comparison with thermal camera videos of the repair process on proxy blades. Combined with the Chapters 3 and 4, the second objective of a detailed open-source GTAW simulation tool has thus been achieved. Further, possible routes for extension of this case that fully account for the aerofoil shape of a turbine blade have been suggested. Finally, the application of \gIF to turbine blade repair shown in this chapter is one of many possible applications. The demonstration of additive manufacture shows this versatility. The use of an arbitrary multiple for the thermal conductivity of the gas phase also shows how \gIF can adapt to a new situation through the reevaluation of already present features.

%% file: chapters/thesis-chapter-8.tex
\lhead{Chapter 8: Conclusions}
\section{General conclusions}
This work has introduced a novel multiphysics solver - \gIF - that was benchmarked and subsequently applied to both ultra-thin-walled tube welding and thin-structure welding. Revisiting the three objectives of this thesis given in the introduction they were (abridged):

\begin{enumerate}
    \item \label{7-item-1} Create a general case for ultra-thin-walled tube welding. 
    \item \label{7-item-2} Create a well documented open source GTAW simulation tool. 
    \item \label{7-item-3} Elucidate and optimize the enthalpy-porosity method for GTAW simulation. 
\end{enumerate}

\noindent The general case for ultra-thin-walled tube welding was identified as a critical `goldilocks zone' for each wall thickness. This `goldilocks zone' was found in both the simulation and experimental work. The experimental work demonstrated that the effect of the current and internal gas flow are orthogonal to each other meaning apt values for both are required for a successful weld. This information was subsequently used in \gIF through mapping the current and internal gas flow to simulation variables namely the heat source power $Q_{Source}$ and the pressure inside the simulated tubing. With this mapping created, a large batch of simulations were performed with different pairings of $Q_{source}$ and internal pressure to create a parameter space of passed and failed welds. This parameter space revealed similar behaviour to the experimental work with an identifiable `goldilocks zone'. The `goldilocks zones' from both the experimental and simulation results were then combined to allow the simulation results to be `translated' into an experimental procedure.

\gIF was then used as a prediction tool for other ultra-thin-walled tubing. The experimental parameters used for orbital welding on the \SI{2.275}{\milli\meter} outer diameter, \SI{300}{\micro\meter} wall thickness (at the point of weld) titanium tubing were used as a basis for the procedures on similar tubes except with wall thicknesses ranging from \SI{250}{\micro\meter} to \SI{350}{\micro\meter}. These results suggested that the required inner pressure for the different wall thicknesses was roughly constant but that less heat was required to penetrate the thinner walls and more heat required to penetrate the thicker walls. The simulation results from \gIF for this hypothetical tubing were then interpreted to suggest inner pressures and average current to complete the experimental procedures. These values were deliberately given with low precision to illustrate that they are guidelines and experimental `tinkering' will be required.  

Whilst it was not possible to test the veracity of these predictions experimentally, they are an obvious starting point for future work. On this point, whilst the higher heat input for thicker tube wall results make intuitive sense, this is not the case for the inner pressure. Some of the experimental results and the experimental experience of the author suggests that slightly higher pressures should be required for thicker walls where higher currents are required. However, short of robust experimental evidence, it is concluded conservatively that similar inner pressures can be used for a variety of wall thicknesses.

The results also suggest that to extend these predictions to ultra-thin-walled tubing with other diameters requires simply to ensure the `goldilocks zone' is identified. With the flattened tubing it is difficult to represent tubing of other diameters so this `goldilocks zone' should - according to the simulation results - only depend on the wall thickness. Due to the limited experimental work, it is unclear whether this is the case however, in principal, the results suggest that it is. In a similar vein, the physical mechanism of what causes a tube to break can only be suggested as the pressure pushing against the buttressing force is unknown. Given that gravity was not included for reasons explained in Chapter 5, in the simulation procedures would fail when the surface tension was overcome. It is the view of the author that this is probably also what causes experimental ultra-thin-walled tube welding to fail.

For the second objective, as covered in Chapter 1 the available computational solvers for GTAW simulation were insufficient for simulating ultra-thin-walled tube welding. Therefore, the first objective required initially creating a custom solver. This led to the novel solver detailed in Chapter 3 and the git repository that includes both \gIF and all the case files featured in this thesis. Compared to the other models detailed in the introduction, \gIF achieves all of the target features. Specifically, it is open source, models three phases, features melting and solidification, uses a custom moving volumetric GTAW heat source and is extensively benchmarked. In fact, the extensive benchmarking in Chapter 4 is a key uniqueness within \gIF as it demonstrates how \gIF can be used for other materials processing problems. This transferability is further illustrated through the application of \gIF to thin-structure welding in Chapter 5. This included an initial introduction of a liquid metal source term allowing additive manufacture to be simulated.

Finally, the above paragraphs illustrate that the enthalpy-porosity method was successfully applied to create a GTAW simulation. A key insight into this method was presented in Chapter 4. Here, the effect of the commonly used Darcy constant (also referred to as the momentum sink in the literature) was quantified through assessing its value against the quality of the results produced using that value. The role of the Darcy constant is to ensure the solid region remains stationary. Given the change in the amount of solid region within the domain is determined (computationally) by the phase change constant, it follows that this quality assessment also investigate the pairing of the Darcy constant with a phase change constant. It was found that an optimum values for these constants were \num{8e8} for the Darcy constant and 10 for the phase change constant. Whilst many pairings achieved strong results, the optimum values exhibited `stability' with neighbouring values achieving similarly high results. This stability is important in \gIF as the ultra-thin-walled tube welding process is predicted thus it is important to have confidence that the results will not be radically different with a slightly different case. However, it should also be noted that many pairings achieved high results which explains the variety of choices for the Darcy constant seen in the literature. An additional smaller insight was the efficacy of a class base volumetric GTAW heat source for enthalpy-porosity based solvers. This implementation allows the modelling of multiple heat source to be achieved through simply adding additional $Q_{GTAW}$ terms to the right hand side of the energy equation. In fact, this can be repeated \emph{n} times which may prove useful for particular applications. Further, the additive manufacturing shown in Chapter 6 illustrates the versatility of these type of sources as the volumetric heat source is converted into a volumetric liquid metal source.

\section{Novelty assessment}
In addition to achieving the objectives, the present work also successfully achieved the novelty outlined in Section \ref{2-sec-novelty}. The literature review highlighted previous work and found that there was a research gap for experimental research that involved orbital GTAW on ultra-thin-walled titanium tubing. This research gap was fulfilled through the experimental work presented in Chapter 5. Further, the literature review also identified a research gap for a well benchmarked, three phase, ultra-thin-walled tube welding simulation complete with melting and solidification and a moving GTAW source term. The \gIF solver presented in Chapter 4, benchmarked in Chapter 5 and applied in Chapter 6 successfully fulfilled this niche. It is thus concluded that the present work presents a substantial novel contribution to the research on ultra-thin-walled tube welding. Abridged versions of the tables of the research gaps identified in Chapter 2 are shown in Table \ref{8-tab-exp-summary} and Table \ref{8-tab-model-summary}.

\begin{table}[h]
\caption{\label{8-tab-exp-summary}Summary of the previous experimental work covered in Chapter 2 that was closest to the target experiment.} 
\centering
\begin{tabular}{l c c c c}
\toprule
Investigation & Geometry & Material & Process & Wall thickness \\
\hline
Target\rule{0pt}{2.6ex} & Tube & Titanium & Orbital GTAW & $\leq$ \SI{500}{\micro\meter} \\
\hline
de Garcia et al.\rule{0pt}{2.6ex} \cite{garcia2010advances} & \ding{51} & \ding{51} & \ding{51} & \ding{55} \\
Kumar et al. \cite{kumar2010} & \ding{51} & \ding{51} & \ding{51} & \ding{55} \\
Liu et al. \cite{LiuThinWallTube} & \ding{51} & \ding{55} & \ding{51} & \ding{51} \\
\hline
Chapter 6 results\rule{0pt}{2.6ex} & \ding{51} & \ding{51} & \ding{51} & \ding{51} \\ 
\bottomrule
\end{tabular}
\end{table}

\begin{table}[h]
\caption{\label{8-tab-model-summary}Summary of the previous simulation work covered in Chapter 2 that was closest to the target simulation.} 
\centering
\begin{tabular}{l c c c c c}
\toprule
Model & \begin{tabular}{@{}c@{}}Bench-\\marks\\\end{tabular} & \begin{tabular}{@{}c@{}}Three\\phase\\\end{tabular} & \begin{tabular}{@{}c@{}}Moving\\GTAW source\\\end{tabular} & \begin{tabular}{@{}c@{}}Melting \&\\ Solidification\\\end{tabular} &\begin{tabular}{@{}c@{}}Geometry\\\end{tabular} \\
\hline
Target\rule{0pt}{2.6ex} & $\geq 5$ & \ding{51} & \ding{51} & \ding{51} & Tube \\
\hline
Traidia\rule{0pt}{2.6ex} \cite{traidiaThesis} & 0 & \ding{51} & \ding{51} & \ding{51} & Plane \\
Wang \cite{WangCoupled} & 0 & \ding{51} & \ding{55} & \ding{51} & Plane \\
Saldi \cite{saldi2012marangoni} & $6$ & \ding{51} & \ding{55} & \ding{51} & Plane \\
\hline
\gIF \rule{0pt}{2.6ex} & $8$ & \ding{51} & \ding{51} & \ding{51} & Tube \\
\bottomrule
\end{tabular}
\end{table}

\section{Future work}
The clear next step is to experimentally test the general case predictions made by \emph{gtawFoam}. This should be both at the \SI{2.275}{\milli\meter} OD, \SI{160}{\micro\meter} titanium tubing used for the ATLAS ITk and for tubing with different dimensions. For the ATLAS ITk tubing, the welding process should be run with all of the mapped parameters that \gIF was run with. This should enable effective validation and tuning of the \gIF results. With this, the predictions for the procedure for other tube wall thicknesses should be tested. Given the titanium tubing is made to order, realistically the way to do with would be CNC milling a thicker titanium tube sleeve and inserting the original ATLAS tube inside. This is not ideal but it is very feasible. The consideration for what wall thickness to use will depend on what tooling is available for the orbital weld head; this will have to be worked around.        

A possible consideration for improving the general case would be to use a cylindrical simulation domain rather than a flattened domain. In the present work, a flattened domain was found to be superior given the time and computational limitations. Although, in principal a cylindrical domain should be better as it will take into account the curved surface of the tube. Further, changing from a linearly translating heat source into a rotationally translating heat source is theoretically performed through correctly applying $sine$ and $cosine$. An initial implementation of this was already performed by the author but it was hampered by a bug whereby the heat source would rotate around a different but parallel axis to the central axis of the cylindrical domain. If this issue was fixed then a `toy' model for this could be evaluated. Upon evaluation, the next step would be to implement radiation. A good place to start for this radiation implementation is the inbuilt one in \emph{chtMultiRegionFoam}. Another feature that could be included is the Lorentz force. This is actually already implemented in \gIF but needs to be integrated with the volumetric heat source. 

More broadly, to investigate tube welding further a CWM approach could be taken which would enable the role of residual stresses and distortions to be investigated. Given how \gIF focuses on the fluid mechanics of the molten metal there is no feature that could be added to investigate this the same way the AM source term was. However in principle as \gIF records the temperature at each time step this information could be fed into a CWM model.

\begin{figure}[!htbp]
\centering
\includegraphics[width=10cm]{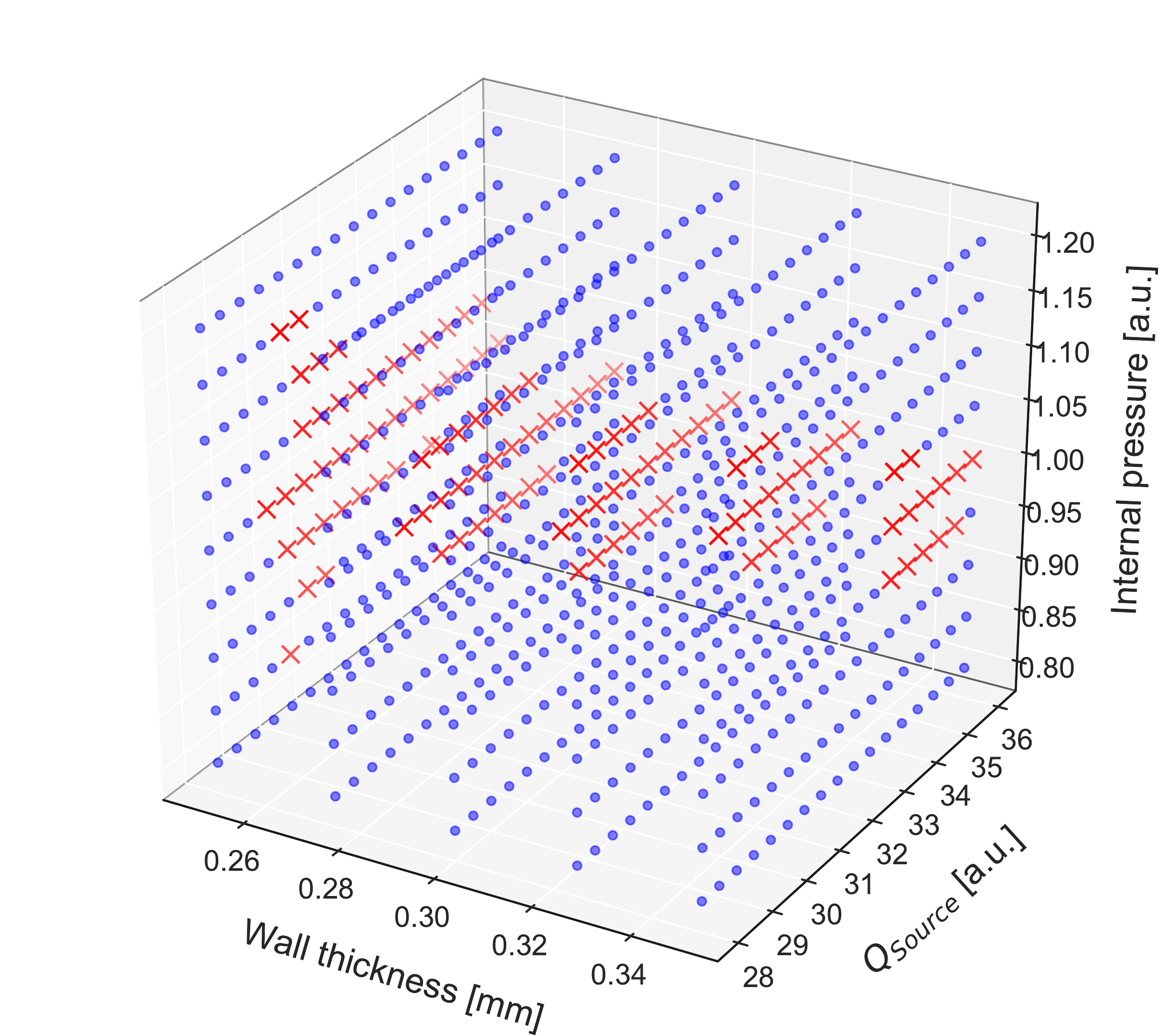}
\caption{All tube simulations presented in Chapter 5 scored as either successful (red cross) or fail (blue circle).}
\label{7-fig-sim-stacked}
\end{figure}

Another possibility for future work is to treat the identification of the `goldilocks zone' as a classification problem. All of the failure modes (burst, collapsed and lack of penetration) could simply be categorized as fail to increase the available data. With enough data, a learning algorithm could potentially be used to predict whether procedures will pass or fail. As an illustration, Figure \ref{7-fig-sim-stacked} shows the simulation results from Chapter 5 as a full parameter space with successful and failed welds. Given that there is only experimental data for tubing with \SI{300}{\micro\meter} thick walls, it is unwise to extrapolate too much with this data. However, if the aforementioned experimental verification was performed, this would be the logical next step.   

The overall understanding of the welding process used for ultra-thin-walled titanium tubing could be greatly enhanced through taking arc pressure and heat input measurements from the welding arc. As covered in Chapter 5, there is a well known technique for performing this involving a water cooled copper anode complete with thermocouples and an inbuilt cavity for arc pressure measurements. If the exact heat input and arc pressure were known, it would further elucidate the tube welding process. This set up could then be expanded to further tune the heat source term in \emph{gtawFoam}. Combining heat input measurements with the results from large amount of experimental cases from the literature would enable a more accurate formula for the heat source constant $C_C$ to be established. In fact, even without a heat input measurements extra experimental cases from the literature for tuning $C_C$ should be added to the eight used in this thesis.  

Beyond tube welding, there are a few wider goals that should be attempted. Chapter 6 provides the first steps for this showing that \gIF can be used for a blade repair case. The immediate next step in this is to run the case on an aerofoil shape - ideally one from an actual physical blade where the repair process could be compared directly against. Further, rerecording the thermal camera footage used for the evaluation in Chapter 6 with the express purpose of evaluating \gIF and combing this footage with a metallographic examination would enable accurate tuning of the blade repair case. Another goal would be to extend the novel optimization technique presented in Chapter 4 to deal with other computational constants and even thermophysical constants. As demonstrated throughout this thesis, the simulation results can change wildly depending on the values used for various constants. A researcher is incentivized to pick the constants that work best for their application but these constants may not be universally the best for all applications. Applying a systematic open source search over a wide variety of applications may show that there are values for these constants that are in fact universal.

%% file: appendicies/thesis-appendix-glossary.tex
\pagestyle{plain}
\begin{longtable}{l l l}
\caption{Definitions of the symbols used in this thesis. Symbols with an extra numerical subscript refer to a specific phase. For example, $k_{2}$ is the thermal conductivity for the liquid phase ($\alpha_2$).}
\label{ap-tab-glossary-symbols} \\
\toprule
Symbol & Description & Units\\
\hline
$\alpha_1$\rule{0pt}{2.6ex} & Gas Phase Fraction &\\
$\alpha_2$ & Liquid Phase Fraction &\\
$\alpha_3$ & Solid Phase Fraction &\\
$\beta$ & Volumetric Thermal Expansion Coefficient & \si{\per\kelvin} \\
$C_d$ & Darcy Computational Constant & None \\
$C_{pc}$ & Phase Change Computational Constant & None \\
$c_{i}$ & Goldak Heat Source Ellipsoid Length & \si{\metre} \\
$c_{p}$ & Specific Heat Capacity & \si{\metre\squared\per\second\squared\per\kelvin} \\
$c_{p, a}$ & Specific Heat Capacity Coefficient $a$ & \si{\metre\squared\per\second\squared\per\kelvin} \\
$c_{p, b}$ & Specific Heat Capacity Coefficient $b$ & \si{\metre\squared\per\second\squared\per\kelvin\squared} \\
$c_{p, c}$ & Specific Heat Capacity Coefficient $c$ & \si{\metre\squared\per\second\squared\per\kelvin\cubed} \\
$\epsilon$ & Emissivity & None \\
$\epsilon_{ij}$ & Rate Of Strain Tensor & \si{\per\second} \\
$\eta$ & Efficiency & None \\
$\vec{F}$ & General External Force & None \\
$g$ & Gravity & \si{\metre\per\second\squared} \\ 
$h_{conv}$ & Convective Heat Transfer Coefficient & \si{\watt\per\metre\squared\per\kelvin} \\ 
$I$ & Current & \si{\ampere} \\
$k$ & Thermal Conductivity & \si{\kilogram\metre\per\second\cubed\per\kelvin} \\ 
$k_{a}$ & Thermal Conductivity Coefficient $a$ & \si{\kilogram\metre\per\second\cubed\per\kelvin} \\ 
$k_{b}$ & Thermal Conductivity Coefficient $b$ & \si{\kilogram\metre\per\second\cubed\per\kelvin\squared} \\
$k_{c}$ & Thermal Conductivity Coefficient $c$ & \si{\kilogram\metre\per\second\cubed\per\kelvin\cubed} \\
$\mu$ & Dynamic Viscosity & \si{\kilogram\per\meter\per\second} \\
$\nu$ & Dynamic Viscosity & \si{\meter\squared\per\second} \\
$P$ & Power & \si{\watt} \\
$\phi$ & General Fluid Property & \si{\per\kilogram} \\ 
$T_{melt}$ & Melting Point & \si{\kelvin} \\
$T_{ref}$ & Reference Temperature & \si{\kelvin} \\
$t$ & Time & \si{\second} \\
$\tau_{ij}$ & Deviatoric Stress Tensor & \si{\kilogram\per\meter\per\second\squared} \\
$\hat{U}$ & Velocity Vector & \si{\meter\per\second} \\ 
$\vec{U}$ & Bulk Velocity Vector & \si{\meter\per\second} \\ 
$\textbf{U}$ & Cell Face Flux & \si{\meter\cubed\per\second} \\ 
$L_f$ & Latent Heat of Fusion & \si{\metre\squared\per\second\squared} \\
$\nu$ & Kinematic Viscosity & \si{\metre\squared\per\second} \\
$\rho$ & Density & \si{\kilogram\per\metre\cubed} \\
$\rho_{a}$ & Density Coefficient $a$ & \si{\kilogram\per\metre\cubed\per\kelvin} \\
$\rho_{b}$ & Density Coefficient $b$ & \si{\kilogram\per\metre\cubed\per\kelvin\squared} \\
$\rho_{c}$ & Density Coefficient $c$ & \si{\kilogram\per\metre\cubed\per\kelvin\cubed} \\
$\rho_{d}$ & Density Coefficient $d$ & \si{\kilogram\per\metre\cubed\per\kelvin\tothe{4}} \\
$\boldsymbol{\rho U}$ & Cell Face Mass Flux & \si{\kilogram\per\second} \\
$\sigma$ & Stefan-Boltzmann Constant & \si{\kilogram\per\metre\cubed\per\kelvin\tothe{4}} \\
$\sigma_{ij}$ & Normal Stress Tensor & \si{\kilogram\per\meter\per\second\squared} \\
$V$ & Voltage & \si{\volt} \\
\bottomrule
\end{longtable}

\begin{longtable}{l l}
\caption{Definitions of the acronyms used in this thesis.}
\label{ap-tab-glossary-acronyms} \\
\toprule
Acronyms & Full words\\
\hline
AM\rule{0pt}{2.6ex} & Additive Manufacturing \\
CP-2 & Commercially Pure Grade 2 Titanium \\
CWM & Computational Welding Mechanics \\
EBW & Electron Beam Welding \\
FEM & Finite Element Method \\
FFP & Fixed Flux Pressure \\
FV & Fixed Value \\
FVM & Finite Volume Method \\
GTAW & Gas Tungsten Arc Welding \\
IO & Inlet Outlet \\
LBW & Laser Beam Welding \\
NS & No Slip \\
PIOV & Pressure Inlet Outlet Velocity \\
TP & Total Pressure \\
WPM & Weld Process Modelling \\
ZG & Zero Gradient \\
\bottomrule
\end{longtable}

%% file: appendicies/thesis-appendix-cases.tex
\pagestyle{plain}
\section{Toy Gallium Melting Benchmark}
\begin{longtable}{l l l l}
\caption{Thermophysical properties used for the toy gallium melting case. Values taken from \cite{thermoPhysProp} unless noted else wise.}
\label{ap-tab-toy-ga-TPP} \\
\toprule
Phase & Property & Value & Units\\
\hline
$\alpha_1$\rule{0pt}{2.6ex} & Density, $\rho_1$ & 6092 & \si{\kilogram\per\metre\cubed} \\
 & Specific Heat Capacity, $c_{p,1}$ & 381.5 & \si{\metre\squared\per\second\squared\per\kelvin} \\
 & Thermal Conductivity, $k_1$ & 32 & \si{\kilogram\metre\per\second\cubed\per\kelvin} \\
 & Kinematic Viscosity, $\nu_1$ & \num{2.97e-07} & \si{\metre\squared\per\second} \\
\hline
$\alpha_2$\rule{0pt}{2.6ex} & Density, $\rho_2$ & 6092 & \si{\kilogram\per\metre\cubed} \\
 & Specific Heat Capacity, $c_{p,2}$ & 381.5 & \si{\metre\squared\per\second\squared\per\kelvin} \\
 & Volumetric Thermal Expansion Coefficient, $\beta$ & \num{1.2e-4} & \si{\per\kelvin} \\
 & Thermal Conductivity, $k_2$ & 32 & \si{\kilogram\metre\per\second\cubed\per\kelvin} \\ 
 & Melting Point, $T_{melt}$ & 302.93 & \si{\kelvin} \\
 & Reference Temperature, $T_{ref}$ & 302.78 & \si{\kelvin} \\
 & Latent Heat of Fusion, $L_f$ & \num{8.016e4} & \si{\metre\squared\per\second\squared} \\
 & Kinematic Viscosity, $\nu_2$ & \num{2.97e-07} & \si{\metre\squared\per\second} \\
%\hline
%All & Specific Heat Capacity & 381.5 & \si{\kilogram\per\metre\cubed} \\
\bottomrule
\end{longtable}
 
\begin{table}[hb]
\caption{\label{ap-tab-toy-ga-domain}Domain size and mesh cells for the toy gallium melting case.}
\centering
\begin{tabular}{l l l}
\toprule
Direction & Length [\si{\centi\meter}] & Mesh Cells \\
\hline
x & 8.89 & 84 \\
y & 2 & 1 \\
z & 6.35 & 64 \\
\bottomrule \\
\end{tabular}    
\end{table}

\newpage
\section{Toy Welding Benchmark}
\begin{longtable}{l l l l}
\caption{Thermophysical properties used for the toy weld case. Values taken from \cite{thermoPhysProp} unless noted else wise.}
\label{ap-tab-toy-weld-TPP} \\
\toprule
Phase & Property & Value & Units\\
\hline
$\alpha_1$\rule{0pt}{2.6ex} & Density, $\rho_1$ & 8052 & \si{\kilogram\per\metre\cubed} \\
 & Specific Heat Capacity, $c_{p,1}$ & 381.5 & \si{\metre\squared\per\second\squared\per\kelvin} \\
 & Thermal Conductivity, $k_1$ & 32 & \si{\kilogram\metre\per\second\cubed\per\kelvin} \\
 & Kinematic Viscosity, $\nu_1$ & \num{2.97e-07} & \si{\metre\squared\per\second} \\
\hline
$\alpha_2$\rule{0pt}{2.6ex} & Density, $\rho_2$ & 8052 & \si{\kilogram\per\metre\cubed} \\
 & Specific Heat Capacity, $c_{p,2}$ & 381.5 & \si{\metre\squared\per\second\squared\per\kelvin} \\
 & Volumetric Thermal Expansion Coefficient, $\beta$ & \num{1.2e-4} & \si{\per\kelvin} \\
 & Thermal Conductivity, $k_2$ & 32 & \si{\kilogram\metre\per\second\cubed\per\kelvin} \\ 
 & Melting Point, $T_{melt}$ & 302.93 & \si{\kelvin} \\
 & Reference Temperature, $T_{ref}$ & 302.78 & \si{\kelvin} \\
 & Latent Heat of Fusion, $L_f$ & \num{8.016e4} & \si{\metre\squared\per\second\squared} \\
 & Kinematic Viscosity, $\nu_2$ & \num{2.97e-07} & \si{\metre\squared\per\second} \\
%\hline
%All & Specific Heat Capacity & 381.5 & \si{\kilogram\per\metre\cubed} \\
\bottomrule
\end{longtable}

\begin{table}[hb]
\caption{\label{ap-tab-toy-weld-domain}Domain size and mesh cells for the toy weld case. Values taken from \cite{thermoPhysProp} unless noted else wise.}
\centering
\begin{tabular}{l l l}
\toprule
Direction & Length [\si{\centi\meter}] & Mesh Cells \\
\hline
x & 2 & 50 \\
y & \num{2e-2} & 1 \\
z & 2 & 50 \\
\bottomrule \\
\end{tabular}    
\end{table}

\newpage
\section{Three Phase Gallium Melting Benchmark}
\begin{longtable}{l l l l}
\caption{Thermophysical properties used for the three phase gallium melting case. Values taken from \cite{thermoPhysProp} unless noted else wise.}
\label{ap-tab-three-phase-TPP} \\
\toprule
Phase & Property & Value & Units\\
\hline
$\alpha_1$\rule{0pt}{2.6ex} & Density, $\rho_1$ & 1.1 & \si{\kilogram\per\metre\cubed} \\
 & Specific Heat Capacity, $c_{p,1}$ & 1000 & \si{\metre\squared\per\second\squared\per\kelvin} \\
 & Thermal Conductivity, $k_1$ & 0.02 & \si{\kilogram\metre\per\second\cubed\per\kelvin} \\
 & Kinematic Viscosity, $\nu_1$ & \num{1.48e-5} & \si{\metre\squared\per\second} \\
\hline
$\alpha_2$\rule{0pt}{2.6ex} & Density, $\rho_2$ & 6093 & \si{\kilogram\per\metre\cubed} \\
 & Volumetric Thermal Expansion Coefficient, $\beta$ & \num{1.2e-4} & \si{\per\kelvin} \\
 & Specific Heat Capacity, $c_{p,2}$ & 381 & \si{\metre\squared\per\second\squared\per\kelvin} \\
 & Thermal Conductivity, $k_3$ & 32 & \si{\kilogram\metre\per\second\cubed\per\kelvin} \\ & Melting Point, $T_{melt}$ & 302.93 & \si{\kelvin} \\
 & Reference Temperature, $T_{ref}$ & 302.93 & \si{\kelvin} \\
 & Latent Heat of Fusion, $L_f$ & \num{8.016e4} & \si{\metre\squared\per\second\squared} \\
 & Kinematic Viscosity, $\nu_2$ & \num{2.97e-7} & \si{\metre\squared\per\second} \\
\hline
$\alpha_3$\rule{0pt}{2.6ex} & Density, $\rho_3$ & 5910 & \si{\kilogram\per\metre\cubed} \\
 & Specific Heat Capacity, $c_{p.3}$ & 385 & \si{\metre\squared\per\second\squared\per\kelvin} \\
 & Thermal Conductivity, $k_3$ & 30 & \si{\kilogram\metre\per\second\cubed\per\kelvin} \\
 & Kinematic Viscosity, $\nu_3$ & \num{3.06e-7} & \si{\metre\squared\per\second} \\
%\hline
%All & Specific Heat Capacity & 381.5 & \si{\kilogram\per\metre\cubed} \\
\bottomrule
\end{longtable}

\begin{table}[hb]
\caption{\label{ap-tab-three-phase-domain}Domain size and mesh cells for the three phase gallium melting case. Values taken from \cite{thermoPhysProp} unless noted else wise.}
\centering
\begin{tabular}{l l l}
\toprule
Direction & Length [\si{\centi\meter}] & Mesh Cells \\
\hline
x & 8.89 & 84 \\
y & 2 & 1 \\
z & 7 & 71 \\
\bottomrule \\
\end{tabular}    
\end{table}

\newpage
\section{Tin Melting and Solidification Benchmark}
\begin{longtable}{l l l l}
\caption{Thermophysical properties used for the melting and solidification of tin case. Values taken from \cite{thermoPhysProp} unless noted else wise.}
\label{ap-tab-tin-TPP} \\
\toprule
Phase & Property & Value & Units\\
\hline
$\alpha_2$\rule{0pt}{2.6ex} & Density, $\rho_2$ & 6980 & \si{\kilogram\per\metre\cubed} \\
 & Volumetric Thermal Expansion Coefficient, $\beta$ & \num{1.06e-4} & \si{\per\kelvin} \\
 & Specific Heat Capacity, $c_{p,2}$ & 228.3885 & \si{\metre\squared\per\second\squared\per\kelvin} \\
 & Thermal Conductivity Base, $k_2$ & 30 & \si{\kilogram\metre\per\second\cubed\per\kelvin} \\
 & Thermal Conductivity Coefficient $a$, $k_{2a}$ & 10.204 & \si{\kilogram\metre\per\second\cubed\per\kelvin} \\ 
 & Thermal Conductivity Coefficient $b$, $k_{2b}$ & \num{32.063} & \si{\kilogram\metre\per\second\cubed\per\kelvin\squared} \\
 & Thermal Conductivity Coefficient $c$, $k_{2c}$ & \num{5.686} & \si{\kilogram\metre\per\second\cubed\per\kelvin\cubed} \\
 & Melting Point, $T_{melt}$ & 505.1 & \si{\kelvin} \\
 & Reference Temperature, $T_{ref}$ & 505.1 & \si{\kelvin} \\
 & Latent Heat of Fusion, $L_f$ & \num{5.9219e4} & \si{\metre\squared\per\second\squared} \\
 & Kinematic Viscosity, $\nu_2$ & \num{2.81e-7} & \si{\metre\squared\per\second} \\
\hline
$\alpha_3$\rule{0pt}{2.6ex} & Density, $\rho_3$ & 7500 & \si{\kilogram\per\metre\cubed} \\
 & Specific Heat Capacity, $c_{p.3}$ & 228.3885 & \si{\metre\squared\per\second\squared\per\kelvin} \\
 & Thermal Conductivity, $k_3$ & 59.5 & \si{\kilogram\metre\per\second\cubed\per\kelvin} \\
 & Kinematic Viscosity, $\nu_3$ & \num{3.06e-7} & \si{\metre\squared\per\second} \\
%\hline
%All & Specific Heat Capacity & 381.5 & \si{\kilogram\per\metre\cubed} \\
\bottomrule
\end{longtable}

\begin{table}[hb]
\caption{\label{ap-tab-tin-domain}Domain size and mesh cells for the melting and solidification of tin case. Values taken from \cite{thermoPhysProp} unless noted else wise.}
\centering
\begin{tabular}{l l l}
\toprule
Direction & Length [\si{\centi\meter}] & Mesh Cells \\
\hline
x & 9 & 90 \\
y & 2 & 1 \\
z & 7 & 70 \\
\bottomrule \\
\end{tabular}    
\end{table}

\newpage
\section{Ice Solidification Benchmark}
\begin{longtable}{l l l l}
\caption{Thermophysical properties used for the ice solidification case. Values taken from \cite{thermoPhysProp} unless noted else wise. Note the density coefficients are given fully in Chapter 4 but are rounded here to fit in the table.}
\label{ap-tab-ice-TPP} \\
\toprule
Phase & Property & Value & Units\\
\hline
$\alpha_2$\rule{0pt}{2.6ex} & Density Base, $\rho_2$ & 999.84 & \si{\kilogram\per\metre\cubed} \\
 & Density Coefficient $a$, $\rho_{2, a}$ & \num{6.7327e-2} & \si{\kilogram\per\metre\cubed\per\kelvin} \\
 & Density Coefficient $b$, $\rho_{2, b}$ & \num{-8.945e-3} & \si{\kilogram\per\metre\cubed\per\kelvin\squared} \\
 & Density Coefficient $c$, $\rho_{2, c}$ & \num{8.7846e-5} & \si{\kilogram\per\metre\cubed\per\kelvin\cubed} \\
 & Density Coefficient $d$, $\rho_{2, d}$ & \num{6.6214e-7} & \si{\kilogram\per\metre\cubed\per\kelvin\tothe{4}} \\
 & Volumetric Thermal Expansion Coefficient, $\beta$ & \num{2.1e-4} & \si{\per\kelvin} \\
 & Specific Heat Capacity, $c_{p,2}$ & 4187 & \si{\metre\squared\per\second\squared\per\kelvin} \\
 & Thermal Conductivity , $k_2$ & 0.6 & \si{\kilogram\metre\per\second\cubed\per\kelvin} \\
 & Melting Point, $T_{melt}$ & 273.15 & \si{\kelvin} \\
 & Reference Temperature, $T_{ref}$ & 273.15 & \si{\kelvin} \\
 & Latent Heat of Fusion, $L_f$ & \num{3.35e5} & \si{\metre\squared\per\second\squared} \\
 & Kinematic Viscosity, $\nu_2$ & \num{1e-6} & \si{\metre\squared\per\second} \\
\hline
$\alpha_3$\rule{0pt}{2.6ex} & Density, $\rho_3$ & 916.8 & \si{\kilogram\per\metre\cubed} \\
 & Specific Heat Capacity, $c_{p.3}$ & 2116 & \si{\metre\squared\per\second\squared\per\kelvin} \\
 & Thermal Conductivity, $k_3$ & 2.26 & \si{\kilogram\metre\per\second\cubed\per\kelvin} \\
 & Kinematic Viscosity, $\nu_3$ & \num{3.06e-7} & \si{\metre\squared\per\second} \\
%\hline
%All & Specific Heat Capacity & 381.5 & \si{\kilogram\per\metre\cubed} \\
\bottomrule
\end{longtable}

\begin{table}[hb]
\caption{\label{ap-tab-ice-domain}Domain size and mesh cells for the ice solidification case. Values taken from \cite{thermoPhysProp} unless noted else wise.}
\centering
\begin{tabular}{l l l}
\toprule
Direction & Length [\si{\centi\meter}] & Mesh Cells \\
\hline
x & 3.8 & 76 \\
y & 0.1 & 1 \\
z & 3.8 & 76 \\
\bottomrule \\
\end{tabular}    
\end{table}

\newpage
\section{Aluminium Flow Benchmark}
\begin{longtable}{l l l l}
\caption{Thermophysical properties used for the aluminium flow case. Values taken from \cite{thermoPhysProp} unless noted else wise.}
\label{ap-tab-aluminium-TPP} \\
\toprule
Phase & Property & Value & Units\\
\hline
$\alpha_1$\rule{0pt}{2.6ex} & Density, $\rho_1$ & 1.1 & \si{\kilogram\per\metre\cubed} \\
 & Specific Heat Capacity, $c_{p,1}$ & 1000 & \si{\metre\squared\per\second\squared\per\kelvin} \\
 & Thermal Conductivity, $k_1$ & 0.02 & \si{\kilogram\metre\per\second\cubed\per\kelvin} \\
 & Kinematic Viscosity, $\nu_1$ & \num{1.48e-5} & \si{\metre\squared\per\second} \\
\hline
$\alpha_2$\rule{0pt}{2.6ex} & Density, $\rho_2$ & 2385 & \si{\kilogram\per\metre\cubed} \\
 & Volumetric Thermal Expansion Coefficient, $\beta$ & \num{1.17e-4} & \si{\per\kelvin} \\
 & Specific Heat Capacity, $c_{p,2}$ & 1080 & \si{\metre\squared\per\second\squared\per\kelvin} \\
 & Thermal Conductivity, $k_3$ & 94.03 & \si{\kilogram\metre\per\second\cubed\per\kelvin} \\ 
 & Melting Point, $T_{melt}$ & 933.5 & \si{\kelvin} \\
 & Reference Temperature, $T_{ref}$ & 933.5 & \si{\kelvin} \\
 & Latent Heat of Fusion, $L_f$ & \num{3.97e5} & \si{\metre\squared\per\second\squared} \\
 & Kinematic Viscosity, $\nu_2$ & \num{5.798e-7} & \si{\metre\squared\per\second} \\
%\hline
%All & Specific Heat Capacity & 381.5 & \si{\kilogram\per\metre\cubed} \\
\bottomrule
\end{longtable}

\begin{table}[hb]
\caption{\label{ap-tab-aluminium-domain}Domain size and mesh cells for the aluminium flow case. Values taken from \cite{thermoPhysProp} unless noted else wise.}
\centering
\begin{tabular}{l l l}
\toprule
Direction & Length [\si{\centi\meter}] & Mesh Cells \\
\hline
x & 2 & 50 \\
y & \num{2e-2} & 1 \\
z & 2 & 50 \\
\bottomrule \\
\end{tabular}    
\end{table}

\newpage
\section{Bismuth Flow Benchmark}
\begin{longtable}{l l l l}
\caption{Thermophysical properties used for the bismuth flow case. Values taken from \cite{thermoPhysProp} unless noted else wise.}
\label{ap-tab-bi-TPP} \\
\toprule
Phase & Property & Value & Units\\
\hline
$\alpha_1$\rule{0pt}{2.6ex} & Density, $\rho_1$ & 1.6337 & \si{\kilogram\per\metre\cubed} \\
 & Specific Heat Capacity, $c_{p,1}$ & 520 & \si{\metre\squared\per\second\squared\per\kelvin} \\
 & Thermal Conductivity, $k_1$ & 0.017 & \si{\kilogram\metre\per\second\cubed\per\kelvin} \\
 & Kinematic Viscosity, $\nu_1$ & \num{1.38e-5} & \si{\metre\squared\per\second} \\
\hline
$\alpha_2$\rule{0pt}{2.6ex} & Density, $\rho_2$ & 9780 & \si{\kilogram\per\metre\cubed} \\
 & Volumetric Thermal Expansion Coefficient, $\beta$ & 0 & \si{\per\kelvin} \\
 & Specific Heat Capacity, $c_{p,2}$ & 123 & \si{\metre\squared\per\second\squared\per\kelvin} \\
 & Thermal Conductivity, $k_2$ & 10.35 & \si{\kilogram\metre\per\second\cubed\per\kelvin} \\ 
 & Melting Point, $T_{melt}$ & 544.55 & \si{\kelvin} \\
 & Reference Temperature, $T_{ref}$ & 544.55 & \si{\kelvin} \\
 & Latent Heat of Fusion, $L_f$ & \num{4.46e4} & \si{\metre\squared\per\second\squared} \\
 & Kinematic Viscosity, $\nu_2$ & \num{1.636e-07} & \si{\metre\squared\per\second} \\
 & Surface Tension, $\sigma$ & 0.0378 & \si{\kilogram\per\second\squared} \\
 & Surface tension to temperature Coefficient, $\frac{d \sigma}{d T}$ & \num{-4.03e-5} & \si{\kilogram\per\second\squared\per\kelvin} \\
\hline
$\alpha_3$\rule{0pt}{2.6ex} & Density, $\rho_3$ & 9780 & \si{\kilogram\per\metre\cubed} \\
 & Specific Heat Capacity, $c_{p.3}$ & 123 & \si{\metre\squared\per\second\squared\per\kelvin} \\
 & Thermal Conductivity, $k_3$ & 10.35 & \si{\kilogram\metre\per\second\cubed\per\kelvin} \\
 & Kinematic Viscosity, $\nu_3$ & \num{1.636e-07} & \si{\metre\squared\per\second} \\
%\hline
%All & Specific Heat Capacity & 381.5 & \si{\kilogram\per\metre\cubed} \\
\bottomrule
\end{longtable}

\begin{table}[hb]
\caption{\label{ap-tab-bi-domain}Domain size and mesh cells for the bismuth flow case. Values taken from \cite{thermoPhysProp} unless noted else wise.}
\centering
\begin{tabular}{l l l}
\toprule
Direction & Length [\si{\centi\meter}] & Mesh Cells \\
\hline
x & 15 & 120 \\
y & 0.2 & 1 \\
z & 5 & 40 \\
\bottomrule \\
\end{tabular}    
\end{table}

\newpage
\section{Base Welding Benchmark}
\begin{longtable}{l l l l}
\caption{Thermophysical properties used for the base welding case.}
\label{ap-tab-weld-TPP} \\
\toprule
Phase & Property & Value & Units\\
\hline
$\alpha_2$\rule{0pt}{2.6ex} & Density, $\rho_2$ & 8065 & \si{\kilogram\per\metre\cubed} \\
 & Volumetric Thermal Expansion Coefficient, $\beta$ & \num{1.96e-5} & \si{\per\kelvin} \\
 & Specific Heat Capacity, $c_{p,2}$ & 620 & \si{\metre\squared\per\second\squared\per\kelvin} \\
 & Thermal Conductivity, $k_2$ & 25 & \si{\kilogram\metre\per\second\cubed\per\kelvin} \\ 
 & Melting Point, $T_{melt}$ & 1705.15 & \si{\kelvin} \\
 & Reference Temperature, $T_{ref}$ & 1705.15 & \si{\kelvin} \\
 & Latent Heat of Fusion, $L_f$ & \num{2.6e5} & \si{\metre\squared\per\second\squared} \\
 & Kinematic Viscosity, $\nu_2$ & \num{2.97e-7} & \si{\metre\squared\per\second} \\
\hline
$\alpha_3$\rule{0pt}{2.6ex} & Density, $\rho_3$ & 8052 & \si{\kilogram\per\metre\cubed} \\
 & Specific Heat Capacity, $c_{p.3}$ & 530 & \si{\metre\squared\per\second\squared\per\kelvin} \\
 & Thermal Conductivity, $k_3$ & 40 & \si{\kilogram\metre\per\second\cubed\per\kelvin} \\
 & Kinematic Viscosity, $\nu_3$ & \num{2.97e-7} & \si{\metre\squared\per\second} \\
%\hline
%All & Specific Heat Capacity & 381.5 & \si{\kilogram\per\metre\cubed} \\
\bottomrule
\end{longtable}

\begin{table}[hb]
\caption{\label{ap-tab-weld-domain}Domain size and mesh cells for the base welding case. Values taken from \cite{thermoPhysProp} unless noted else wise.}
\centering
\begin{tabular}{l l l}
\toprule
Direction & Length [\si{\centi\meter}] & Mesh Cells \\
\hline
x & 4 & 120 \\
y & 5 & 50 \\
z & 0.6 & 18 \\
\bottomrule \\
\end{tabular}    
\end{table}

\newpage
\section{Independent Welding Benchmark}
\begin{longtable}{l l l l}
\caption{Thermophysical properties used for the independent welding case. Values taken from \cite{thermoPhysProp} unless noted else wise.}
\label{ap-tab-independent-TPP} \\
\toprule
Phase & Property & Value & Units\\
\hline
$\alpha_2$\rule{0pt}{2.6ex} & Density, $\rho_2$ & 8065 & \si{\kilogram\per\metre\cubed} \\
 & Volumetric Thermal Expansion Coefficient, $\beta$ & \num{1.96e-5} & \si{\per\kelvin} \\
 & Specific Heat Capacity, $c_{p,2}$ & 800 & \si{\metre\squared\per\second\squared\per\kelvin} \\
 & Thermal Conductivity Base, $k_2$ & 24 & \si{\kilogram\metre\per\second\cubed\per\kelvin} \\ 
 & Thermal Conductivity Coefficient $a$, $k_{2a}$ & 10.204 & \si{\kilogram\metre\per\second\cubed\per\kelvin} \\ 
 & Thermal Conductivity Coefficient $b$, $k_{2b}$ & \num{15.4e-3} & \si{\kilogram\metre\per\second\cubed\per\kelvin\squared} \\
 & Thermal Conductivity Coefficient $c$, $k_{2c}$ & \num{-7e-7} & \si{\kilogram\metre\per\second\cubed\per\kelvin\cubed} \\
 & Melting Point, $T_{melt}$ & 1705.15 & \si{\kelvin} \\
 & Reference Temperature, $T_{ref}$ & 1705.15 & \si{\kelvin} \\
 & Latent Heat of Fusion, $L_f$ & \num{2.6e5} & \si{\metre\squared\per\second\squared} \\
 & Kinematic Viscosity, $\nu_2$ & \num{2.97e-7} & \si{\metre\squared\per\second} \\
\hline
$\alpha_3$\rule{0pt}{2.6ex} & Density, $\rho_3$ & 8052 & \si{\kilogram\per\metre\cubed} \\
 & Specific Heat Capacity, $c_{p.3}$ & 490 & \si{\metre\squared\per\second\squared\per\kelvin} \\
 & Thermal Conductivity Base, $k_3$ & 40 & \si{\kilogram\metre\per\second\cubed\per\kelvin} \\ 
 & Thermal Conductivity Coefficient $a$, $k_{3a}$ & 6.6 & \si{\kilogram\metre\per\second\cubed\per\kelvin} \\ 
 & Thermal Conductivity Coefficient $b$, $k_{3b}$ & \num{12.14e-3} & \si{\kilogram\metre\per\second\cubed\per\kelvin\squared} \\
 & Kinematic Viscosity, $\nu_3$ & \num{2.97e-7} & \si{\metre\squared\per\second} \\
%\hline
%All & Specific Heat Capacity & 381.5 & \si{\kilogram\per\metre\cubed} \\
\bottomrule
\end{longtable}

\begin{table}[hb]
\caption{\label{ap-tab-independent-domain}Domain size and mesh cells for the independent welding case. Values taken from \cite{thermoPhysProp} unless noted else wise.}
\centering
\begin{tabular}{l l l}
\toprule
Direction & Length [\si{\centi\meter}] & Mesh Cells \\
\hline
x & 6 & 120 \\
y & 4 & 40 \\
z & 1.2 & 24 \\
\bottomrule \\
\end{tabular}    
\end{table}

%% file: appendicies/thesis-appendix-extraction.tex
\pagestyle{plain}
\begin{figure}[!htb]
\centering
\includegraphics[width=14.5cm]{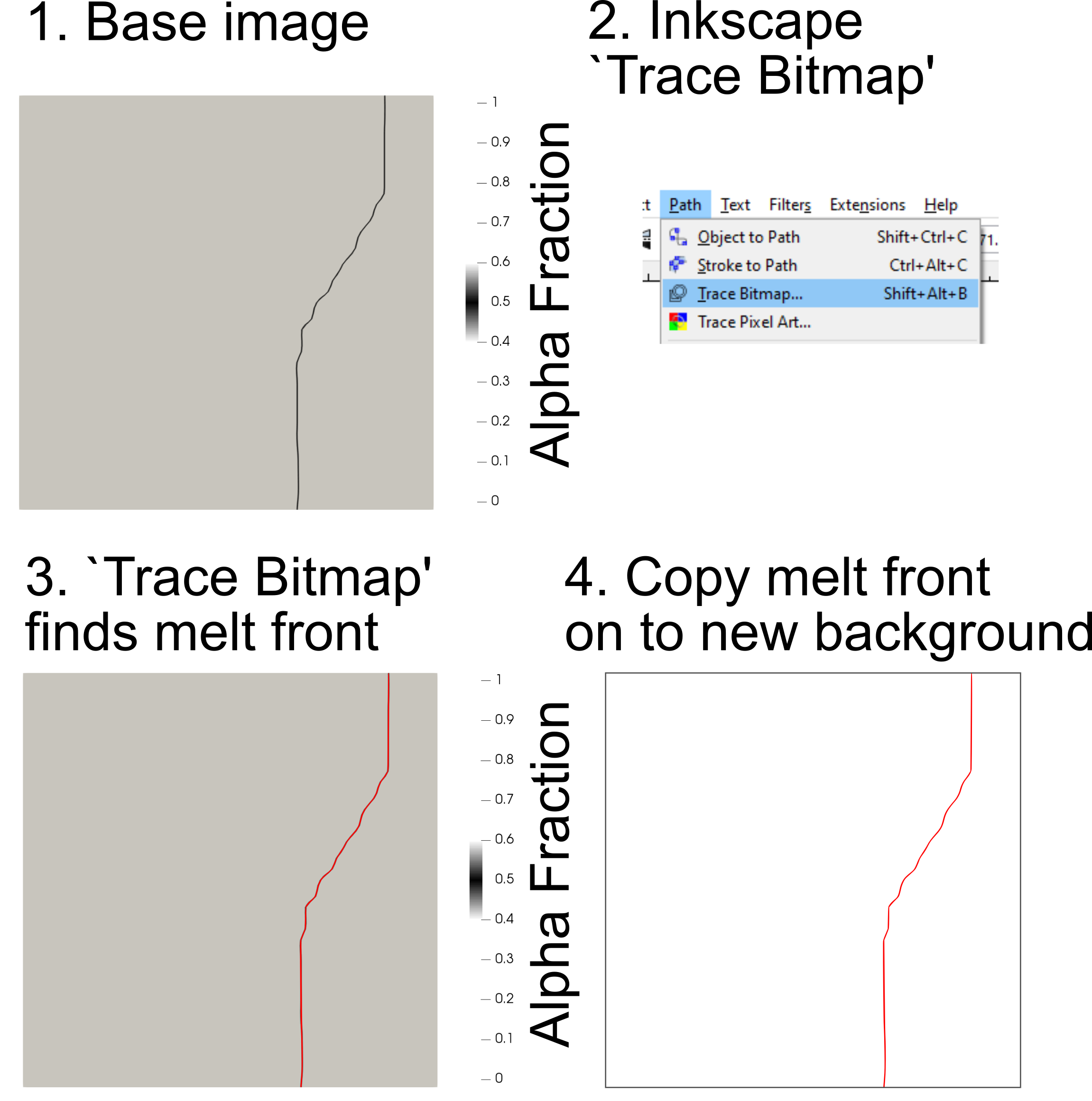}
\caption{Method for melt front extraction.}
\label{ap-fig-melt-front-extraciton}
\end{figure}

%% file: appendicies/thesis-appendix-incomplete.tex
\pagestyle{plain}
\begin{figure}[!htb]
\centering
\includegraphics[width=14.5cm]{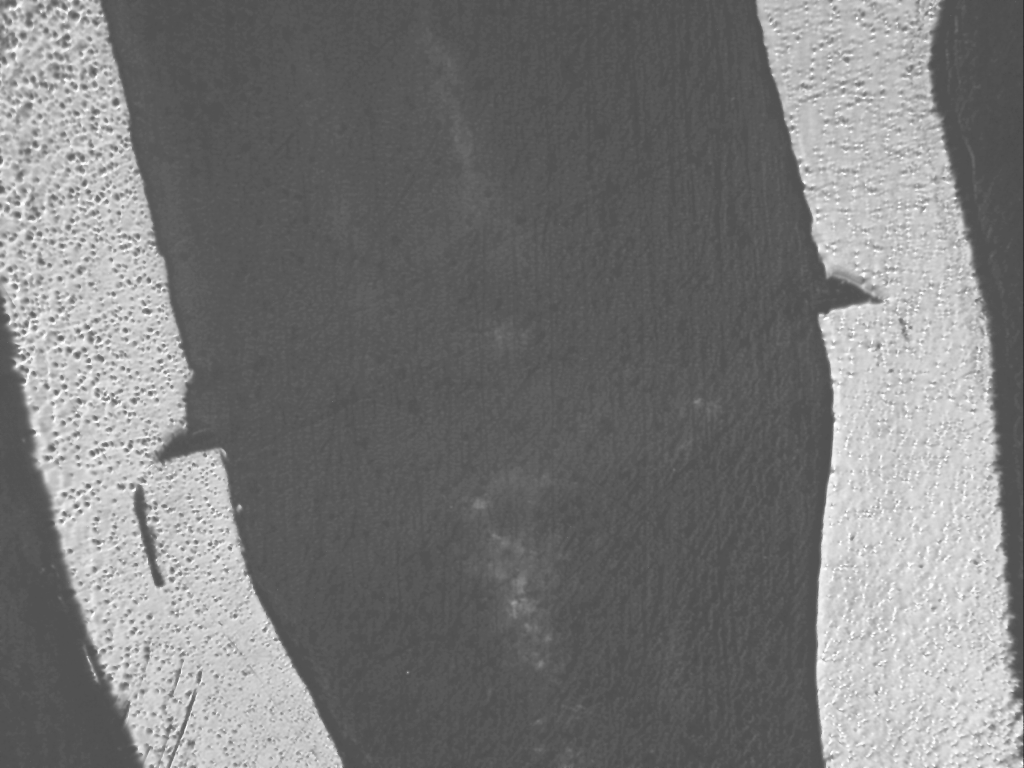}
\caption{Image of a welded tube interior with incomplete fusion.}
\label{ap-fig-incomplete-fusion}
\end{figure}

%% file: appendicies/thesis-appendix-code.tex
\pagestyle{plain}
%\begin{minipage}{\linewidth}
%\noindent
%\begin{lstlisting}[caption = {Implementation of equation \ref{3-energy-equation}}, label = {3-code-T-eqn}]
%fvScalarMatrix TEqn
%(
%    fvm::ddt(rhoCp, T)
%    + fvm::div(rhoPhiCp, T)
%    - fvm::Sp(fvc::ddt(rhoCp) + fvc::div(rhoPhiCp), T)
%    - fvm::laplacian(kEff, T)
%    ==
%    - S_latent
%    + qSource
%);
%\end{lstlisting}
%\end{minipage}

\begin{minipage}{\linewidth}
\begin{lstlisting}[caption = {Code snipping showing the implementation of equation \ref{3-energy-equation} within OpenFOAM.}, label = {3-code-T-eqn}]
fvScalarMatrix TEqn
(
    fvm::ddt(rhoCp, T)
    + fvm::div(rhoPhiCp, T)
    - fvm::Sp(fvc::ddt(rhoCp) + fvc::div(rhoPhiCp), T)
    - fvm::laplacian(kEff, T)
    ==
    - S_latent
    + qSource
);  
\end{lstlisting}
\end{minipage}

\begin{minipage}{\linewidth}
\begin{lstlisting}[caption = {Snippet of member function used to define the geometric field. Equation \ref{3-expX-eq} is used to create \emph{expX}, \emph{expY} and \emph{expZ} (for the y and z-direction) are implemented in a similar manner and combined with \emph{expX} to create the returned field \emph{expTot}. Ellipsis indicate omitted code.}, label = {3-code-expX}]
const volVectorField& cellCentre = U_.mesh().C();
...
volScalarField expX
(
    exp(C_cut*(sqr(cellCentre.component(vector::X)
    - (sourcePos.component(vector::X))
    - (sourceVel.component(vector::X)*(currentTime  - sourcePauseTime.value()))) /sqr(sourceOmega)))
);
...
volScalarField expTot(expX*expY*expZ);

return tmp<volScalarField>
(
    new volScalarField
    ("gaussField",expTot)
);  
\end{lstlisting}
\end{minipage}

\begin{minipage}{\linewidth}
\begin{lstlisting}[caption = {Member function to calculate $c_p(T)$ for both NIST and ASM formulations.}, label = {3-code-cpT-memberfunction}]
Foam::tmp<Foam::volScalarField>
Foam::gtawIncompressibleThreePhaseMixture::cp3TDepVSF() 
const {
scalar shomateDiv = 1;
            
    if (Shomate_)
        shomateDiv = 1000;
            
const volScalarField thirdTerm(
    cp3c_*pow(max(T_, smallT_)/shomateDiv, -2));
            
const volScalarField forthTerm(
    cp3d_*pow(max(T_, smallT_)/shomateDiv, 2));
            
const volScalarField fifthTerm(
    cp3e_*pow(max(T_, smallT_)/shomateDiv, 3));
            
    return tmp<volScalarField>(
    new volScalarField(
    "cp3TDepVSF",
    cp3a_ + cp3b_*((T_)/shomateDiv) + thirdTerm + forthTerm + fifthTerm));
}
\end{lstlisting}
\end{minipage}